\documentclass{masterthesis}

\usepackage[utf8x]{inputenc}
\usepackage{hyperref} 
\usepackage{import}
\usepackage{amssymb}
\usepackage{amsfonts}
\usepackage{amsmath}
\usepackage{agda}
\usepackage{is}
\DeclareUnicodeCharacter{8759}{::}
\DeclareUnicodeCharacter{8718}{$\blacksquare$}
\DeclareUnicodeCharacter{923}{$\Lambda$}

\newif\ifsubmit
\submittrue

\ifsubmit
\newcommand{\EZ}[1]{{#1}} 
\newcommand{\FrD}[1]{{#1}}
\newcommand{\LC}[1]{{#1}}
\newcommand{\EZComm}[1]{} 
\newcommand{\FDComm}[1]{}
\newcommand{\LCComm}[1]{}
\else
\newcommand{\EZ}[1]{\textcolor{blue}{#1}} 
\newcommand{\FrD}[1]{\textcolor{orange}{#1}} 
\newcommand{\LC}[1]{\textcolor{purple}{#1}} 
\newcommand{\EZComm}[1]{{\scriptsize\textcolor{blue}{[\bf{Elena: }#1}]}}
\newcommand{\FDComm}[1]{{\scriptsize\textcolor{orange}{[\bf{Francesco: }#1}]}}
\newcommand{\LCComm}[1]{{\scriptsize\textcolor{purple}{[\bf{Luca: }#1}]}}
\fi

\begin{document}

\title{Flexible Coinduction in Agda}

\author{Luca Ciccone}

\advisor{Elena Zucca, Francesco Dagnino}
\examiner{Eugenio Moggi}

\maketitle

\begin{center}
\large{\textbf{Acknowledgements}}
\end{center}
\vspace{0.75cm}

First of all, I would like to express my gratitude to my thesis supervisors Elena Zucca and Francesco Dagnino of DIBRIS (University of Genoa). Elena's courses during the bachelor and the master degree helped me finding the topics of my interests in computer science.\\

I would like to thank the master degree staff with a particular focus on Elena Zucca and Viviana Mascardi of DIBRIS (University of Genoa) that allowed me to find my right path during the master degree and gave me the possibility to enrich my preparation for the next years of research.\\

I would like also to thank my family for their constant encouragement during these many years of studies.\\

In particular, a special thanks goes to my parents Marirosa and Renato for their daily support especially in all the difficult moments.\\

I am also grateful to my uncle Roberto for having introduced me to my main cultural interests that I deepened every day for more than ten years.\\

Finally I would like to thank a place, Caldasio (AL), for its incredible inspirational energy and for its huge bag of memories especially about my grandparents.

\tableofcontents

\AgdaNoSpaceAroundCode{}

\begin{abstract}
Theorem provers are tools that help users to write machine readable proofs. Some of this tools are also interactive. \LC{The need of such softwares is increasing since they provide proofs that are more certified than the hand written ones}. Agda is based on type theory and on the propositions-as-types correspondence and has a Haskell-like syntax. This means that a proof of a statement is turned into a function. Inference systems are a way of defining inductive and coinductive predicates and induction and coinduction principles are provided to help proving their correctness with respect to a given specification in terms of soundness and completeness. Generalized inference systems deal with predicates whose inductive and coinductive interpretations do not provide the expected set of judgments. In this case inference systems are enriched by corules that are rules that can be applied at infinite depth in a proof tree. \LC{Induction and coinduction principles cannot be used in case of generalized inference systems and the bounded coinduction one has been proposed.}
We first present how Agda supports inductive and coinductive types highlighting the fact that data structures and predicates are defined using the same constructs.\LC{ Then we move to the main topic of this thesis, which is investigating how generalized inference systems can be implemented and how their correctness can be proved}. \LC{ All this work is explained by paradigmatic examples in order to emphasize the way of formalizing in Agda theoretical notions}. At last we study temporal operators of linear temporal logic that are an interesting case study because on the one hand they generalize the examples shown in the previous chapters and on the other hand they find application in static verification of concurrent systems.
\end{abstract}

\chapter*{Introduction}

Nowadays proof assistants are becoming \LC{popular and more usable}. \EZ{When proving the correctness of {huge software systems}, especially in concurrent scenarios,} proofs are hard to be {checked} by hand and so we ask support to {software tools}. By using a theorem prover, proofs are written in a precise syntax, and the tool checks their correctness, so users are guaranteed that their reasoning is correct. 
Agda is a tool of this kind; this means that, on the one hand, it is a programming language with a Haskell-like syntax,  on the other hand \textit{programs} are \textit{proofs}. 

The aim of this thesis is to investigate the use of Agda to \FrD{express and reason on predicates defined by} \emph{inference systems with corules} \cite{AnconaDZ17,Dagnino19}.  
Inference systems  \cite{Aczel77,LeroyGrall09} are a widely-used formalism to define predicates by means of \emph{meta-rules}. 
The user which defines a predicate through an inference system aims at proving that the definition is correct (that is, \emph{sound} and \emph{complete}) with respect to an expected meaning (\emph{specification}).
In this context, Agda can be used to express both the inference system and the specification, and to formally check soundness and completeness proofs.

Inference systems \EZ{with corules generalize (standard) inference systems, which, depending on the chosen interpretation, define either \emph{inductive} or \emph{coinductive} predicates.}
In both cases a canonical proof technique is  available: \FrD{ the \emph{induction principle} to prove soundness of an inductive predicate, and the \emph{coinduction principle} to prove the completeness of a coinductive predicate. }
In some cases, \FrD{however,}  neither the inductive nor the coinductive interpretation of an inference system correspond to the expected meaning. Indeed, many predicates cannot be defined inductively on possibly infinite structures, and, on the other hand, the coinductive interpretation often allows the derivation of non valid judgments. In these cases, corules can be used to filter the coinductive interpretation, so that only desired judgments are derived. \emph{Bounded coinduction}  \cite{AnconaDZ17,Dagnino19}, a generalization of standard coinduction, is the canonical proof technique to show completeness of an inference system with corules with respect to a specification.

\EZ{To express in Agda predicates defined by inference systems with corules, and formally check their correctness, we have to face two challenges.
 First, whereas inductive and coinductive predicates can be directly translated into an Agda type, which has in turn an inductive or coinductive semantics, respectively, for predicates which are neither inductive nor purely coinductive}, Agda has no built-in support.  We show that a predicate definition through an inference system with corules leads to a pair of Agda types\EZ{, notably a coinductive type which internally uses an inductive type.} \EZ{Second, whereas Agda inductive and coinductive types implicitly provide their corresponding induction and coinduction principles, the bounded coinduction principle needs to be explicitly expressed and proved.}
 
 \EZ{Note that the objective of the thesis is \emph{not} to implement the meta-theory of inference systems with corules in Agda. This would be possibile and is interesting in itself, but in this case, in order to obtain a given inference system, the user would have to provide parameters to a generic definition, losing the ``syntactic'' presentation of the specific instance. Instead, the aim here is to provide guidelines to a user to write inference systems with corules in the most natural way as Agda types, and to prove their correctness against a specification.} 

%

\LC{The thesis is structured as follows}.

\begin{description}

\item[\refToChapter{agda}] We introduce the most important features of Agda that will be used throughout the thesis. In particular, we illustrate the constructs for defining inductive and coinductive types, and discuss two different schema for representing possibly infinite structures.

\item[\refToChapter{inductive}] We show how constructs for defining inductive data types can be exploited to implement inductive predicates and how to write their soundness and completeness proofs. In particular we use the \textbf{induction principle} to prove the soundness of inductive predicates.

\item[\refToChapter{coinductive}] Analogously for coinductive predicates. We use the \textbf{coinduction principle} to prove the completeness of this kind of predicates. 

\item[\refToChapter{flex}] We consider predicates defined by inference systems with corules, which are the main focus of the thesis. 
As anticipated, we show that a predicate definition through a generalized inference system leads to a pair of Agda types, notably a coinductive type which internally uses an inductive type.
Moreover, we {express and prove in Agda} the \textbf{bounded coinduction principle} to prove the completeness of this kind of predicates.  

\item[\refToChapter{temp-op}] To investigate predicate \FrD{schema} more general with respect to the previous chapters, we consider an interesting case study: operators of linear temporal logic. First we consider the two basic operators \textit{Eventually} (also called \textit{F}) and \textit{Always} (also called \textit{G}), which can be seen as a generalization of the predicates considered in \refToChapter{inductive} and \refToChapter{coinductive}, respectively. Then we consider the operator \textit{InfinitelyOften} which \EZ{could} be defined by stratification of the inference systems for the two basic operators. We show that there is an alternative definition  through a generalized inference system and we apply the techniques developed in \refToChapter{flex} in order to prove its correctness.

\item[\refToChapter{conclu}] \EZ{We summarize the contribution of the thesis, notably drawing the methodological guidelines to implement inference systems with corules obtained from our investigation. Moreover, we describe directions for further work.}
\end{description}

Finally, the Appendix contains auxiliary Agda definitions.

The presentation is driven by examples. Notably, in each of the Chapters 2-4 we consider a paradigmatic example of predicate. 
For what concerns the underlying data structure, we focus on \emph{lists}. Indeed, the example of lists is simple but powerful enough  to illustrate 
the differences between \emph{finite} data structures (finite lists), \emph{infinite-only} data structures (streams), and \emph{possibly infinite} data structures (possibly infinite lists, also called \emph{colists}).  
Moreover, (possibly infinite) lists are the underlying model of linear temporal logic considered in \refToChapter{temp-op}.

For colists, we use in the thesis two different approaches of implementation in Agda. The standard library of Agda provides an interesting implementation of such data type that hides the difficulties of dealing with infinite structures. Indeed, the syntax is the same as for finite lists, through the use of \emph{thunks} (suspended computations). On the one hand, this is an help for the users, on the other hand to understand the resulting code the meaning of the additional involved types needs to be known. We propose a different implementation of colists that does not use the mechanism of thunks. \LC{Actually the two approaches find application in \EZ{other} coinductive data structures, e.g., binary infinite trees (see \refToSection{trees}).} In \refToChapter{inductive}, \refToChapter{coinductive}, and for the first two operators in \refToChapter{temp-op} we adopt both approaches with the aim of comparing them.  For generalized inference systems we use only the simpler approach of library colists.

\chapter{Agda}\label{chapter:agda}

Agda is both a dependently typed functional programming language and a proof assistant. Here we are interested in this second feature{, that is, we use}  Agda to build machine readable formal {proofs}.
What allows us to state that a program is actually a proof is the Curry-Howard correspondence, a key notion in proof theory.
Roughly, this means that a type $T$ can be considered as a \textit{property}, while a term $t$ of type $T$ is a \textit{proof} that $T$ holds under a set of hypotheses.\EZComm{non si capisce il ruolo delle ipotesi}
In particular, a \textit{dependent type} $T (\x_1, ...,\x_n)$, where $\x_1, ..., \x_n$ have type $T_1, ..., T_n$, {respectively,} represents a predicate on the types $T_1, ..., T_n$.
For instance, we can express  that $\x$ is an element of a list $\xs$ 
by a dependent type $\textit{member}(\x, \xs)$.
Agda also includes type constructors to encode logical connectives and quantifiers, thus providing a support for the full (intuitionistic) first-order logic. It also supports both inductive and coinductive data types and provides language facilities to write programs using them.
Finally, Agda is a \emph{total} language; this means that it has a termination checker that verifies whether {functions terminate on any argument, and} all patterns are matched.
\EZComm{in generale questa presentazione di Agda si potrebbe allungare un pochino}

For what concerns the usage of Agda, the syntax of programs is Haskell-like, and Emacs is provided in the default \EZComm{si chiama cos\`i?} package as interface.
At the beginning a few default modules can be used such as natural numbers and the equality type \LC{(see \cite{Agda} for the documentation)}.
Advanced modules can be found in the standard library that do not come with the default \EZComm{si chiama cos\`i?} package. 

{In the following sections we will first illustrate Agda data types and coinductive records, and then discuss how to use these features to represent possibly infinite structures. Finally, we will provide a quick summary of equality properties needed in the following. }

{Throughout the thesis, w}e will limit the code to the essential for the sake of {brevity}. Hence, additional auxiliary modules will be only described to understand their meaning; their code can be found in \refToSection{agda-modules}. {We will also omit to list the imported modules.}

\section{Data types}

{Agda provides support for defining inductive structures through \emph{data types}}. 
The {paradigmatic} example are the Peano's numbers. 
%
{The following is their definition} in Agda, along with the sum function.

\begin{center}
\begin{code}%
\>[2]\AgdaKeyword{data}\AgdaSpace{}%
\AgdaDatatype{ℕ}\AgdaSpace{}%
\AgdaSymbol{:}\AgdaSpace{}%
\AgdaPrimitiveType{Set}\AgdaSpace{}%
\AgdaKeyword{where}\<%
\\
\>[2][@{}l@{\AgdaIndent{0}}]%
\>[4]\AgdaInductiveConstructor{zero}%
\>[10]\AgdaSymbol{:}\AgdaSpace{}%
\AgdaDatatype{ℕ}\<%
\\
\>[4]\AgdaInductiveConstructor{suc}%
\>[10]\AgdaSymbol{:}\AgdaSpace{}%
\AgdaDatatype{ℕ}\AgdaSpace{}%
\AgdaSymbol{→}\AgdaSpace{}%
\AgdaDatatype{ℕ}\<%
\\
\\[\AgdaEmptyExtraSkip]%
\>[2]\AgdaOperator{\AgdaFunction{\AgdaUnderscore{}+\AgdaUnderscore{}}}\AgdaSpace{}%
\AgdaSymbol{:}\AgdaSpace{}%
\AgdaDatatype{ℕ}\AgdaSpace{}%
\AgdaSymbol{→}\AgdaSpace{}%
\AgdaDatatype{ℕ}\AgdaSpace{}%
\AgdaSymbol{→}\AgdaSpace{}%
\AgdaDatatype{ℕ}\<%
\\
\>[2]\AgdaInductiveConstructor{zero}%
\>[9]\AgdaOperator{\AgdaFunction{+}}\AgdaSpace{}%
\AgdaBound{n}\AgdaSpace{}%
\AgdaSymbol{=}\AgdaSpace{}%
\AgdaBound{n}\<%
\\
\>[2]\AgdaInductiveConstructor{suc}\AgdaSpace{}%
\AgdaBound{m}%
\>[9]\AgdaOperator{\AgdaFunction{+}}\AgdaSpace{}%
\AgdaBound{n}\AgdaSpace{}%
\AgdaSymbol{=}\AgdaSpace{}%
\AgdaInductiveConstructor{suc}\AgdaSpace{}%
\AgdaSymbol{(}\AgdaBound{m}\AgdaSpace{}%
\AgdaOperator{\AgdaFunction{+}}\AgdaSpace{}%
\AgdaBound{n}\AgdaSymbol{)}\<%
\end{code}

\end{center}

{As the example shows, Agda data types are introduced by a \agda{data} declaration, including the name and type of the data type, as well as the constructors, in this case \agda{zero} and \agda{suc},  and their types.}
{Here} $\Set$ is the ``type of types''. 

%
%
%

{Data types} can be also parameterized, the typical example is the data type of (finite) lists which is defined {as follows} inside the {built-in} modules.

\begin{center}
\begin{code}%
\>[2]\AgdaKeyword{data}\AgdaSpace{}%
\AgdaDatatype{List}\AgdaSpace{}%
\AgdaSymbol{\{}\AgdaBound{a}\AgdaSymbol{\}}\AgdaSpace{}%
\AgdaSymbol{(}\AgdaBound{A}\AgdaSpace{}%
\AgdaSymbol{:}\AgdaSpace{}%
\AgdaPrimitiveType{Set}\AgdaSpace{}%
\AgdaBound{a}\AgdaSymbol{)}\AgdaSpace{}%
\AgdaSymbol{:}\AgdaSpace{}%
\AgdaPrimitiveType{Set}\AgdaSpace{}%
\AgdaBound{a}\AgdaSpace{}%
\AgdaKeyword{where}\<%
\\
\>[2][@{}l@{\AgdaIndent{0}}]%
\>[4]\AgdaInductiveConstructor{[]}%
\>[8]\AgdaSymbol{:}\AgdaSpace{}%
\AgdaDatatype{List}\AgdaSpace{}%
\AgdaBound{A}\<%
\\
\>[4]\AgdaOperator{\AgdaInductiveConstructor{\AgdaUnderscore{}::\AgdaUnderscore{}}}\AgdaSpace{}%
\AgdaSymbol{:}\AgdaSpace{}%
\AgdaSymbol{(}\AgdaBound{x}\AgdaSpace{}%
\AgdaSymbol{:}\AgdaSpace{}%
\AgdaBound{A}\AgdaSymbol{)}\AgdaSpace{}%
\AgdaSymbol{(}\AgdaBound{xs}\AgdaSpace{}%
\AgdaSymbol{:}\AgdaSpace{}%
\AgdaDatatype{List}\AgdaSpace{}%
\AgdaBound{A}\AgdaSymbol{)}\AgdaSpace{}%
\AgdaSymbol{→}\AgdaSpace{}%
\AgdaDatatype{List}\AgdaSpace{}%
\AgdaBound{A}\<%
\end{code}

\end{center}

The list constructors are the usual ones, and $A$ is the type parameter. {The parameter $a$ denotes the \emph{level of the universe}, generally present in built-in modules to make code fully general. Elements of type \agda{Level} behave like numbers with the same functions. As an example, we need a level parameter \LC{inside the elimination rule of the disjunction type in \refToSection{agda-modules}} to allow the function to return types (like $\agda{N}$ that has type \agda{Set}) rather than values (like \agda{zero} that has type \agda{N}).}

{Also note} that $a$ is given between curly braces and without an explicit type. Curly braces tell Agda that the parameter is \emph{implicit}, thus Agda has to set it by itself. 
Moreover, Agda has to infer the type of $a$, which is \textit{Level}. This syntax can be adopted not only in data definitions but also in function definitions.

If we consider Agda as a programming language, given a data type definition we can define functions by pattern matching, that is, \emph{inductively}. For instance, the sum of natural numbers above is defined by pattern matching on the first argument.
An analogous example is the function that concatenates two lists.

\begin{center}
\begin{code}%
\>[0]\AgdaOperator{\AgdaFunction{\AgdaUnderscore{}\#\AgdaUnderscore{}}}\AgdaSpace{}%
\AgdaSymbol{:}\AgdaSpace{}%
\AgdaSymbol{\{}\AgdaBound{A}\AgdaSpace{}%
\AgdaSymbol{:}\AgdaSpace{}%
\AgdaPrimitiveType{Set}\AgdaSymbol{\}}\AgdaSpace{}%
\AgdaSymbol{→}\AgdaSpace{}%
\AgdaDatatype{List}\AgdaSpace{}%
\AgdaBound{A}\AgdaSpace{}%
\AgdaSymbol{→}\AgdaSpace{}%
\AgdaDatatype{List}\AgdaSpace{}%
\AgdaBound{A}\AgdaSpace{}%
\AgdaSymbol{→}\AgdaSpace{}%
\AgdaDatatype{List}\AgdaSpace{}%
\AgdaBound{A}\<%
\\
\>[0]\AgdaInductiveConstructor{[]}\AgdaSpace{}%
\AgdaOperator{\AgdaFunction{\#}}\AgdaSpace{}%
\AgdaBound{ys}\AgdaSpace{}%
\AgdaSymbol{=}\AgdaSpace{}%
\AgdaBound{ys}\<%
\\
\>[0]\AgdaSymbol{(}\AgdaBound{x}\AgdaSpace{}%
\AgdaOperator{\AgdaInductiveConstructor{∷}}\AgdaSpace{}%
\AgdaBound{xs}\AgdaSymbol{)}\AgdaSpace{}%
\AgdaOperator{\AgdaFunction{\#}}\AgdaSpace{}%
\AgdaBound{ys}\AgdaSpace{}%
\AgdaSymbol{=}\AgdaSpace{}%
\AgdaBound{x}\AgdaSpace{}%
\AgdaOperator{\AgdaInductiveConstructor{∷}}\AgdaSpace{}%
\AgdaSymbol{(}\AgdaBound{xs}\AgdaSpace{}%
\AgdaOperator{\AgdaFunction{\#}}\AgdaSpace{}%
\AgdaBound{ys}\AgdaSymbol{)}\<%
\end{code}
\end{center}
 
{Finally,} we show the pointwise sum {of two lists,} that needs pattern matching on both {arguments}.

\begin{center}
\begin{code}%
\>[0]\AgdaFunction{sum}\AgdaSpace{}%
\AgdaSymbol{:}\AgdaSpace{}%
\AgdaDatatype{List}\AgdaSpace{}%
\AgdaDatatype{Nat}\AgdaSpace{}%
\AgdaSymbol{→}\AgdaSpace{}%
\AgdaDatatype{List}\AgdaSpace{}%
\AgdaDatatype{Nat}\AgdaSpace{}%
\AgdaSymbol{→}\AgdaSpace{}%
\AgdaDatatype{List}\AgdaSpace{}%
\AgdaDatatype{Nat}\<%
\\
\>[0]\AgdaFunction{sum}\AgdaSpace{}%
\AgdaInductiveConstructor{[]}\AgdaSpace{}%
\AgdaBound{ys}\AgdaSpace{}%
\AgdaSymbol{=}\AgdaSpace{}%
\AgdaBound{ys}\<%
\\
\>[0]\AgdaFunction{sum}\AgdaSpace{}%
\AgdaSymbol{(}\AgdaBound{x}\AgdaSpace{}%
\AgdaOperator{\AgdaInductiveConstructor{∷}}\AgdaSpace{}%
\AgdaBound{xs}\AgdaSymbol{)}\AgdaSpace{}%
\AgdaInductiveConstructor{[]}\AgdaSpace{}%
\AgdaSymbol{=}\AgdaSpace{}%
\AgdaBound{x}\AgdaSpace{}%
\AgdaOperator{\AgdaInductiveConstructor{∷}}\AgdaSpace{}%
\AgdaBound{xs}\<%
\\
\>[0]\AgdaFunction{sum}\AgdaSpace{}%
\AgdaSymbol{(}\AgdaBound{x}\AgdaSpace{}%
\AgdaOperator{\AgdaInductiveConstructor{∷}}\AgdaSpace{}%
\AgdaBound{xs}\AgdaSymbol{)}\AgdaSpace{}%
\AgdaSymbol{(}\AgdaBound{y}\AgdaSpace{}%
\AgdaOperator{\AgdaInductiveConstructor{∷}}\AgdaSpace{}%
\AgdaBound{ys}\AgdaSymbol{)}\AgdaSpace{}%
\AgdaSymbol{=}\AgdaSpace{}%
\AgdaSymbol{(}\AgdaBound{x}\AgdaSpace{}%
\AgdaOperator{\AgdaPrimitive{+}}\AgdaSpace{}%
\AgdaBound{y}\AgdaSymbol{)}\AgdaSpace{}%
\AgdaOperator{\AgdaInductiveConstructor{∷}}\AgdaSpace{}%
\AgdaFunction{sum}\AgdaSpace{}%
\AgdaBound{xs}\AgdaSpace{}%
\AgdaBound{ys}\<%
\end{code}

\end{center}

In the same way, through \agda{data} declarations we can inductively define \emph{predicates}: a predicate $P$ over $S$ is a data type $S \rightarrow \Set$.
Here below are two examples: $\xs\ \agda{reverseOf}\ \ys$ holds if $\xs$ is the reverse of $\ys$ and 
$\xs \subseteq \ys$ if $\xs$ is a sublist of $\ys$.

\begin{center}
\begin{code}%
\>[0]\AgdaKeyword{data}\AgdaSpace{}%
\AgdaOperator{\AgdaDatatype{\AgdaUnderscore{}reverseOf\AgdaUnderscore{}}}\AgdaSpace{}%
\AgdaSymbol{\{}\AgdaBound{A}\AgdaSpace{}%
\AgdaSymbol{:}\AgdaSpace{}%
\AgdaPrimitiveType{Set}\AgdaSymbol{\}}\AgdaSpace{}%
\AgdaSymbol{:}\AgdaSpace{}%
\AgdaDatatype{List}\AgdaSpace{}%
\AgdaBound{A}\AgdaSpace{}%
\AgdaSymbol{→}\AgdaSpace{}%
\AgdaDatatype{List}\AgdaSpace{}%
\AgdaBound{A}\AgdaSpace{}%
\AgdaSymbol{→}\AgdaSpace{}%
\AgdaPrimitiveType{Set}\AgdaSpace{}%
\AgdaKeyword{where}\<%
\\
\>[0][@{}l@{\AgdaIndent{0}}]%
\>[2]\AgdaInductiveConstructor{rev{-}Λ}\AgdaSpace{}%
\AgdaSymbol{:}\AgdaSpace{}%
\AgdaInductiveConstructor{[]}\AgdaSpace{}%
\AgdaOperator{\AgdaDatatype{reverseOf}}\AgdaSpace{}%
\AgdaInductiveConstructor{[]}\<%
\\
\>[2]\AgdaInductiveConstructor{rev{-}t}\AgdaSpace{}%
\AgdaSymbol{:}\AgdaSpace{}%
\AgdaSymbol{\{}\AgdaBound{x}\AgdaSpace{}%
\AgdaSymbol{:}\AgdaSpace{}%
\AgdaBound{A}\AgdaSymbol{\}\{}\AgdaBound{xs}\AgdaSpace{}%
\AgdaBound{ys}\AgdaSpace{}%
\AgdaSymbol{:}\AgdaSpace{}%
\AgdaDatatype{List}\AgdaSpace{}%
\AgdaBound{A}\AgdaSymbol{\}}\AgdaSpace{}%
\AgdaSymbol{→}\AgdaSpace{}%
\AgdaBound{xs}\AgdaSpace{}%
\AgdaOperator{\AgdaDatatype{reverseOf}}\AgdaSpace{}%
\AgdaBound{ys}\AgdaSpace{}%
\AgdaSymbol{→}%
\>[51]\AgdaSymbol{(}\AgdaBound{x}\AgdaSpace{}%
\AgdaOperator{\AgdaInductiveConstructor{∷}}\AgdaSpace{}%
\AgdaBound{xs}\AgdaSymbol{)}\AgdaSpace{}%
\AgdaOperator{\AgdaDatatype{reverseOf}}%
\>[70]\AgdaSymbol{(}\AgdaBound{ys}\AgdaSpace{}%
\AgdaOperator{\AgdaFunction{\#}}\AgdaSpace{}%
\AgdaSymbol{(}\AgdaBound{x}\AgdaSpace{}%
\AgdaOperator{\AgdaInductiveConstructor{∷}}\AgdaSpace{}%
\AgdaInductiveConstructor{[]}\AgdaSymbol{))}\<%
\\
\>[0]\<%
\\
\>[0]\AgdaKeyword{data}\AgdaSpace{}%
\AgdaOperator{\AgdaDatatype{\AgdaUnderscore{}⊆\AgdaUnderscore{}}}%
\>[10]\AgdaSymbol{\{}\AgdaBound{A}\AgdaSpace{}%
\AgdaSymbol{:}\AgdaSpace{}%
\AgdaPrimitiveType{Set}\AgdaSymbol{\}}\AgdaSpace{}%
\AgdaSymbol{:}\AgdaSpace{}%
\AgdaDatatype{List}\AgdaSpace{}%
\AgdaBound{A}\AgdaSpace{}%
\AgdaSymbol{→}\AgdaSpace{}%
\AgdaDatatype{List}\AgdaSpace{}%
\AgdaBound{A}\AgdaSpace{}%
\AgdaSymbol{→}\AgdaSpace{}%
\AgdaPrimitiveType{Set}\AgdaSpace{}%
\AgdaKeyword{where}\<%
\\
\>[0][@{}l@{\AgdaIndent{0}}]%
\>[2]\AgdaInductiveConstructor{sub{-}Λ}\AgdaSpace{}%
\AgdaSymbol{:}%
\>[11]\AgdaInductiveConstructor{[]}\AgdaSpace{}%
\AgdaOperator{\AgdaDatatype{⊆}}\AgdaSpace{}%
\AgdaInductiveConstructor{[]}\<%
\\
\>[2]\AgdaInductiveConstructor{sub{-}right}\AgdaSpace{}%
\AgdaSymbol{:}\AgdaSpace{}%
\AgdaSymbol{\{}\AgdaBound{n}\AgdaSpace{}%
\AgdaSymbol{:}\AgdaSpace{}%
\AgdaBound{A}\AgdaSymbol{\}\{}\AgdaBound{ys}\AgdaSpace{}%
\AgdaBound{xs}\AgdaSpace{}%
\AgdaSymbol{:}\AgdaSpace{}%
\AgdaDatatype{List}\AgdaSpace{}%
\AgdaBound{A}\AgdaSymbol{\}}\AgdaSpace{}%
\AgdaSymbol{→}\AgdaSpace{}%
\AgdaBound{ys}\AgdaSpace{}%
\AgdaOperator{\AgdaDatatype{⊆}}\AgdaSpace{}%
\AgdaBound{xs}\AgdaSpace{}%
\AgdaSymbol{→}\AgdaSpace{}%
\AgdaBound{ys}\AgdaSpace{}%
\AgdaOperator{\AgdaDatatype{⊆}}\AgdaSpace{}%
\AgdaSymbol{(}\AgdaBound{n}\AgdaSpace{}%
\AgdaOperator{\AgdaInductiveConstructor{∷}}\AgdaSpace{}%
\AgdaBound{xs}\AgdaSymbol{)}\<%
\\
\>[2]\AgdaInductiveConstructor{sub{-}ind}\AgdaSpace{}%
\AgdaSymbol{:}\AgdaSpace{}%
\AgdaSymbol{\{}\AgdaBound{n}\AgdaSpace{}%
\AgdaSymbol{:}\AgdaSpace{}%
\AgdaBound{A}\AgdaSpace{}%
\AgdaSymbol{\}\{}\AgdaBound{ys}\AgdaSpace{}%
\AgdaBound{xs}\AgdaSpace{}%
\AgdaSymbol{:}\AgdaSpace{}%
\AgdaDatatype{List}\AgdaSpace{}%
\AgdaBound{A}\AgdaSymbol{\}}\AgdaSpace{}%
\AgdaSymbol{→}\AgdaSpace{}%
\AgdaBound{ys}\AgdaSpace{}%
\AgdaOperator{\AgdaDatatype{⊆}}\AgdaSpace{}%
\AgdaBound{xs}\AgdaSpace{}%
\AgdaSymbol{→}\AgdaSpace{}%
\AgdaSymbol{(}\AgdaBound{n}\AgdaSpace{}%
\AgdaOperator{\AgdaInductiveConstructor{∷}}\AgdaSpace{}%
\AgdaBound{ys}\AgdaSymbol{)}\AgdaSpace{}%
\AgdaOperator{\AgdaDatatype{⊆}}\AgdaSpace{}%
\AgdaSymbol{(}\AgdaBound{n}\AgdaSpace{}%
\AgdaOperator{\AgdaInductiveConstructor{∷}}\AgdaSpace{}%
\AgdaBound{xs}\AgdaSymbol{)}\<%
\end{code}
\end{center}

From the proof assistant point of view, having defined a predicate inductively, as in the examples above, we can do \emph{proofs  by induction} on the definition, as will be shown in \refToChapter{inductive}.

\section{Coinductive records}

Agda supports also the definition of coinductive structures through the {\agda{coinductive} variant of the \agda{record} declaration}. {First we show an example of (standard) record type.} 

\begin{center}
\begin{code}%
\>[2]\AgdaKeyword{record}\AgdaSpace{}%
\AgdaRecord{Pair}\AgdaSpace{}%
\AgdaSymbol{(}\AgdaBound{A}\AgdaSpace{}%
\AgdaBound{B}\AgdaSpace{}%
\AgdaSymbol{:}\AgdaSpace{}%
\AgdaPrimitiveType{Set}\AgdaSymbol{)}\AgdaSpace{}%
\AgdaSymbol{:}\AgdaSpace{}%
\AgdaPrimitiveType{Set}\AgdaSpace{}%
\AgdaKeyword{where}\<%
\\
\>[2][@{}l@{\AgdaIndent{0}}]%
\>[4]\AgdaKeyword{field}\<%
\\
\>[4][@{}l@{\AgdaIndent{0}}]%
\>[6]\AgdaField{fst}\AgdaSpace{}%
\AgdaSymbol{:}\AgdaSpace{}%
\AgdaBound{A}\<%
\\
\>[6]\AgdaField{snd}\AgdaSpace{}%
\AgdaSymbol{:}\AgdaSpace{}%
\AgdaBound{B}\<%
\end{code}

\end{center}

{Such declaration introduces} the projections \agda{Pair.fst} and \agda{Pair.snd}. 
In order to define a pair, the syntax is \agda{record \{ fst = a, snd = b \}} where $a$ and $b$ have type $A$ and $B$ respectively. Alternatively, a constructor can be added in the record declaration before the fields, e.g., \agda{constructor Pair ( \_ , \_ )}, where {the underscore denotes as usual the position of argument.}

{To support infinite structures, in the last version of Agda the \agda{coinductive} variant has been introduced}. The simplest example is the {type of the infinite lists, also called \emph{streams}}.

\begin{center}
\begin{code}%
\>[0]\AgdaKeyword{record}\AgdaSpace{}%
\AgdaRecord{MyStream}\AgdaSpace{}%
\AgdaSymbol{(}\AgdaBound{A}\AgdaSpace{}%
\AgdaSymbol{:}\AgdaSpace{}%
\AgdaPrimitiveType{Set}\AgdaSymbol{)}\AgdaSpace{}%
\AgdaSymbol{:}\AgdaSpace{}%
\AgdaPrimitiveType{Set}\AgdaSpace{}%
\AgdaKeyword{where}\<%
\\
\>[0][@{}l@{\AgdaIndent{0}}]%
\>[2]\AgdaKeyword{coinductive}\<%
\\
\>[2]\AgdaKeyword{field}\<%
\\
\>[2][@{}l@{\AgdaIndent{0}}]%
\>[4]\AgdaField{hd}\AgdaSpace{}%
\AgdaSymbol{:}\AgdaSpace{}%
\AgdaBound{A}\<%
\\
\>[4]\AgdaField{tl}\AgdaSpace{}%
\AgdaSymbol{:}\AgdaSpace{}%
\AgdaRecord{MyStream}\AgdaSpace{}%
\AgdaBound{A}\<%
\end{code}

\end{center}

{Analogously to finite lists, the type is parameterized on the type $A$ of the elements}. 
{Using a coinductive record, we represent a data structure through the \emph{observations} which can be made on it. For instance, a stream is completely determined by its \emph{head} and its \emph{tail}, which is a stream in turn.} Turning from finite lists to streams, functions can no longer be defined inductively (by pattern matching). Instead, Agda allows us to use \emph{copattern matching} \cite{CopAbel}. That is, we must {specify} how the result of the function will be observed. This is shown below on the example of the pointwise sum of two streams {(compare with the version for finite lists)}. 

\begin{center}
\begin{code}%
\>[0]\AgdaFunction{sum}\AgdaSpace{}%
\AgdaSymbol{:}\AgdaSpace{}%
\AgdaRecord{MyStream}\AgdaSpace{}%
\AgdaDatatype{Nat}\AgdaSpace{}%
\AgdaSymbol{→}\AgdaSpace{}%
\AgdaRecord{MyStream}\AgdaSpace{}%
\AgdaDatatype{Nat}\AgdaSpace{}%
\AgdaSymbol{→}\AgdaSpace{}%
\AgdaRecord{MyStream}\AgdaSpace{}%
\AgdaDatatype{Nat}\<%
\\
\\[\AgdaEmptyExtraSkip]%
\>[0]\AgdaField{MyStream.hd}\AgdaSpace{}%
\AgdaSymbol{(}\AgdaFunction{sum}\AgdaSpace{}%
\AgdaBound{a}\AgdaSpace{}%
\AgdaBound{b}\AgdaSymbol{)}\AgdaSpace{}%
\AgdaSymbol{=}\AgdaSpace{}%
\AgdaSymbol{(}\AgdaField{MyStream.hd}\AgdaSpace{}%
\AgdaBound{a}\AgdaSymbol{)}\AgdaSpace{}%
\AgdaOperator{\AgdaPrimitive{+}}\AgdaSpace{}%
\AgdaSymbol{(}\AgdaField{MyStream.hd}\AgdaSpace{}%
\AgdaBound{b}\AgdaSymbol{)}\<%
\\
\\[\AgdaEmptyExtraSkip]%
\>[0]\AgdaField{MyStream.tl}\AgdaSpace{}%
\AgdaSymbol{(}\AgdaFunction{sum}\AgdaSpace{}%
\AgdaBound{a}\AgdaSpace{}%
\AgdaBound{b}\AgdaSymbol{)}\AgdaSpace{}%
\AgdaSymbol{=}\AgdaSpace{}%
\AgdaFunction{sum}\AgdaSpace{}%
\AgdaSymbol{(}\AgdaField{MyStream.tl}\AgdaSpace{}%
\AgdaBound{a}\AgdaSymbol{)}\AgdaSpace{}%
\AgdaSymbol{(}\AgdaField{MyStream.tl}\AgdaSpace{}%
\AgdaBound{b}\AgdaSymbol{)}\<%
\end{code}

\end{center}

where the sum of natural numbers is that presented before.

\section{Possibly infinite structures}\label{sect:colists}

{In the previous sections we provided finite and infinite lists as examples of Agda definitions of inductive and coinductive structures, respectively, and we illustrated \textit{pattern} and \textit{copattern matching}.} \LC{We investigate now how to represent in Agda \emph{possibly infinite structures},  that is, structures which can be either finite or infinite.
Analogously to the previous sections, we can again refer to lists.}
Possibly infinite lists are also called \emph{colists}, and implemented in the standard library by using the mechanism of \emph{thunks}. We describe this implementation, and then 
propose a different implementation which does not require such additional concept. In the following sections we will compare the two versions{, notably discussing differences in the way proofs are driven}.
Clearly the two patterns can be used for all kinds of possibly infinite data structures.

The following is the standard library definition.

\begin{center}
\centering
\begin{code}%
\>[0]\AgdaKeyword{data}\AgdaSpace{}%
\AgdaDatatype{Colist}\AgdaSpace{}%
\AgdaSymbol{\{}\AgdaBound{a}\AgdaSymbol{\}}\AgdaSpace{}%
\AgdaSymbol{(}\AgdaBound{A}\AgdaSpace{}%
\AgdaSymbol{:}\AgdaSpace{}%
\AgdaPrimitiveType{Set}\AgdaSpace{}%
\AgdaBound{a}\AgdaSymbol{)}\AgdaSpace{}%
\AgdaSymbol{(}\AgdaBound{i}\AgdaSpace{}%
\AgdaSymbol{:}\AgdaSpace{}%
\AgdaPostulate{Size}\AgdaSymbol{)}\AgdaSpace{}%
\AgdaSymbol{:}\AgdaSpace{}%
\AgdaPrimitiveType{Set}\AgdaSpace{}%
\AgdaBound{a}\AgdaSpace{}%
\AgdaKeyword{where}\<%
\\
\>[0][@{}l@{\AgdaIndent{0}}]%
\>[2]\AgdaInductiveConstructor{[]}%
\>[6]\AgdaSymbol{:}\AgdaSpace{}%
\AgdaDatatype{Colist}\AgdaSpace{}%
\AgdaBound{A}\AgdaSpace{}%
\AgdaBound{i}\<%
\\
\>[2]\AgdaOperator{\AgdaInductiveConstructor{\AgdaUnderscore{}∷\AgdaUnderscore{}}}\AgdaSpace{}%
\AgdaSymbol{:}\AgdaSpace{}%
\AgdaBound{A}\AgdaSpace{}%
\AgdaSymbol{→}\AgdaSpace{}%
\AgdaRecord{Thunk}\AgdaSpace{}%
\AgdaSymbol{(}\AgdaDatatype{Colist}\AgdaSpace{}%
\AgdaBound{A}\AgdaSymbol{)}\AgdaSpace{}%
\AgdaBound{i}\AgdaSpace{}%
\AgdaSymbol{→}\AgdaSpace{}%
\AgdaDatatype{Colist}\AgdaSpace{}%
\AgdaBound{A}\AgdaSpace{}%
\AgdaBound{i}\<%
\end{code}

\end{center} 

This code takes advantage from {the mechanism of thunks} in order to follow the same scheme used for {finite} lists. That is, a \agda{data} declaration is used even though the data to be represented can be infinite. 
\begin{itemize}
\item The \agda{Size} parameter \LC{\cite{SizeAbel, SizeHughes}} represents {an} approximation level and can be $\infty$. {Sizes can also be used to help the termination checker by tracking the depth of data structures.} 

\item \agda{Thunk} is the type of suspended computations, used to simulate laziness. When the computation {should be} \emph{forced} (e.g., to get the tail of a colist) the termination checker has to know that the size of the data will decrease. This {can be better understood by looking at} the implementation:
\end{itemize}  

\begin{center}
\centering
\begin{code}%
\>[0]\AgdaKeyword{record}\AgdaSpace{}%
\AgdaRecord{Thunk}\AgdaSpace{}%
\AgdaSymbol{\{}\AgdaBound{ℓ}\AgdaSymbol{\}}\AgdaSpace{}%
\AgdaSymbol{(}\AgdaBound{F}\AgdaSpace{}%
\AgdaSymbol{:}\AgdaSpace{}%
\AgdaPostulate{Size}\AgdaSpace{}%
\AgdaSymbol{→}\AgdaSpace{}%
\AgdaPrimitiveType{Set}\AgdaSpace{}%
\AgdaBound{ℓ}\AgdaSymbol{)}\AgdaSpace{}%
\AgdaSymbol{(}\AgdaBound{i}\AgdaSpace{}%
\AgdaSymbol{:}\AgdaSpace{}%
\AgdaPostulate{Size}\AgdaSymbol{)}\AgdaSpace{}%
\AgdaSymbol{:}\AgdaSpace{}%
\AgdaPrimitiveType{Set}\AgdaSpace{}%
\AgdaBound{ℓ}\AgdaSpace{}%
\AgdaKeyword{where}\<%
\\
\>[0][@{}l@{\AgdaIndent{0}}]%
\>[2]\AgdaKeyword{coinductive}\<%
\\
\>[2]\AgdaKeyword{field}\AgdaSpace{}%
\AgdaField{force}\AgdaSpace{}%
\AgdaSymbol{:}\AgdaSpace{}%
\AgdaSymbol{\{}\AgdaBound{j}\AgdaSpace{}%
\AgdaSymbol{:}\AgdaSpace{}%
\AgdaOperator{\AgdaPostulate{Size<}}\AgdaSpace{}%
\AgdaBound{i}\AgdaSymbol{\}}\AgdaSpace{}%
\AgdaSymbol{→}\AgdaSpace{}%
\AgdaBound{F}\AgdaSpace{}%
\AgdaBound{j}\<%
\end{code} 

\end{center}

A thunk {is a coinductive record with only one field, intuitively the suspended computation.} {It takes as argument} a function from \agda{Size} to \Set, {\agda{Colist $A$}} can be considered as an example. When the field is accessed the computation is forced {and} implicitly the size decreases. The type \agda{Size $< i$} represents all the sizes that are smaller than $i$.

Our {alternative implementation, instead, follows the pattern used for coinductive types, that is, is based on coinductive records, as in the example of streams above. However, in order to represent a structure which can be \emph{either finite or infinite}, we  use a coinductive record with \emph{a unique field} representing the whole observation which can be made on the structure. This field will have typically a variant type, since this observation could take different shapes. In the example of colists, if the colist is non-empty we can observe the pair consisting of head and tail, as for streams, otherwise nothing.}

\begin{center}
\centering
\begin{code}%
\>[0]\AgdaKeyword{record}\AgdaSpace{}%
\AgdaRecord{MyColist}%
\>[17]\AgdaSymbol{(}\AgdaBound{A}\AgdaSpace{}%
\AgdaSymbol{:}\AgdaSpace{}%
\AgdaPrimitiveType{Set}\AgdaSymbol{)}\AgdaSpace{}%
\AgdaSymbol{:}\AgdaSpace{}%
\AgdaPrimitiveType{Set}\AgdaSpace{}%
\AgdaKeyword{where}\<%
\\
\>[0][@{}l@{\AgdaIndent{0}}]%
\>[2]\AgdaKeyword{constructor}\AgdaSpace{}%
\AgdaOperator{\AgdaCoinductiveConstructor{CoL\AgdaUnderscore{}}}\<%
\\
\>[2]\AgdaKeyword{coinductive}\<%
\\
\>[2]\AgdaKeyword{field}\<%
\\
\>[2][@{}l@{\AgdaIndent{0}}]%
\>[4]\AgdaField{list}\AgdaSpace{}%
\AgdaSymbol{:}\AgdaSpace{}%
\AgdaDatatype{Maybe}\AgdaSpace{}%
\AgdaSymbol{(}\AgdaBound{A}\AgdaSpace{}%
\AgdaOperator{\AgdaDatatype{×}}\AgdaSpace{}%
\AgdaRecord{MyColist}\AgdaSpace{}%
\AgdaBound{A}\AgdaSymbol{)}\<%
\end{code}

\end{center}

{This choice is implemented using \agda{Maybe}, taken from the standard library, which is, as in Haskell, the type which encapsulates optional values, with the two standard} constructors \agda{nothing} and \agda{just}.
The \agda{$\times$} denotes the product type.

{Comparing the two implementations, it is clear that to define functions and predicates on top of the implementation using \agda{data} and thunks we will use pattern matching, whereas copattern matching will be needed for the second. This will be illustrated in the following sections. }

As said above, these two approaches can be ``canonically'' applied every time we deal with possibly infinite structures. For instance, applying the two techniques to streams we obtain two implementations which are both different from that shown in the previous section.

\begin{center}
\centering
\begin{code}%
\>[0]\AgdaKeyword{data}\AgdaSpace{}%
\AgdaDatatype{Stream}\AgdaSpace{}%
\AgdaSymbol{\{}\AgdaBound{ℓ}\AgdaSymbol{\}}\AgdaSpace{}%
\AgdaSymbol{(}\AgdaBound{A}\AgdaSpace{}%
\AgdaSymbol{:}\AgdaSpace{}%
\AgdaPrimitiveType{Set}\AgdaSpace{}%
\AgdaBound{ℓ}\AgdaSymbol{)}\AgdaSpace{}%
\AgdaSymbol{(}\AgdaBound{i}\AgdaSpace{}%
\AgdaSymbol{:}\AgdaSpace{}%
\AgdaPostulate{Size}\AgdaSymbol{)}\AgdaSpace{}%
\AgdaSymbol{:}\AgdaSpace{}%
\AgdaPrimitiveType{Set}\AgdaSpace{}%
\AgdaBound{ℓ}\AgdaSpace{}%
\AgdaKeyword{where}\<%
\\
\>[0][@{}l@{\AgdaIndent{0}}]%
\>[2]\AgdaOperator{\AgdaInductiveConstructor{\AgdaUnderscore{}∷\AgdaUnderscore{}}}\AgdaSpace{}%
\AgdaSymbol{:}\AgdaSpace{}%
\AgdaBound{A}\AgdaSpace{}%
\AgdaSymbol{→}\AgdaSpace{}%
\AgdaRecord{Thunk}\AgdaSpace{}%
\AgdaSymbol{(}\AgdaDatatype{Stream}\AgdaSpace{}%
\AgdaBound{A}\AgdaSymbol{)}\AgdaSpace{}%
\AgdaBound{i}\AgdaSpace{}%
\AgdaSymbol{→}\AgdaSpace{}%
\AgdaDatatype{Stream}\AgdaSpace{}%
\AgdaBound{A}\AgdaSpace{}%
\AgdaBound{i}\<%
\\
\\
\>[0]\AgdaKeyword{record}\AgdaSpace{}%
\AgdaOperator{\AgdaRecord{MyStream\AgdaUnderscore{}bis}}\AgdaSpace{}%
\AgdaSymbol{(}\AgdaBound{A}\AgdaSpace{}%
\AgdaSymbol{:}\AgdaSpace{}%
\AgdaPrimitiveType{Set}\AgdaSymbol{)}\AgdaSpace{}%
\AgdaSymbol{:}\AgdaSpace{}%
\AgdaPrimitiveType{Set}\AgdaSpace{}%
\AgdaKeyword{where}\<%
\\
\>[0][@{}l@{\AgdaIndent{0}}]%
\>[2]\AgdaKeyword{coinductive}\<%
\\
\>[2]\AgdaKeyword{field}\<%
\\
\>[2][@{}l@{\AgdaIndent{0}}]%
\>[4]\AgdaField{stream}\AgdaSpace{}%
\AgdaSymbol{:}\AgdaSpace{}%
\AgdaBound{A}\AgdaSpace{}%
\AgdaOperator{\AgdaDatatype{×}}\AgdaSpace{}%
\AgdaSymbol{(}\AgdaOperator{\AgdaRecord{MyStream\AgdaUnderscore{}bis}}\AgdaSpace{}%
\AgdaBound{A}\AgdaSymbol{)}\<%
\end{code}

\end{center}

The first version is the standard library code and is very similar to the implementation with thunks of colists,  without the empty case.
The second version uses a coinductive record with a single field which is the pair of head and tail (always observable for a stream). 
The version that we provided initially is an ad-hoc solution, that is, does not follow either canonical technique.

\section{Equality properties}\label{sect:equality}

The definition of equality in the built-in \agda{Equality} module is reported below.

\begin{center}
\begin{code}%
\>[0]\AgdaKeyword{data}\AgdaSpace{}%
\AgdaOperator{\AgdaDatatype{\AgdaUnderscore{}≡\AgdaUnderscore{}}}\AgdaSpace{}%
\AgdaSymbol{\{}\AgdaBound{a}\AgdaSymbol{\}}\AgdaSpace{}%
\AgdaSymbol{\{}\AgdaBound{A}\AgdaSpace{}%
\AgdaSymbol{:}\AgdaSpace{}%
\AgdaPrimitiveType{Set}\AgdaSpace{}%
\AgdaBound{a}\AgdaSymbol{\}}\AgdaSpace{}%
\AgdaSymbol{(}\AgdaBound{x}\AgdaSpace{}%
\AgdaSymbol{:}\AgdaSpace{}%
\AgdaBound{A}\AgdaSymbol{)}\AgdaSpace{}%
\AgdaSymbol{:}\AgdaSpace{}%
\AgdaBound{A}\AgdaSpace{}%
\AgdaSymbol{→}\AgdaSpace{}%
\AgdaPrimitiveType{Set}\AgdaSpace{}%
\AgdaBound{a}\AgdaSpace{}%
\AgdaKeyword{where}\<%
\\
\>[0][@{}l@{\AgdaIndent{0}}]%
\>[2]\AgdaKeyword{instance}\AgdaSpace{}%
\AgdaInductiveConstructor{refl}\AgdaSpace{}%
\AgdaSymbol{:}\AgdaSpace{}%
\AgdaBound{x}\AgdaSpace{}%
\AgdaOperator{\AgdaDatatype{≡}}\AgdaSpace{}%
\AgdaBound{x}\<%
\end{code}

\end{center}

Intuitively, given $\x, \y$ of type $A$, $\x \equiv \y$ is the proof that they are equal. The constructor \agda{refl} of the data type stands for the reflexive property: $\x$ is equal to itself.

We defined an additional module with other equality properties which will be heavily used in the next chapters; we only list here the properties, the code is reported in \refToSection{agda-modules}. 

In Agda each property is expressed as a function that takes proofs as arguments and returns a proof as result. Let $\x, \y, \z$ be elements of type $A$, $P$ a predicate $A \rightarrow \Set$, and $f$ a function $A \rightarrow B$. For each property we report the corresponding Agda constructor which will be used in proofs in the following.
\begin{description}
\item[Symmetry (\agda{sym})] if  $\x \equiv \y$ then $\y \equiv \x$.
\item[Transitivity (\agda{trans})] if $\x \equiv \y$ and $\y \equiv \z$ then $\x \equiv \z$. 
\item[Congruence (\agda{cong})] if $\x \equiv \y$ then $f \x \equiv f \y$
\item[Substitution (\agda{subst})] if $\x \equiv y$ and $P$ holds on $\x$ then $P$ holds on $\y$
\end{description}

\chapter{Inductive reasoning}
\label{chapter:inductive}
We begin this {chapter} by recalling {the basic} notions about {\textit{inference systems}.} 

\section{Induction}
{Assume a \emph{universe} $\universe$ whose elements are called \emph{judgments}.}
An \textit{inference system} $\is$ consists of a set of \textit{inference rules}, which are pairs $\Rule {\prem} {\co}$, with $\prem \subseteq \universe$ the set of \textit{premises}, {and} $c \in \universe$ the \textit{consequence}. Intuitively, given a rule, if the premises hold then the consequence should hold {as well}.
A rule of the form $\Rule {\emptyset} {\co}$ is called \textit{axiom}. {In this case,} since the set of premises is empty, $c$ must necessarily hold. 
In order to define {the rules of an inference system in a finitary way}, \textit{meta-rules} can be used along with meta-variables and side-conditions. 
{An inference system defines a subset of the judgments, which can be equivalently seen as a predicate over the universe.}


If we want to {formally define such a} predicate, a formal semantics {of inference systems} has to be introduced. {To this end, we provide the following definitions.}
\begin{itemize}
\item A set $\Spec \subseteq \universe$ is \textit{closed} if, for each rule $\Rule {\prem} {\co} \in \is$, $\prem \subseteq \Spec$ implies that $\co \in \Spec$.
\item A set $\Spec \subseteq \universe$ is \textit{consistent} if, for all $\x \subseteq \Spec$, there exists a rule $\Rule {\prem} {\x} \in \is$ such that $\prem \subseteq \Spec$.
\end{itemize}

{The \emph{inductive interpretation} of $\is$, denoted $\Ind{\is}$, is the smallest closed set, that is, the intersection of all closed sets, and the \emph{coinductive interpretation} of $\is$, denoted $\CoInd{\is}$, is the largest consistent set, that is, the union of all consistent sets.}

{The inductive and coinductive interpretation can also be characterized in terms of proof trees, as follows.}

Let $\is$ be an inference system, a \textit{proof tree} in $\is$ is a tree whose nodes are judgments in $\universe$ and there is a node $c$ with children $\prem$ only if there exists a rule $\Rule {\prem} {\co}$. 
The \textit{inductive interpretation} of $\is$ is the set of judgments having a {finite}\footnote{{Under the common assumption that the set of premises of all the rules are finite, otherwise we should say a finite depth tree.}} proof tree{, whereas the coinductive interpretation is the set of judgment having a (finite or infinite) proof tree}. {In this chapter we describe proof techniques for predicates defined as the inductive interpretation of an inference system, shortly, \emph{inductively defined predicates}.}

Usually an \textit{expected semantics} or \textit{specification} $\Spec\subseteq\universe$ is available when a predicate is defined through an inference system $\is${, and we want} to prove that $\Spec$ and $\Ind{\is}$ agree. 
{That is}, we want to prove that $\Ind{\is} \subseteq \Spec$ and $\Spec \subseteq \Ind{\is}$. These two properties are called{, respectively,} \textit{soundness} and \textit{completeness} of $\Ind{\is}$ with respect to the \mbox{specification $\Spec$}.

{To prove the} \textit{soundness} of an inductive predicate we can use the \emph{induction principle}. 

\paragraph*{Induction Principle} If a set $\Spec\subseteq\universe$ is closed, then $\Ind{\is} \subseteq \Spec$.
{The proof is immediate since $\Ind{\is}$ is the smallest closed set by definition. Proving that $\Spec$ is closed amounts to show that, for each rule $\Rule {\prem} {\co}$ of the inference system, if $\prem\subseteq\Spec$ then $\co\in\Spec$.}

{On the other hand, to prove completeness of an inductive predicate there is no canonical technique, hence, for each concrete case, we must find an ``ad-hoc'' technique, typically some other form of induction.}
 
\section{Example}
{Now we provide an example of inductive predicate on lists, together with its soundness and completeness proofs. Then we will discuss the Agda implementation.}

\newcommand{\Seq}[1]{\textit{Seq}(#1)}

\EZ{In this example, the universe can be either the set $A^\star$ of finite lists, or the set $A^\omega$ of streams, or the set $A^\star+A^\omega$ of colists (possibly infinite lists).}

{We will use the following notations on lists: $\Lambda$ denotes the empty list, $\x{:}\xs$ the list with head $\x$ and tail $\xs$, $\get{\xs}{i}$ the $i$-th element of list $\xs$, if any (assuming that the first element has index $0$).}

{The following inference system defines the predicate \textit{memberOf}, where} $\memberOf {\x} {\xs}$ is expected to hold if $\x$ is an element of the list $\xs$.

\begin{small}
\begin{quote}
$\NamedRule{mem-h} {} {\memberOf {\x} {x{:}\xs}} \Space
\NamedRule{mem-t} {\memberOf {\x} {\xs}} {\memberOf {\x} {\y{:}\xs}}$
\end{quote}
\end{small}

\begin{itemize}
\item The (meta-)axiom \refToRule{mem-h} states that the head of a list is an element of the list
\item The (meta-)rule \refToRule{mem-t} states that an element of the tail of a list is an element of the list.  
\end{itemize}

The specification of the predicate is $\Spec = \{ {\Pair{\x}{\xs}} \mid \exists i. \get{\xs}{i}=\x \}$. 

\begin{statement}
The inductive definition of \textit{memberOf} is sound with respect to its specification.
\end{statement}
\begin{proof} We have to prove that, if $\memberOf{\x}{\xs}\in\Ind{\is}$, then there exists $i$ such that $\get{\xs}{i}=\x$. The proof is by induction on the definition of $\memberOf{\x}{\xs}$. That is, 
we have to show that, for each (meta-)rule, if the premises are in $\Spec$, then the consequence is in $\Spec$ as well.
\begin{description}
\item[\refToRule{mem-h}] We have to prove that there exists $i$ such that $\get{\x{:}\xs}{i}=\x$. This is true for $i=0$.
\item[\refToRule{mem-t}] Assume that $\Pair{\x} {\xs} \in \Spec${, that is, there exists $i$ such that $\get{\xs}{i}=\x$}. We have to prove that, for all $\y$, $\Pair{\x} {\y{:}\xs} \in \Spec$. This is true taking as index  $i+1$.
\end{description}
\end{proof}

\begin{statement}
The inductive definition of \textit{memberOf} is complete with respect to its specification.
\end{statement}
\begin{proof}
We have to prove that, if $\Pair {\x} {\xs} \in \Spec$, that is, there exists $i$ such that $\get{\xs}{i}=\x$, then $\memberOf {\x} {\xs}$ can be derived.
The proof is by arithmetic induction on $i$. Note that $\xs$ cannot be empty, hence it is of shape $y{:}\ys$. 
\begin{description}
\item[$i=0$] We have $\x\equiv y$, hence the judgment can be derived by axiom \refToRule{mem-h}. 
\item[$i+1$] We have that $\x$ belongs to $\ys$ in position $i$. By inductive hypothesis the judgment $\memberOf{\x}{\ys}$ can be derived, hence $\memberOf {\x} {\xs}$ can be derived by rule \refToRule{mem-t}.
\end{description}
\end{proof}

{Note that\EZ{, as anticipated,} the inductively defined predicate turns out to be sound and complete even considering (possibly) infinite lists.}

\section{Agda implementation}
We start from the simple case of finite lists. 

\begin{center}
\begin{code}%
\>[0]\AgdaKeyword{data}\AgdaSpace{}%
\AgdaOperator{\AgdaDatatype{\AgdaUnderscore{}memberOf\AgdaUnderscore{}}}\AgdaSpace{}%
\AgdaSymbol{\{}\AgdaBound{A}\AgdaSpace{}%
\AgdaSymbol{:}\AgdaSpace{}%
\AgdaPrimitiveType{Set}\AgdaSymbol{\}}\AgdaSpace{}%
\AgdaSymbol{:}\AgdaSpace{}%
\AgdaBound{A}\AgdaSpace{}%
\AgdaSymbol{→}\AgdaSpace{}%
\AgdaDatatype{List}\AgdaSpace{}%
\AgdaBound{A}\AgdaSpace{}%
\AgdaSymbol{→}\AgdaSpace{}%
\AgdaPrimitiveType{Set}\AgdaSpace{}%
\AgdaKeyword{where}\<%
\\
\>[0][@{}l@{\AgdaIndent{0}}]%
\>[2]\AgdaInductiveConstructor{mem{-}h}\AgdaSpace{}%
\AgdaSymbol{:}\AgdaSpace{}%
\AgdaSymbol{\{}\AgdaBound{x}\AgdaSpace{}%
\AgdaSymbol{:}\AgdaSpace{}%
\AgdaBound{A}\AgdaSymbol{\}}\AgdaSpace{}%
\AgdaSymbol{→}\AgdaSpace{}%
\AgdaSymbol{\{}\AgdaBound{xs}\AgdaSpace{}%
\AgdaSymbol{:}\AgdaSpace{}%
\AgdaDatatype{List}\AgdaSpace{}%
\AgdaBound{A}\AgdaSymbol{\}}\AgdaSpace{}%
\AgdaSymbol{→}\AgdaSpace{}%
\AgdaBound{x}\AgdaSpace{}%
\AgdaOperator{\AgdaDatatype{memberOf}}\AgdaSpace{}%
\AgdaSymbol{(}\AgdaBound{x}\AgdaSpace{}%
\AgdaOperator{\AgdaInductiveConstructor{∷}}\AgdaSpace{}%
\AgdaBound{xs}\AgdaSymbol{)}\<%
\\
\>[2]\AgdaInductiveConstructor{mem{-}t}\AgdaSpace{}%
\AgdaSymbol{:}\AgdaSpace{}%
\AgdaSymbol{\{}\AgdaBound{x}\AgdaSpace{}%
\AgdaBound{y}\AgdaSpace{}%
\AgdaSymbol{:}\AgdaSpace{}%
\AgdaBound{A}\AgdaSymbol{\}}\AgdaSpace{}%
\AgdaSymbol{→}\AgdaSpace{}%
\AgdaSymbol{\{}\AgdaBound{xs}\AgdaSpace{}%
\AgdaSymbol{:}\AgdaSpace{}%
\AgdaDatatype{List}\AgdaSpace{}%
\AgdaBound{A}\AgdaSymbol{\}}\AgdaSpace{}%
\AgdaSymbol{→}\AgdaSpace{}%
\AgdaBound{x}\AgdaSpace{}%
\AgdaOperator{\AgdaDatatype{memberOf}}\AgdaSpace{}%
\AgdaBound{xs}\AgdaSpace{}%
\AgdaSymbol{→}\AgdaSpace{}%
\AgdaBound{x}\AgdaSpace{}%
\AgdaOperator{\AgdaDatatype{memberOf}}\AgdaSpace{}%
\AgdaSymbol{(}\AgdaBound{y}\AgdaSpace{}%
\AgdaOperator{\AgdaInductiveConstructor{∷}}\AgdaSpace{}%
\AgdaBound{xs}\AgdaSymbol{)}\<%
\end{code}

\end{center}

As said in \refToChapter{agda}, \EZ{predicates are represented as (dependent) types. In the example, $\x\ \agda{memberOf}\ \xs$ is a type, and an element of type $\x\ \agda{memberOf}\ \xs$ {is} the proof that $\x$ is an element of the list $\xs$. In particular, for}
{an inductive predicate, we use} the \agda{data} construct. {The two constructors of proofs exactly correspond to the two meta-rules of the inference system}. 
\EZ{For example, given a proof that $\x$ is an element of $\xs$, the constructor \agda{mem-t} gives a proof that $\x$ is an element of $\y{:}\xs$.}
{Note} that some input arguments are between curly braces meaning that they are \emph{implicit}, as discussed  in \refToChapter{agda}.
\agda{List} is the module taken from the built-in library described in the previous chapter, parameterized over an arbitrary type. 

Now we move to infinite lists, that is, streams, and for this case we also show the soundness and completeness proofs.
{To this end, we have to express in Agda the specification $\Spec = \{ \Pair{\x}{\xs} \mid \exists i. \get{\xs}{i}=\x \}$.} 

\begin{center}
\begin{code}%
\>[0]\AgdaFunction{S}\AgdaSpace{}%
\AgdaSymbol{:}\AgdaSpace{}%
\AgdaSymbol{\{}\AgdaBound{A}\AgdaSpace{}%
\AgdaSymbol{:}\AgdaSpace{}%
\AgdaPrimitiveType{Set}\AgdaSymbol{\}(}\AgdaBound{x}\AgdaSpace{}%
\AgdaSymbol{:}\AgdaSpace{}%
\AgdaBound{A}\AgdaSymbol{)(}\AgdaBound{xs}\AgdaSpace{}%
\AgdaSymbol{:}\AgdaSpace{}%
\AgdaRecord{MyStream}\AgdaSpace{}%
\AgdaBound{A}\AgdaSymbol{)}\AgdaSpace{}%
\AgdaSymbol{→}\AgdaSpace{}%
\AgdaPrimitiveType{Set}\<%
\\
\>[0]\AgdaFunction{S}\AgdaSpace{}%
\AgdaBound{x}\AgdaSpace{}%
\AgdaBound{xs}\AgdaSpace{}%
\AgdaSymbol{=}\AgdaSpace{}%
\AgdaDatatype{∃}\AgdaSpace{}%
\AgdaDatatype{Nat}\AgdaSpace{}%
\AgdaSymbol{(λ}\AgdaSpace{}%
\AgdaBound{i}\AgdaSpace{}%
\AgdaSymbol{→}\AgdaSpace{}%
\AgdaFunction{get}\AgdaSpace{}%
\AgdaBound{xs}\AgdaSpace{}%
\AgdaBound{i}\AgdaSpace{}%
\AgdaOperator{\AgdaDatatype{≡}}\AgdaSpace{}%
\AgdaBound{x}\AgdaSymbol{)}\<%
\end{code}

\end{center}

{We need} the auxiliary function \agda{get} that returns the $i$-th element of a stream, defined in \refToSection{get}. \EZ{Moreover, we need the existential quantifier, defined in \refToSection{exists},}
 encoded as a pair whose elements are:
\begin{itemize}
\item the \emph{witness}, in this case a natural number
\item the \emph{proof} that the predicate actually holds for the witness. 
\end{itemize} 

We show below the Agda code for the predicate \textit{memberOf} on streams, and the corresponding proofs.

\begin{center}
\begin{code}%
\>[0]\AgdaKeyword{data}\AgdaSpace{}%
\AgdaOperator{\AgdaDatatype{\AgdaUnderscore{}memberOf\AgdaUnderscore{}}}\AgdaSpace{}%
\AgdaSymbol{\{}\AgdaBound{A}\AgdaSpace{}%
\AgdaSymbol{:}\AgdaSpace{}%
\AgdaPrimitiveType{Set}\AgdaSymbol{\}(}\AgdaBound{x}\AgdaSpace{}%
\AgdaSymbol{:}\AgdaSpace{}%
\AgdaBound{A}\AgdaSymbol{)(}\AgdaBound{xs}\AgdaSpace{}%
\AgdaSymbol{:}\AgdaSpace{}%
\AgdaRecord{MyStream}\AgdaSpace{}%
\AgdaBound{A}\AgdaSymbol{)}\AgdaSpace{}%
\AgdaSymbol{:}\AgdaSpace{}%
\AgdaPrimitiveType{Set}\AgdaSpace{}%
\AgdaKeyword{where}\<%
\\
\>[0][@{}l@{\AgdaIndent{0}}]%
\>[2]\AgdaInductiveConstructor{mem{-}h}\AgdaSpace{}%
\AgdaSymbol{:}\AgdaSpace{}%
\AgdaBound{x}\AgdaSpace{}%
\AgdaOperator{\AgdaDatatype{≡}}\AgdaSpace{}%
\AgdaSymbol{(}\AgdaField{MyStream.hd}\AgdaSpace{}%
\AgdaBound{xs}\AgdaSymbol{)}\AgdaSpace{}%
\AgdaSymbol{→}\AgdaSpace{}%
\AgdaBound{x}\AgdaSpace{}%
\AgdaOperator{\AgdaDatatype{memberOf}}\AgdaSpace{}%
\AgdaBound{xs}\<%
\\
\>[2]\AgdaInductiveConstructor{mem{-}t}\AgdaSpace{}%
\AgdaSymbol{:}\AgdaSpace{}%
\AgdaBound{x}\AgdaSpace{}%
\AgdaOperator{\AgdaDatatype{memberOf}}\AgdaSpace{}%
\AgdaSymbol{(}\AgdaField{MyStream.tl}\AgdaSpace{}%
\AgdaBound{xs}\AgdaSymbol{)}\AgdaSpace{}%
\AgdaSymbol{→}%
\>[41]\AgdaBound{x}\AgdaSpace{}%
\AgdaOperator{\AgdaDatatype{memberOf}}\AgdaSpace{}%
\AgdaBound{xs}\<%
\\
\\[\AgdaEmptyExtraSkip]%
\>[0]\AgdaFunction{mem{-}sound}\AgdaSpace{}%
\AgdaSymbol{:}%
\>[66I]\AgdaSymbol{\{}\AgdaBound{A}\AgdaSpace{}%
\AgdaSymbol{:}\AgdaSpace{}%
\AgdaPrimitiveType{Set}\AgdaSymbol{\}\{}\AgdaBound{x}\AgdaSpace{}%
\AgdaSymbol{:}\AgdaSpace{}%
\AgdaBound{A}\AgdaSymbol{\}\{}\AgdaBound{xs}\AgdaSpace{}%
\AgdaSymbol{:}\AgdaSpace{}%
\AgdaRecord{MyStream}\AgdaSpace{}%
\AgdaBound{A}\AgdaSymbol{\}}\AgdaSpace{}%
\AgdaSymbol{→}\<%
\\
\>[66I][@{}l@{\AgdaIndent{0}}]%
\>[13]\AgdaBound{x}\AgdaSpace{}%
\AgdaOperator{\AgdaDatatype{memberOf}}\AgdaSpace{}%
\AgdaBound{xs}\AgdaSpace{}%
\AgdaSymbol{→}\AgdaSpace{}%
\AgdaDatatype{∃}\AgdaSpace{}%
\AgdaDatatype{Nat}\AgdaSpace{}%
\AgdaSymbol{(λ}\AgdaSpace{}%
\AgdaBound{i}\AgdaSpace{}%
\AgdaSymbol{→}\AgdaSpace{}%
\AgdaSymbol{(}\AgdaFunction{get}\AgdaSpace{}%
\AgdaBound{xs}\AgdaSpace{}%
\AgdaBound{i}\AgdaSymbol{)}\AgdaSpace{}%
\AgdaOperator{\AgdaDatatype{≡}}\AgdaSpace{}%
\AgdaBound{x}\AgdaSymbol{)}\<%
\\
\>[0]\AgdaFunction{mem{-}sound}\AgdaSpace{}%
\AgdaSymbol{(}\AgdaInductiveConstructor{mem{-}h}\AgdaSpace{}%
\AgdaBound{eq}\AgdaSymbol{)}\AgdaSpace{}%
\AgdaSymbol{=}\AgdaSpace{}%
\AgdaOperator{\AgdaInductiveConstructor{<}}\AgdaSpace{}%
\AgdaInductiveConstructor{zero}\AgdaSpace{}%
\AgdaOperator{\AgdaInductiveConstructor{,}}\AgdaSpace{}%
\AgdaFunction{sym}\AgdaSpace{}%
\AgdaBound{eq}\AgdaSpace{}%
\AgdaOperator{\AgdaInductiveConstructor{>}}\<%
\\
\>[0]\AgdaFunction{mem{-}sound}\AgdaSpace{}%
\AgdaSymbol{(}\AgdaInductiveConstructor{mem{-}t}%
\>[98I]\AgdaBound{mem}\AgdaSymbol{)}\AgdaSpace{}%
\AgdaSymbol{=}\<%
\\
\>[.]\<[98I]%
\>[17]\AgdaOperator{\AgdaInductiveConstructor{<}}\<%
\\
\>[17][@{}l@{\AgdaIndent{0}}]%
\>[19]\AgdaSymbol{(}\AgdaInductiveConstructor{suc}\AgdaSpace{}%
\AgdaSymbol{(}\AgdaFunction{witness}\AgdaSpace{}%
\AgdaSymbol{(}\AgdaFunction{mem{-}sound}\AgdaSpace{}%
\AgdaBound{mem}\AgdaSymbol{)))}\AgdaSpace{}%
\AgdaOperator{\AgdaInductiveConstructor{,}}\<%
\\
\>[19]\AgdaFunction{proof}\AgdaSpace{}%
\AgdaSymbol{(}\AgdaFunction{mem{-}sound}\AgdaSpace{}%
\AgdaBound{mem}\AgdaSymbol{)}\<%
\\
\>[17]\AgdaOperator{\AgdaInductiveConstructor{>}}\<%
\\
\\[\AgdaEmptyExtraSkip]%
\\[\AgdaEmptyExtraSkip]%
\>[0]\AgdaFunction{mem{-}compl}\AgdaSpace{}%
\AgdaSymbol{:}\AgdaSpace{}%
\AgdaSymbol{(}\AgdaBound{xs}\AgdaSpace{}%
\AgdaSymbol{:}\AgdaSpace{}%
\AgdaRecord{MyStream}\AgdaSpace{}%
\AgdaDatatype{Nat}\AgdaSymbol{)}\AgdaSpace{}%
\AgdaSymbol{→}\AgdaSpace{}%
\AgdaSymbol{(}\AgdaBound{i}\AgdaSpace{}%
\AgdaSymbol{:}\AgdaSpace{}%
\AgdaDatatype{Nat}\AgdaSymbol{)}\AgdaSpace{}%
\AgdaSymbol{→}\AgdaSpace{}%
\AgdaSymbol{(}\AgdaFunction{get}\AgdaSpace{}%
\AgdaBound{xs}\AgdaSpace{}%
\AgdaBound{i}\AgdaSymbol{)}\AgdaSpace{}%
\AgdaOperator{\AgdaDatatype{memberOf}}\AgdaSpace{}%
\AgdaBound{xs}\<%
\\
\>[0]\AgdaFunction{mem{-}compl}\AgdaSpace{}%
\AgdaBound{xs}\AgdaSpace{}%
\AgdaInductiveConstructor{zero}\AgdaSpace{}%
\AgdaSymbol{=}\AgdaSpace{}%
\AgdaInductiveConstructor{mem{-}h}\AgdaSpace{}%
\AgdaInductiveConstructor{refl}\<%
\\
\>[0]\AgdaFunction{mem{-}compl}\AgdaSpace{}%
\AgdaBound{xs}\AgdaSpace{}%
\AgdaSymbol{(}\AgdaInductiveConstructor{suc}\AgdaSpace{}%
\AgdaBound{i}\AgdaSymbol{)}\AgdaSpace{}%
\AgdaSymbol{=}\AgdaSpace{}%
\AgdaInductiveConstructor{mem{-}t}\AgdaSpace{}%
\AgdaSymbol{(}\AgdaFunction{mem{-}compl}\AgdaSpace{}%
\AgdaSymbol{(}\AgdaField{MyStream.tl}\AgdaSpace{}%
\AgdaBound{xs}\AgdaSymbol{)}\AgdaSpace{}%
\AgdaBound{i}\AgdaSymbol{)}\<%
\end{code}

\end{center}

{Note that the definition of the predicate} is slightly different from the one for lists, {indeed} here we {cannot rely on pattern-matching, but must explicitly require $\x$ to be equal to the head of the list.}\EZComm{tentare di spiegare perch\'e} {Equality $\equiv$ is provided in the equality module, reported in \refToSection{agda-modules}, which} must be imported.

{In Agda, the soundness proof is a function \agda{mem-sound} which, given a proof of $\memberOf{\x}{\xs}$, returns a proof of $\agda{S}\ \x\ \xs$. The proof follows the schema of the hand-written proof, which is by induction on the definition of the predicate. That is, in Agda, by pattern-matching on the constructors of $\memberOf{\x}{\xs}$. Note that the fact that the proof is by induction corresponds, in Agda, to have a recursive call on smaller arguments. In the base case of the soundness proof, the variable \textit{eq} is the proof that the element $\x$ is equal to the head of the stream, as required in  in the data type definition. On the other hand, to prove that the specification holds we need the proof that the element of the stream at position zero is equal to $\x$, which is the symmetrical equality with respect to \textit{eq}. Hence, to get the right equality, we need \agda{sym} from the equality properties module \LC{\refToSection{agda-modules}}.}


Also the completeness proof follows the schema shown before, since it is by induction on the position of the element. When it is not the first, \agda{mem-t} requires the proof {for the tail and the previous position}, and this leads to a recursive call.

Note that in the case of streams the function \agda{get} always returns an element, since the data structure is infinite.

{Finally, we consider possibly infinite lists (colists). In this case,} we can use both the techniques presented in the previous chapter. We illustrate both solutions and discuss the differences.  First, we use the colists defined in the standard library.

\begin{center}
\begin{code}%
\>[0]\AgdaKeyword{data}\AgdaSpace{}%
\AgdaOperator{\AgdaDatatype{\AgdaUnderscore{}memberOf\AgdaUnderscore{}}}\AgdaSpace{}%
\AgdaSymbol{\{}\AgdaBound{A}\AgdaSpace{}%
\AgdaSymbol{:}\AgdaSpace{}%
\AgdaPrimitiveType{Set}\AgdaSymbol{\}}\AgdaSpace{}%
\AgdaSymbol{:}\AgdaSpace{}%
\AgdaBound{A}\AgdaSpace{}%
\AgdaSymbol{→}\AgdaSpace{}%
\AgdaDatatype{Colist}\AgdaSpace{}%
\AgdaBound{A}\AgdaSpace{}%
\AgdaPostulate{∞}\AgdaSpace{}%
\AgdaSymbol{→}\AgdaSpace{}%
\AgdaPrimitiveType{Set}\AgdaSpace{}%
\AgdaKeyword{where}\<%
\\
\>[0][@{}l@{\AgdaIndent{0}}]%
\>[2]\AgdaInductiveConstructor{mem{-}h}\AgdaSpace{}%
\AgdaSymbol{:}\AgdaSpace{}%
\AgdaSymbol{∀}\AgdaSpace{}%
\AgdaSymbol{\{}\AgdaBound{x}\AgdaSpace{}%
\AgdaBound{xs}\AgdaSymbol{\}}\AgdaSpace{}%
\AgdaSymbol{→}\AgdaSpace{}%
\AgdaBound{x}\AgdaSpace{}%
\AgdaOperator{\AgdaDatatype{memberOf}}\AgdaSpace{}%
\AgdaSymbol{(}\AgdaSpace{}%
\AgdaBound{x}\AgdaSpace{}%
\AgdaOperator{\AgdaInductiveConstructor{∷}}\AgdaSpace{}%
\AgdaBound{xs}\AgdaSpace{}%
\AgdaSymbol{)}\<%
\\
\>[2]\AgdaInductiveConstructor{mem{-}t}\AgdaSpace{}%
\AgdaSymbol{:}\AgdaSpace{}%
\AgdaSymbol{∀}\AgdaSpace{}%
\AgdaSymbol{\{}\AgdaBound{x}\AgdaSpace{}%
\AgdaBound{y}\AgdaSpace{}%
\AgdaBound{xs}\AgdaSymbol{\}}\AgdaSpace{}%
\AgdaSymbol{→}\AgdaSpace{}%
\AgdaBound{x}\AgdaSpace{}%
\AgdaOperator{\AgdaDatatype{memberOf}}\AgdaSpace{}%
\AgdaSymbol{(}\AgdaField{Thunk.force}\AgdaSpace{}%
\AgdaBound{xs}\AgdaSymbol{)}\AgdaSpace{}%
\AgdaSymbol{→}\AgdaSpace{}%
\AgdaBound{x}\AgdaSpace{}%
\AgdaOperator{\AgdaDatatype{memberOf}}\AgdaSpace{}%
\AgdaSymbol{(}\AgdaBound{y}\AgdaSpace{}%
\AgdaOperator{\AgdaInductiveConstructor{∷}}\AgdaSpace{}%
\AgdaBound{xs}\AgdaSymbol{)}\<%
\end{code}

\end{center}

{In this case, since we use thunks (suspended computations), the definition of the predicate follows the same schema used for finite lists.}
\LC{The \agda{Size} parameter of colists is set to $\infty$ because we are dealing with lists that are not approximated.}\EZComm{non si capisce} 
{Correspondingly, in the second constructor of the predicate (corresponding to the second meta-rule), the projection \agda{force} must be called since the type requires a \agda{Colist} rather than a \agda{Thunk}.} Implicit parameters are written in a more compact way; notably, their types {are inferred} by Agda thanks to the usage of $\forall$. For example, in the constructor \agda{mem-h}, type $A$ will be inferred for $\x$, and type \agda{Thunk Colist A} for $\xs$, from the fact that the operator $::$ for colists of type $A$ requires two elements with such types.
The predicate is inductive, thus the data type does not depend on \agda{Size}; all the parameters of these types are related to the colist.
The specification can be written as:

\begin{center}
\begin{code}%
\>[0]\AgdaFunction{S}\AgdaSpace{}%
\AgdaSymbol{:}\AgdaSpace{}%
\AgdaSymbol{\{}\AgdaBound{A}\AgdaSpace{}%
\AgdaSymbol{:}\AgdaSpace{}%
\AgdaPrimitiveType{Set}\AgdaSymbol{\}(}\AgdaBound{x}\AgdaSpace{}%
\AgdaSymbol{:}\AgdaSpace{}%
\AgdaBound{A}\AgdaSymbol{)(}\AgdaBound{xs}\AgdaSpace{}%
\AgdaSymbol{:}\AgdaSpace{}%
\AgdaDatatype{Colist}\AgdaSpace{}%
\AgdaBound{A}\AgdaSpace{}%
\AgdaPostulate{∞}\AgdaSymbol{)}\AgdaSpace{}%
\AgdaSymbol{→}\AgdaSpace{}%
\AgdaPrimitiveType{Set}\<%
\\
\>[0]\AgdaFunction{S}\AgdaSpace{}%
\AgdaBound{x}\AgdaSpace{}%
\AgdaBound{xs}\AgdaSpace{}%
\AgdaSymbol{=}\AgdaSpace{}%
\AgdaDatatype{∃}\AgdaSpace{}%
\AgdaDatatype{Nat}\AgdaSpace{}%
\AgdaSymbol{(λ}\AgdaSpace{}%
\AgdaBound{i}\AgdaSpace{}%
\AgdaSymbol{→}\AgdaSpace{}%
\AgdaFunction{lookup}\AgdaSpace{}%
\AgdaBound{i}\AgdaSpace{}%
\AgdaBound{xs}\AgdaSpace{}%
\AgdaOperator{\AgdaDatatype{≡}}\AgdaSpace{}%
\AgdaInductiveConstructor{just}\AgdaSpace{}%
\AgdaBound{x}\AgdaSymbol{)}\<%
\end{code}

\end{center}

{Here} \agda{lookup} is the library version of \agda{get}, and \agda{just} is the constructor of the standard library type \agda{Maybe}. The latter must be used since now the lists are not infinite-only and there are cases in which \agda{nothing} (the other constructor) must be returned. 
Recall that in the predicate on streams there was an  {explicit condition} inside the first constructor {requiring}  the element at position zero to be actually the head of the list. This difference will be better highlighted when we will deal with the second approach to colists.
 
{We show the soundness proof.}
 
\begin{center}
\begin{code}
\>[0]\AgdaFunction{mem{-}sound}\AgdaSpace{}%
\AgdaSymbol{:}\AgdaSpace{}%
\AgdaSymbol{\{}\AgdaBound{A}\AgdaSpace{}%
\AgdaSymbol{:}\AgdaSpace{}%
\AgdaPrimitiveType{Set}\AgdaSymbol{\}\{}\AgdaBound{x}%
\>[175I]\AgdaSymbol{:}\AgdaSpace{}%
\AgdaBound{A}\AgdaSymbol{\}\{}\AgdaBound{xs}\AgdaSpace{}%
\AgdaSymbol{:}\AgdaSpace{}%
\AgdaDatatype{Colist}\AgdaSpace{}%
\AgdaBound{A}\AgdaSpace{}%
\AgdaPostulate{∞}\AgdaSymbol{\}}\AgdaSpace{}%
\AgdaSymbol{→}\AgdaSpace{}%
\AgdaBound{x}\AgdaSpace{}%
\AgdaOperator{\AgdaDatatype{memberOf}}\AgdaSpace{}%
\AgdaBound{xs}\AgdaSpace{}%
\AgdaSymbol{→}\<%
\\
\>[.]\<[175I]%
\>[24]\AgdaDatatype{∃}\AgdaSpace{}%
\AgdaDatatype{Nat}\AgdaSpace{}%
\AgdaSymbol{(λ}\AgdaSpace{}%
\AgdaBound{i}\AgdaSpace{}%
\AgdaSymbol{→}\AgdaSpace{}%
\AgdaFunction{lookup}\AgdaSpace{}%
\AgdaBound{i}\AgdaSpace{}%
\AgdaBound{xs}\AgdaSpace{}%
\AgdaOperator{\AgdaDatatype{≡}}\AgdaSpace{}%
\AgdaInductiveConstructor{just}\AgdaSpace{}%
\AgdaBound{x}\AgdaSymbol{)}\<%
\\
\>[0]\<%
\\
\>[0]\AgdaFunction{mem{-}sound}\AgdaSpace{}%
\AgdaInductiveConstructor{mem{-}h}\AgdaSpace{}%
\AgdaSymbol{=}\AgdaSpace{}%
\AgdaOperator{\AgdaInductiveConstructor{<}}\AgdaSpace{}%
\AgdaInductiveConstructor{zero}\AgdaSpace{}%
\AgdaOperator{\AgdaInductiveConstructor{,}}\AgdaSpace{}%
\AgdaInductiveConstructor{refl}\AgdaSpace{}%
\AgdaOperator{\AgdaInductiveConstructor{>}}\<%
\\
\>[0]\AgdaFunction{mem{-}sound}\AgdaSpace{}%
\AgdaSymbol{(}\AgdaInductiveConstructor{mem{-}t}\AgdaSpace{}%
\AgdaBound{mem}\AgdaSymbol{)}\AgdaSpace{}%
\AgdaSymbol{=}%
\>[206I]\AgdaOperator{\AgdaInductiveConstructor{<}}\<%
\\
\>[.]\<[206I]%
\>[24]\AgdaInductiveConstructor{suc}\AgdaSpace{}%
\AgdaSymbol{(}\AgdaFunction{witness}\AgdaSpace{}%
\AgdaSymbol{(}\AgdaFunction{mem{-}sound}\AgdaSpace{}%
\AgdaBound{mem}\AgdaSymbol{))}\AgdaSpace{}%
\AgdaOperator{\AgdaInductiveConstructor{,}}\<%
\\
\>[24]\AgdaFunction{proof}\AgdaSpace{}%
\AgdaSymbol{(}\AgdaFunction{mem{-}sound}\AgdaSpace{}%
\AgdaBound{mem}\AgdaSymbol{)}\<%
\\
\>[24]\AgdaOperator{\AgdaInductiveConstructor{>}}\<%
\end{code}
\end{center}

The proof is again by induction on the definition of \textit{memberOf}.
{The proof of completeness is the following.}

\begin{center}
\begin{code}
\>[0]\AgdaFunction{mem{-}compl}\AgdaSpace{}%
\AgdaSymbol{:}\AgdaSpace{}%
\AgdaSymbol{\{}\AgdaBound{A}\AgdaSpace{}%
\AgdaSymbol{:}%
\>[216I]\AgdaPrimitiveType{Set}\AgdaSymbol{\}\{}\AgdaBound{ys}\AgdaSpace{}%
\AgdaSymbol{:}\AgdaSpace{}%
\AgdaDatatype{Colist}\AgdaSpace{}%
\AgdaBound{A}\AgdaSpace{}%
\AgdaPostulate{∞}\AgdaSymbol{\}\{}\AgdaBound{i}\AgdaSpace{}%
\AgdaSymbol{:}\AgdaSpace{}%
\AgdaDatatype{Nat}\AgdaSymbol{\}\{}\AgdaBound{x}\AgdaSpace{}%
\AgdaSymbol{:}\AgdaSpace{}%
\AgdaBound{A}\AgdaSymbol{\}}\AgdaSpace{}%
\AgdaSymbol{→}\<%
\\
\>[216I][@{}l@{\AgdaIndent{0}}]%
\>[24]\AgdaFunction{lookup}\AgdaSpace{}%
\AgdaBound{i}\AgdaSpace{}%
\AgdaBound{ys}\AgdaSpace{}%
\AgdaOperator{\AgdaDatatype{≡}}\AgdaSpace{}%
\AgdaInductiveConstructor{just}\AgdaSpace{}%
\AgdaBound{x}\AgdaSpace{}%
\AgdaSymbol{→}\AgdaSpace{}%
\AgdaBound{x}\AgdaSpace{}%
\AgdaOperator{\AgdaDatatype{memberOf}}\AgdaSpace{}%
\AgdaBound{ys}\<%
\\
\>[0]\<%
\\
\>[0]\AgdaFunction{mem{-}compl}\AgdaSpace{}%
\AgdaSymbol{\{}\AgdaArgument{ys}\AgdaSpace{}%
\AgdaSymbol{=}\AgdaSpace{}%
\AgdaInductiveConstructor{[]}\AgdaSymbol{\}}\AgdaSpace{}%
\AgdaSymbol{\{}\AgdaBound{i}\AgdaSymbol{\}}\AgdaSpace{}%
\AgdaSymbol{()}\<%
\\
\>[0]\AgdaFunction{mem{-}compl}\AgdaSpace{}%
\AgdaSymbol{\{}\AgdaArgument{ys}\AgdaSpace{}%
\AgdaSymbol{=}\AgdaSpace{}%
\AgdaBound{x}\AgdaSpace{}%
\AgdaOperator{\AgdaInductiveConstructor{∷}}\AgdaSpace{}%
\AgdaBound{xs}\AgdaSymbol{\}}\AgdaSpace{}%
\AgdaSymbol{\{}\AgdaInductiveConstructor{zero}\AgdaSymbol{\}}\AgdaSpace{}%
\AgdaInductiveConstructor{refl}\AgdaSpace{}%
\AgdaSymbol{=}\AgdaSpace{}%
\AgdaInductiveConstructor{mem{-}h}\<%
\\
\>[0]\AgdaFunction{mem{-}compl}\AgdaSpace{}%
\AgdaSymbol{\{}\AgdaArgument{ys}\AgdaSpace{}%
\AgdaSymbol{=}\AgdaSpace{}%
\AgdaBound{x}\AgdaSpace{}%
\AgdaOperator{\AgdaInductiveConstructor{∷}}\AgdaSpace{}%
\AgdaBound{xs}\AgdaSymbol{\}}\AgdaSpace{}%
\AgdaSymbol{\{}\AgdaInductiveConstructor{suc}\AgdaSpace{}%
\AgdaBound{i}\AgdaSymbol{\}}\AgdaSpace{}%
\AgdaBound{eq}\AgdaSpace{}%
\AgdaSymbol{=}\AgdaSpace{}%
\AgdaInductiveConstructor{mem{-}t}\AgdaSpace{}%
\AgdaSymbol{(}\AgdaFunction{mem{-}compl}\AgdaSpace{}%
\AgdaSymbol{\{}\AgdaArgument{ys}\AgdaSpace{}%
\AgdaSymbol{=}\AgdaSpace{}%
\AgdaField{Thunk.force}\AgdaSpace{}%
\AgdaBound{xs}\AgdaSymbol{\}}\AgdaSpace{}%
\AgdaBound{eq}\AgdaSymbol{)}\<%
\end{code}
\end{center}

{Also in this case the Agda proof follows the schema of the hand-written proof. However, now we must} consider the case that the list is empty, and we get an absurd. The absurd is automatically inferred by Agda if the proof is written interactively. Then, for a non-empty list,
the proof is by induction over the position. The arguments to the recursive call are provided in order to make the boxing and unboxing of the {colists} inside \agda{Thunk} easier to understand.

{Now we show the code that uses the second approach to colists, that is, conductive records.}

\begin{center}
\begin{code}%
\>[0]\AgdaKeyword{data}\AgdaSpace{}%
\AgdaOperator{\AgdaDatatype{\AgdaUnderscore{}memberOf\AgdaUnderscore{}}}\AgdaSpace{}%
\AgdaSymbol{\{}\AgdaBound{A}\AgdaSpace{}%
\AgdaSymbol{:}\AgdaSpace{}%
\AgdaPrimitiveType{Set}\AgdaSymbol{\}(}\AgdaBound{x}\AgdaSpace{}%
\AgdaSymbol{:}\AgdaSpace{}%
\AgdaBound{A}\AgdaSymbol{)(}\AgdaBound{ys}\AgdaSpace{}%
\AgdaSymbol{:}\AgdaSpace{}%
\AgdaRecord{MyColist}\AgdaSpace{}%
\AgdaBound{A}\AgdaSymbol{)}\AgdaSpace{}%
\AgdaSymbol{:}\AgdaSpace{}%
\AgdaPrimitiveType{Set}\AgdaSpace{}%
\AgdaKeyword{where}\<%
\\
\>[0][@{}l@{\AgdaIndent{0}}]%
\>[2]\AgdaInductiveConstructor{mem{-}h}\AgdaSpace{}%
\AgdaSymbol{:\{}\AgdaBound{xs}\AgdaSpace{}%
\AgdaSymbol{:}\AgdaSpace{}%
\AgdaRecord{MyColist}\AgdaSpace{}%
\AgdaBound{A}\AgdaSymbol{\}}\AgdaSpace{}%
\AgdaSymbol{→}\AgdaSpace{}%
\AgdaSymbol{(}\AgdaField{MyColist.list}\AgdaSpace{}%
\AgdaBound{ys}\AgdaSymbol{)}\AgdaSpace{}%
\AgdaOperator{\AgdaDatatype{≡}}\AgdaSpace{}%
\AgdaSymbol{(}\AgdaInductiveConstructor{just}\AgdaSpace{}%
\AgdaOperator{\AgdaInductiveConstructor{⟨}}\AgdaSpace{}%
\AgdaBound{x}\AgdaSpace{}%
\AgdaOperator{\AgdaInductiveConstructor{,}}\AgdaSpace{}%
\AgdaBound{xs}\AgdaSpace{}%
\AgdaOperator{\AgdaInductiveConstructor{⟩}}\AgdaSymbol{)}\AgdaSpace{}%
\AgdaSymbol{→}\AgdaSpace{}%
\AgdaBound{x}\AgdaSpace{}%
\AgdaOperator{\AgdaDatatype{memberOf}}\AgdaSpace{}%
\AgdaBound{ys}\<%
\\
\>[2]\AgdaInductiveConstructor{mem{-}t}%
\>[62I]\AgdaSymbol{:}\AgdaSpace{}%
\AgdaSymbol{\{}\AgdaBound{y}\AgdaSpace{}%
\AgdaSymbol{:}\AgdaSpace{}%
\AgdaBound{A}\AgdaSymbol{\}\{}\AgdaBound{xs}\AgdaSpace{}%
\AgdaSymbol{:}\AgdaSpace{}%
\AgdaRecord{MyColist}\AgdaSpace{}%
\AgdaBound{A}\AgdaSymbol{\}}\AgdaSpace{}%
\AgdaSymbol{→}\<%
\\
\>[.]\<[62I]%
\>[8]\AgdaSymbol{(}\AgdaField{MyColist.list}\AgdaSpace{}%
\AgdaBound{ys}\AgdaSymbol{)}\AgdaSpace{}%
\AgdaOperator{\AgdaDatatype{≡}}\AgdaSpace{}%
\AgdaSymbol{(}\AgdaInductiveConstructor{just}\AgdaSpace{}%
\AgdaOperator{\AgdaInductiveConstructor{⟨}}\AgdaSpace{}%
\AgdaBound{y}\AgdaSpace{}%
\AgdaOperator{\AgdaInductiveConstructor{,}}\AgdaSpace{}%
\AgdaBound{xs}\AgdaSpace{}%
\AgdaOperator{\AgdaInductiveConstructor{⟩}}\AgdaSymbol{)}\AgdaSpace{}%
\AgdaSymbol{→}\AgdaSpace{}%
\AgdaBound{x}\AgdaSpace{}%
\AgdaOperator{\AgdaDatatype{memberOf}}\AgdaSpace{}%
\AgdaBound{xs}\AgdaSpace{}%
\AgdaSymbol{→}\AgdaSpace{}%
\AgdaBound{x}\AgdaSpace{}%
\AgdaOperator{\AgdaDatatype{memberOf}}\AgdaSpace{}%
\AgdaBound{ys}\<%
\end{code}

\end{center}

The definition of \textit{memberOf} {is similar to that} presented for streams. Also in this case the {explicit} condition that tells how the colist is built is necessary. 
The code above seems to be a good alternative with respect to the library colists since there is no usage of additional data structures (\agda{Thunk}, \agda{Size}). On the other hand, the proofs are much harder, since they are based on how the list is observed {rather than} on how it is built. This has a huge impact, since now there is the need of additional equality proofs, and management of the absurd cases. Thus, we listed them in \LC{\refToSection{mycolisteq}}.\EZComm{ma che cosa?}

In this approach, the specification is defined as:

\begin{center}
\begin{code}%
\>[0]\AgdaFunction{S}\AgdaSpace{}%
\AgdaSymbol{:}\AgdaSpace{}%
\AgdaSymbol{\{}\AgdaBound{A}\AgdaSpace{}%
\AgdaSymbol{:}\AgdaSpace{}%
\AgdaPrimitiveType{Set}\AgdaSymbol{\}(}\AgdaBound{x}\AgdaSpace{}%
\AgdaSymbol{:}\AgdaSpace{}%
\AgdaBound{A}\AgdaSymbol{)(}\AgdaBound{xs}\AgdaSpace{}%
\AgdaSymbol{:}\AgdaSpace{}%
\AgdaRecord{MyColist}\AgdaSpace{}%
\AgdaBound{A}\AgdaSymbol{)}\AgdaSpace{}%
\AgdaSymbol{→}\AgdaSpace{}%
\AgdaPrimitiveType{Set}\<%
\\
\>[0]\AgdaFunction{S}\AgdaSpace{}%
\AgdaBound{x}\AgdaSpace{}%
\AgdaBound{xs}\AgdaSpace{}%
\AgdaSymbol{=}\AgdaSpace{}%
\AgdaDatatype{∃}\AgdaSpace{}%
\AgdaDatatype{Nat}\AgdaSpace{}%
\AgdaSymbol{(λ}\AgdaSpace{}%
\AgdaBound{i}\AgdaSpace{}%
\AgdaSymbol{→}\AgdaSpace{}%
\AgdaFunction{get}\AgdaSpace{}%
\AgdaBound{xs}\AgdaSpace{}%
\AgdaBound{i}\AgdaSpace{}%
\AgdaOperator{\AgdaDatatype{≡}}\AgdaSpace{}%
\AgdaInductiveConstructor{just}\AgdaSpace{}%
\AgdaBound{x}\AgdaSymbol{)}\<%
\end{code}

\end{center}

where \agda{get} is the function analogous to \agda{lookup} that we defined for colists. \LC{Its code is reported in \refToSection{get}}.

The following is the proof of soundness.

\begin{center}
\begin{code}
\>[0]\AgdaFunction{mem{-}sound}\AgdaSpace{}%
\AgdaSymbol{:}\AgdaSpace{}%
\AgdaSymbol{\{}\AgdaBound{A}\AgdaSpace{}%
\AgdaSymbol{:}\AgdaSpace{}%
\AgdaPrimitiveType{Set}\AgdaSymbol{\}\{}\AgdaBound{x}%
\>[90I]\AgdaSymbol{:}\AgdaSpace{}%
\AgdaBound{A}\AgdaSymbol{\}\{}\AgdaBound{xs}\AgdaSpace{}%
\AgdaSymbol{:}\AgdaSpace{}%
\AgdaRecord{MyColist}\AgdaSpace{}%
\AgdaBound{A}\AgdaSymbol{\}}\AgdaSpace{}%
\AgdaSymbol{→}\<%
\\
\>[.]\<[90I]%
\>[24]\AgdaBound{x}\AgdaSpace{}%
\AgdaOperator{\AgdaDatatype{memberOf}}\AgdaSpace{}%
\AgdaBound{xs}\AgdaSpace{}%
\AgdaSymbol{→}\AgdaSpace{}%
\AgdaDatatype{∃}\AgdaSpace{}%
\AgdaDatatype{Nat}\AgdaSpace{}%
\AgdaSymbol{(λ}\AgdaSpace{}%
\AgdaBound{i}\AgdaSpace{}%
\AgdaSymbol{→}\AgdaSpace{}%
\AgdaSymbol{(}\AgdaFunction{get}\AgdaSpace{}%
\AgdaBound{xs}\AgdaSpace{}%
\AgdaBound{i}\AgdaSymbol{)}\AgdaSpace{}%
\AgdaOperator{\AgdaDatatype{≡}}\AgdaSpace{}%
\AgdaInductiveConstructor{just}\AgdaSpace{}%
\AgdaBound{x}\AgdaSymbol{)}\<%
\\
\>[0]\<%
\\
\>[0]\AgdaFunction{mem{-}sound}\AgdaSpace{}%
\AgdaSymbol{\{}\AgdaArgument{xs}\AgdaSpace{}%
\AgdaSymbol{=}\AgdaSpace{}%
\AgdaBound{xs}\AgdaSymbol{\}}\AgdaSpace{}%
\AgdaSymbol{(}\AgdaInductiveConstructor{mem{-}h}\AgdaSpace{}%
\AgdaBound{eq}\AgdaSymbol{)}\AgdaSpace{}%
\AgdaSymbol{=}\AgdaSpace{}%
\AgdaOperator{\AgdaInductiveConstructor{<}}\AgdaSpace{}%
\AgdaInductiveConstructor{zero}\AgdaSpace{}%
\AgdaOperator{\AgdaInductiveConstructor{,}}\AgdaSpace{}%
\AgdaFunction{get{-}eq{-}0}\AgdaSpace{}%
\AgdaSymbol{\{}\AgdaArgument{l}\AgdaSpace{}%
\AgdaSymbol{=}\AgdaSpace{}%
\AgdaBound{xs}\AgdaSymbol{\}}\AgdaSpace{}%
\AgdaBound{eq}\AgdaSpace{}%
\AgdaOperator{\AgdaInductiveConstructor{>}}\<%
\\
\>[0]\AgdaFunction{mem{-}sound}\AgdaSpace{}%
\AgdaSymbol{\{}\AgdaArgument{xs}\AgdaSpace{}%
\AgdaSymbol{=}\AgdaSpace{}%
\AgdaBound{xs}\AgdaSymbol{\}}%
\>[128I]\AgdaSymbol{(}\AgdaInductiveConstructor{mem{-}t}\AgdaSpace{}%
\AgdaBound{eq}\AgdaSpace{}%
\AgdaBound{mem}\AgdaSymbol{)}\AgdaSpace{}%
\AgdaSymbol{=}\AgdaSpace{}%
\AgdaKeyword{let}\AgdaSpace{}%
\AgdaBound{ex{-}p}\AgdaSpace{}%
\AgdaSymbol{=}\AgdaSpace{}%
\AgdaFunction{mem{-}sound}\AgdaSpace{}%
\AgdaBound{mem}\<%
\\
\>[128I][@{}l@{\AgdaIndent{0}}]%
\>[24]\AgdaKeyword{in}\AgdaSpace{}%
\AgdaOperator{\AgdaInductiveConstructor{<}}\AgdaSpace{}%
\AgdaInductiveConstructor{suc}\AgdaSpace{}%
\AgdaSymbol{(}\AgdaFunction{witness}\AgdaSpace{}%
\AgdaBound{ex{-}p}\AgdaSymbol{)}\AgdaSpace{}%
\AgdaOperator{\AgdaInductiveConstructor{,}}\AgdaSpace{}%
\AgdaFunction{get{-}eq{-}tl}\AgdaSpace{}%
\AgdaSymbol{\{}\AgdaArgument{l}\AgdaSpace{}%
\AgdaSymbol{=}\AgdaSpace{}%
\AgdaBound{xs}\AgdaSymbol{\}}\AgdaSpace{}%
\AgdaBound{eq}\AgdaSpace{}%
\AgdaSymbol{(}\AgdaFunction{proof}\AgdaSpace{}%
\AgdaBound{ex{-}p}\AgdaSymbol{)}\AgdaSpace{}%
\AgdaOperator{\AgdaInductiveConstructor{>}}\<%
\end{code}
\end{center}

In the first case, it is necessary to explicitly state that the element at position zero is the head of the list. Here \agda{eq} is the proof that the element is the head of the list and \agda{get-eq-0} returns the proof that it is the element at position zero. Some implicit parameters are passed to avoid warnings. In the second case a similar proof is needed to relate \textit{eq}, which states how the list is built, to the $i$-th element.

The proof of completeness is by induction on the existential pair (to access the position) and shows the usage of \agda{inspect}.

\begin{center}
\begin{code}
\>[0]\AgdaFunction{mem{-}compl}\AgdaSpace{}%
\AgdaSymbol{:}\AgdaSpace{}%
\AgdaSymbol{\{}\AgdaBound{A}\AgdaSpace{}%
\AgdaSymbol{:}\AgdaSpace{}%
\AgdaPrimitiveType{Set}\AgdaSymbol{\}\{}\AgdaBound{x}%
\>[154I]\AgdaSymbol{:}\AgdaSpace{}%
\AgdaBound{A}\AgdaSymbol{\}(}\AgdaBound{xs}\AgdaSpace{}%
\AgdaSymbol{:}\AgdaSpace{}%
\AgdaRecord{MyColist}\AgdaSpace{}%
\AgdaBound{A}\AgdaSymbol{)}\AgdaSpace{}%
\AgdaSymbol{→}\<%
\\
\>[.]\<[154I]%
\>[24]\AgdaDatatype{∃}\AgdaSpace{}%
\AgdaDatatype{Nat}\AgdaSpace{}%
\AgdaSymbol{(λ}\AgdaSpace{}%
\AgdaBound{i}\AgdaSpace{}%
\AgdaSymbol{→}\AgdaSpace{}%
\AgdaSymbol{(}\AgdaFunction{get}\AgdaSpace{}%
\AgdaBound{xs}\AgdaSpace{}%
\AgdaBound{i}\AgdaSymbol{)}\AgdaSpace{}%
\AgdaOperator{\AgdaDatatype{≡}}\AgdaSpace{}%
\AgdaInductiveConstructor{just}\AgdaSpace{}%
\AgdaBound{x}\AgdaSymbol{)}\AgdaSpace{}%
\AgdaSymbol{→}\AgdaSpace{}%
\AgdaBound{x}\AgdaSpace{}%
\AgdaOperator{\AgdaDatatype{memberOf}}\AgdaSpace{}%
\AgdaBound{xs}\<%
\\
\\[\AgdaEmptyExtraSkip]%
\>[0]\AgdaFunction{mem{-}compl}\AgdaSpace{}%
\AgdaSymbol{\{}\AgdaArgument{x}\AgdaSpace{}%
\AgdaSymbol{=}\AgdaSpace{}%
\AgdaBound{x}\AgdaSymbol{\}}\AgdaSpace{}%
\AgdaBound{xs}\AgdaSpace{}%
\AgdaOperator{\AgdaInductiveConstructor{<}}\AgdaSpace{}%
\AgdaInductiveConstructor{zero}\AgdaSpace{}%
\AgdaOperator{\AgdaInductiveConstructor{,}}\AgdaSpace{}%
\AgdaBound{p}\AgdaSpace{}%
\AgdaOperator{\AgdaInductiveConstructor{>}}\AgdaSpace{}%
\AgdaKeyword{with}\AgdaSpace{}%
\AgdaFunction{inspect}\AgdaSpace{}%
\AgdaSymbol{(}\AgdaField{MyColist.list}\AgdaSpace{}%
\AgdaBound{xs}\AgdaSymbol{)}\<%
\\
\>[0]\AgdaSymbol{...}\AgdaSpace{}%
\AgdaSymbol{|}\AgdaSpace{}%
\AgdaInductiveConstructor{nothing}\AgdaSpace{}%
\AgdaOperator{\AgdaInductiveConstructor{with≡}}\AgdaSpace{}%
\AgdaBound{eq}\AgdaSpace{}%
\AgdaSymbol{=}\AgdaSpace{}%
\AgdaFunction{⊥{-}elim}\AgdaSpace{}%
\AgdaSymbol{(}\AgdaFunction{maybe{-}abs{-}2}\AgdaSpace{}%
\AgdaSymbol{\{}\AgdaArgument{x}\AgdaSpace{}%
\AgdaSymbol{=}\AgdaSpace{}%
\AgdaBound{x}\AgdaSymbol{\}}\AgdaSpace{}%
\AgdaBound{p}\AgdaSymbol{)}\<%
\\
\>[0]\AgdaSymbol{...}\AgdaSpace{}%
\AgdaSymbol{|}\AgdaSpace{}%
\AgdaInductiveConstructor{just}\AgdaSpace{}%
\AgdaOperator{\AgdaInductiveConstructor{⟨}}\AgdaSpace{}%
\AgdaBound{y}\AgdaSpace{}%
\AgdaOperator{\AgdaInductiveConstructor{,}}\AgdaSpace{}%
\AgdaBound{ys}\AgdaSpace{}%
\AgdaOperator{\AgdaInductiveConstructor{⟩}}\AgdaSpace{}%
\AgdaOperator{\AgdaInductiveConstructor{with≡}}\AgdaSpace{}%
\AgdaBound{eq}\AgdaSpace{}%
\AgdaSymbol{=}\AgdaSpace{}%
\AgdaKeyword{let}\AgdaSpace{}%
\AgdaBound{eq{-}h{-}x}\AgdaSpace{}%
\AgdaSymbol{=}\AgdaSpace{}%
\AgdaFunction{just{-}elim}\AgdaSpace{}%
\AgdaBound{p}\AgdaSpace{}%
\AgdaKeyword{in}\<%
\\
\>[0][@{}l@{\AgdaIndent{0}}]%
\>[2]\AgdaInductiveConstructor{mem{-}h}\AgdaSpace{}%
\AgdaSymbol{\{}\AgdaArgument{xs}\AgdaSpace{}%
\AgdaSymbol{=}\AgdaSpace{}%
\AgdaBound{ys}\AgdaSymbol{\}}\AgdaSpace{}%
\AgdaSymbol{(}\AgdaOperator{\AgdaFunction{begin}}\<%
\\
\>[2][@{}l@{\AgdaIndent{0}}]%
\>[4]\AgdaField{MyColist.list}\AgdaSpace{}%
\AgdaBound{xs}\<%
\\
\>[4]\AgdaOperator{\AgdaFunction{≡⟨}}\AgdaSpace{}%
\AgdaBound{eq}\AgdaSpace{}%
\AgdaOperator{\AgdaFunction{⟩}}\<%
\\
\>[4]\AgdaInductiveConstructor{just}\AgdaSpace{}%
\AgdaOperator{\AgdaInductiveConstructor{⟨}}\AgdaSpace{}%
\AgdaBound{y}\AgdaSpace{}%
\AgdaOperator{\AgdaInductiveConstructor{,}}\AgdaSpace{}%
\AgdaBound{ys}\AgdaSpace{}%
\AgdaOperator{\AgdaInductiveConstructor{⟩}}\<%
\\
\>[4]\AgdaOperator{\AgdaFunction{≡⟨}}\AgdaSpace{}%
\AgdaFunction{cong}\AgdaSpace{}%
\AgdaSymbol{(λ}\AgdaSpace{}%
\AgdaBound{z}\AgdaSpace{}%
\AgdaSymbol{→}\AgdaSpace{}%
\AgdaInductiveConstructor{just}\AgdaSpace{}%
\AgdaOperator{\AgdaInductiveConstructor{⟨}}\AgdaSpace{}%
\AgdaBound{z}\AgdaSpace{}%
\AgdaOperator{\AgdaInductiveConstructor{,}}\AgdaSpace{}%
\AgdaBound{ys}\AgdaSpace{}%
\AgdaOperator{\AgdaInductiveConstructor{⟩}}\AgdaSymbol{)}\AgdaSpace{}%
\AgdaBound{eq{-}h{-}x}\AgdaSpace{}%
\AgdaOperator{\AgdaFunction{⟩}}\<%
\\
\>[4]\AgdaInductiveConstructor{just}\AgdaSpace{}%
\AgdaOperator{\AgdaInductiveConstructor{⟨}}\AgdaSpace{}%
\AgdaBound{x}\AgdaSpace{}%
\AgdaOperator{\AgdaInductiveConstructor{,}}\AgdaSpace{}%
\AgdaBound{ys}\AgdaSpace{}%
\AgdaOperator{\AgdaInductiveConstructor{⟩}}\<%
\\
\>[2]\AgdaOperator{\AgdaFunction{∎}}\AgdaSymbol{)}\<%
\\
\\[\AgdaEmptyExtraSkip]%
\>[0]\AgdaFunction{mem{-}compl}\AgdaSpace{}%
\AgdaBound{xs}\AgdaSpace{}%
\AgdaOperator{\AgdaInductiveConstructor{<}}\AgdaSpace{}%
\AgdaInductiveConstructor{suc}\AgdaSpace{}%
\AgdaBound{i}\AgdaSpace{}%
\AgdaOperator{\AgdaInductiveConstructor{,}}\AgdaSpace{}%
\AgdaBound{p}\AgdaSpace{}%
\AgdaOperator{\AgdaInductiveConstructor{>}}\AgdaSpace{}%
\AgdaKeyword{with}\AgdaSpace{}%
\AgdaFunction{inspect}\AgdaSpace{}%
\AgdaSymbol{(}\AgdaField{MyColist.list}\AgdaSpace{}%
\AgdaBound{xs}\AgdaSymbol{)}\<%
\\
\>[0]\AgdaSymbol{...}\AgdaSpace{}%
\AgdaSymbol{|}\AgdaSpace{}%
\AgdaInductiveConstructor{nothing}\AgdaSpace{}%
\AgdaOperator{\AgdaInductiveConstructor{with≡}}\AgdaSpace{}%
\AgdaBound{eq}\AgdaSpace{}%
\AgdaSymbol{=}%
\>[27]\AgdaFunction{⊥{-}elim}\AgdaSpace{}%
\AgdaSymbol{(}\AgdaFunction{maybe{-}abs{-}2}\AgdaSpace{}%
\AgdaBound{p}\AgdaSymbol{)}\<%
\\
\>[0]\AgdaSymbol{...}\AgdaSpace{}%
\AgdaSymbol{|}\AgdaSpace{}%
\AgdaInductiveConstructor{just}\AgdaSpace{}%
\AgdaOperator{\AgdaInductiveConstructor{⟨}}\AgdaSpace{}%
\AgdaSymbol{\AgdaUnderscore{}}\AgdaSpace{}%
\AgdaOperator{\AgdaInductiveConstructor{,}}\AgdaSpace{}%
\AgdaBound{ys}\AgdaSpace{}%
\AgdaOperator{\AgdaInductiveConstructor{⟩}}\AgdaSpace{}%
\AgdaOperator{\AgdaInductiveConstructor{with≡}}\AgdaSpace{}%
\AgdaBound{eq}\AgdaSpace{}%
\AgdaSymbol{=}\AgdaSpace{}%
\AgdaInductiveConstructor{mem{-}t}\AgdaSpace{}%
\AgdaBound{eq}\AgdaSpace{}%
\AgdaSymbol{(}\AgdaFunction{mem{-}compl}\AgdaSpace{}%
\AgdaBound{ys}\AgdaSpace{}%
\AgdaOperator{\AgdaInductiveConstructor{<}}\AgdaSpace{}%
\AgdaBound{i}\AgdaSpace{}%
\AgdaOperator{\AgdaInductiveConstructor{,}}\AgdaSpace{}%
\AgdaBound{p}\AgdaSpace{}%
\AgdaOperator{\AgdaInductiveConstructor{>}}\AgdaSymbol{)}\<%
\end{code}
 \end{center}

The construct \agda{with} allows one to reason by pattern-matching on structures where standard pattern-matching is not possibile, such as coinductive records. 
Indeed, pattern-matching is done on the result of a function applied to the structure (e.g., a record projection). The function \agda{inspect}, defined in \LC{\refToSection{agda-modules}}, is needed because when generalizing over a term $t$ using \agda{with} we lose the connection between $t$ and the new variables.\EZComm{non si capisce}

In the proof, first of all two absurd cases must be handled: \agda{maybe-abs-2} checks whether an object of type \agda{Maybe} is built using different constructors and it returns $\bot$ in this case. Then everything can be returned by the elimination rule (see \refToSection{agda-modules} for more details).
Then, the cases corresponding to the constructors of natural numbers are considered.
\begin{description}
\item[\agda{zero}] In this case, \agda{mem-h} requires the proof that the list has \agda{x} and \agda{ys} as head and tail,  respectively, while \agda{eq} is the proof that \agda{y} is equal to the head. The proof is obtained through the equality reasoning module that allows one to make chains of equalities without defining each of them separately.
\item[\agda{suc i}] The proof is obtained by recursion.
\end{description}

The predicate \textit{memberOf} analyzed so far can be defined inductively independently from the data structure over which it is defined. 
We conclude this chapter by considering an example where this is not true: the predicate \textit{allPos}, expected to hold on a list \agda{xs} of natural numbers when all the elements are (strictly) positive.
The inference system is the following:

\begin{small}
\begin{quote}
$\NamedRule {allp-$\Lambda$}{} {\allPos {\Lambda}} \Space
\NamedRule {allp-t}{\allPos{\xs}} {\allPos {\x{:}\xs}} \x > 0$
\end{quote}
\end{small}

The axiom states that the predicate holds on an empty list, while the second rule that it holds on a non-empty list if it holds on the tail and the head is positive.

In this case, the inductive interpretation works well on finite lists. However, given an infinite list $\xs$ of all positive elements, $\allPos{\xs}$ cannot be derived, since there is no finite proof tree for this judgment. 

The specification of the predicate is $\Spec = \{ \xs \mid \forall \x. \memberOf {\x} {\xs}\ \mbox{implies}\ \x > 0 \}$. Note that {we can use} \textit{memberOf} as auxiliary predicate since we already proved its correctness before. {That is, the above specification is equivalent to the following: $\{ \xs\mid \forall \x.(\exists i. \get{\xs}{i}=\x\ \mbox{implies}\ x > 0) \}$. This allows us to reason by induction on the definition of \textit{memberOf} rather than by arithmetic induction.}

\begin{statement}
The inductive definition of \textit{allPos} is sound with respect to its specification.
\end{statement}

\begin{proof}
We have to prove that, if $\allPos{\xs}\in\Ind{\is}$, then $\memberOf {\x} {\xs}\ \mbox{implies}\ \x > 0$. The proof is by induction on the definition of $\allPos{\xs}$.
\begin{description}
\item[\refToRule{allp-$\Lambda$}] The thesis trivially holds.
\item[\refToRule{allp-t}] Assume that $\xs\in\Spec$, that is, all the elements of $\xs$ are positive. We have to prove that $\x{:}\xs\in\Spec$ holds as well, and this is true since $\x>0$  from the side condition.  
\end{description}
\end{proof}

\begin{statement}
{Considering only finite lists,} the inductive definition of \textit{allPos} is {complete} with respect to its specification.
\end{statement}

\begin{proof} We have to prove that, if $\xs\in\Spec$, that is, all the elements of $\xs$ are positive, then $\allPos{\xs}$ can be derived.
The proof is by induction on the (finite) list. 
\begin{description}
\item[$\Lambda$] The judgment $\allPos{\Lambda}$ can be derived by rule \agda{allp-$\Lambda$}.
\item[$\x{:}\xs$] Since all the elements of $\x{:}\xs$ are positive, we know that $\xs\in\Spec$, hence the judgment $\allPos{\xs}$ can be derived by inductive hypothesis. Thus, the judgment $\allPos{\x{:}\xs}$ can be derived by rule \agda{allp-t}.
\end{description}
\end{proof}

Now we move to Agda; the predicate is implemented using \agda{data} since it is interpreted inductively.

\begin{center}
\begin{code}%

\>[0]\AgdaKeyword{data}\AgdaSpace{}%
\AgdaOperator{\AgdaDatatype{\AgdaUnderscore{}>\AgdaUnderscore{}}}\AgdaSpace{}%
\AgdaSymbol{:}\AgdaSpace{}%
\AgdaDatatype{Nat}\AgdaSpace{}%
\AgdaSymbol{→}\AgdaSpace{}%
\AgdaDatatype{Nat}\AgdaSpace{}%
\AgdaSymbol{→}\AgdaSpace{}%
\AgdaPrimitiveType{Set}\AgdaSpace{}%
\AgdaKeyword{where}\<%
\\
\>[0][@{}l@{\AgdaIndent{0}}]%
\>[2]\AgdaInductiveConstructor{g{-}zero}\AgdaSpace{}%
\AgdaSymbol{:}\AgdaSpace{}%
\AgdaSymbol{\{}\AgdaBound{x}\AgdaSpace{}%
\AgdaSymbol{:}\AgdaSpace{}%
\AgdaDatatype{Nat}\AgdaSymbol{\}}\AgdaSpace{}%
\AgdaSymbol{→}\AgdaSpace{}%
\AgdaInductiveConstructor{suc}%
\>[28]\AgdaBound{x}\AgdaSpace{}%
\AgdaOperator{\AgdaDatatype{>}}\AgdaSpace{}%
\AgdaInductiveConstructor{zero}\<%
\\
\>[2]\AgdaInductiveConstructor{g{-}suc}\AgdaSpace{}%
\AgdaSymbol{:}\AgdaSpace{}%
\AgdaSymbol{\{}\AgdaBound{x}\AgdaSpace{}%
\AgdaBound{y}\AgdaSpace{}%
\AgdaSymbol{:}\AgdaSpace{}%
\AgdaDatatype{Nat}\AgdaSymbol{\}}\AgdaSpace{}%
\AgdaSymbol{→}\AgdaSpace{}%
\AgdaBound{x}\AgdaSpace{}%
\AgdaOperator{\AgdaDatatype{>}}\AgdaSpace{}%
\AgdaBound{y}\AgdaSpace{}%
\AgdaSymbol{→}\AgdaSpace{}%
\AgdaInductiveConstructor{suc}\AgdaSpace{}%
\AgdaBound{x}\AgdaSpace{}%
\AgdaOperator{\AgdaDatatype{>}}\AgdaSpace{}%
\AgdaInductiveConstructor{suc}\AgdaSpace{}%
\AgdaBound{y}\<%
\\
\\[\AgdaEmptyExtraSkip]%
\>[0]\AgdaKeyword{data}\AgdaSpace{}%
\AgdaOperator{\AgdaDatatype{allPos\AgdaUnderscore{}}}\AgdaSpace{}%
\AgdaSymbol{:}\AgdaSpace{}%
\AgdaDatatype{List}\AgdaSpace{}%
\AgdaDatatype{Nat}\AgdaSpace{}%
\AgdaSymbol{→}\AgdaSpace{}%
\AgdaPrimitiveType{Set}\AgdaSpace{}%
\AgdaKeyword{where}\<%
\\
\>[0][@{}l@{\AgdaIndent{0}}]%
\>[2]\AgdaInductiveConstructor{allp{-}Λ}\AgdaSpace{}%
\AgdaSymbol{:}\AgdaSpace{}%
\AgdaOperator{\AgdaDatatype{allPos}}\AgdaSpace{}%
\AgdaInductiveConstructor{[]}\<%
\\
\>[2]\AgdaInductiveConstructor{allp{-}t}\AgdaSpace{}%
\AgdaSymbol{:}\AgdaSpace{}%
\AgdaSymbol{forall}\AgdaSpace{}%
\AgdaSymbol{\{}\AgdaBound{x}\AgdaSpace{}%
\AgdaSymbol{:}\AgdaSpace{}%
\AgdaDatatype{Nat}\AgdaSymbol{\}\{}\AgdaBound{xs}\AgdaSpace{}%
\AgdaSymbol{:}\AgdaSpace{}%
\AgdaDatatype{List}\AgdaSpace{}%
\AgdaDatatype{Nat}\AgdaSymbol{\}}\AgdaSpace{}%
\AgdaSymbol{→}\AgdaSpace{}%
\AgdaOperator{\AgdaDatatype{allPos}}\AgdaSpace{}%
\AgdaBound{xs}\AgdaSpace{}%
\AgdaSymbol{→}\AgdaSpace{}%
\AgdaBound{x}\AgdaSpace{}%
\AgdaOperator{\AgdaDatatype{>}}\AgdaSpace{}%
\AgdaInductiveConstructor{zero}\AgdaSpace{}%
\AgdaSymbol{→}\AgdaSpace{}%
\AgdaOperator{\AgdaDatatype{allPos}}\AgdaSpace{}%
\AgdaSymbol{(}\AgdaBound{x}%
\>[82]\AgdaOperator{\AgdaInductiveConstructor{∷}}\AgdaSpace{}%
\AgdaBound{xs}\AgdaSymbol{)}\<%
\end{code}

\end{center}

{The predicate $>$ is in turn an inductive predicate, hence implemented in Agda by \agda{data}, defined by two meta-rules.}
{For instance, an} element of type $1 > 0$ {is} the proof that 1 is greater than 0 and it is {constructed by} \agda{g-zero}.

In the definition of \agda{allPos}, we add in the inductive constructor the side condition that $\x$ is positive, analogously to the inference system.

\begin{center}
\begin{code}%
\>[0]\AgdaFunction{allp{-}sound}\AgdaSpace{}%
\AgdaSymbol{:}\AgdaSpace{}%
\AgdaSymbol{\{}\AgdaBound{xs}\AgdaSpace{}%
\AgdaSymbol{:}\AgdaSpace{}%
\AgdaDatatype{List}\AgdaSpace{}%
\AgdaDatatype{Nat}\AgdaSymbol{\}}\AgdaSpace{}%
\AgdaSymbol{→}\AgdaSpace{}%
\AgdaOperator{\AgdaDatatype{allPos}}\AgdaSpace{}%
\AgdaBound{xs}\AgdaSpace{}%
\AgdaSymbol{→}\AgdaSpace{}%
\AgdaSymbol{(\{}\AgdaBound{n}\AgdaSpace{}%
\AgdaSymbol{:}\AgdaSpace{}%
\AgdaDatatype{Nat}\AgdaSymbol{\}}\AgdaSpace{}%
\AgdaSymbol{→}\AgdaSpace{}%
\AgdaBound{n}\AgdaSpace{}%
\AgdaOperator{\AgdaDatatype{memberOf}}\AgdaSpace{}%
\AgdaBound{xs}\AgdaSpace{}%
\AgdaSymbol{→}\AgdaSpace{}%
\AgdaBound{n}\AgdaSpace{}%
\AgdaOperator{\AgdaDatatype{>}}\AgdaSpace{}%
\AgdaInductiveConstructor{zero}\AgdaSymbol{)}\<%
\\
\>[0]\AgdaFunction{allp{-}sound}\AgdaSpace{}%
\AgdaInductiveConstructor{allp{-}Λ}\AgdaSpace{}%
\AgdaSymbol{()}\<%
\\
\>[0]\AgdaFunction{allp{-}sound}\AgdaSpace{}%
\AgdaSymbol{(}\AgdaInductiveConstructor{allp{-}t}\AgdaSpace{}%
\AgdaBound{allp}\AgdaSpace{}%
\AgdaBound{gr}\AgdaSymbol{)}\AgdaSpace{}%
\AgdaInductiveConstructor{mem{-}h}\AgdaSpace{}%
\AgdaSymbol{=}\AgdaSpace{}%
\AgdaBound{gr}\<%
\\
\>[0]\AgdaFunction{allp{-}sound}\AgdaSpace{}%
\AgdaSymbol{(}\AgdaInductiveConstructor{allp{-}t}\AgdaSpace{}%
\AgdaBound{allp}\AgdaSpace{}%
\AgdaBound{gr}\AgdaSymbol{)}\AgdaSpace{}%
\AgdaSymbol{(}\AgdaInductiveConstructor{mem{-}t}\AgdaSpace{}%
\AgdaBound{mem}\AgdaSymbol{)}\AgdaSpace{}%
\AgdaSymbol{=}\AgdaSpace{}%
\AgdaFunction{allp{-}sound}\AgdaSpace{}%
\AgdaBound{allp}\AgdaSpace{}%
\AgdaBound{mem}\<%
\\
\\[\AgdaEmptyExtraSkip]%
\>[0]\AgdaFunction{allp{-}compl}\AgdaSpace{}%
\AgdaSymbol{:}%
\>[14]\AgdaSymbol{\{}\AgdaBound{xs}\AgdaSpace{}%
\AgdaSymbol{:}\AgdaSpace{}%
\AgdaDatatype{List}\AgdaSpace{}%
\AgdaDatatype{Nat}\AgdaSymbol{\}}\AgdaSpace{}%
\AgdaSymbol{→}\AgdaSpace{}%
\AgdaSymbol{(\{}\AgdaBound{n}\AgdaSpace{}%
\AgdaSymbol{:}\AgdaSpace{}%
\AgdaDatatype{Nat}\AgdaSymbol{\}}\AgdaSpace{}%
\AgdaSymbol{→}\AgdaSpace{}%
\AgdaBound{n}\AgdaSpace{}%
\AgdaOperator{\AgdaDatatype{memberOf}}\AgdaSpace{}%
\AgdaBound{xs}\AgdaSpace{}%
\AgdaSymbol{→}\AgdaSpace{}%
\AgdaBound{n}\AgdaSpace{}%
\AgdaOperator{\AgdaDatatype{>}}\AgdaSpace{}%
\AgdaNumber{0}\AgdaSymbol{)}\AgdaSpace{}%
\AgdaSymbol{→}\AgdaSpace{}%
\AgdaOperator{\AgdaDatatype{allPos}}\AgdaSpace{}%
\AgdaBound{xs}\<%
\\
\>[0]\AgdaFunction{allp{-}compl}\AgdaSpace{}%
\AgdaSymbol{\{}\AgdaInductiveConstructor{[]}\AgdaSymbol{\}}\AgdaSpace{}%
\AgdaBound{p}\AgdaSpace{}%
\AgdaSymbol{=}\AgdaSpace{}%
\AgdaInductiveConstructor{allp{-}Λ}\<%
\\
\>[0]\AgdaFunction{allp{-}compl}\AgdaSpace{}%
\AgdaSymbol{\{}\AgdaBound{x}\AgdaSpace{}%
\AgdaOperator{\AgdaInductiveConstructor{∷}}\AgdaSpace{}%
\AgdaBound{xs}\AgdaSymbol{\}}\AgdaSpace{}%
\AgdaBound{p}\AgdaSpace{}%
\AgdaSymbol{=}\AgdaSpace{}%
\AgdaInductiveConstructor{allp{-}t}\AgdaSpace{}%
\AgdaSymbol{(}\AgdaFunction{allp{-}compl}\AgdaSpace{}%
\AgdaSymbol{(λ}\AgdaSpace{}%
\AgdaBound{m}\AgdaSpace{}%
\AgdaSymbol{→}\AgdaSpace{}%
\AgdaBound{p}\AgdaSpace{}%
\AgdaSymbol{(}\AgdaInductiveConstructor{mem{-}t}\AgdaSpace{}%
\AgdaBound{m}\AgdaSymbol{)))}\AgdaSpace{}%
\AgdaSymbol{(}\AgdaBound{p}\AgdaSpace{}%
\AgdaInductiveConstructor{mem{-}h}\AgdaSymbol{)}\<%
\end{code}

\end{center}

The soundness proof is done by pattern matching on the constructors of \textit{allPos} and then on those of \textit{memberOf}, that is, by induction on its definition.\EZComm{nella prova su carta questo \`e implicito}. The first case leads to the absurd, denoted $()$: indeed, \agda{allp-$\Lambda$} implies that the list is empty, and there is no valid constructor of \textit{memberOf}. Agda automatically recognizes this absurd if we try to interactively split \EZComm{oscuro} over \textit{memberOf}.

The proof of completeness reflects the one we provided with the inference system and it is by induction on how the list is built.
When recursively calling \agda{complete-allpos} it is necessary to restrict $p$ to the tail $xs$ of the list. In fact $p$ is a function that, for each element in $l$, returns the proof that it is positive. The recursive call requires a similar function that acts only on the elements of the tail. The type of $m$ is not reported but it is quite clear that it identifies the elements of $xs$ according to the inputs of the constructor of \textit{allPos}. Thus, from $p$ we get the proofs for all the numbers in $x{:}xs$ and so in $xs$ and we explicitly say that the output of the new function is $p$ applied to $m$ with the additional constructor \agda{mem-t} to avoid involving the head of $x{:}xs$.

We presented a situation in which the inductive interpretation of \textit{allPos} works well. Clearly if the reference data structure is infinite, the proof that its elements are all positive is infinite as well because the objects at each position have to be checked. In this case the inductive interpretation does not work and we have to interpret the predicate coinductively. In a more detailed way, the inductive predicate is still sound but not complete. In the next chapter we are going to study into detail coinductively interpreted predicates and how Agda support them.

\chapter{Coinductive reasoning}
\label{chapter:coinductive}
{In \refToChapter{inductive} we introduced basic notions about inference systems and we described proof techniques for inductive predicates. In particular we discovered that in Agda the implementation of \textit{memberOf} does not depend on the data structure on which it is defined since it is an inductive predicate. 
In this chapter, we consider predicates defined as the coinductive interpretation of an inference system, shortly \emph{coinductive predicates}.}
\LC{For what concerns Agda, in this case the two approaches that we proposed for implementing colists (see \refToSection{colists} for more details) can be adopted for implementing the predicates too. Clearly this leads to different implementation possibilities according to how we choose to define the universe and the predicate. We decided not to mix the approaches, thus considering also the cases in which the colists and the predicate are implemented in the same way.}

\section{Coinduction} We recall that the coinductive interpretation of an inference system $\is$, denoted $\CoInd{\is}$, is the largest consistent set, that is, the union of all consistent sets, or, in proof-theoretic terms, the set of judgments which have a (finite or infinite) proof tree. In this case, we have a canonical technique to prove \textit{completeness}, that is, the \emph{coinduction principle}.

\paragraph*{Coinduction Principle} If a set $\Spec\subseteq\universe$ is consistent, then $\Spec\subseteq\CoInd{\is}$.
{The proof is immediate since $\CoInd{\is}$ is the largest consistent set by definition. Proving that $\Spec$ is consistent amounts to show that, for each $\co\in\Spec$, there is a rule $\Rule {\prem} {\co}$ of the inference system such that $\prem\subseteq\Spec$.}

{On the other hand, to prove soundness of a coinductive predicate there is no canonical technique, hence, for each concrete case, we must find an ``ad-hoc'' technique.}

\section{Example}{As paradigmatic example of coinductive predicate,} we recall \textit{allPos}, where $\textit{allPos}(\xs)$ holds if all the elements of $\xs$ are positive. 
{If the previously introduced inference system}
\begin{small}
\begin{quote}
$\NamedRule {allP-$\Lambda$}{} {\allPos {\Lambda}} \Space
\NamedRule {allP-t}{\allPos{\xs}} {\allPos {\x{:}\xs}} \x > 0$
\end{quote}
\end{small}
{is interpreted \emph{coinductively}, then the definition also works for infinite lists, as will be shown in the following.}

The specification of the predicate is as before:
\begin{quote}
$\Spec = \{ \xs \mid \forall \x. \memberOf {\x} {\xs}\ \mbox{implies}\ \x > 0 \}$
\end{quote}

For soundness, since now \textit{allPos} is coinductive, we can no longer reason by induction on its definition.

\begin{statement}
The coinductive definition of \textit{allPos} is sound with respect to its specification.
\end{statement}
\begin{proof}
We have to prove that, if $\allPos{\xs}\in\CoInd{\is}$, then $\memberOf{\x}{\xs}$ implies that $\x > 0$ holds. If $\memberOf {\x} {\xs}$ holds, then $\xs$ cannot be empty, hence $\xs=\y{:}\ys$. Hence, to derive that $\xs\in\CoInd{\is}$, we have used rule \refToRule{allPos-t}, thus $\y>0$ and $\ys\in\CoInd{\is}$.
The proof is by induction on the definition of $\memberOf{\x}{\xs}$. 
\begin{description}
\item[\refToRule{mem-h}] We have $\x=\y$, hence $\x>0$ holds by the side condition of \refToRule{allP-t}.
\item[\refToRule{mem-t}] We have $\memberOf{\x}{\ys}$, and $\x>0$ by the inductive hypothesis. 
\end{description}
\end{proof}

Completeness is no longer restricted to finite lists, and can now be proved by coinduction on the definition of \textit{allPos}.

\begin{statement}
The coinductive definition of \textit{allPos} is complete with respect to its specification.
\end{statement} 

\begin{proof}
We have to prove that, if $\xs\in\Spec$, that is, all the elements of $\xs$ are positive, then $\allPos{\xs}$ can be derived. 
The proof is by coinduction on the definition of $\allPos{\xs}$. That is, we have to show that $\Spec$ is consistent: if $\xs$ is in $\Spec$, then it is the consequence of a rule with premises which are in $\Spec$ as well. We consider two cases.
\begin{description}
\item If $\xs \in \Spec$ and $\xs$ is empty, then it is the consequence of \refToRule{allP-$\Lambda$}.
\item If $\ys\in\Spec$ and $\ys = \x{:}\xs$, then it is the consequence of \refToRule{allP-t} with premise $\xs$, and we know that $\xs\in\Spec$. 
\end{description}
\end{proof}

\section{Agda implementation}We show now the Agda implementation of the predicate on streams, and the corresponding soundness and completeness proofs.

\begin{center}
\begin{code}
\>[0]\AgdaKeyword{record}\AgdaSpace{}%
\AgdaRecord{allPos}\AgdaSpace{}%
\AgdaSymbol{(}\AgdaBound{xs}\AgdaSpace{}%
\AgdaSymbol{:}\AgdaSpace{}%
\AgdaRecord{MyStream}\AgdaSpace{}%
\AgdaDatatype{Nat}\AgdaSymbol{)}\AgdaSpace{}%
\AgdaSymbol{:}\AgdaSpace{}%
\AgdaPrimitiveType{Set}\AgdaSpace{}%
\AgdaKeyword{where}\<%
\\
\>[0][@{}l@{\AgdaIndent{0}}]%
\>[2]\AgdaKeyword{coinductive}\<%
\\
\>[2]\AgdaKeyword{field}\<%
\\
\>[2][@{}l@{\AgdaIndent{0}}]%
\>[4]\AgdaField{h}\AgdaSpace{}%
\AgdaSymbol{:}\AgdaSpace{}%
\AgdaSymbol{(}\AgdaField{MyStream.hd}\AgdaSpace{}%
\AgdaBound{xs}\AgdaSymbol{)}\AgdaSpace{}%
\AgdaOperator{\AgdaDatatype{>}}\AgdaSpace{}%
\AgdaInductiveConstructor{zero}\<%
\\
\>[4]\AgdaField{t}\AgdaSpace{}%
\AgdaSymbol{:}\AgdaSpace{}%
\AgdaRecord{allPos}\AgdaSpace{}%
\AgdaSymbol{(}\AgdaField{MyStream.tl}\AgdaSpace{}%
\AgdaBound{xs}\AgdaSymbol{)}\<%
\\
\\[\AgdaEmptyExtraSkip]%
\\[\AgdaEmptyExtraSkip]%
\>[0]\AgdaFunction{allp{-}sound}\AgdaSpace{}%
\AgdaSymbol{:}\AgdaSpace{}%
\AgdaSymbol{\{}\AgdaBound{xs}\AgdaSpace{}%
\AgdaSymbol{:}\AgdaSpace{}%
\AgdaRecord{MyStream}%
\>[35I]\AgdaDatatype{Nat}\AgdaSymbol{\}}\AgdaSpace{}%
\AgdaSymbol{→}\AgdaSpace{}%
\AgdaRecord{allPos}\AgdaSpace{}%
\AgdaBound{xs}\AgdaSpace{}%
\AgdaSymbol{→}\<%
\\
\>[35I][@{}l@{\AgdaIndent{0}}]%
\>[29]\AgdaSymbol{(\{}\AgdaBound{n}\AgdaSpace{}%
\AgdaSymbol{:}\AgdaSpace{}%
\AgdaDatatype{Nat}\AgdaSymbol{\}}\AgdaSpace{}%
\AgdaSymbol{→}\AgdaSpace{}%
\AgdaBound{n}\AgdaSpace{}%
\AgdaOperator{\AgdaDatatype{memberOf}}\AgdaSpace{}%
\AgdaBound{xs}\AgdaSpace{}%
\AgdaSymbol{→}\AgdaSpace{}%
\AgdaBound{n}\AgdaSpace{}%
\AgdaOperator{\AgdaDatatype{>}}\AgdaSpace{}%
\AgdaInductiveConstructor{zero}\AgdaSymbol{)}\<%
\\
\>[0]\<%
\\
\>[0]\AgdaFunction{allp{-}sound}\AgdaSpace{}%
\AgdaBound{ap}\AgdaSpace{}%
\AgdaSymbol{(}\AgdaInductiveConstructor{mem{-}h}\AgdaSpace{}%
\AgdaInductiveConstructor{refl}\AgdaSymbol{)}\AgdaSpace{}%
\AgdaSymbol{=}\AgdaSpace{}%
\AgdaField{allPos.h}\AgdaSpace{}%
\AgdaBound{ap}\<%
\\
\>[0]\AgdaFunction{allp{-}sound}\AgdaSpace{}%
\AgdaBound{ap}\AgdaSpace{}%
\AgdaSymbol{(}\AgdaInductiveConstructor{mem{-}t}\AgdaSpace{}%
\AgdaBound{mem}\AgdaSymbol{)}\AgdaSpace{}%
\AgdaSymbol{=}\AgdaSpace{}%
\AgdaFunction{allp{-}sound}\AgdaSpace{}%
\AgdaSymbol{(}\AgdaField{allPos.t}\AgdaSpace{}%
\AgdaBound{ap}\AgdaSymbol{)}\AgdaSpace{}%
\AgdaBound{mem}\<%
\\
\\[\AgdaEmptyExtraSkip]%
\\[\AgdaEmptyExtraSkip]%
\>[0]\AgdaFunction{allp{-}compl}\AgdaSpace{}%
\AgdaSymbol{:}%
\>[14]\AgdaSymbol{(}\AgdaBound{xs}\AgdaSpace{}%
\AgdaSymbol{:}\AgdaSpace{}%
\AgdaRecord{MyStream}%
\>[67I]\AgdaDatatype{Nat}\AgdaSymbol{)}\AgdaSpace{}%
\AgdaSymbol{→}\<%
\\
\>[67I][@{}l@{\AgdaIndent{0}}]%
\>[30]\AgdaSymbol{(\{}\AgdaBound{n}\AgdaSpace{}%
\AgdaSymbol{:}\AgdaSpace{}%
\AgdaDatatype{Nat}\AgdaSymbol{\}}\AgdaSpace{}%
\AgdaSymbol{→}\AgdaSpace{}%
\AgdaBound{n}\AgdaSpace{}%
\AgdaOperator{\AgdaDatatype{memberOf}}\AgdaSpace{}%
\AgdaBound{xs}\AgdaSpace{}%
\AgdaSymbol{→}\AgdaSpace{}%
\AgdaBound{n}\AgdaSpace{}%
\AgdaOperator{\AgdaDatatype{>}}\AgdaSpace{}%
\AgdaNumber{0}\AgdaSymbol{)}\AgdaSpace{}%
\AgdaSymbol{→}\AgdaSpace{}%
\AgdaRecord{allPos}\AgdaSpace{}%
\AgdaBound{xs}\<%
\\
\>[0]\<%
\\
\>[0]\AgdaField{allPos.h}\AgdaSpace{}%
\AgdaSymbol{(}\AgdaFunction{allp{-}compl}\AgdaSpace{}%
\AgdaBound{xs}\AgdaSpace{}%
\AgdaBound{p}\AgdaSymbol{)}\AgdaSpace{}%
\AgdaSymbol{=}\AgdaSpace{}%
\AgdaBound{p}\AgdaSpace{}%
\AgdaSymbol{(}\AgdaInductiveConstructor{mem{-}h}\AgdaSpace{}%
\AgdaInductiveConstructor{refl}\AgdaSymbol{)}\<%
\\
\>[0]\AgdaField{allPos.t}\AgdaSpace{}%
\AgdaSymbol{(}\AgdaFunction{allp{-}compl}\AgdaSpace{}%
\AgdaBound{xs}\AgdaSpace{}%
\AgdaBound{p}\AgdaSymbol{)}\AgdaSpace{}%
\AgdaSymbol{=}\AgdaSpace{}%
\AgdaFunction{allp{-}compl}\AgdaSpace{}%
\AgdaSymbol{(}\AgdaField{MyStream.tl}\AgdaSpace{}%
\AgdaBound{xs}\AgdaSymbol{)}\AgdaSpace{}%
\AgdaSymbol{λ}\AgdaSpace{}%
\AgdaBound{m}\AgdaSpace{}%
\AgdaSymbol{→}\AgdaSpace{}%
\AgdaBound{p}\AgdaSpace{}%
\AgdaSymbol{(}\AgdaInductiveConstructor{mem{-}t}\AgdaSpace{}%
\AgdaBound{m}\AgdaSymbol{)}\<%
\end{code}

\end{center}

{In this case, the coinductive} predicate is defined using a coinductive record analogously to streams. While the soundness proof is by induction on the definition of \textit{memberOf}, the completeness proof shows an example of \textit{copattern matching}.

{We consider now colists.} In this case, \textit{allPos} can be represented by the two different strategies illustrated in \refToSection{colists} for possibly infinite data structures, that is, either by the colists of the standard library, ot through coinductive records.
The proofs will differ depending on the chosen implementation. 

{We first show the representation with library colists.}

\begin{center}
\begin{code}%
\>[0]\AgdaKeyword{data}\AgdaSpace{}%
\AgdaDatatype{allPos}\AgdaSpace{}%
\AgdaSymbol{:}\AgdaSpace{}%
\AgdaDatatype{Colist}\AgdaSpace{}%
\AgdaDatatype{Nat}\AgdaSpace{}%
\AgdaPostulate{∞}\AgdaSpace{}%
\AgdaSymbol{→}\AgdaSpace{}%
\AgdaPostulate{Size}\AgdaSpace{}%
\AgdaSymbol{→}\AgdaSpace{}%
\AgdaPrimitiveType{Set}\AgdaSpace{}%
\AgdaKeyword{where}\<%
\\
\>[0][@{}l@{\AgdaIndent{0}}]%
\>[2]\AgdaInductiveConstructor{allp{-}Λ}\AgdaSpace{}%
\AgdaSymbol{:}\AgdaSpace{}%
\AgdaSymbol{∀}\AgdaSpace{}%
\AgdaSymbol{\{}\AgdaBound{i}\AgdaSymbol{\}}\AgdaSpace{}%
\AgdaSymbol{→}%
\>[20]\AgdaDatatype{allPos}\AgdaSpace{}%
\AgdaInductiveConstructor{[]}\AgdaSpace{}%
\AgdaBound{i}\<%
\\
\>[2]\AgdaInductiveConstructor{allp{-}t}\AgdaSpace{}%
\AgdaSymbol{:}\AgdaSpace{}%
\AgdaSymbol{∀}\AgdaSpace{}%
\AgdaSymbol{\{}\AgdaBound{x}\AgdaSpace{}%
\AgdaBound{i}\AgdaSpace{}%
\AgdaBound{xs}\AgdaSymbol{\}}\AgdaSpace{}%
\AgdaSymbol{→}\AgdaSpace{}%
\AgdaBound{x}\AgdaSpace{}%
\AgdaOperator{\AgdaDatatype{>}}\AgdaSpace{}%
\AgdaInductiveConstructor{zero}\AgdaSpace{}%
\AgdaSymbol{→}\AgdaSpace{}%
\AgdaRecord{Thunk}\AgdaSpace{}%
\AgdaSymbol{(}\AgdaDatatype{allPos}\AgdaSpace{}%
\AgdaSymbol{(}\AgdaField{Thunk.force}\AgdaSpace{}%
\AgdaBound{xs}\AgdaSymbol{))}\AgdaSpace{}%
\AgdaBound{i}\AgdaSpace{}%
\AgdaSymbol{→}\AgdaSpace{}%
\AgdaDatatype{allPos}\AgdaSpace{}%
\AgdaSymbol{(}\AgdaBound{x}\AgdaSpace{}%
\AgdaOperator{\AgdaInductiveConstructor{∷}}\AgdaSpace{}%
\AgdaBound{xs}\AgdaSymbol{)}\AgdaSpace{}%
\AgdaBound{i}\<%
\end{code}

\end{center}

\EZComm{frase oscura da discutere}\LC{The predicate is coinductive thus it depends on \agda{Size}, which is different from the colists one because it identifies the approximations of the predicate. This leads also to the introduction of \agda{Thunk} for what concerns the second rule \refToRule{allp-t}. Notice that the structure of the type of the predicate recalls the structure of the colists. In fact the proof that \textit{allPos} holds on the tail must be boxed in a suspended computation.}

\begin{center}
\begin{code}
\>[0]\AgdaFunction{allp{-}sound}\AgdaSpace{}%
\AgdaSymbol{:}\AgdaSpace{}%
\AgdaSymbol{\{}\AgdaBound{xs}\AgdaSpace{}%
\AgdaSymbol{:}\AgdaSpace{}%
\AgdaDatatype{Colist}%
\>[57I]\AgdaDatatype{Nat}\AgdaSpace{}%
\AgdaPostulate{∞}\AgdaSymbol{\}}\AgdaSpace{}%
\AgdaSymbol{→}\AgdaSpace{}%
\AgdaSymbol{(∀}\AgdaSpace{}%
\AgdaSymbol{\{}\AgdaBound{i}\AgdaSymbol{\}}\AgdaSpace{}%
\AgdaSymbol{→}\AgdaSpace{}%
\AgdaDatatype{allPos}\AgdaSpace{}%
\AgdaBound{xs}\AgdaSpace{}%
\AgdaBound{i}\AgdaSymbol{)}\AgdaSpace{}%
\AgdaSymbol{→}\<%
\\
\>[57I][@{}l@{\AgdaIndent{0}}]%
\>[29]\AgdaSymbol{(\{}\AgdaBound{n}\AgdaSpace{}%
\AgdaSymbol{:}\AgdaSpace{}%
\AgdaDatatype{Nat}\AgdaSymbol{\}}\AgdaSpace{}%
\AgdaSymbol{→}\AgdaSpace{}%
\AgdaBound{n}\AgdaSpace{}%
\AgdaOperator{\AgdaDatatype{memberOf}}\AgdaSpace{}%
\AgdaBound{xs}\AgdaSpace{}%
\AgdaSymbol{→}\AgdaSpace{}%
\AgdaBound{n}\AgdaSpace{}%
\AgdaOperator{\AgdaDatatype{>}}\AgdaSpace{}%
\AgdaInductiveConstructor{zero}\AgdaSymbol{)}\<%
\\
\\[\AgdaEmptyExtraSkip]%
\>[0]\AgdaFunction{allp{-}sound}\AgdaSpace{}%
\AgdaBound{ap}\AgdaSpace{}%
\AgdaInductiveConstructor{mem{-}h}\AgdaSpace{}%
\AgdaKeyword{with}\AgdaSpace{}%
\AgdaBound{ap}\<%
\\
\>[0]\AgdaSymbol{...}\AgdaSpace{}%
\AgdaSymbol{|}\AgdaSpace{}%
\AgdaInductiveConstructor{allp{-}t}\AgdaSpace{}%
\AgdaBound{x>0}\AgdaSpace{}%
\AgdaBound{ap{-}xs}\AgdaSpace{}%
\AgdaSymbol{=}\AgdaSpace{}%
\AgdaBound{x>0}\<%
\\
\>[0]\AgdaFunction{allp{-}sound}\AgdaSpace{}%
\AgdaBound{ap}\AgdaSpace{}%
\AgdaSymbol{(}\AgdaInductiveConstructor{mem{-}t}\AgdaSpace{}%
\AgdaBound{mem}\AgdaSymbol{)}\AgdaSpace{}%
\AgdaKeyword{with}\AgdaSpace{}%
\AgdaBound{ap}\<%
\\
\>[0]\AgdaSymbol{...}\AgdaSpace{}%
\AgdaSymbol{|}\AgdaSpace{}%
\AgdaInductiveConstructor{allp{-}t}\AgdaSpace{}%
\AgdaBound{x>0}\AgdaSpace{}%
\AgdaBound{ap{-}xs}\AgdaSpace{}%
\AgdaSymbol{=}\AgdaSpace{}%
\AgdaFunction{allp{-}sound}\AgdaSpace{}%
\AgdaSymbol{(}\AgdaField{Thunk.force}\AgdaSpace{}%
\AgdaBound{ap{-}xs}\AgdaSymbol{)}\AgdaSpace{}%
\AgdaBound{mem}\<%
\end{code}
\end{center}

Note that the hypothesis that \textit{allPos} holds is required for all sizes \textit{i}. \LC{Using the universally quantified \agda{Size} we obtain a predicate \agda{Colist Nat $\infty$ $\rightarrow$ Set} which does not depend on \agda{Size} anymore}. The proof is by induction the rules of \textit{memberOf}. \LC{Agda knows that the colist cannot be empty, and indeed there are no absurd cases.}

\begin{center}
\begin{code}%
\>[0]\AgdaFunction{allp{-}compl}\AgdaSpace{}%
\AgdaSymbol{:}\AgdaSpace{}%
\AgdaSymbol{\{}\AgdaBound{xs}\AgdaSpace{}%
\AgdaSymbol{:}\AgdaSpace{}%
\AgdaDatatype{Colist}\AgdaSpace{}%
\AgdaDatatype{Nat}%
\>[106I]\AgdaPostulate{∞}\AgdaSymbol{\}}\AgdaSpace{}%
\AgdaSymbol{→}\AgdaSpace{}%
\AgdaSymbol{(∀}\AgdaSpace{}%
\AgdaSymbol{\{}\AgdaBound{n}\AgdaSymbol{\}}\AgdaSpace{}%
\AgdaSymbol{→}\AgdaSpace{}%
\AgdaBound{n}\AgdaSpace{}%
\AgdaOperator{\AgdaDatatype{memberOf}}\AgdaSpace{}%
\AgdaBound{xs}\AgdaSpace{}%
\AgdaSymbol{→}\AgdaSpace{}%
\AgdaBound{n}\AgdaSpace{}%
\AgdaOperator{\AgdaDatatype{>}}\AgdaSpace{}%
\AgdaNumber{0}\AgdaSymbol{)}\AgdaSpace{}%
\AgdaSymbol{→}\<%
\\
\>[.]\<[106I]%
\>[30]\AgdaSymbol{(∀}\AgdaSpace{}%
\AgdaSymbol{\{}\AgdaBound{i}\AgdaSymbol{\}}\AgdaSpace{}%
\AgdaSymbol{→}\AgdaSpace{}%
\AgdaDatatype{allPos}\AgdaSpace{}%
\AgdaBound{xs}\AgdaSpace{}%
\AgdaBound{i}\AgdaSymbol{)}\<%
\\
\>[0]\<%
\\
\>[0]\AgdaFunction{allp{-}compl}\AgdaSpace{}%
\AgdaSymbol{\{}\AgdaInductiveConstructor{[]}\AgdaSymbol{\}}\AgdaSpace{}%
\AgdaBound{f}\AgdaSpace{}%
\AgdaSymbol{=}\AgdaSpace{}%
\AgdaInductiveConstructor{allp{-}Λ}\<%
\\
\>[0]\AgdaFunction{allp{-}compl}\AgdaSpace{}%
\AgdaSymbol{\{}\AgdaBound{x}\AgdaSpace{}%
\AgdaOperator{\AgdaInductiveConstructor{∷}}\AgdaSpace{}%
\AgdaBound{xs}\AgdaSymbol{\}}\AgdaSpace{}%
\AgdaBound{f}\AgdaSpace{}%
\AgdaSymbol{=}\AgdaSpace{}%
\AgdaInductiveConstructor{allp{-}t}%
\>[134I]\AgdaSymbol{(}\AgdaBound{f}\AgdaSpace{}%
\AgdaInductiveConstructor{mem{-}h}\AgdaSymbol{)}\<%
\\
\>[.]\<[134I]%
\>[31]\AgdaSymbol{(λ}\AgdaSpace{}%
\AgdaKeyword{where}\AgdaSpace{}%
\AgdaSymbol{.}\AgdaField{force}\AgdaSpace{}%
\AgdaSymbol{→}\AgdaSpace{}%
\AgdaFunction{allp{-}compl}\AgdaSpace{}%
\AgdaSymbol{(λ}\AgdaSpace{}%
\AgdaBound{m}\AgdaSpace{}%
\AgdaSymbol{→}\AgdaSpace{}%
\AgdaBound{f}\AgdaSpace{}%
\AgdaSymbol{(}\AgdaInductiveConstructor{mem{-}t}\AgdaSpace{}%
\AgdaBound{m}\AgdaSymbol{)))}\<%
\end{code}
\end{center}

\EZComm{non capisco}\LC{The proof that \textit{allPos} holds for $xs$ that is obtained by recursion must be boxed inside a \agda{Thunk} due to the required argument type of the constructor \refToRule{allp-t} as discussed before. 
We adopted the syntax \agda{$\lambda$ where .force $\rightarrow$ \_} for defining a new \agda{Thunk} record, which is also used in the libraries.}

The representation through a coinductive record is as follows.

\begin{center}
\begin{code}%
\>[0]\AgdaKeyword{record}\AgdaSpace{}%
\AgdaRecord{allPos}\AgdaSpace{}%
\AgdaSymbol{(}\AgdaBound{xs}\AgdaSpace{}%
\AgdaSymbol{:}\AgdaSpace{}%
\AgdaRecord{MyColist}\AgdaSpace{}%
\AgdaDatatype{Nat}\AgdaSymbol{)}\AgdaSpace{}%
\AgdaSymbol{:}\AgdaSpace{}%
\AgdaPrimitiveType{Set}\AgdaSpace{}%
\AgdaKeyword{where}\<%
\\
\>[0][@{}l@{\AgdaIndent{0}}]%
\>[2]\AgdaKeyword{constructor}\AgdaSpace{}%
\AgdaOperator{\AgdaCoinductiveConstructor{AllP\AgdaUnderscore{}}}\<%
\\
\>[2]\AgdaKeyword{coinductive}\<%
\\
\>[2]\AgdaKeyword{field}\<%
\\
\>[2][@{}l@{\AgdaIndent{0}}]%
\>[4]\AgdaField{list}\AgdaSpace{}%
\AgdaSymbol{:}\AgdaSpace{}%
\AgdaSymbol{(}\AgdaField{MyColist.list}\AgdaSpace{}%
\AgdaBound{xs}\AgdaSpace{}%
\AgdaOperator{\AgdaDatatype{≡}}\AgdaSpace{}%
\AgdaInductiveConstructor{nothing}\AgdaSymbol{)}\AgdaSpace{}%
\AgdaOperator{\AgdaDatatype{∨}}\<%
\\
\>[4][@{}l@{\AgdaIndent{0}}]%
\>[6]\AgdaSymbol{(}\AgdaDatatype{∃}\AgdaSpace{}%
\AgdaSymbol{(}\AgdaDatatype{Nat}\AgdaSpace{}%
\AgdaOperator{\AgdaDatatype{×}}\AgdaSpace{}%
\AgdaRecord{MyColist}\AgdaSpace{}%
\AgdaDatatype{Nat}\AgdaSymbol{)}\AgdaSpace{}%
\AgdaSymbol{(λ}\AgdaSpace{}%
\AgdaBound{c}\AgdaSpace{}%
\AgdaSymbol{→}\AgdaSpace{}%
\AgdaSymbol{((}\AgdaField{MyColist.list}\AgdaSpace{}%
\AgdaBound{xs}\AgdaSymbol{)}\AgdaSpace{}%
\AgdaOperator{\AgdaDatatype{≡}}\AgdaSpace{}%
\AgdaSymbol{(}\AgdaInductiveConstructor{just}\AgdaSpace{}%
\AgdaBound{c}\AgdaSymbol{))}\AgdaSpace{}%
\AgdaOperator{\AgdaDatatype{∧}}\<%
\\
\>[6]\AgdaSymbol{((}\AgdaFunction{∧{-}left}\AgdaSpace{}%
\AgdaBound{c}\AgdaSpace{}%
\AgdaOperator{\AgdaDatatype{>}}\AgdaSpace{}%
\AgdaInductiveConstructor{zero}\AgdaSymbol{)}\AgdaSpace{}%
\AgdaOperator{\AgdaDatatype{∧}}\AgdaSpace{}%
\AgdaSymbol{(}\AgdaRecord{allPos}\AgdaSpace{}%
\AgdaSymbol{(}\AgdaFunction{∧{-}right}\AgdaSpace{}%
\AgdaBound{c}\AgdaSymbol{)))))}\<%
\end{code}

\end{center}

\LC{
where \agda{$\land$} is the conjunction type. Since \agda{$\land$} and \agda{$\times$} can be implemented in the same way, we wrote \agda{$\times$} as an alias for the conjunction type. They share the projections \agda{$\land$-left} and \agda{$\land$-right} (see \refToSection{agda-modules} for more details).} 
The unique field represents the two cases of an empty or non-empty list through the data type $\lor$ whose constructors are the injections \agda{in-l} and \agda{in-r} (more details in \refToSection{agda-modules}). In the second case, \EZComm{non capisco}\LC{ the proof that the list is not empty is needed to say that the head is positive and the predicate holds for the tail. In details, if the list is not empty it is observed as a couple \textit{c} boxed inside \agda{Maybe} because we refer to colists implemented as records. Then the two projections of the product type allow to extract the information to state that the head is greater than zero and that \textit{allPos} holds on the tail.}

\begin{center}
\begin{code}
\>[0]\AgdaFunction{allp{-}sound}\AgdaSpace{}%
\AgdaSymbol{:}\AgdaSpace{}%
\AgdaSymbol{\{}\AgdaBound{xs}\AgdaSpace{}%
\AgdaSymbol{:}\AgdaSpace{}%
\AgdaRecord{MyColist}\AgdaSpace{}%
\AgdaDatatype{Nat}\AgdaSymbol{\}}\AgdaSpace{}%
\AgdaSymbol{→}\AgdaSpace{}%
\AgdaRecord{allPos}\AgdaSpace{}%
\AgdaBound{xs}\AgdaSpace{}%
\AgdaSymbol{→}\AgdaSpace{}%
\AgdaSymbol{(\{}\AgdaBound{n}\AgdaSpace{}%
\AgdaSymbol{:}\AgdaSpace{}%
\AgdaDatatype{Nat}\AgdaSymbol{\}}\AgdaSpace{}%
\AgdaSymbol{→}\AgdaSpace{}%
\AgdaBound{n}\AgdaSpace{}%
\AgdaOperator{\AgdaDatatype{memberOf}}\AgdaSpace{}%
\AgdaBound{xs}\AgdaSpace{}%
\AgdaSymbol{→}\AgdaSpace{}%
\AgdaBound{n}\AgdaSpace{}%
\AgdaOperator{\AgdaDatatype{>}}\AgdaSpace{}%
\AgdaInductiveConstructor{zero}\AgdaSymbol{)}\<%
\\
\\[\AgdaEmptyExtraSkip]%
\>[0]\AgdaFunction{allp{-}sound}\AgdaSpace{}%
\AgdaSymbol{\{}\AgdaBound{xs}\AgdaSymbol{\}}\AgdaSpace{}%
\AgdaBound{ap}\AgdaSpace{}%
\AgdaSymbol{(}\AgdaInductiveConstructor{mem{-}h}\AgdaSpace{}%
\AgdaBound{eq}\AgdaSymbol{)}\AgdaSpace{}%
\AgdaKeyword{with}\AgdaSpace{}%
\AgdaField{allPos.list}\AgdaSpace{}%
\AgdaBound{ap}\<%
\\
\>[0]\AgdaSymbol{...}\AgdaSpace{}%
\AgdaSymbol{|}\AgdaSpace{}%
\AgdaInductiveConstructor{inl}\AgdaSpace{}%
\AgdaBound{p}\AgdaSpace{}%
\AgdaSymbol{=}\AgdaSpace{}%
\AgdaFunction{⊥{-}elim}\AgdaSpace{}%
\AgdaSymbol{(}\AgdaFunction{mycolist{-}abs}\AgdaSpace{}%
\AgdaSymbol{\{}\AgdaArgument{l}\AgdaSpace{}%
\AgdaSymbol{=}\AgdaSpace{}%
\AgdaBound{xs}\AgdaSymbol{\}}\AgdaSpace{}%
\AgdaBound{p}\AgdaSpace{}%
\AgdaBound{eq}\AgdaSymbol{)}\<%
\\
\>[0]\AgdaSymbol{...}%
\>[168I]\AgdaSymbol{|}\AgdaSpace{}%
\AgdaInductiveConstructor{inr}\AgdaSpace{}%
\AgdaOperator{\AgdaInductiveConstructor{<}}\AgdaSpace{}%
\AgdaOperator{\AgdaInductiveConstructor{⟨}}\AgdaSpace{}%
\AgdaBound{x}\AgdaSpace{}%
\AgdaOperator{\AgdaInductiveConstructor{,}}\AgdaSpace{}%
\AgdaSymbol{\AgdaUnderscore{}}\AgdaSpace{}%
\AgdaOperator{\AgdaInductiveConstructor{⟩}}\AgdaSpace{}%
\AgdaOperator{\AgdaInductiveConstructor{,}}\AgdaSpace{}%
\AgdaOperator{\AgdaInductiveConstructor{⟨}}\AgdaSpace{}%
\AgdaBound{eq₁}\AgdaSpace{}%
\AgdaOperator{\AgdaInductiveConstructor{,}}\AgdaSpace{}%
\AgdaOperator{\AgdaInductiveConstructor{⟨}}\AgdaSpace{}%
\AgdaBound{x{-}pos}\AgdaSpace{}%
\AgdaOperator{\AgdaInductiveConstructor{,}}\AgdaSpace{}%
\AgdaSymbol{\AgdaUnderscore{}}\AgdaSpace{}%
\AgdaOperator{\AgdaInductiveConstructor{⟩}}\AgdaSpace{}%
\AgdaOperator{\AgdaInductiveConstructor{⟩}}\AgdaSpace{}%
\AgdaOperator{\AgdaInductiveConstructor{>}}\AgdaSpace{}%
\AgdaSymbol{=}\AgdaSpace{}%
\AgdaFunction{subst}\<%
\\
\>[.]\<[168I]%
\>[4]\AgdaSymbol{(λ}\AgdaSpace{}%
\AgdaBound{n}\AgdaSpace{}%
\AgdaSymbol{→}\AgdaSpace{}%
\AgdaBound{n}\AgdaSpace{}%
\AgdaOperator{\AgdaDatatype{>}}\AgdaSpace{}%
\AgdaInductiveConstructor{zero}\AgdaSymbol{)}\<%
\\
\>[4]\AgdaSymbol{(}\AgdaFunction{eq2sx}\AgdaSpace{}%
\AgdaSymbol{(}\AgdaFunction{just{-}elim}\AgdaSpace{}%
\AgdaSymbol{(}\AgdaFunction{trans}\AgdaSpace{}%
\AgdaSymbol{(}\AgdaFunction{sym}\AgdaSpace{}%
\AgdaBound{eq₁}\AgdaSymbol{)}\AgdaSpace{}%
\AgdaBound{eq}\AgdaSymbol{))}\AgdaSpace{}%
\AgdaSymbol{)}\<%
\\
\>[4]\AgdaBound{x{-}pos}\<%
\\
\\[\AgdaEmptyExtraSkip]%
\>[0]\AgdaFunction{allp{-}sound}\AgdaSpace{}%
\AgdaSymbol{\{}\AgdaBound{xs}\AgdaSymbol{\}}\AgdaSpace{}%
\AgdaBound{ap}\AgdaSpace{}%
\AgdaSymbol{(}\AgdaInductiveConstructor{mem{-}t}\AgdaSpace{}%
\AgdaBound{eq}\AgdaSpace{}%
\AgdaBound{mem}\AgdaSymbol{)}\AgdaSpace{}%
\AgdaKeyword{with}\AgdaSpace{}%
\AgdaField{allPos.list}\AgdaSpace{}%
\AgdaBound{ap}\<%
\\
\>[0]\AgdaSymbol{...}\AgdaSpace{}%
\AgdaSymbol{|}\AgdaSpace{}%
\AgdaInductiveConstructor{inl}\AgdaSpace{}%
\AgdaBound{p}\AgdaSpace{}%
\AgdaSymbol{=}\AgdaSpace{}%
\AgdaFunction{⊥{-}elim}\AgdaSpace{}%
\AgdaSymbol{(}\AgdaFunction{mycolist{-}abs}\AgdaSpace{}%
\AgdaSymbol{\{}\AgdaArgument{l}\AgdaSpace{}%
\AgdaSymbol{=}\AgdaSpace{}%
\AgdaBound{xs}\AgdaSymbol{\}}\AgdaSpace{}%
\AgdaBound{p}\AgdaSpace{}%
\AgdaBound{eq}\AgdaSymbol{)}\<%
\\
\>[0]\AgdaSymbol{...}%
\>[219I]\AgdaSymbol{|}\AgdaSpace{}%
\AgdaInductiveConstructor{inr}\AgdaSpace{}%
\AgdaOperator{\AgdaInductiveConstructor{<}}\AgdaSpace{}%
\AgdaOperator{\AgdaInductiveConstructor{⟨}}\AgdaSpace{}%
\AgdaSymbol{\AgdaUnderscore{}}\AgdaSpace{}%
\AgdaOperator{\AgdaInductiveConstructor{,}}\AgdaSpace{}%
\AgdaBound{ys}\AgdaSpace{}%
\AgdaOperator{\AgdaInductiveConstructor{⟩}}\AgdaSpace{}%
\AgdaOperator{\AgdaInductiveConstructor{,}}\AgdaSpace{}%
\AgdaOperator{\AgdaInductiveConstructor{⟨}}\AgdaSpace{}%
\AgdaBound{eq₁}\AgdaSpace{}%
\AgdaOperator{\AgdaInductiveConstructor{,}}\AgdaSpace{}%
\AgdaOperator{\AgdaInductiveConstructor{⟨}}\AgdaSpace{}%
\AgdaSymbol{\AgdaUnderscore{}}\AgdaSpace{}%
\AgdaOperator{\AgdaInductiveConstructor{,}}\AgdaSpace{}%
\AgdaBound{ap{-}ys}\AgdaSpace{}%
\AgdaOperator{\AgdaInductiveConstructor{⟩}}\AgdaSpace{}%
\AgdaOperator{\AgdaInductiveConstructor{⟩}}\AgdaSpace{}%
\AgdaOperator{\AgdaInductiveConstructor{>}}\AgdaSpace{}%
\AgdaSymbol{=}\AgdaSpace{}%
\AgdaFunction{allp{-}sound}\<%
\\
\>[.]\<[219I]%
\>[4]\AgdaSymbol{(}\AgdaFunction{subst}\<%
\\
\>[4][@{}l@{\AgdaIndent{0}}]%
\>[6]\AgdaSymbol{(λ}\AgdaSpace{}%
\AgdaBound{v}\AgdaSpace{}%
\AgdaSymbol{→}\AgdaSpace{}%
\AgdaRecord{allPos}\AgdaSpace{}%
\AgdaBound{v}\AgdaSymbol{)}\<%
\\
\>[6]\AgdaSymbol{(}\AgdaFunction{eq2dx}\AgdaSpace{}%
\AgdaSymbol{(}\AgdaFunction{just{-}elim}\AgdaSpace{}%
\AgdaSymbol{(}\AgdaFunction{trans}\AgdaSpace{}%
\AgdaSymbol{(}\AgdaFunction{sym}\AgdaSpace{}%
\AgdaBound{eq₁}\AgdaSymbol{)}\AgdaSpace{}%
\AgdaBound{eq}\AgdaSymbol{)))}\<%
\\
\>[6]\AgdaBound{ap{-}ys}\AgdaSymbol{)}\AgdaSpace{}%
\AgdaBound{mem}\<%
\end{code}
\end{center}

The soundness proof contains two absurd cases and they are solved using the $\bot$-elimination rule\EZComm{riferimento?}. Notice that in this case they are not automatically recognized by Agda, since we are reasoning by observations, thus \agda{()} cannot be used even interactively. Then the proof is by induction on the rules of \textit{memberOf}.

\begin{itemize}
\item \agda{mem-h}: \textit{x-pos} is the proof that $\x$ is positive \EZComm{non capisco} but the condition on the head referenced by \textit{eq} is required. \LC{Notice that there are two equality proofs (\textit{eq} and \textit{eq$_{1}$}) that tell how the list is built in terms of couples, but the one we have to refer to is \textit{eq} because the second one is generated by the usage of \agda{with}. An intermediate result shows that the involved pairs are equal and thus the the first elements are equal as well. By substitution the proof that the head is positive is obtained. Roughly speaking we start from the fact that a list is observed in two ways and we prove that they are equivalent; so if a predicate holds on one head, it also holds on the other. }
\item \agda{mem-t}: \textit{ap-ys} is the proof that \textit{allPos} holds for $\ys$, but, as above, the proof \textit{eq} that it holds for the tail  is required. Again the crucial point is that the pairs are equal, hence the second elements are equal as well. The solution is again obtained by substitution.
\end{itemize}
It should be clear that the main challenge in writing the code above is to relate the variables in the definition of the predicate with those available in the proofs. 

\begin{center}
\begin{code}%
\>[0]\AgdaFunction{allp{-}compl}\AgdaSpace{}%
\AgdaSymbol{:}%
\>[14]\AgdaSymbol{(}\AgdaBound{xs}\AgdaSpace{}%
\AgdaSymbol{:}\AgdaSpace{}%
\AgdaRecord{MyColist}\AgdaSpace{}%
\AgdaDatatype{Nat}\AgdaSymbol{)}\AgdaSpace{}%
\AgdaSymbol{→}\AgdaSpace{}%
\AgdaSymbol{(\{}\AgdaBound{n}\AgdaSpace{}%
\AgdaSymbol{:}\AgdaSpace{}%
\AgdaDatatype{Nat}\AgdaSymbol{\}}\AgdaSpace{}%
\AgdaSymbol{→}\AgdaSpace{}%
\AgdaBound{n}\AgdaSpace{}%
\AgdaOperator{\AgdaDatatype{memberOf}}\AgdaSpace{}%
\AgdaBound{xs}\AgdaSpace{}%
\AgdaSymbol{→}\AgdaSpace{}%
\AgdaBound{n}\AgdaSpace{}%
\AgdaOperator{\AgdaDatatype{>}}\AgdaSpace{}%
\AgdaNumber{0}\AgdaSymbol{)}\AgdaSpace{}%
\AgdaSymbol{→}\AgdaSpace{}%
\AgdaRecord{allPos}\AgdaSpace{}%
\AgdaBound{xs}\<%
\\
\\[\AgdaEmptyExtraSkip]%
\>[0]\AgdaField{allPos.list}\AgdaSpace{}%
\AgdaSymbol{(}\AgdaFunction{allp{-}compl}\AgdaSpace{}%
\AgdaBound{xs}\AgdaSpace{}%
\AgdaBound{f}\AgdaSymbol{)}\AgdaSpace{}%
\AgdaKeyword{with}\AgdaSpace{}%
\AgdaFunction{inspect}\AgdaSpace{}%
\AgdaSymbol{(}\AgdaField{MyColist.list}\AgdaSpace{}%
\AgdaBound{xs}\AgdaSymbol{)}\<%
\\
\>[0]\AgdaSymbol{...}\AgdaSpace{}%
\AgdaSymbol{|}\AgdaSpace{}%
\AgdaInductiveConstructor{nothing}\AgdaSpace{}%
\AgdaOperator{\AgdaInductiveConstructor{with≡}}\AgdaSpace{}%
\AgdaBound{eq}\AgdaSpace{}%
\AgdaSymbol{=}\AgdaSpace{}%
\AgdaInductiveConstructor{inl}\AgdaSpace{}%
\AgdaBound{eq}\<%
\\
\>[0]\AgdaSymbol{...}%
\>[283I]\AgdaSymbol{|}\AgdaSpace{}%
\AgdaInductiveConstructor{just}\AgdaSpace{}%
\AgdaOperator{\AgdaInductiveConstructor{⟨}}\AgdaSpace{}%
\AgdaBound{y}\AgdaSpace{}%
\AgdaOperator{\AgdaInductiveConstructor{,}}\AgdaSpace{}%
\AgdaBound{ys}\AgdaSpace{}%
\AgdaOperator{\AgdaInductiveConstructor{⟩}}\AgdaSpace{}%
\AgdaOperator{\AgdaInductiveConstructor{with≡}}\AgdaSpace{}%
\AgdaBound{eq}\AgdaSpace{}%
\AgdaSymbol{=}\AgdaSpace{}%
\AgdaInductiveConstructor{inr}\AgdaSpace{}%
\AgdaOperator{\AgdaInductiveConstructor{<}}\<%
\\
\>[.]\<[283I]%
\>[4]\AgdaOperator{\AgdaInductiveConstructor{⟨}}\AgdaSpace{}%
\AgdaBound{y}\AgdaSpace{}%
\AgdaOperator{\AgdaInductiveConstructor{,}}\AgdaSpace{}%
\AgdaBound{ys}\AgdaSpace{}%
\AgdaOperator{\AgdaInductiveConstructor{⟩}}\AgdaSpace{}%
\AgdaOperator{\AgdaInductiveConstructor{,}}\<%
\\
\>[4]\AgdaOperator{\AgdaInductiveConstructor{⟨}}\AgdaSpace{}%
\AgdaBound{eq}\AgdaSpace{}%
\AgdaOperator{\AgdaInductiveConstructor{,}}\AgdaSpace{}%
\AgdaOperator{\AgdaInductiveConstructor{⟨}}\AgdaSpace{}%
\AgdaBound{f}\AgdaSpace{}%
\AgdaSymbol{(}\AgdaInductiveConstructor{mem{-}h}\AgdaSpace{}%
\AgdaBound{eq}\AgdaSymbol{)}\AgdaSpace{}%
\AgdaOperator{\AgdaInductiveConstructor{,}}\AgdaSpace{}%
\AgdaFunction{allp{-}compl}\AgdaSpace{}%
\AgdaBound{ys}\AgdaSpace{}%
\AgdaSymbol{(λ}\AgdaSpace{}%
\AgdaBound{m}\AgdaSpace{}%
\AgdaSymbol{→}\AgdaSpace{}%
\AgdaBound{f}\AgdaSpace{}%
\AgdaSymbol{(}\AgdaInductiveConstructor{mem{-}t}\AgdaSpace{}%
\AgdaBound{eq}\AgdaSpace{}%
\AgdaBound{m}\AgdaSymbol{))}\AgdaSpace{}%
\AgdaOperator{\AgdaInductiveConstructor{⟩}}\AgdaSpace{}%
\AgdaOperator{\AgdaInductiveConstructor{⟩}}\<%
\\
\>[4]\AgdaOperator{\AgdaInductiveConstructor{>}}\<%
\end{code}
\end{center}

The completeness proof must be written using copattern matching, since it returns a coinductive record. It is simpler than the previous proofs since there is no need of additional equalites.
Indeed, in this case the \agda{inspect} function is sufficient to obtain the correct ingredients to define a record of type \textit{allPos}. The empty case is easy: \textit{eq} is the proof that the {field} is \agda{nothing}.
In the second case, \textit{eq} is used in different positions to show that it is the only way in which the list is built. \EZComm{non capisco}Thanks to this fact no absurd cases are present.  

To summarize, to use the approach based on coinductive records some challenges have to be faced, \LC{analogous to those described in \refToChapter{inductive}\EZComm{\`e giusto?} when we wrote the proofs of \textit{memberOf} for the same type of colists, again since we reason by observations. 
In this case, things are even harder; we also introduced variables that are existentially quantified in the field of the coinductive record. \EZComm{frase molto oscura} They are inside the pair \textit{c} that identifies how the colist is built such that the proofs that the head is positive and \textit{allPos} holds on the tail refer to them. The \agda{inspect} construct has an important role in the proofs since it allows to bring the information on the structure of the reference colist. On the other hand the usage of the \textit{Singleton} module leads to situations in which a list is observed in different ways. So the main difficulty is to relate the variables provided by \agda{inspect} with those inside the definition of the coinductive record type.}
This fact becomes more evident in predicates on more than one colist. 
{To illustrate this, we provide an implementation of binary relations on} colists using the two approaches.

First the implementation using library colists:

\begin{center}
\begin{code}%
\>[0]\AgdaKeyword{data}\AgdaSpace{}%
\AgdaDatatype{cBinRel}\AgdaSpace{}%
\AgdaSymbol{\{}\AgdaBound{A}\AgdaSpace{}%
\AgdaBound{B}\AgdaSpace{}%
\AgdaSymbol{:}\AgdaSpace{}%
\AgdaPrimitiveType{Set}\AgdaSymbol{\}}\AgdaSpace{}%
\AgdaSymbol{:}\AgdaSpace{}%
\AgdaSymbol{(}\AgdaBound{A}\AgdaSpace{}%
\AgdaSymbol{→}\AgdaSpace{}%
\AgdaBound{B}\AgdaSpace{}%
\AgdaSymbol{→}\AgdaSpace{}%
\AgdaPrimitiveType{Set}\AgdaSymbol{)}\AgdaSpace{}%
\AgdaSymbol{→}\<%
\\
\>[0][@{}l@{\AgdaIndent{0}}]%
\>[2]\AgdaDatatype{Colist}\AgdaSpace{}%
\AgdaBound{A}\AgdaSpace{}%
\AgdaPostulate{∞}\AgdaSpace{}%
\AgdaSymbol{→}\AgdaSpace{}%
\AgdaDatatype{Colist}\AgdaSpace{}%
\AgdaBound{B}\AgdaSpace{}%
\AgdaPostulate{∞}\AgdaSpace{}%
\AgdaSymbol{→}\AgdaSpace{}%
\AgdaPostulate{Size}\AgdaSpace{}%
\AgdaSymbol{→}\AgdaSpace{}%
\AgdaPrimitiveType{Set}\AgdaSpace{}%
\AgdaKeyword{where}\<%
\\
\>[2][@{}l@{\AgdaIndent{0}}]%
\>[4]\AgdaInductiveConstructor{bin{-}null}\AgdaSpace{}%
\AgdaSymbol{:}\AgdaSpace{}%
\AgdaSymbol{∀}\AgdaSpace{}%
\AgdaSymbol{\{}\AgdaBound{R}\AgdaSpace{}%
\AgdaSymbol{:}\AgdaSpace{}%
\AgdaBound{A}\AgdaSpace{}%
\AgdaSymbol{→}\AgdaSpace{}%
\AgdaBound{B}\AgdaSpace{}%
\AgdaSymbol{→}\AgdaSpace{}%
\AgdaPrimitiveType{Set}\AgdaSymbol{\}\{}\AgdaBound{i}\AgdaSymbol{\}}\AgdaSpace{}%
\AgdaSymbol{→}\AgdaSpace{}%
\AgdaDatatype{cBinRel}\AgdaSpace{}%
\AgdaBound{R}\AgdaSpace{}%
\AgdaInductiveConstructor{[]}\AgdaSpace{}%
\AgdaInductiveConstructor{[]}\AgdaSpace{}%
\AgdaBound{i}\<%
\\
\>[4]\AgdaInductiveConstructor{bin{-}i}\AgdaSpace{}%
\AgdaSymbol{:}%
\>[49I]\AgdaSymbol{∀}\AgdaSpace{}%
\AgdaSymbol{\{}\AgdaBound{R}\AgdaSpace{}%
\AgdaSymbol{:}\AgdaSpace{}%
\AgdaBound{A}\AgdaSpace{}%
\AgdaSymbol{→}\AgdaSpace{}%
\AgdaBound{B}\AgdaSpace{}%
\AgdaSymbol{→}\AgdaSpace{}%
\AgdaPrimitiveType{Set}\AgdaSymbol{\}\{}\AgdaBound{x}\AgdaSpace{}%
\AgdaBound{y}\AgdaSpace{}%
\AgdaBound{xs}\AgdaSpace{}%
\AgdaBound{ys}\AgdaSpace{}%
\AgdaBound{i}\AgdaSymbol{\}}\AgdaSpace{}%
\AgdaSymbol{→}\<%
\\
\>[.]\<[49I]%
\>[12]\AgdaBound{R}\AgdaSpace{}%
\AgdaBound{x}\AgdaSpace{}%
\AgdaBound{y}\AgdaSpace{}%
\AgdaSymbol{→}\<%
\\
\>[12]\AgdaRecord{Thunk}\AgdaSpace{}%
\AgdaSymbol{(}\AgdaDatatype{cBinRel}\AgdaSpace{}%
\AgdaBound{R}\AgdaSpace{}%
\AgdaSymbol{(}\AgdaField{Thunk.force}\AgdaSpace{}%
\AgdaBound{xs}\AgdaSymbol{)}\AgdaSpace{}%
\AgdaSymbol{(}\AgdaField{Thunk.force}\AgdaSpace{}%
\AgdaBound{ys}\AgdaSymbol{))}\AgdaSpace{}%
\AgdaBound{i}\AgdaSpace{}%
\AgdaSymbol{→}\<%
\\
\>[12]\AgdaDatatype{cBinRel}\AgdaSpace{}%
\AgdaBound{R}\AgdaSpace{}%
\AgdaSymbol{(}\AgdaBound{x}\AgdaSpace{}%
\AgdaOperator{\AgdaInductiveConstructor{∷}}\AgdaSpace{}%
\AgdaBound{xs}\AgdaSymbol{)}\AgdaSpace{}%
\AgdaSymbol{(}\AgdaBound{y}\AgdaSpace{}%
\AgdaOperator{\AgdaInductiveConstructor{∷}}\AgdaSpace{}%
\AgdaBound{ys}\AgdaSymbol{)}\AgdaSpace{}%
\AgdaBound{i}\<%
\end{code}

\end{center}

and then the implementation based on coinductive records:

\begin{center}
\begin{code}%
\>[0]\AgdaKeyword{record}\AgdaSpace{}%
\AgdaRecord{sBinRel}\AgdaSpace{}%
\AgdaSymbol{\{}\AgdaBound{A}\AgdaSpace{}%
\AgdaBound{B}\AgdaSpace{}%
\AgdaSymbol{:}\AgdaSpace{}%
\AgdaPrimitiveType{Set}\AgdaSymbol{\}}\AgdaSpace{}%
\AgdaSymbol{(}\AgdaBound{R}\AgdaSpace{}%
\AgdaSymbol{:}\AgdaSpace{}%
\AgdaBound{A}\AgdaSpace{}%
\AgdaSymbol{→}\AgdaSpace{}%
\AgdaBound{B}\AgdaSpace{}%
\AgdaSymbol{→}\AgdaSpace{}%
\AgdaPrimitiveType{Set}\AgdaSymbol{)}\<%
\\
\>[0][@{}l@{\AgdaIndent{0}}]%
\>[2]\AgdaSymbol{(}\AgdaBound{xs}\AgdaSpace{}%
\AgdaSymbol{:}\AgdaSpace{}%
\AgdaRecord{MyColist}\AgdaSpace{}%
\AgdaBound{A}\AgdaSymbol{)}\AgdaSpace{}%
\AgdaSymbol{(}\AgdaBound{ys}\AgdaSpace{}%
\AgdaSymbol{:}\AgdaSpace{}%
\AgdaRecord{MyColist}\AgdaSpace{}%
\AgdaBound{B}\AgdaSymbol{)}\AgdaSpace{}%
\AgdaSymbol{:}\AgdaSpace{}%
\AgdaPrimitiveType{Set}\AgdaSpace{}%
\AgdaKeyword{where}\<%
\\
\>[2]\AgdaKeyword{coinductive}\<%
\\
\>[2]\AgdaKeyword{field}\<%
\\
\>[2][@{}l@{\AgdaIndent{0}}]%
\>[4]\AgdaField{list}\AgdaSpace{}%
\AgdaSymbol{:}\AgdaSpace{}%
\AgdaSymbol{((}\AgdaField{MyColist.list}\AgdaSpace{}%
\AgdaBound{xs}\AgdaSpace{}%
\AgdaOperator{\AgdaDatatype{≡}}\AgdaSpace{}%
\AgdaInductiveConstructor{nothing}\AgdaSymbol{)}\AgdaSpace{}%
\AgdaOperator{\AgdaDatatype{∧}}\AgdaSpace{}%
\AgdaSymbol{(}\AgdaField{MyColist.list}\AgdaSpace{}%
\AgdaBound{ys}\AgdaSpace{}%
\AgdaOperator{\AgdaDatatype{≡}}\AgdaSpace{}%
\AgdaInductiveConstructor{nothing}\AgdaSymbol{))}\AgdaSpace{}%
\AgdaOperator{\AgdaDatatype{∨}}\<%
\\
\>[4][@{}l@{\AgdaIndent{0}}]%
\>[6]\AgdaDatatype{∃}\AgdaSpace{}%
\AgdaSymbol{((}\AgdaBound{A}\AgdaSpace{}%
\AgdaOperator{\AgdaDatatype{×}}\AgdaSpace{}%
\AgdaRecord{MyColist}\AgdaSpace{}%
\AgdaBound{A}\AgdaSymbol{)}\AgdaSpace{}%
\AgdaOperator{\AgdaDatatype{×}}\AgdaSpace{}%
\AgdaSymbol{(}\AgdaBound{B}\AgdaSpace{}%
\AgdaOperator{\AgdaDatatype{×}}\AgdaSpace{}%
\AgdaRecord{MyColist}\AgdaSpace{}%
\AgdaBound{B}\AgdaSymbol{))}\<%
\\
\>[6]\AgdaSymbol{(λ}\AgdaSpace{}%
\AgdaBound{c}\AgdaSpace{}%
\AgdaSymbol{→}\AgdaSpace{}%
\AgdaSymbol{((}\AgdaField{MyColist.list}\AgdaSpace{}%
\AgdaBound{xs}\AgdaSpace{}%
\AgdaOperator{\AgdaDatatype{≡}}\AgdaSpace{}%
\AgdaInductiveConstructor{just}\AgdaSpace{}%
\AgdaSymbol{(}\AgdaFunction{∧{-}left}\AgdaSpace{}%
\AgdaBound{c}\AgdaSymbol{))}\AgdaSpace{}%
\AgdaOperator{\AgdaDatatype{∧}}\<%
\\
\>[6]\AgdaSymbol{(}\AgdaField{MyColist.list}\AgdaSpace{}%
\AgdaBound{ys}\AgdaSpace{}%
\AgdaOperator{\AgdaDatatype{≡}}\AgdaSpace{}%
\AgdaInductiveConstructor{just}\AgdaSpace{}%
\AgdaSymbol{(}\AgdaFunction{∧{-}right}\AgdaSpace{}%
\AgdaBound{c}\AgdaSymbol{)))}\AgdaSpace{}%
\AgdaOperator{\AgdaDatatype{∧}}\<%
\\
\>[6]\AgdaSymbol{((}\AgdaBound{R}\AgdaSpace{}%
\AgdaSymbol{(}\AgdaFunction{∧{-}left}\AgdaSpace{}%
\AgdaSymbol{(}\AgdaFunction{∧{-}left}\AgdaSpace{}%
\AgdaBound{c}\AgdaSymbol{))}\AgdaSpace{}%
\AgdaSymbol{(}\AgdaFunction{∧{-}left}\AgdaSpace{}%
\AgdaSymbol{(}\AgdaFunction{∧{-}right}\AgdaSpace{}%
\AgdaBound{c}\AgdaSymbol{)))}\AgdaSpace{}%
\AgdaOperator{\AgdaDatatype{∧}}\<%
\\
\>[6]\AgdaSymbol{(}\AgdaRecord{sBinRel}\AgdaSpace{}%
\AgdaBound{R}\AgdaSpace{}%
\AgdaSymbol{(}\AgdaFunction{∧{-}right}\AgdaSpace{}%
\AgdaSymbol{(}\AgdaFunction{∧{-}left}\AgdaSpace{}%
\AgdaBound{c}\AgdaSymbol{))}\AgdaSpace{}%
\AgdaSymbol{(}\AgdaFunction{∧{-}right}\AgdaSpace{}%
\AgdaSymbol{(}\AgdaFunction{∧{-}right}\AgdaSpace{}%
\AgdaBound{c}\AgdaSymbol{)))))}\<%
\end{code}

\end{center}

The schema is similar to that for \textit{allPos}, except that here a generic binary predicate is considered.\LC{\EZComm{non capisco}The type is necessarily coinductive by definition. In fact it depends on a binary relation \textit{R} and it holds if \textit{R} holds element-wise. For example, $l_{1} \leq l_{2}$ if all their elements are in a $\leq$ relation; in this case, the two lists must have the same type. If we forget about the need of many projections that are simply required inside a big product type, the main points are the new existentially quantified variables that tell how both lists are built.
This allows us to conclude that the representation of colists through coinductive records cannot be easily scaled. }

\chapter{\EZ{Reasoning with corules}}
\label{chapter:flex}
So far we dealt with inductive and coinductive predicates. However, sometimes the intended meaning of a predicate cannot be expressed through these techniques. 

{To show the problem, let us consider the following inference system for the predicate \textit{maxElem}, where $\maxElem{\x}{\xs}$ holds if $\x$ is the maximal element of $\xs$.}

\begin{small}
\begin{quote}
$\NamedRule{max-h}{}{\maxElem {\x} {\x{:}\Lambda}}$ \Space
$\NamedRule{max-t}{\maxElem {\y} {\xs}} {\maxElem {z} {\x{:}\xs}}$\Space $z = \Max{\x}{\y}$
\end{quote}
\end{small}

{In order to compute this predicate on a list, all its elements have to be inspected. Thus, the inductive interpretation cannot work on infinite lists. 
On the other hand, the coinductive interpretation turns out to be unsatisfactory as well. To see this consider, for instance, the infinite list defined by the equation $\xs = 1:2:\xs$. It is easy to see that  the intuitively valid judgment 
$\maxElem{2}{\xs}$ can be derived, since it has an infinite proof tree, constructed by infinitely many instantiations of rule \refToRule{max-t}. However, the judgment $\maxElem {3} {\xs}$, wrong with respect to the intended meaning, can be derived analogously. Actually, the judgment $\maxElem {\x} {\xs}$  can be derived for each $\x$ which is greater or equal than the maximum.}

In summary, neither the inductive nor the coinductive interpretation works on infinite lists. \emph{Inference systems with corules}, or \emph{generalized inference systems} \cite{AnconaDZ17,Dagnino19}, have been proposed to deal with this kind of situations. We provide here a short introduction.

\section{Flexible coinduction} An \emph{inference system with corules}, or \emph{generalized inference system}, is a pair $\Pair{\is}{\cois}$ where $\is$ and $\cois$ are inference systems, whose elements are called \emph{rules} and \emph{corules}\footnote{{In the original formulation in \cite{AnconaDZ17,Dagnino19} only \emph{coaxioms} were considered.}}, respectively.

Analogously to rules, the meaning of corules is to derive a consequence from the premises. 
However, they can only be used in a special way, described in the following.

Given an inference system with corules $\Pair{\is}{\cois}$, $\is\cup\cois$ is the (standard) inference system where corules can be used as rules as well. 
Moreover, given a subset $S$ of the universe, $\Restricted{\is}{S}$ denotes the inference system obtained from $\is$ by keeping only rules with 
{consequence} in $S$.
Then, the interpretation $\Generated{\is}{\cois}$ of an inference system with corules $\Pair{\is}{\cois}$ is defined as follows.\footnote{\textit{Gen} stands for ``interpretation generated by the corules''. In \cite{AnconaDZ17,Dagnino19} it is shown that it corresponds to {taking} a fixed point which is, {in general}, neither the least, nor the greatest.} 
\begin{quote}
 $\Generated{\is}{\cois}=\CoInd{\Restricted{\is}{\Ind{\is\cup\cois}}}$
 \end{quote}
 
 That is, first we consider the inference system $\is\cup\cois$, and we take its inductive interpretation $\Ind{\is\cup\cois}$. Then, we take the coinductive interpretation of the inference system obtained from $\is$ by {keeping only rules with consequence in} $\Ind{\is\cup\cois}$.
 
In proof-theoretic terms, $\Generated{\is}{\cois}$ is the set of judgments which have an arbitrary (finite or infinite) proof tree in $\is$, whose nodes all have a finite proof tree in $\is\cup\cois$. 

Note that the inductive and coinductive interpretation of {$\is$} are special cases, obtained by taking as set of corules the empty set, and the set $\{\Rule{\emptyset}{\co}\mid\co\in\universe\}$, respectively

When a predicate is defined by an inference system of corules, we have a canonical technique to prove \emph{completeness}, that is, the \emph{bounded coinduction principle}, which is a generalization of the standard coinduction principle.

\paragraph*{Bounded coinduction principle} If a set $\Spec$ satisfies the following two conditions:
\begin{description}
\item[Boundedness] $\Spec \subseteq \Ind{\is\cup\cois}$
\item[Consistency] $\Spec$ is consistent with respect to $\is$
\end{description}
then $\Spec\subseteq\Generated{\is}{\cois}$. 

The proof can be found in \cite{AnconaDZ17,Dagnino19}, and is a direct consequence of the definition. Proving that $\Spec$ is consistent, as for the standard coinduction principle, amounts to show that, for each $\co\in\Spec$, there is a rule $\Rule {\prem} {\co}$ of the inference system such that $\prem\subseteq\Spec$. Proving that $\Spec$ is bounded means proving completeness of the inference system extended by corules, interpreted inductively, with respect to  $\Spec$. Hence, there is no canonical technique, and for each concrete case we must find an ``ad-hoc'' technique.

The standard coinduction principle can be obtained when $\cois=\{\Rule{\emptyset}{\co}\mid\co\in\universe\}$; for this particular case the first condition trivially holds.

\section{Example}Let us now consider again the \textit{maxElem} example. The following is an inference system with corules which correctly defines the predicate. Corules are written with a thicker line.

\begin{small}
\begin{quote}
$\NamedRule{max-h}{}{\maxElem {\x} {\x{:}\Lambda}}$ \Space
$\NamedRule{max-t}{\maxElem {\y} {\xs}} {\maxElem {\z} {\x{:}\xs}}$\Space $\z =\Max{\x}{\y}$\\[4ex]
$\NamedCoRule{co-max-h} {} {\maxElem {\x} {\x{:}\xs}}$
\end{quote}
\end{small}

In order to prove the correctness of such definition we show the specification $\Spec$ of the predicate.

\begin{center}
$\Spec = \{ \Pair{m}{\xs}\mid \memberOf {m} {\xs}\ \mbox{and, for all \x,}\ \memberOf {\x} {\xs}\ \mbox{implies}\ m=\Max{m}{\x}\}$
\end{center}


\begin{statement}
The definition of \textit{maxElem} by the inference system with corules is sound with respect to its specification.
\end{statement}
\begin{proof}
We have to prove that $\Generated{\is}{\cois}\subseteq\Spec$. That is, assuming that $\maxElem {m} {\xs}\in\Generated{\is}{\cois}$, we have to prove the following: 
\begin{enumerate}
\item $m$ is an element of $\xs$, that is $\memberOf {m} {\xs}$ holds.
\item for all $\x$ such that $\memberOf {\x} {\xs}$ holds, $m=\Max{m}{\x}$.
\end{enumerate}
We prove separately the two facts, since different techniques are needed.\\
To prove (1), since $\Generated{\is}{\cois}\subseteq\Ind{\is\cup\cois}$ holds by definition, we can reason by induction on the definition of $\Ind{\is\cup\cois}$.
\begin{description}
\item[\refToRule{max-h}-\refToRule{co-max-h}] In both cases we have $\maxElem{\x}{\x{:}\xs}$, hence the maximum $\x$ is the head of the list, that is, $\memberOf {\x} {\x{:}\xs}$ holds by rule \refToRule{mem-h}.
\item[\refToRule{max-t}] We have $\maxElem {\z} {\x{:}\xs}$, \maxElem {\y} {\xs}, and $z = \Max{\x}{\y}$. By inductive hypothesis we know that $\memberOf {\y} {\xs}$ holds, and we have to prove that $\memberOf {\z} {\x{:}\xs}$ holds as well. There are two cases:
\begin{itemize}
\item If $\z = \x$ then the maximum is again the head of the list, that is, $\memberOf {\x} {\x{:}\xs}$ holds by rule \refToRule{mem-h}.
\item If $\z = \y$ then the maximum is an element of the tail by inductive hypothesis, and thus it is an element of the list. That is, $\memberOf {\y} {\x{:}\xs}$ holds by rule \refToRule{mem-t}.
\end{itemize}
\end{description}

To prove (2), we can reason by induction over the definition of $\memberOf {\x} {\xs}$.
\begin{description}
\item[\refToRule {mem-h}] We have $\memberOf{\x}{\x{:}\xs}$, and $\maxElem {m} {\x{:}\xs}$, and we have to prove that $m=\Max{m}{\x}$. By cases on the rule used to derive $\maxElem {m} {\x{:}\xs}$.
\begin{description}
\item[\refToRule{max-h}] We have $\xs=\Lambda$, and $m=\x$, hence $m=\Max{m}{m}$ trivially holds.
\item[\refToRule{max-t}] We have  $\maxElem{n}{\xs}$, and $m = \Max {\x} {n}$ by the side condition, hence $m = \Max {m} {\x}$.\EZComm{non mi sembra servisse distinguere i due casi}
\end{description}

\item[\refToRule {mem-t}] We have $\memberOf{\x}{\y{:}\xs}$, $\memberOf{\x}{\xs}$ and $\maxElem {m} {\y{:}\xs}$, and we have to prove that $m=\Max{m}{\x}$. 
To derive $\maxElem {m} {\y{:}\xs}$, we cannot have applied \refToRule{max-h} since $\memberOf{\x}{\xs}$ implies that $\xs$ is not empty. 
Hence we have applied \refToRule{max-t}, and $\maxElem {n} {\xs}$ holds, with $m = \Max {\y} {n}$. From $\maxElem {n} {\xs}$ and $\memberOf{\x}{\xs}$, by inductive hypothesis we have that $n = \Max {n} {\x}$, hence, since $m = \Max {\y} {n}$, we get the thesis $m=\Max{m}{\x}$.\EZComm{idem}
\end{description}
\end{proof}

\begin{statement}
The definition of \textit{maxElem} by the inference system with corules is complete with respect to its specification.
\end{statement}

To prove the completeness statement above, we separately prove \textit{boundedness} and \textit{consistency} of $\Spec$.

\begin{statement}
\label{stmt:max-bound}
$\Spec$ is bounded with respect to the inference system with corules of \textit{maxElem}.
\end{statement}
\begin{proof}
We have to prove that $\Spec \subseteq \Ind{\is\cup\cois}$, that is, for each $\Pair{m}{\xs}\in\Spec$, the judgment $\maxElem{m}{\xs}$ can be derived in $\is\cup\cois$. 
Since $\Pair{m}{\xs}\in\Spec$ implies $\memberOf{m}{\xs}$,  the proof is by induction on the definition of $\memberOf{m}{\xs}$.
\begin{description}
\item[\refToRule{mem-h}] We have $\memberOf{m}{m{:}\xs}$, and the judgment $\maxElem{m}{m{:}\xs}$  can be derived by \refToRule{co-max-h}.
\item[\refToRule{mem-t}] We have $\memberOf{m}{\y{:}\xs}$, and $\memberOf{m}{\xs}$. By inductive hypothesis the judgment $\maxElem {m} {\xs}$ can be derived in $\is\cup\cois$.
Moreover, since $\Pair{m}{\y{:}\xs}\in\Spec$, we know that $m=\Max{m}{\y}$. Hence we can apply rule \refToRule{max-t} and get that the judgment $\maxElem{m}{\y{:}\xs}$ can be derived as well.
\end{description}
\end{proof}

\begin{statement}
\label{stmt:max-cons}
$\Spec$ is consistent with respect to $\is$.
\end{statement}
\begin{proof}
We have to prove that, for each $\Pair{m}{\xs}\in\Spec$, $\maxElem{m}{\xs}$ is the consequence of a rule with premises which are in $\Spec$ as well.
Since $\Pair{m}{\xs}\in\Spec$, $\xs$ cannot be empty, hence it is of shape $\x{:}\ys$. We consider two cases.\EZComm{non mi sembra servisse distinguere ulteriormente il secondo caso}
\begin{itemize}
\item If $\ys=\Lambda$, hence $m=\x$, then $\maxElem {m} {m{:}\Lambda}$ is the consequence of \refToRule{max-h}.
\item Otherwise, since $\xs$ has a maximum element $m$, $\ys$ has a maximum element as well, say, $n$ (this implication could be formally proved by absurd), and $m=\Max{\x}{n}$. Hence, $\maxElem{m}{\xs}$ is the consequence of \refToRule{max-t} with premise $\maxElem{n}{\ys}$.
\end{itemize}
\end{proof}

Translating these notions into Agda is not as simple as before because we have to mix all the builtin constructs that we presented before. \LC{According to the comparison between the two approaches to potentially infinte structures we made in the end of \refToChapter{coinductive},} we decided to show the implementations of \textit{maxElem} only for standard libray colists. 

\section{Agda implementation: predicate}
We begin by showing the implementation of the inference system along with the ingredients that will be used inside proofs.

\begin{center}
\begin{code}
\>[0]\AgdaKeyword{data}\AgdaSpace{}%
\AgdaOperator{\AgdaDatatype{\AgdaUnderscore{}maxElem{-}ind\AgdaUnderscore{}}}%
\>[20]\AgdaSymbol{:}\AgdaSpace{}%
\AgdaDatatype{Nat}\AgdaSpace{}%
\AgdaSymbol{→}\AgdaSpace{}%
\AgdaDatatype{Colist}\AgdaSpace{}%
\AgdaDatatype{Nat}\AgdaSpace{}%
\AgdaPostulate{∞}\AgdaSpace{}%
\AgdaSymbol{→}\AgdaSpace{}%
\AgdaPrimitiveType{Set}\AgdaSpace{}%
\AgdaKeyword{where}\<%
\\
\>[0][@{}l@{\AgdaIndent{0}}]%
\>[2]\AgdaInductiveConstructor{max{-}h{-}ind}\AgdaSpace{}%
\AgdaSymbol{:}\AgdaSpace{}%
\AgdaSymbol{∀}\AgdaSpace{}%
\AgdaSymbol{\{}\AgdaBound{x}\AgdaSpace{}%
\AgdaBound{xs}\AgdaSymbol{\}}\AgdaSpace{}%
\AgdaSymbol{→}\AgdaSpace{}%
\AgdaField{Thunk.force}\AgdaSpace{}%
\AgdaBound{xs}\AgdaSpace{}%
\AgdaOperator{\AgdaDatatype{≡}}\AgdaSpace{}%
\AgdaInductiveConstructor{[]}\AgdaSpace{}%
\AgdaSymbol{→}\AgdaSpace{}%
\AgdaBound{x}\AgdaSpace{}%
\AgdaOperator{\AgdaDatatype{maxElem{-}ind}}\AgdaSpace{}%
\AgdaSymbol{(}\AgdaBound{x}\AgdaSpace{}%
\AgdaOperator{\AgdaInductiveConstructor{∷}}\AgdaSpace{}%
\AgdaBound{xs}\AgdaSymbol{)}\<%
\\
\>[2]\AgdaInductiveConstructor{max{-}t{-}ind}\AgdaSpace{}%
\AgdaSymbol{:}\AgdaSpace{}%
\AgdaSymbol{∀}\AgdaSpace{}%
\AgdaSymbol{\{}\AgdaBound{n}%
\>[57I]\AgdaBound{x}\AgdaSpace{}%
\AgdaBound{z}\AgdaSpace{}%
\AgdaBound{xs}\AgdaSymbol{\}}\AgdaSpace{}%
\AgdaSymbol{→}\AgdaSpace{}%
\AgdaBound{n}\AgdaSpace{}%
\AgdaOperator{\AgdaDatatype{maxElem{-}ind}}\AgdaSpace{}%
\AgdaSymbol{(}\AgdaField{Thunk.force}\AgdaSpace{}%
\AgdaBound{xs}\AgdaSymbol{)}\AgdaSpace{}%
\AgdaSymbol{→}\<%
\\
\>[.]\<[57I]%
\>[19]\AgdaBound{z}\AgdaSpace{}%
\AgdaOperator{\AgdaDatatype{≡}}\AgdaSpace{}%
\AgdaFunction{max}\AgdaSpace{}%
\AgdaBound{n}\AgdaSpace{}%
\AgdaBound{x}\AgdaSpace{}%
\AgdaSymbol{→}\AgdaSpace{}%
\AgdaBound{z}\AgdaSpace{}%
\AgdaOperator{\AgdaDatatype{maxElem{-}ind}}\AgdaSpace{}%
\AgdaSymbol{(}\AgdaBound{x}\AgdaSpace{}%
\AgdaOperator{\AgdaInductiveConstructor{∷}}\AgdaSpace{}%
\AgdaBound{xs}\AgdaSymbol{)}\<%
\\
\>[2]\AgdaInductiveConstructor{co{-}max{-}h}\AgdaSpace{}%
\AgdaSymbol{:}\AgdaSpace{}%
\AgdaSymbol{∀}\AgdaSpace{}%
\AgdaSymbol{\{}\AgdaBound{x}\AgdaSpace{}%
\AgdaBound{xs}\AgdaSymbol{\}}\AgdaSpace{}%
\AgdaSymbol{→}\AgdaSpace{}%
\AgdaBound{x}\AgdaSpace{}%
\AgdaOperator{\AgdaDatatype{maxElem{-}ind}}\AgdaSpace{}%
\AgdaSymbol{(}\AgdaBound{x}\AgdaSpace{}%
\AgdaOperator{\AgdaInductiveConstructor{∷}}\AgdaSpace{}%
\AgdaBound{xs}\AgdaSymbol{)}\<%
\\
\>[0]\<%
\\
\>[0]\AgdaKeyword{data}\AgdaSpace{}%
\AgdaOperator{\AgdaDatatype{\AgdaUnderscore{}maxElem\AgdaUnderscore{}}}\AgdaSpace{}%
\AgdaSymbol{:}\AgdaSpace{}%
\AgdaDatatype{Nat}\AgdaSpace{}%
\AgdaSymbol{→}\AgdaSpace{}%
\AgdaDatatype{Colist}\AgdaSpace{}%
\AgdaDatatype{Nat}\AgdaSpace{}%
\AgdaPostulate{∞}\AgdaSpace{}%
\AgdaSymbol{→}\AgdaSpace{}%
\AgdaPostulate{Size}\AgdaSpace{}%
\AgdaSymbol{→}\AgdaSpace{}%
\AgdaPrimitiveType{Set}\AgdaSpace{}%
\AgdaKeyword{where}\<%
\\
\>[0][@{}l@{\AgdaIndent{0}}]%
\>[2]\AgdaInductiveConstructor{max{-}h}\AgdaSpace{}%
\AgdaSymbol{:}\AgdaSpace{}%
\AgdaSymbol{∀}\AgdaSpace{}%
\AgdaSymbol{\{}\AgdaBound{x}\AgdaSpace{}%
\AgdaBound{xs}\AgdaSpace{}%
\AgdaBound{i}\AgdaSymbol{\}}\AgdaSpace{}%
\AgdaSymbol{→}\AgdaSpace{}%
\AgdaField{Thunk.force}\AgdaSpace{}%
\AgdaBound{xs}\AgdaSpace{}%
\AgdaOperator{\AgdaDatatype{≡}}\AgdaSpace{}%
\AgdaInductiveConstructor{[]}\AgdaSpace{}%
\AgdaSymbol{→}\AgdaSpace{}%
\AgdaSymbol{(}\AgdaBound{x}\AgdaSpace{}%
\AgdaOperator{\AgdaDatatype{maxElem}}\AgdaSpace{}%
\AgdaSymbol{(}\AgdaBound{x}\AgdaSpace{}%
\AgdaOperator{\AgdaInductiveConstructor{∷}}\AgdaSpace{}%
\AgdaBound{xs}\AgdaSymbol{))}\AgdaSpace{}%
\AgdaBound{i}\<%
\\
\>[2]\AgdaInductiveConstructor{max{-}t}\AgdaSpace{}%
\AgdaSymbol{:}\AgdaSpace{}%
\AgdaSymbol{∀}\AgdaSpace{}%
\AgdaSymbol{\{}\AgdaBound{n}\AgdaSpace{}%
\AgdaBound{x}\AgdaSpace{}%
\AgdaBound{z}%
\>[120I]\AgdaBound{xs}\AgdaSpace{}%
\AgdaBound{i}\AgdaSymbol{\}}\AgdaSpace{}%
\AgdaSymbol{→}\AgdaSpace{}%
\AgdaRecord{Thunk}\AgdaSpace{}%
\AgdaSymbol{(}\AgdaBound{n}\AgdaSpace{}%
\AgdaOperator{\AgdaDatatype{maxElem}}\AgdaSpace{}%
\AgdaSymbol{(}\AgdaField{Thunk.force}\AgdaSpace{}%
\AgdaBound{xs}\AgdaSymbol{))}\AgdaSpace{}%
\AgdaBound{i}\AgdaSpace{}%
\AgdaSymbol{→}\<%
\\
\>[120I][@{}l@{\AgdaIndent{0}}]%
\>[20]\AgdaBound{z}\AgdaSpace{}%
\AgdaOperator{\AgdaDatatype{≡}}\AgdaSpace{}%
\AgdaFunction{max}\AgdaSpace{}%
\AgdaBound{n}\AgdaSpace{}%
\AgdaBound{x}\AgdaSpace{}%
\AgdaSymbol{→}\AgdaSpace{}%
\AgdaBound{z}\AgdaSpace{}%
\AgdaOperator{\AgdaDatatype{maxElem{-}ind}}\AgdaSpace{}%
\AgdaSymbol{(}\AgdaBound{x}\AgdaSpace{}%
\AgdaOperator{\AgdaInductiveConstructor{∷}}\AgdaSpace{}%
\AgdaBound{xs}\AgdaSymbol{)}\AgdaSpace{}%
\AgdaSymbol{→}\AgdaSpace{}%
\AgdaSymbol{(}\AgdaBound{z}\AgdaSpace{}%
\AgdaOperator{\AgdaDatatype{maxElem}}\AgdaSpace{}%
\AgdaSymbol{(}\AgdaBound{x}\AgdaSpace{}%
\AgdaOperator{\AgdaInductiveConstructor{∷}}\AgdaSpace{}%
\AgdaBound{xs}\AgdaSymbol{))}\AgdaSpace{}%
\AgdaBound{i}\<%
\\
\\[\AgdaEmptyExtraSkip]%
\>[0]\AgdaKeyword{data}\AgdaSpace{}%
\AgdaDatatype{Step}\AgdaSpace{}%
\AgdaSymbol{(}\AgdaBound{S}\AgdaSpace{}%
\AgdaSymbol{:}\AgdaSpace{}%
\AgdaDatatype{Nat}\AgdaSpace{}%
\AgdaSymbol{→}\AgdaSpace{}%
\AgdaDatatype{Colist}\AgdaSpace{}%
\AgdaDatatype{Nat}\AgdaSpace{}%
\AgdaPostulate{∞}\AgdaSpace{}%
\AgdaSymbol{→}\AgdaSpace{}%
\AgdaPrimitiveType{Set}\AgdaSymbol{)}\AgdaSpace{}%
\AgdaSymbol{:}\AgdaSpace{}%
\AgdaDatatype{Nat}\AgdaSpace{}%
\AgdaSymbol{→}\AgdaSpace{}%
\AgdaDatatype{Colist}\AgdaSpace{}%
\AgdaDatatype{Nat}\AgdaSpace{}%
\AgdaPostulate{∞}\AgdaSpace{}%
\AgdaSymbol{→}\AgdaSpace{}%
\AgdaPrimitiveType{Set}\AgdaSpace{}%
\AgdaKeyword{where}\<%
\\
\>[0][@{}l@{\AgdaIndent{0}}]%
\>[2]\AgdaInductiveConstructor{max{-}h}\AgdaSpace{}%
\AgdaSymbol{:}\AgdaSpace{}%
\AgdaSymbol{∀}\AgdaSpace{}%
\AgdaSymbol{\{}\AgdaBound{x}\AgdaSpace{}%
\AgdaBound{xs}\AgdaSymbol{\}}\AgdaSpace{}%
\AgdaSymbol{→}\AgdaSpace{}%
\AgdaField{Thunk.force}\AgdaSpace{}%
\AgdaBound{xs}\AgdaSpace{}%
\AgdaOperator{\AgdaDatatype{≡}}\AgdaSpace{}%
\AgdaInductiveConstructor{[]}\AgdaSpace{}%
\AgdaSymbol{→}\AgdaSpace{}%
\AgdaDatatype{Step}\AgdaSpace{}%
\AgdaBound{S}\AgdaSpace{}%
\AgdaBound{x}\AgdaSpace{}%
\AgdaSymbol{(}\AgdaBound{x}\AgdaSpace{}%
\AgdaOperator{\AgdaInductiveConstructor{∷}}\AgdaSpace{}%
\AgdaBound{xs}\AgdaSymbol{)}\<%
\\
\>[2]\AgdaInductiveConstructor{max{-}t}\AgdaSpace{}%
\AgdaSymbol{:}\AgdaSpace{}%
\AgdaSymbol{∀}\AgdaSpace{}%
\AgdaSymbol{\{}\AgdaBound{n}\AgdaSpace{}%
\AgdaBound{x}\AgdaSpace{}%
\AgdaBound{z}%
\>[187I]\AgdaBound{xs}\AgdaSymbol{\}}\AgdaSpace{}%
\AgdaSymbol{→}\AgdaSpace{}%
\AgdaBound{S}\AgdaSpace{}%
\AgdaBound{n}\AgdaSpace{}%
\AgdaSymbol{(}\AgdaField{Thunk.force}\AgdaSpace{}%
\AgdaBound{xs}\AgdaSymbol{)}\AgdaSpace{}%
\AgdaSymbol{→}\<%
\\
\>[187I][@{}l@{\AgdaIndent{0}}]%
\>[20]\AgdaBound{z}\AgdaSpace{}%
\AgdaOperator{\AgdaDatatype{≡}}\AgdaSpace{}%
\AgdaFunction{max}\AgdaSpace{}%
\AgdaBound{n}\AgdaSpace{}%
\AgdaBound{x}\AgdaSpace{}%
\AgdaSymbol{→}\AgdaSpace{}%
\AgdaDatatype{Step}\AgdaSpace{}%
\AgdaBound{S}\AgdaSpace{}%
\AgdaBound{z}\AgdaSpace{}%
\AgdaSymbol{(}\AgdaBound{x}\AgdaSpace{}%
\AgdaOperator{\AgdaInductiveConstructor{∷}}\AgdaSpace{}%
\AgdaBound{xs}\AgdaSymbol{)}\<%
\\
\\[\AgdaEmptyExtraSkip]%
\>[0]\AgdaFunction{S}\AgdaSpace{}%
\AgdaFunction{S{-}in}\AgdaSpace{}%
\AgdaFunction{S{-}gr}\AgdaSpace{}%
\AgdaSymbol{:}\AgdaSpace{}%
\AgdaDatatype{Nat}\AgdaSpace{}%
\AgdaSymbol{→}\AgdaSpace{}%
\AgdaDatatype{Colist}\AgdaSpace{}%
\AgdaDatatype{Nat}\AgdaSpace{}%
\AgdaPostulate{∞}\AgdaSpace{}%
\AgdaSymbol{→}\AgdaSpace{}%
\AgdaPrimitiveType{Set}\<%
\\
\>[0]\AgdaFunction{S{-}in}\AgdaSpace{}%
\AgdaBound{n}\AgdaSpace{}%
\AgdaBound{xs}\AgdaSpace{}%
\AgdaSymbol{=}\AgdaSpace{}%
\AgdaBound{n}\AgdaSpace{}%
\AgdaOperator{\AgdaDatatype{memberOf}}\AgdaSpace{}%
\AgdaBound{xs}\<%
\\
\>[0]\AgdaFunction{S{-}gr}\AgdaSpace{}%
\AgdaBound{n}\AgdaSpace{}%
\AgdaBound{xs}\AgdaSpace{}%
\AgdaSymbol{=}\AgdaSpace{}%
\AgdaSymbol{(∀}\AgdaSpace{}%
\AgdaSymbol{\{}\AgdaBound{i}\AgdaSymbol{\}}\AgdaSpace{}%
\AgdaSymbol{→}\AgdaSpace{}%
\AgdaBound{i}\AgdaSpace{}%
\AgdaOperator{\AgdaDatatype{memberOf}}\AgdaSpace{}%
\AgdaBound{xs}\AgdaSpace{}%
\AgdaSymbol{→}\AgdaSpace{}%
\AgdaBound{n}\AgdaSpace{}%
\AgdaOperator{\AgdaDatatype{≡}}\AgdaSpace{}%
\AgdaFunction{max}\AgdaSpace{}%
\AgdaBound{n}\AgdaSpace{}%
\AgdaBound{i}\AgdaSpace{}%
\AgdaSymbol{)}\<%
\\
\>[0]\AgdaFunction{S}\AgdaSpace{}%
\AgdaBound{n}\AgdaSpace{}%
\AgdaBound{xs}\AgdaSpace{}%
\AgdaSymbol{=}\AgdaSpace{}%
\AgdaSymbol{(}\AgdaFunction{S{-}in}\AgdaSpace{}%
\AgdaBound{n}\AgdaSpace{}%
\AgdaBound{xs}\AgdaSymbol{)}\AgdaSpace{}%
\AgdaOperator{\AgdaDatatype{∧}}\AgdaSpace{}%
\AgdaSymbol{(}\AgdaFunction{S{-}gr}\AgdaSpace{}%
\AgdaBound{n}\AgdaSpace{}%
\AgdaBound{xs}\AgdaSymbol{)}\<%
\end{code}
\end{center}

\begin{itemize}
\item \agda{maxElem-ind} represents the implementation of $\Pair\is\cois$ inductively interpreted. In fact there are two rules that correspond to those in $\is$ and a coaxiom that is inside $\cois$.
\item \agda{maxElem} is the implementation of interpretation generated by coaxiom $\Generated{\is}{\cois}$. Notice that an element of this type represents a proof tree (that can be well-founded or not) such that each node has a finite proof tree in $\Pair\is\cois$. This last fact is implemented by requiring a proof of type \agda{maxElem-ind} inside the second constructor.
\item \agda{Step} is a type that will be used to prove \textit{consistency} of the specification. Actually it says that there exists a rule in the inference system such that the premises are in $\Spec$. Clearly the structure of this type recalls the one of \agda{maxElem} in order to refer to its rules and it is not recursive. 
\item \agda{S} is the implementation of the specification. As we discussed before, $\Spec$ consist of two parts where the first denotes the membership of the maximum inside the reference list while the second says that the maximum it is greater than the other elements of the list.
\textit{S-in} and \textit{S-gr} are the implementations of this two properties.
\end{itemize}
The side conditions are written in terms of \agda{max} which is a binary function that returns the maximum between two numbers. Its implementation is showed in \refToSection{agda-modules} along with its properties.
\LC{Notice that for translating predicates defined through generalized inference systems we need two types, in this case \agda{maxElem-ind} and \agda{maxElem}. Since now the correspondence between the names of the rules and the constructors is not obvious, we decided to name those constructors in the inductive type using the names in the inference systems followed by \agda{-ind}. Also the corules are translated into constructors of the inductive type and are identified by the prefix \agda{co}.}

\section{Agda implementation: soundness}
If we want to prove the soundness of the predicate, the proof can be divided in two parts as we explained in the theoretical proof. Now we show the proof that the maximum is inside the list by considering the predicate \textit{memberOf}. 

\begin{center}
\begin{code}
\>[0]\AgdaFunction{meof2me}\AgdaSpace{}%
\AgdaSymbol{:}\AgdaSpace{}%
\AgdaSymbol{∀\{}\AgdaBound{m}\AgdaSpace{}%
\AgdaBound{ys}\AgdaSymbol{\}}\AgdaSpace{}%
\AgdaSymbol{→}\AgdaSpace{}%
\AgdaSymbol{(∀}\AgdaSpace{}%
\AgdaSymbol{\{}\AgdaBound{i}\AgdaSymbol{\}}\AgdaSpace{}%
\AgdaSymbol{→}\AgdaSpace{}%
\AgdaSymbol{(}\AgdaBound{m}\AgdaSpace{}%
\AgdaOperator{\AgdaDatatype{maxElem}}\AgdaSpace{}%
\AgdaBound{ys}\AgdaSymbol{)}\AgdaSpace{}%
\AgdaBound{i}\AgdaSymbol{)}\AgdaSpace{}%
\AgdaSymbol{→}\AgdaSpace{}%
\AgdaBound{m}\AgdaSpace{}%
\AgdaOperator{\AgdaDatatype{maxElem{-}ind}}\AgdaSpace{}%
\AgdaBound{ys}\<%
\\
\\[\AgdaEmptyExtraSkip]%
\>[0]\AgdaFunction{meof2me}\AgdaSpace{}%
\AgdaSymbol{\{}\AgdaBound{m}\AgdaSymbol{\}}\AgdaSpace{}%
\AgdaSymbol{\{}\AgdaBound{ys}\AgdaSymbol{\}}\AgdaSpace{}%
\AgdaBound{meof}\AgdaSpace{}%
\AgdaKeyword{with}\AgdaSpace{}%
\AgdaBound{meof}\<%
\\
\>[0]\AgdaFunction{meof2me}\AgdaSpace{}%
\AgdaSymbol{\{}\AgdaBound{m}\AgdaSymbol{\}}\AgdaSpace{}%
\AgdaSymbol{\{}\AgdaSymbol{\AgdaDottedPattern{.(}}\AgdaBound{\AgdaDottedPattern{m}}\AgdaSpace{}%
\AgdaOperator{\AgdaInductiveConstructor{\AgdaDottedPattern{∷}}}\AgdaSpace{}%
\AgdaSymbol{\AgdaDottedPattern{\AgdaUnderscore{})}}\AgdaSymbol{\}}\AgdaSpace{}%
\AgdaBound{meof}\AgdaSpace{}%
\AgdaSymbol{|}\AgdaSpace{}%
\AgdaInductiveConstructor{max{-}h}\AgdaSpace{}%
\AgdaBound{eq}\AgdaSpace{}%
\AgdaSymbol{=}\AgdaSpace{}%
\AgdaInductiveConstructor{max{-}h{-}ind}\AgdaSpace{}%
\AgdaBound{eq}\<%
\\
\>[0]\AgdaFunction{meof2me}\AgdaSpace{}%
\AgdaSymbol{\{}\AgdaBound{m}\AgdaSymbol{\}}\AgdaSpace{}%
\AgdaSymbol{\{\AgdaUnderscore{}}\AgdaSpace{}%
\AgdaOperator{\AgdaInductiveConstructor{∷}}\AgdaSpace{}%
\AgdaBound{xs}\AgdaSymbol{\}}\AgdaSpace{}%
\AgdaBound{meof}\AgdaSpace{}%
\AgdaSymbol{|}\AgdaSpace{}%
\AgdaInductiveConstructor{max{-}t}\AgdaSpace{}%
\AgdaBound{meof{-}xs}\AgdaSpace{}%
\AgdaBound{z}\AgdaSpace{}%
\AgdaBound{me{-}xs}\AgdaSpace{}%
\AgdaSymbol{=}\AgdaSpace{}%
\AgdaBound{me{-}xs}\<%
\\
\\[\AgdaEmptyExtraSkip]%
\\[\AgdaEmptyExtraSkip]%
\>[0]\AgdaFunction{me2memb}\AgdaSpace{}%
\AgdaSymbol{:}\AgdaSpace{}%
\AgdaSymbol{∀}\AgdaSpace{}%
\AgdaSymbol{\{}\AgdaBound{m}\AgdaSpace{}%
\AgdaBound{ys}\AgdaSymbol{\}}\AgdaSpace{}%
\AgdaSymbol{→}\AgdaSpace{}%
\AgdaBound{m}\AgdaSpace{}%
\AgdaOperator{\AgdaDatatype{maxElem{-}ind}}\AgdaSpace{}%
\AgdaBound{ys}\AgdaSpace{}%
\AgdaSymbol{→}\AgdaSpace{}%
\AgdaBound{m}\AgdaSpace{}%
\AgdaOperator{\AgdaDatatype{memberOf}}\AgdaSpace{}%
\AgdaBound{ys}\<%
\\
\\[\AgdaEmptyExtraSkip]%
\>[0]\AgdaFunction{me2memb}\AgdaSpace{}%
\AgdaSymbol{\{}\AgdaBound{m}\AgdaSymbol{\}}\AgdaSpace{}%
\AgdaSymbol{\{}\AgdaSymbol{\AgdaDottedPattern{.(}}\AgdaBound{\AgdaDottedPattern{m}}\AgdaSpace{}%
\AgdaOperator{\AgdaInductiveConstructor{\AgdaDottedPattern{∷}}}\AgdaSpace{}%
\AgdaSymbol{\AgdaDottedPattern{\AgdaUnderscore{})}}\AgdaSymbol{\}}\AgdaSpace{}%
\AgdaSymbol{(}\AgdaInductiveConstructor{max{-}h{-}ind}\AgdaSpace{}%
\AgdaBound{x}\AgdaSymbol{)}\AgdaSpace{}%
\AgdaSymbol{=}\AgdaSpace{}%
\AgdaInductiveConstructor{mem{-}h}\<%
\\
\>[0]\AgdaFunction{me2memb}\AgdaSpace{}%
\AgdaSymbol{\{}\AgdaBound{m}\AgdaSymbol{\}}\AgdaSpace{}%
\AgdaSymbol{\{}\AgdaBound{x}\AgdaSpace{}%
\AgdaOperator{\AgdaInductiveConstructor{∷}}\AgdaSpace{}%
\AgdaBound{xs}\AgdaSymbol{\}}\AgdaSpace{}%
\AgdaSymbol{(}\AgdaInductiveConstructor{max{-}t{-}ind}\AgdaSpace{}%
\AgdaBound{me}\AgdaSpace{}%
\AgdaBound{z{-}max}\AgdaSymbol{)}\AgdaSpace{}%
\AgdaKeyword{with}\AgdaSpace{}%
\AgdaFunction{max{-}refl}\AgdaSpace{}%
\AgdaBound{z{-}max}\<%
\\
\>[0]\AgdaFunction{me2memb}\AgdaSpace{}%
\AgdaSymbol{\{}\AgdaBound{m}\AgdaSymbol{\}}\AgdaSpace{}%
\AgdaSymbol{\{}\AgdaBound{x}\AgdaSpace{}%
\AgdaOperator{\AgdaInductiveConstructor{∷}}%
\>[343I]\AgdaBound{xs}\AgdaSymbol{\}}\AgdaSpace{}%
\AgdaSymbol{(}\AgdaInductiveConstructor{max{-}t{-}ind}\AgdaSpace{}%
\AgdaBound{me}\AgdaSpace{}%
\AgdaBound{z{-}max}\AgdaSymbol{)}\AgdaSpace{}%
\AgdaSymbol{|}\AgdaSpace{}%
\AgdaInductiveConstructor{inl}\AgdaSpace{}%
\AgdaBound{eq}\AgdaSpace{}%
\AgdaSymbol{=}\AgdaSpace{}%
\AgdaInductiveConstructor{mem{-}t}\<%
\\
\>[.]\<[343I]%
\>[17]\AgdaSymbol{(}\AgdaFunction{subst}\AgdaSpace{}%
\AgdaSymbol{(λ}\AgdaSpace{}%
\AgdaBound{t}\AgdaSpace{}%
\AgdaSymbol{→}\AgdaSpace{}%
\AgdaBound{t}\AgdaSpace{}%
\AgdaOperator{\AgdaDatatype{memberOf}}\AgdaSpace{}%
\AgdaSymbol{(}\AgdaBound{xs}\AgdaSpace{}%
\AgdaSymbol{.}\AgdaField{force}\AgdaSymbol{))}\AgdaSpace{}%
\AgdaSymbol{(}\AgdaFunction{sym}\AgdaSpace{}%
\AgdaBound{eq}\AgdaSymbol{)}\AgdaSpace{}%
\AgdaSymbol{(}\AgdaFunction{me2memb}\AgdaSpace{}%
\AgdaBound{me}\AgdaSymbol{))}\<%
\\
\>[0]\AgdaFunction{me2memb}\AgdaSpace{}%
\AgdaSymbol{\{}\AgdaSymbol{\AgdaDottedPattern{.}}\AgdaBound{\AgdaDottedPattern{x}}\AgdaSymbol{\}}\AgdaSpace{}%
\AgdaSymbol{\{}\AgdaBound{x}\AgdaSpace{}%
\AgdaOperator{\AgdaInductiveConstructor{∷}}\AgdaSpace{}%
\AgdaBound{xs}\AgdaSymbol{\}}\AgdaSpace{}%
\AgdaSymbol{(}\AgdaInductiveConstructor{max{-}t{-}ind}\AgdaSpace{}%
\AgdaBound{me}\AgdaSpace{}%
\AgdaBound{z{-}max}\AgdaSymbol{)}\AgdaSpace{}%
\AgdaSymbol{|}\AgdaSpace{}%
\AgdaInductiveConstructor{inr}\AgdaSpace{}%
\AgdaInductiveConstructor{refl}\AgdaSpace{}%
\AgdaSymbol{=}\AgdaSpace{}%
\AgdaInductiveConstructor{mem{-}h}\<%
\\
\>[0]\AgdaFunction{me2memb}\AgdaSpace{}%
\AgdaSymbol{\{}\AgdaBound{m}\AgdaSymbol{\}}\AgdaSpace{}%
\AgdaSymbol{\{}\AgdaSymbol{\AgdaDottedPattern{.(}}\AgdaBound{\AgdaDottedPattern{m}}\AgdaSpace{}%
\AgdaOperator{\AgdaInductiveConstructor{\AgdaDottedPattern{∷}}}\AgdaSpace{}%
\AgdaSymbol{\AgdaDottedPattern{\AgdaUnderscore{})}}\AgdaSymbol{\}}\AgdaSpace{}%
\AgdaInductiveConstructor{co{-}max{-}h}\AgdaSpace{}%
\AgdaSymbol{=}\AgdaSpace{}%
\AgdaInductiveConstructor{mem{-}h}\<%
\\
\\[\AgdaEmptyExtraSkip]%
\\[\AgdaEmptyExtraSkip]%
\>[0]\AgdaFunction{me{-}in}\AgdaSpace{}%
\AgdaSymbol{:}\AgdaSpace{}%
\AgdaSymbol{∀\{}\AgdaBound{m}\AgdaSpace{}%
\AgdaBound{ys}\AgdaSymbol{\}}\AgdaSpace{}%
\AgdaSymbol{→}\AgdaSpace{}%
\AgdaSymbol{(∀}\AgdaSpace{}%
\AgdaSymbol{\{}\AgdaBound{i}\AgdaSymbol{\}}\AgdaSpace{}%
\AgdaSymbol{→}\AgdaSpace{}%
\AgdaSymbol{(}\AgdaBound{m}\AgdaSpace{}%
\AgdaOperator{\AgdaDatatype{maxElem}}\AgdaSpace{}%
\AgdaBound{ys}\AgdaSymbol{)}\AgdaSpace{}%
\AgdaBound{i}\AgdaSymbol{)}\AgdaSpace{}%
\AgdaSymbol{→}\AgdaSpace{}%
\AgdaBound{m}\AgdaSpace{}%
\AgdaOperator{\AgdaDatatype{memberOf}}\AgdaSpace{}%
\AgdaBound{ys}\<%
\\
\>[0]\AgdaFunction{me{-}in}\AgdaSpace{}%
\AgdaBound{meof}\AgdaSpace{}%
\AgdaSymbol{=}\AgdaSpace{}%
\AgdaFunction{me2memb}\AgdaSpace{}%
\AgdaSymbol{(}\AgdaFunction{meof2me}\AgdaSpace{}%
\AgdaBound{meof}\AgdaSymbol{)}\<%
\end{code}
\end{center}

The proof is by induction of the rules of $\Pair\is\cois$. In order to do that we have to get first the finite proof in $\Pair\is\cois$, that is an element of type \agda{maxElem-ind}. This is done through the function \textit{meof2me}. \textit{me2memb} is the function that allows us to obtain the membership proof and you can see that it is very similar to that presented in the theoretical part. In fact pattern matching on the input element leads to the previously analyzed situations:
\begin{itemize}
\item If the axiom or the coaxiom are used, then the membership proof is built using the axiom \refToRule{mem-h} of \textit{memberOf} because the maximum is actually the head.
\item If the rule \refToRule{mem-t} is used then we have to make a distinction between the case in which the max is equal to the max of the tail and the case case in which it is the new element. Clearly this second case says again that the maximum is the head. On the other hand we have the inductive hypothesis that tells that the maximum of the tail is inside the tail (through the recursive call to \agda{me2memb}) but also that $m$ is equal to that maximum. The proof is by substitution.
\end{itemize}
From this last consideration it can be noticed that Agda does not know that the maximum is equal to the previous one because the \textit{max} function is called inside the constructor and this leads to the loss of connection between the inputs and the output. \agda{max-refl} is a property that basically says that if $z = \Max {x} {y}$ then $z$ must be equal to $x$ or to $y$; the statement is given in \refToSection{agda-modules} and the output term uses the disjunction module (this is why we meet the constructors \agda{inl} and \agda{inr}). Precisely, the variable $eq$ is the needed proof to apply the substitution.
\agda{xs .force} is a more compact way to write \agda{Thunk.force xs}. 
We proceed by showing the proof that the maximum is actually the maximum between all the elements of the list and itself.

\begin{center}
\begin{code}
\>[0]\AgdaFunction{me{-}gr}\AgdaSpace{}%
\AgdaSymbol{:}\AgdaSpace{}%
\AgdaSymbol{∀\{}\AgdaBound{n}\AgdaSpace{}%
\AgdaBound{ys}\AgdaSymbol{\}}\AgdaSpace{}%
\AgdaSymbol{→}\AgdaSpace{}%
\AgdaSymbol{(∀}%
\>[407I]\AgdaSymbol{\{}\AgdaBound{i}\AgdaSymbol{\}}\AgdaSpace{}%
\AgdaSymbol{→}\AgdaSpace{}%
\AgdaSymbol{(}\AgdaBound{n}\AgdaSpace{}%
\AgdaOperator{\AgdaDatatype{maxElem}}\AgdaSpace{}%
\AgdaBound{ys}\AgdaSymbol{)}\AgdaSpace{}%
\AgdaBound{i}\AgdaSymbol{)}\AgdaSpace{}%
\AgdaSymbol{→}\<%
\\
\>[.]\<[407I]%
\>[21]\AgdaSymbol{(∀}\AgdaSpace{}%
\AgdaSymbol{\{}\AgdaBound{i}\AgdaSymbol{\}}\AgdaSpace{}%
\AgdaSymbol{→}\AgdaSpace{}%
\AgdaBound{i}\AgdaSpace{}%
\AgdaOperator{\AgdaDatatype{memberOf}}\AgdaSpace{}%
\AgdaBound{ys}\AgdaSpace{}%
\AgdaSymbol{→}\AgdaSpace{}%
\AgdaBound{n}\AgdaSpace{}%
\AgdaOperator{\AgdaDatatype{≡}}\AgdaSpace{}%
\AgdaFunction{max}\AgdaSpace{}%
\AgdaBound{n}\AgdaSpace{}%
\AgdaBound{i}\AgdaSymbol{)}\<%
\\
\\[\AgdaEmptyExtraSkip]%
\>[0]\AgdaFunction{me{-}gr}\AgdaSpace{}%
\AgdaSymbol{\{}\AgdaBound{n}\AgdaSymbol{\}}\AgdaSpace{}%
\AgdaSymbol{\{}\AgdaSymbol{\AgdaDottedPattern{.(\AgdaUnderscore{}}}\AgdaSpace{}%
\AgdaOperator{\AgdaInductiveConstructor{\AgdaDottedPattern{∷}}}\AgdaSpace{}%
\AgdaSymbol{\AgdaDottedPattern{\AgdaUnderscore{})}}\AgdaSymbol{\}}\AgdaSpace{}%
\AgdaBound{meof}\AgdaSpace{}%
\AgdaInductiveConstructor{mem{-}h}\AgdaSpace{}%
\AgdaKeyword{with}\AgdaSpace{}%
\AgdaBound{meof}\<%
\\
\>[0]\AgdaFunction{me{-}gr}\AgdaSpace{}%
\AgdaSymbol{\{\AgdaUnderscore{}\}}\AgdaSpace{}%
\AgdaSymbol{\{}\AgdaSymbol{\AgdaDottedPattern{.(\AgdaUnderscore{}}}\AgdaSpace{}%
\AgdaOperator{\AgdaInductiveConstructor{\AgdaDottedPattern{∷}}}\AgdaSpace{}%
\AgdaSymbol{\AgdaDottedPattern{\AgdaUnderscore{})}}\AgdaSymbol{\}}\AgdaSpace{}%
\AgdaBound{meof}\AgdaSpace{}%
\AgdaInductiveConstructor{mem{-}h}\AgdaSpace{}%
\AgdaSymbol{|}\AgdaSpace{}%
\AgdaInductiveConstructor{max{-}h}\AgdaSpace{}%
\AgdaBound{eq}\AgdaSpace{}%
\AgdaSymbol{=}\AgdaSpace{}%
\AgdaFunction{max{-}eq}\<%
\\
\>[0]\AgdaFunction{me{-}gr}\AgdaSpace{}%
\AgdaSymbol{\{}\AgdaBound{n}\AgdaSymbol{\}}\AgdaSpace{}%
\AgdaSymbol{\{}\AgdaBound{x}\AgdaSpace{}%
\AgdaOperator{\AgdaInductiveConstructor{∷}}\AgdaSpace{}%
\AgdaBound{xs}\AgdaSymbol{\}}\AgdaSpace{}%
\AgdaBound{meof}\AgdaSpace{}%
\AgdaInductiveConstructor{mem{-}h}\AgdaSpace{}%
\AgdaSymbol{|}\AgdaSpace{}%
\AgdaInductiveConstructor{max{-}t}\AgdaSpace{}%
\AgdaBound{meof{-}xs}\AgdaSpace{}%
\AgdaBound{z}\AgdaSpace{}%
\AgdaSymbol{\AgdaUnderscore{}}\AgdaSpace{}%
\AgdaKeyword{with}\AgdaSpace{}%
\AgdaFunction{max{-}refl}\AgdaSpace{}%
\AgdaSymbol{\{}\AgdaBound{n}\AgdaSymbol{\}\{}\AgdaArgument{y}\AgdaSpace{}%
\AgdaSymbol{=}\AgdaSpace{}%
\AgdaBound{x}\AgdaSymbol{\}}\AgdaSpace{}%
\AgdaBound{z}\<%
\\
\>[0]\AgdaFunction{me{-}gr}\AgdaSpace{}%
\AgdaSymbol{\{}\AgdaBound{n}\AgdaSymbol{\}}%
\>[462I]\AgdaSymbol{\{}\AgdaBound{x}\AgdaSpace{}%
\AgdaOperator{\AgdaInductiveConstructor{∷}}\AgdaSpace{}%
\AgdaBound{xs}\AgdaSymbol{\}}\AgdaSpace{}%
\AgdaBound{meof}\AgdaSpace{}%
\AgdaInductiveConstructor{mem{-}h}\AgdaSpace{}%
\AgdaSymbol{|}\AgdaSpace{}%
\AgdaInductiveConstructor{max{-}t}\AgdaSpace{}%
\AgdaBound{meof{-}xs}\AgdaSpace{}%
\AgdaBound{z}\AgdaSpace{}%
\AgdaSymbol{\AgdaUnderscore{}}\AgdaSpace{}%
\AgdaSymbol{|}\AgdaSpace{}%
\AgdaInductiveConstructor{inl}\AgdaSpace{}%
\AgdaBound{eq}\AgdaSpace{}%
\AgdaSymbol{=}\<%
\\
\>[.]\<[462I]%
\>[10]\AgdaFunction{subst}\AgdaSpace{}%
\AgdaSymbol{(λ}\AgdaSpace{}%
\AgdaBound{t}\AgdaSpace{}%
\AgdaSymbol{→}\AgdaSpace{}%
\AgdaBound{n}\AgdaSpace{}%
\AgdaOperator{\AgdaDatatype{≡}}\AgdaSpace{}%
\AgdaFunction{max}\AgdaSpace{}%
\AgdaBound{t}\AgdaSpace{}%
\AgdaBound{x}\AgdaSymbol{)}\AgdaSpace{}%
\AgdaSymbol{(}\AgdaFunction{sym}\AgdaSpace{}%
\AgdaBound{eq}\AgdaSymbol{)}\AgdaSpace{}%
\AgdaBound{z}\<%
\\
\>[0]\AgdaFunction{me{-}gr}\AgdaSpace{}%
\AgdaSymbol{\{}\AgdaBound{n}\AgdaSymbol{\}}\AgdaSpace{}%
\AgdaSymbol{\{}\AgdaSymbol{\AgdaDottedPattern{.(\AgdaUnderscore{}}}\AgdaSpace{}%
\AgdaOperator{\AgdaInductiveConstructor{\AgdaDottedPattern{∷}}}\AgdaSpace{}%
\AgdaSymbol{\AgdaDottedPattern{\AgdaUnderscore{})}}\AgdaSymbol{\}}\AgdaSpace{}%
\AgdaBound{meof}\AgdaSpace{}%
\AgdaInductiveConstructor{mem{-}h}\AgdaSpace{}%
\AgdaSymbol{|}\AgdaSpace{}%
\AgdaInductiveConstructor{max{-}t}\AgdaSpace{}%
\AgdaBound{meof{-}xs}\AgdaSpace{}%
\AgdaBound{z}\AgdaSpace{}%
\AgdaSymbol{\AgdaUnderscore{}}\AgdaSpace{}%
\AgdaSymbol{|}\AgdaSpace{}%
\AgdaInductiveConstructor{inr}\AgdaSpace{}%
\AgdaInductiveConstructor{refl}\AgdaSpace{}%
\AgdaSymbol{=}\AgdaSpace{}%
\AgdaFunction{max{-}eq}\<%
\\
\\[\AgdaEmptyExtraSkip]%
\>[0]\AgdaFunction{me{-}gr}\AgdaSpace{}%
\AgdaSymbol{\{}\AgdaBound{n}\AgdaSymbol{\}}\AgdaSpace{}%
\AgdaSymbol{\{}\AgdaSymbol{\AgdaDottedPattern{.(\AgdaUnderscore{}}}\AgdaSpace{}%
\AgdaOperator{\AgdaInductiveConstructor{\AgdaDottedPattern{∷}}}\AgdaSpace{}%
\AgdaSymbol{\AgdaDottedPattern{\AgdaUnderscore{})}}\AgdaSymbol{\}}\AgdaSpace{}%
\AgdaBound{meof}\AgdaSpace{}%
\AgdaSymbol{(}\AgdaInductiveConstructor{mem{-}t}\AgdaSpace{}%
\AgdaBound{mem}\AgdaSymbol{)}\AgdaSpace{}%
\AgdaKeyword{with}\AgdaSpace{}%
\AgdaBound{meof}\<%
\\
\>[0]\AgdaFunction{me{-}gr}\AgdaSpace{}%
\AgdaSymbol{\{\AgdaUnderscore{}\}}\AgdaSpace{}%
\AgdaSymbol{\{}\AgdaSymbol{\AgdaDottedPattern{.(\AgdaUnderscore{}}}\AgdaSpace{}%
\AgdaOperator{\AgdaInductiveConstructor{\AgdaDottedPattern{∷}}}\AgdaSpace{}%
\AgdaSymbol{\AgdaDottedPattern{\AgdaUnderscore{})}}\AgdaSymbol{\}}\AgdaSpace{}%
\AgdaBound{meof}\AgdaSpace{}%
\AgdaSymbol{(}\AgdaInductiveConstructor{mem{-}t}\AgdaSpace{}%
\AgdaBound{mem}\AgdaSymbol{)}\AgdaSpace{}%
\AgdaSymbol{|}\AgdaSpace{}%
\AgdaInductiveConstructor{max{-}h}\AgdaSpace{}%
\AgdaBound{eq}\AgdaSpace{}%
\AgdaSymbol{=}\AgdaSpace{}%
\AgdaFunction{⊥{-}elim}\AgdaSpace{}%
\AgdaSymbol{(}\AgdaFunction{abs}\AgdaSpace{}%
\AgdaBound{mem}\AgdaSpace{}%
\AgdaBound{eq}\AgdaSymbol{)}\<%
\\
\>[0]\AgdaFunction{me{-}gr}\AgdaSpace{}%
\AgdaSymbol{\{}\AgdaBound{n}\AgdaSymbol{\}}\AgdaSpace{}%
\AgdaSymbol{\{}\AgdaBound{x}\AgdaSpace{}%
\AgdaOperator{\AgdaInductiveConstructor{∷}}\AgdaSpace{}%
\AgdaBound{xs}\AgdaSymbol{\}}\AgdaSpace{}%
\AgdaBound{meof}\AgdaSpace{}%
\AgdaSymbol{(}\AgdaInductiveConstructor{mem{-}t}\AgdaSpace{}%
\AgdaBound{mem}\AgdaSymbol{)}\AgdaSpace{}%
\AgdaSymbol{|}\AgdaSpace{}%
\AgdaInductiveConstructor{max{-}t}\AgdaSpace{}%
\AgdaBound{meof{-}xs}\AgdaSpace{}%
\AgdaBound{z}\AgdaSpace{}%
\AgdaSymbol{\AgdaUnderscore{}}\AgdaSpace{}%
\AgdaKeyword{with}\AgdaSpace{}%
\AgdaFunction{max{-}refl}\AgdaSpace{}%
\AgdaSymbol{\{}\AgdaBound{n}\AgdaSymbol{\}\{}\AgdaArgument{y}\AgdaSpace{}%
\AgdaSymbol{=}\AgdaSpace{}%
\AgdaBound{x}\AgdaSymbol{\}}\AgdaSpace{}%
\AgdaBound{z}\<%
\\
\>[0]\AgdaFunction{me{-}gr}\AgdaSpace{}%
\AgdaSymbol{\{}\AgdaBound{n}\AgdaSymbol{\}}%
\>[546I]\AgdaSymbol{\{}\AgdaBound{x}\AgdaSpace{}%
\AgdaOperator{\AgdaInductiveConstructor{∷}}\AgdaSpace{}%
\AgdaBound{xs}\AgdaSymbol{\}}\AgdaSpace{}%
\AgdaBound{meof}\AgdaSpace{}%
\AgdaSymbol{(}\AgdaInductiveConstructor{mem{-}t}\AgdaSpace{}%
\AgdaBound{mem}\AgdaSymbol{)}\AgdaSpace{}%
\AgdaSymbol{|}\AgdaSpace{}%
\AgdaInductiveConstructor{max{-}t}\AgdaSpace{}%
\AgdaBound{meof{-}xs}\AgdaSpace{}%
\AgdaBound{z}\AgdaSpace{}%
\AgdaSymbol{\AgdaUnderscore{}}\AgdaSpace{}%
\AgdaSymbol{|}\AgdaSpace{}%
\AgdaInductiveConstructor{inl}\AgdaSpace{}%
\AgdaInductiveConstructor{refl}\AgdaSpace{}%
\AgdaSymbol{=}\<%
\\
\>[.]\<[546I]%
\>[10]\AgdaFunction{me{-}gr}%
\>[561I]\AgdaSymbol{(λ}\AgdaSpace{}%
\AgdaSymbol{\{}\AgdaBound{i}\AgdaSymbol{\}}\AgdaSpace{}%
\AgdaSymbol{→}\AgdaSpace{}%
\AgdaFunction{subst}\<%
\\
\>[.]\<[561I]%
\>[16]\AgdaSymbol{(λ}\AgdaSpace{}%
\AgdaBound{t}\AgdaSpace{}%
\AgdaSymbol{→}\AgdaSpace{}%
\AgdaSymbol{(}\AgdaBound{t}\AgdaSpace{}%
\AgdaOperator{\AgdaDatatype{maxElem}}\AgdaSpace{}%
\AgdaField{force}\AgdaSpace{}%
\AgdaBound{xs}\AgdaSymbol{)}\AgdaSpace{}%
\AgdaBound{i}\AgdaSymbol{)}\AgdaSpace{}%
\AgdaSymbol{(}\AgdaFunction{sym}\AgdaSpace{}%
\AgdaInductiveConstructor{refl}\AgdaSymbol{)}\AgdaSpace{}%
\AgdaSymbol{(}\AgdaBound{meof{-}xs}\AgdaSpace{}%
\AgdaSymbol{.}\AgdaField{force}\AgdaSymbol{))}\<%
\\
\>[10]\AgdaBound{mem}\<%
\\
\>[0]\AgdaFunction{me{-}gr}\AgdaSpace{}%
\AgdaSymbol{\{}\AgdaBound{n}\AgdaSymbol{\}}%
\>[577I]\AgdaSymbol{\{}\AgdaBound{x}\AgdaSpace{}%
\AgdaOperator{\AgdaInductiveConstructor{∷}}\AgdaSpace{}%
\AgdaBound{xs}\AgdaSymbol{\}}\AgdaSpace{}%
\AgdaBound{meof}\AgdaSpace{}%
\AgdaSymbol{(}\AgdaInductiveConstructor{mem{-}t}\AgdaSpace{}%
\AgdaBound{mem}\AgdaSymbol{)}\AgdaSpace{}%
\AgdaSymbol{|}\AgdaSpace{}%
\AgdaInductiveConstructor{max{-}t}\AgdaSpace{}%
\AgdaBound{meof{-}xs}\AgdaSpace{}%
\AgdaBound{z}\AgdaSpace{}%
\AgdaSymbol{\AgdaUnderscore{}}\AgdaSpace{}%
\AgdaSymbol{|}\AgdaSpace{}%
\AgdaInductiveConstructor{inr}\AgdaSpace{}%
\AgdaInductiveConstructor{refl}\AgdaSpace{}%
\AgdaSymbol{=}\<%
\\
\>[.]\<[577I]%
\>[10]\AgdaFunction{max{-}trans}\AgdaSpace{}%
\AgdaSymbol{(}\AgdaFunction{max{-}comm}\AgdaSpace{}%
\AgdaSymbol{\{}\AgdaArgument{b}\AgdaSpace{}%
\AgdaSymbol{=}\AgdaSpace{}%
\AgdaBound{n}\AgdaSymbol{\}}\AgdaSpace{}%
\AgdaBound{z}\AgdaSymbol{)}\<%
\\
\>[10]\AgdaSymbol{(}\AgdaFunction{me{-}gr}\AgdaSpace{}%
\AgdaSymbol{(}\AgdaBound{meof{-}xs}\AgdaSpace{}%
\AgdaSymbol{.}\AgdaField{force}\AgdaSymbol{)}\AgdaSpace{}%
\AgdaBound{mem}\AgdaSymbol{)}\<%
\end{code}
\end{center}

The proof is by induction over the element of type \textit{memberOf}, roughly speaking on the position of the element inside the list. Again the code follows exactly the scheme we followed in the theoretical part. 
The first part of the code deals with the head of the list and different cases are generated according on how the term of type \agda{maxElem} is built.

\begin{itemize}
\item If the axiom is applied then the only element in the list is the maximum itself. The fact that $n = \Max {n} {n}$ must be proved separately, you can find the code in \refToSection{agda-modules}.
\item If the rule is applied and the maximum is $x$, then the proof is as before.
\item If the rule is applied and the maximum is the maximum of $xs$ (let $m$ be this value), then $z$ contains the proof that $n = \Max {m} {x}$ and again we have to tell Agda that $n$ is equal to $m$. The proof follows by substitution. 
\end{itemize}

Things become a little more complicate when we consider an element which is not the head of the list. In this case, the term of type \agda{maxElem} cannot be built using the axiom; in fact we manage the absurd case through the $\bot$ elimination rule (\refToSection{agda-modules}). \agda{abs} is a function that returns $\bot$ if we give it a proof that an element is inside a list and the proof that the list is empty. 
If the rule is applied:
\begin{itemize}
\item If $n$ is the maximum of $xs$ we can use the inductive hypothesis, that is the recursive call to the function with \agda{meof-xs} as input. This last variable refers to the maximum of $xs$ and Agda again does not know that it is $n$. So we proceed by substitution.
\item If the maximum is $x$ we have to prove that there exists a maximum of $xs$ and, since $x$ is greater than this value, that $x$ is the maximum between itself and all the elements of $xs$. We obtain the maximum of $xs$ by recursively calling the function while \agda{max-trans} returns the proof of the second fact. \agda{max-comm} proves, without spending words on technicalities, that $\Max {a} {b}$ is equal to $\Max {b} {a}$ where $a$ and $b$ are two numbers. The codes of the types of \agda{max-trans} and \agda{max-comm} can be viewed in \refToSection{agda-modules}.
\end{itemize}
Finally we can prove the soundness of \agda{maxElem} with respect to the specification.
Using the two proofs above the code becomes very simple:

\begin{center}
\begin{code}
\>[0]\AgdaFunction{max{-}sound}\AgdaSpace{}%
\AgdaSymbol{:}\AgdaSpace{}%
\AgdaSymbol{\{}\AgdaBound{n}\AgdaSpace{}%
\AgdaSymbol{:}\AgdaSpace{}%
\AgdaDatatype{Nat}\AgdaSymbol{\}\{}\AgdaBound{xs}\AgdaSpace{}%
\AgdaSymbol{:}\AgdaSpace{}%
\AgdaDatatype{Colist}\AgdaSpace{}%
\AgdaDatatype{Nat}\AgdaSpace{}%
\AgdaPostulate{∞}\AgdaSymbol{\}}\AgdaSpace{}%
\AgdaSymbol{→}\AgdaSpace{}%
\AgdaSymbol{(∀}\AgdaSpace{}%
\AgdaSymbol{\{}\AgdaBound{i}\AgdaSymbol{\}}\AgdaSpace{}%
\AgdaSymbol{→}\AgdaSpace{}%
\AgdaSymbol{(}\AgdaBound{n}\AgdaSpace{}%
\AgdaOperator{\AgdaDatatype{maxElem}}\AgdaSpace{}%
\AgdaBound{xs}\AgdaSymbol{)}\AgdaSpace{}%
\AgdaBound{i}\AgdaSymbol{)}\AgdaSpace{}%
\AgdaSymbol{→}\AgdaSpace{}%
\AgdaFunction{S}\AgdaSpace{}%
\AgdaBound{n}\AgdaSpace{}%
\AgdaBound{xs}\<%
\\
\>[0]\AgdaFunction{max{-}sound}\AgdaSpace{}%
\AgdaSymbol{\{}\AgdaBound{n}\AgdaSymbol{\}}\AgdaSpace{}%
\AgdaSymbol{\{}\AgdaBound{xs}\AgdaSymbol{\}}\AgdaSpace{}%
\AgdaBound{me}\AgdaSpace{}%
\AgdaSymbol{=}\AgdaSpace{}%
\AgdaOperator{\AgdaInductiveConstructor{⟨}}\AgdaSpace{}%
\AgdaSymbol{(}\AgdaFunction{me{-}in}\AgdaSpace{}%
\AgdaBound{me}\AgdaSymbol{)}\AgdaSpace{}%
\AgdaOperator{\AgdaInductiveConstructor{,}}\AgdaSpace{}%
\AgdaSymbol{(}\AgdaFunction{me{-}gr}\AgdaSpace{}%
\AgdaBound{me}\AgdaSymbol{)}\AgdaSpace{}%
\AgdaOperator{\AgdaInductiveConstructor{⟩}}\<%
\end{code}
\end{center}

Where \agda{S} is the specification. We use the constructor of the conjunction type because it is used in the code of the specification.
The elements of the couple are exactly the two proofs above applied to the element of type \agda{maxElem}.

\section{Agda implementation: bounded coinduction principle}
Now we can move to completeness; in the theoretical part we presented the \textbf{bounded coinduction principle} that can be used to prove it. First of all we show the implementation and the proof of this principle. 

\begin{center}
\begin{code}
\>[0]\AgdaFunction{bd{-}c}\AgdaSpace{}%
\AgdaSymbol{:}%
\>[632I]\AgdaSymbol{(}\AgdaBound{S}%
\>[633I]\AgdaSymbol{:}\AgdaSpace{}%
\AgdaDatatype{Nat}\AgdaSpace{}%
\AgdaSymbol{→}\AgdaSpace{}%
\AgdaDatatype{Colist}\AgdaSpace{}%
\AgdaDatatype{Nat}\AgdaSpace{}%
\AgdaPostulate{∞}\AgdaSpace{}%
\AgdaSymbol{→}\AgdaSpace{}%
\AgdaPrimitiveType{Set}\AgdaSymbol{)}\AgdaSpace{}%
\AgdaSymbol{→}\<%
\\
\>[.]\<[633I]%
\>[10]\AgdaSymbol{(\{}\AgdaBound{x}\AgdaSpace{}%
\AgdaSymbol{:}\AgdaSpace{}%
\AgdaDatatype{Nat}\AgdaSymbol{\}\{}\AgdaBound{xs}\AgdaSpace{}%
\AgdaSymbol{:}\AgdaSpace{}%
\AgdaDatatype{Colist}\AgdaSpace{}%
\AgdaDatatype{Nat}\AgdaSpace{}%
\AgdaPostulate{∞}\AgdaSymbol{\}}\AgdaSpace{}%
\AgdaSymbol{→}\AgdaSpace{}%
\AgdaBound{S}\AgdaSpace{}%
\AgdaBound{x}\AgdaSpace{}%
\AgdaBound{xs}\AgdaSpace{}%
\AgdaSymbol{→}%
\>[52]\AgdaBound{x}\AgdaSpace{}%
\AgdaOperator{\AgdaDatatype{maxElem{-}ind}}\AgdaSpace{}%
\AgdaBound{xs}\AgdaSymbol{)}\AgdaSpace{}%
\AgdaSymbol{→}\<%
\\
\>[10]\AgdaSymbol{(\{}\AgdaBound{x}\AgdaSpace{}%
\AgdaSymbol{:}\AgdaSpace{}%
\AgdaDatatype{Nat}\AgdaSymbol{\}\{}\AgdaBound{xs}\AgdaSpace{}%
\AgdaSymbol{:}\AgdaSpace{}%
\AgdaDatatype{Colist}\AgdaSpace{}%
\AgdaDatatype{Nat}\AgdaSpace{}%
\AgdaPostulate{∞}\AgdaSymbol{\}}\AgdaSpace{}%
\AgdaSymbol{→}\AgdaSpace{}%
\AgdaBound{S}\AgdaSpace{}%
\AgdaBound{x}\AgdaSpace{}%
\AgdaBound{xs}\AgdaSpace{}%
\AgdaSymbol{→}\AgdaSpace{}%
\AgdaDatatype{Step}\AgdaSpace{}%
\AgdaBound{S}\AgdaSpace{}%
\AgdaBound{x}\AgdaSpace{}%
\AgdaBound{xs}\AgdaSymbol{)}\AgdaSpace{}%
\AgdaSymbol{→}\<%
\\
\>[.]\<[632I]%
\>[7]\AgdaComment{{-}{-}{-}{-}{-}{-}{-}{-}{-}{-}{-}{-}{-}{-}{-}{-}{-}{-}{-}{-}{-}{-}{-}{-}{-}{-}{-}{-}{-}{-}{-}{-}{-}{-}{-}{-}{-}{-}{-}{-}{-}{-}{-}{-}{-}{-}{-}{-}{-}{-}{-}{-}{-}{-}{-}{-}{-}{-}{-}{-}{-}{-}{-}}\<%
\\
\>[7]\AgdaSymbol{(∀}\AgdaSpace{}%
\AgdaSymbol{\{}\AgdaBound{i}\AgdaSpace{}%
\AgdaBound{xs}\AgdaSpace{}%
\AgdaBound{x}\AgdaSymbol{\}}\AgdaSpace{}%
\AgdaSymbol{→}\AgdaSpace{}%
\AgdaBound{S}\AgdaSpace{}%
\AgdaBound{x}\AgdaSpace{}%
\AgdaBound{xs}\AgdaSpace{}%
\AgdaSymbol{→}\AgdaSpace{}%
\AgdaSymbol{(}\AgdaBound{x}\AgdaSpace{}%
\AgdaOperator{\AgdaDatatype{maxElem}}\AgdaSpace{}%
\AgdaBound{xs}\AgdaSymbol{)}\AgdaSpace{}%
\AgdaBound{i}\AgdaSymbol{)}\<%
\\
\\[\AgdaEmptyExtraSkip]%
\>[0]\AgdaFunction{bd{-}c}\AgdaSpace{}%
\AgdaBound{S}\AgdaSpace{}%
\AgdaBound{bd}\AgdaSpace{}%
\AgdaBound{cs}\AgdaSpace{}%
\AgdaBound{Sx}\AgdaSpace{}%
\AgdaKeyword{with}\AgdaSpace{}%
\AgdaBound{cs}\AgdaSpace{}%
\AgdaBound{Sx}\<%
\\
\>[0]\AgdaFunction{bd{-}c}\AgdaSpace{}%
\AgdaBound{S}\AgdaSpace{}%
\AgdaBound{bd}\AgdaSpace{}%
\AgdaBound{cs}\AgdaSpace{}%
\AgdaBound{Sx}\AgdaSpace{}%
\AgdaSymbol{|}\AgdaSpace{}%
\AgdaInductiveConstructor{max{-}h}\AgdaSpace{}%
\AgdaBound{eq}\AgdaSpace{}%
\AgdaSymbol{=}\AgdaSpace{}%
\AgdaInductiveConstructor{max{-}h}\AgdaSpace{}%
\AgdaBound{eq}\<%
\\
\>[0]\AgdaFunction{bd{-}c}\AgdaSpace{}%
\AgdaBound{S}\AgdaSpace{}%
\AgdaBound{bd}\AgdaSpace{}%
\AgdaBound{cs}\AgdaSpace{}%
\AgdaBound{Sx}\AgdaSpace{}%
\AgdaSymbol{|}\AgdaSpace{}%
\AgdaInductiveConstructor{max{-}t}\AgdaSpace{}%
\AgdaBound{Sxs}\AgdaSpace{}%
\AgdaBound{eq}\AgdaSpace{}%
\AgdaSymbol{=}\AgdaSpace{}%
\AgdaInductiveConstructor{max{-}t}\AgdaSpace{}%
\AgdaSymbol{(λ}\AgdaSpace{}%
\AgdaKeyword{where}\AgdaSpace{}%
\AgdaSymbol{.}\AgdaField{force}\AgdaSpace{}%
\AgdaSymbol{→}\AgdaSpace{}%
\AgdaFunction{bd{-}c}\AgdaSpace{}%
\AgdaBound{S}\AgdaSpace{}%
\AgdaBound{bd}\AgdaSpace{}%
\AgdaBound{cs}\AgdaSpace{}%
\AgdaBound{Sxs}\AgdaSymbol{)}\AgdaSpace{}%
\AgdaBound{eq}\AgdaSpace{}%
\AgdaSymbol{(}\AgdaBound{bd}\AgdaSpace{}%
\AgdaBound{Sx}\AgdaSymbol{)}\<%
\end{code}
\end{center}

Notice that the input elements to \agda{bd-c} correspond to the requirements on the specification that we discussed in the theoretical part. We recall that in the previous chapters we did not implement induction and coinduction principles because Agda provides the tools to make proofs on inductive and coinductive types. 

\section{Agda implementation: completeness}
Coming back the completeness, it is clear that two proofs are needed in order to apply \agda{bd-c}.
The first one is the proof that the specification is bounded, that is the judgments have a finite proof tree in $\Ind{\is\cup\cois}$. In Agda this means that we can build a term of type \agda{maxElem-ind}.

\begin{center}
\begin{code}
\>[0]\AgdaFunction{bound{-}aux}\AgdaSpace{}%
\AgdaSymbol{:}\AgdaSpace{}%
\AgdaSymbol{∀}\AgdaSpace{}%
\AgdaSymbol{\{}\AgdaBound{n}\AgdaSpace{}%
\AgdaBound{xs}\AgdaSymbol{\}}\AgdaSpace{}%
\AgdaSymbol{→}\AgdaSpace{}%
\AgdaFunction{S{-}in}\AgdaSpace{}%
\AgdaBound{n}\AgdaSpace{}%
\AgdaBound{xs}\AgdaSpace{}%
\AgdaSymbol{→}\AgdaSpace{}%
\AgdaFunction{S{-}gr}\AgdaSpace{}%
\AgdaBound{n}\AgdaSpace{}%
\AgdaBound{xs}\AgdaSpace{}%
\AgdaSymbol{→}\AgdaSpace{}%
\AgdaBound{n}\AgdaSpace{}%
\AgdaOperator{\AgdaDatatype{maxElem{-}ind}}\AgdaSpace{}%
\AgdaBound{xs}\<%
\\
\>[0]\AgdaFunction{bound{-}aux}\AgdaSpace{}%
\AgdaInductiveConstructor{mem{-}h}\AgdaSpace{}%
\AgdaBound{gr}\AgdaSpace{}%
\AgdaSymbol{=}\AgdaSpace{}%
\AgdaInductiveConstructor{co{-}max{-}h}\<%
\\
\>[0]\AgdaFunction{bound{-}aux}\AgdaSpace{}%
\AgdaSymbol{(}\AgdaInductiveConstructor{mem{-}t}%
\>[744I]\AgdaBound{mem}\AgdaSymbol{)}\AgdaSpace{}%
\AgdaBound{gr}\AgdaSpace{}%
\AgdaSymbol{=}\<%
\\
\>[.]\<[744I]%
\>[17]\AgdaInductiveConstructor{max{-}t{-}ind}\AgdaSpace{}%
\AgdaSymbol{(}\AgdaFunction{bound{-}aux}\AgdaSpace{}%
\AgdaBound{mem}\AgdaSpace{}%
\AgdaSymbol{(λ}\AgdaSpace{}%
\AgdaBound{m}\AgdaSpace{}%
\AgdaSymbol{→}\AgdaSpace{}%
\AgdaBound{gr}\AgdaSpace{}%
\AgdaSymbol{(}\AgdaInductiveConstructor{mem{-}t}\AgdaSpace{}%
\AgdaBound{m}\AgdaSymbol{)))}\AgdaSpace{}%
\AgdaSymbol{(}\AgdaBound{gr}\AgdaSpace{}%
\AgdaInductiveConstructor{mem{-}h}\AgdaSymbol{)}\<%
\\
\\[\AgdaEmptyExtraSkip]%
\>[0]\AgdaFunction{bound}\AgdaSpace{}%
\AgdaSymbol{:\{}\AgdaBound{xs}\AgdaSpace{}%
\AgdaSymbol{:}\AgdaSpace{}%
\AgdaDatatype{Colist}\AgdaSpace{}%
\AgdaDatatype{Nat}\AgdaSpace{}%
\AgdaPostulate{∞}\AgdaSymbol{\}\{}\AgdaBound{n}\AgdaSpace{}%
\AgdaSymbol{:}\AgdaSpace{}%
\AgdaDatatype{Nat}\AgdaSymbol{\}}\AgdaSpace{}%
\AgdaSymbol{→}\AgdaSpace{}%
\AgdaFunction{S}\AgdaSpace{}%
\AgdaBound{n}\AgdaSpace{}%
\AgdaBound{xs}\AgdaSpace{}%
\AgdaSymbol{→}\AgdaSpace{}%
\AgdaBound{n}\AgdaSpace{}%
\AgdaOperator{\AgdaDatatype{maxElem{-}ind}}\AgdaSpace{}%
\AgdaBound{xs}\<%
\\
\>[0]\AgdaFunction{bound}\AgdaSpace{}%
\AgdaBound{Snxs}\AgdaSpace{}%
\AgdaSymbol{=}\AgdaSpace{}%
\AgdaFunction{bound{-}aux}\AgdaSpace{}%
\AgdaSymbol{(}\AgdaFunction{∧{-}left}\AgdaSpace{}%
\AgdaBound{Snxs}\AgdaSymbol{)}\AgdaSpace{}%
\AgdaSymbol{(}\AgdaFunction{∧{-}right}\AgdaSpace{}%
\AgdaBound{Snxs}\AgdaSymbol{)}\<%
\end{code}
\end{center}

The proof is by induction over \textit{memberOf} referred to the maximum. 
If it is the head, then the coaxiom is applied. There is no need of analyzing when the list has more than one element since the axiom is dealt inside the coaxiom. If the maximum is not the head, then we know that is it is the maximum of the tail by inductive hypothesis, that is the recursive call and that it is greater than the head thanks to second part of the specification. Thus, the rule can be applied.
Now we can show the proof that the specification is consistent.

\begin{center}
\begin{code}
\>[0]\AgdaKeyword{postulate}\<%
\\
\>[0][@{}l@{\AgdaIndent{0}}]%
\>[2]\AgdaPostulate{cons{-}aux}\AgdaSpace{}%
\AgdaSymbol{:}\AgdaSpace{}%
\AgdaSymbol{∀}\AgdaSpace{}%
\AgdaSymbol{\{}\AgdaBound{n}\AgdaSpace{}%
\AgdaBound{x}\AgdaSpace{}%
\AgdaBound{y}%
\>[784I]\AgdaBound{xs}\AgdaSpace{}%
\AgdaBound{ys}\AgdaSymbol{\}}\AgdaSpace{}%
\AgdaSymbol{→}\AgdaSpace{}%
\AgdaField{Thunk.force}\AgdaSpace{}%
\AgdaBound{xs}\AgdaSpace{}%
\AgdaOperator{\AgdaDatatype{≡}}\AgdaSpace{}%
\AgdaSymbol{(}\AgdaBound{y}\AgdaSpace{}%
\AgdaOperator{\AgdaInductiveConstructor{∷}}\AgdaSpace{}%
\AgdaBound{ys}\AgdaSymbol{)}\AgdaSpace{}%
\AgdaSymbol{→}\<%
\\
\>[.]\<[784I]%
\>[22]\AgdaFunction{S}\AgdaSpace{}%
\AgdaBound{n}\AgdaSpace{}%
\AgdaSymbol{(}\AgdaBound{x}\AgdaSpace{}%
\AgdaOperator{\AgdaInductiveConstructor{∷}}\AgdaSpace{}%
\AgdaBound{xs}\AgdaSymbol{)}\AgdaSpace{}%
\AgdaSymbol{→}\AgdaSpace{}%
\AgdaDatatype{∃}\AgdaSpace{}%
\AgdaDatatype{Nat}\AgdaSpace{}%
\AgdaSymbol{(λ}\AgdaSpace{}%
\AgdaBound{m}\AgdaSpace{}%
\AgdaSymbol{→}\AgdaSpace{}%
\AgdaFunction{S}\AgdaSpace{}%
\AgdaBound{m}\AgdaSpace{}%
\AgdaSymbol{(}\AgdaBound{y}\AgdaSpace{}%
\AgdaOperator{\AgdaInductiveConstructor{∷}}\AgdaSpace{}%
\AgdaBound{ys}\AgdaSymbol{))}\<%
\\
\\[\AgdaEmptyExtraSkip]%
\>[0]\AgdaFunction{consistent}%
\>[809I]\AgdaSymbol{:\{}\AgdaBound{ys}\AgdaSpace{}%
\AgdaSymbol{:}\AgdaSpace{}%
\AgdaDatatype{Colist}\AgdaSpace{}%
\AgdaDatatype{Nat}\AgdaSpace{}%
\AgdaPostulate{∞}\AgdaSymbol{\}\{}\AgdaBound{n}\AgdaSpace{}%
\AgdaSymbol{:}\AgdaSpace{}%
\AgdaDatatype{Nat}\AgdaSymbol{\}}\AgdaSpace{}%
\AgdaSymbol{→}\<%
\\
\>[.]\<[809I]%
\>[11]\AgdaFunction{S}\AgdaSpace{}%
\AgdaBound{n}\AgdaSpace{}%
\AgdaBound{ys}\AgdaSpace{}%
\AgdaSymbol{→}\AgdaSpace{}%
\AgdaDatatype{Step}\AgdaSpace{}%
\AgdaFunction{S}\AgdaSpace{}%
\AgdaBound{n}\AgdaSpace{}%
\AgdaBound{ys}\<%
\\
\\[\AgdaEmptyExtraSkip]%
\>[0]\AgdaFunction{consistent}\AgdaSpace{}%
\AgdaSymbol{\{}\AgdaInductiveConstructor{[]}\AgdaSymbol{\}}\AgdaSpace{}%
\AgdaSymbol{\{}\AgdaBound{n}\AgdaSymbol{\}}\AgdaSpace{}%
\AgdaOperator{\AgdaInductiveConstructor{⟨}}\AgdaSpace{}%
\AgdaSymbol{()}\AgdaSpace{}%
\AgdaOperator{\AgdaInductiveConstructor{,}}\AgdaSpace{}%
\AgdaBound{gr}\AgdaSpace{}%
\AgdaOperator{\AgdaInductiveConstructor{⟩}}\<%
\\
\>[0]\AgdaFunction{consistent}\AgdaSpace{}%
\AgdaSymbol{\{}\AgdaBound{x}\AgdaSpace{}%
\AgdaOperator{\AgdaInductiveConstructor{∷}}\AgdaSpace{}%
\AgdaBound{xs}\AgdaSymbol{\}}\AgdaSpace{}%
\AgdaSymbol{\{}\AgdaBound{n}\AgdaSymbol{\}}\AgdaSpace{}%
\AgdaBound{Sl}\AgdaSpace{}%
\AgdaKeyword{with}\AgdaSpace{}%
\AgdaFunction{inspect}\AgdaSpace{}%
\AgdaSymbol{(}\AgdaField{Thunk.force}\AgdaSpace{}%
\AgdaBound{xs}\AgdaSymbol{)}\<%
\\
\\[\AgdaEmptyExtraSkip]%
\>[0]\AgdaFunction{consistent}\AgdaSpace{}%
\AgdaSymbol{\{}\AgdaBound{x}\AgdaSpace{}%
\AgdaOperator{\AgdaInductiveConstructor{∷}}\AgdaSpace{}%
\AgdaBound{xs}\AgdaSymbol{\}}\AgdaSpace{}%
\AgdaSymbol{\{}\AgdaSymbol{\AgdaDottedPattern{.}}\AgdaBound{\AgdaDottedPattern{x}}\AgdaSymbol{\}}\AgdaSpace{}%
\AgdaOperator{\AgdaInductiveConstructor{⟨}}\AgdaSpace{}%
\AgdaInductiveConstructor{mem{-}h}\AgdaSpace{}%
\AgdaOperator{\AgdaInductiveConstructor{,}}\AgdaSpace{}%
\AgdaBound{gr}\AgdaSpace{}%
\AgdaOperator{\AgdaInductiveConstructor{⟩}}\AgdaSpace{}%
\AgdaSymbol{|}\AgdaSpace{}%
\AgdaInductiveConstructor{[]}\AgdaSpace{}%
\AgdaOperator{\AgdaInductiveConstructor{with≡}}\AgdaSpace{}%
\AgdaBound{eq}\AgdaSpace{}%
\AgdaSymbol{=}\AgdaSpace{}%
\AgdaInductiveConstructor{max{-}h}\AgdaSpace{}%
\AgdaBound{eq}\<%
\\
\>[0]\AgdaFunction{consistent}\AgdaSpace{}%
\AgdaSymbol{\{}\AgdaBound{x}\AgdaSpace{}%
\AgdaOperator{\AgdaInductiveConstructor{∷}}\AgdaSpace{}%
\AgdaBound{xs}\AgdaSymbol{\}}\AgdaSpace{}%
\AgdaSymbol{\{}\AgdaBound{n}\AgdaSymbol{\}}\AgdaSpace{}%
\AgdaOperator{\AgdaInductiveConstructor{⟨}}\AgdaSpace{}%
\AgdaInductiveConstructor{mem{-}t}\AgdaSpace{}%
\AgdaBound{mem}\AgdaSpace{}%
\AgdaOperator{\AgdaInductiveConstructor{,}}\AgdaSpace{}%
\AgdaBound{gr}\AgdaSpace{}%
\AgdaOperator{\AgdaInductiveConstructor{⟩}}\AgdaSpace{}%
\AgdaSymbol{|}\AgdaSpace{}%
\AgdaInductiveConstructor{[]}\AgdaSpace{}%
\AgdaOperator{\AgdaInductiveConstructor{with≡}}\AgdaSpace{}%
\AgdaBound{eq}\AgdaSpace{}%
\AgdaSymbol{=}\AgdaSpace{}%
\AgdaFunction{⊥{-}elim}\AgdaSpace{}%
\AgdaSymbol{(}\AgdaFunction{abs}\AgdaSpace{}%
\AgdaBound{mem}\AgdaSpace{}%
\AgdaBound{eq}\AgdaSymbol{)}\<%
\\
\\[\AgdaEmptyExtraSkip]%
\>[0]\AgdaFunction{consistent}\AgdaSpace{}%
\AgdaSymbol{\{}\AgdaBound{x}%
\>[876I]\AgdaOperator{\AgdaInductiveConstructor{∷}}\AgdaSpace{}%
\AgdaBound{xs}\AgdaSymbol{\}}\AgdaSpace{}%
\AgdaSymbol{\{}\AgdaSymbol{\AgdaDottedPattern{.}}\AgdaBound{\AgdaDottedPattern{x}}\AgdaSymbol{\}}\AgdaSpace{}%
\AgdaOperator{\AgdaInductiveConstructor{⟨}}\AgdaSpace{}%
\AgdaInductiveConstructor{mem{-}h}\AgdaSpace{}%
\AgdaOperator{\AgdaInductiveConstructor{,}}\AgdaSpace{}%
\AgdaBound{gr}\AgdaSpace{}%
\AgdaOperator{\AgdaInductiveConstructor{⟩}}\AgdaSpace{}%
\AgdaSymbol{|}\AgdaSpace{}%
\AgdaSymbol{(}\AgdaBound{y}\AgdaSpace{}%
\AgdaOperator{\AgdaInductiveConstructor{∷}}\AgdaSpace{}%
\AgdaBound{ys}\AgdaSymbol{)}\AgdaSpace{}%
\AgdaOperator{\AgdaInductiveConstructor{with≡}}\AgdaSpace{}%
\AgdaBound{eq}\AgdaSpace{}%
\AgdaSymbol{=}\<%
\\
\>[.]\<[876I]%
\>[14]\AgdaKeyword{let}\AgdaSpace{}%
\AgdaBound{wit{-}aux}\AgdaSpace{}%
\AgdaSymbol{=}\AgdaSpace{}%
\AgdaFunction{witness}\AgdaSpace{}%
\AgdaSymbol{(}\AgdaPostulate{cons{-}aux}\AgdaSpace{}%
\AgdaBound{eq}\AgdaSpace{}%
\AgdaOperator{\AgdaInductiveConstructor{⟨}}\AgdaSpace{}%
\AgdaInductiveConstructor{mem{-}h}\AgdaSpace{}%
\AgdaOperator{\AgdaInductiveConstructor{,}}\AgdaSpace{}%
\AgdaBound{gr}\AgdaSpace{}%
\AgdaOperator{\AgdaInductiveConstructor{⟩}}\AgdaSymbol{)}\AgdaSpace{}%
\AgdaKeyword{in}\<%
\\
\>[14]\AgdaKeyword{let}\AgdaSpace{}%
\AgdaBound{pr{-}aux}\AgdaSpace{}%
\AgdaSymbol{=}\AgdaSpace{}%
\AgdaFunction{proof}\AgdaSpace{}%
\AgdaSymbol{(}\AgdaPostulate{cons{-}aux}\AgdaSpace{}%
\AgdaBound{eq}\AgdaSpace{}%
\AgdaOperator{\AgdaInductiveConstructor{⟨}}\AgdaSpace{}%
\AgdaInductiveConstructor{mem{-}h}\AgdaSpace{}%
\AgdaOperator{\AgdaInductiveConstructor{,}}\AgdaSpace{}%
\AgdaBound{gr}\AgdaSpace{}%
\AgdaOperator{\AgdaInductiveConstructor{⟩}}\AgdaSymbol{)}\AgdaSpace{}%
\AgdaKeyword{in}\<%
\\
\>[14]\AgdaInductiveConstructor{max{-}t}%
\>[913I]\AgdaSymbol{\{}\AgdaArgument{n}\AgdaSpace{}%
\AgdaSymbol{=}\AgdaSpace{}%
\AgdaBound{wit{-}aux}\AgdaSymbol{\}\{}\AgdaArgument{x}\AgdaSpace{}%
\AgdaSymbol{=}\AgdaSpace{}%
\AgdaBound{x}\AgdaSymbol{\}\{}\AgdaArgument{z}\AgdaSpace{}%
\AgdaSymbol{=}\AgdaSpace{}%
\AgdaBound{x}\AgdaSymbol{\}}\<%
\\
\>[913I][@{}l@{\AgdaIndent{0}}]%
\>[21]\AgdaComment{{-}{-} There exists a maximum in the tail}\<%
\\
\>[21]\AgdaSymbol{(}\AgdaFunction{subst}\AgdaSpace{}%
\AgdaSymbol{(λ}\AgdaSpace{}%
\AgdaBound{ll}\AgdaSpace{}%
\AgdaSymbol{→}\AgdaSpace{}%
\AgdaFunction{S}\AgdaSpace{}%
\AgdaBound{wit{-}aux}\AgdaSpace{}%
\AgdaBound{ll}\AgdaSymbol{)}\AgdaSpace{}%
\AgdaSymbol{(}\AgdaFunction{sym}\AgdaSpace{}%
\AgdaBound{eq}\AgdaSymbol{)}\AgdaSpace{}%
\AgdaBound{pr{-}aux}\AgdaSymbol{)}\<%
\\
\>[21]\AgdaSymbol{(}\AgdaFunction{max{-}comm}\AgdaSpace{}%
\AgdaSymbol{\{}\AgdaArgument{a}\AgdaSpace{}%
\AgdaSymbol{=}\AgdaSpace{}%
\AgdaBound{x}\AgdaSymbol{\}\{}\AgdaArgument{b}\AgdaSpace{}%
\AgdaSymbol{=}\AgdaSpace{}%
\AgdaBound{wit{-}aux}\AgdaSymbol{\}\{}\AgdaArgument{c}\AgdaSpace{}%
\AgdaSymbol{=}\AgdaSpace{}%
\AgdaBound{x}\AgdaSymbol{\}}\AgdaSpace{}%
\AgdaSymbol{(}\AgdaBound{gr}\AgdaSpace{}%
\AgdaSymbol{(}\AgdaInductiveConstructor{mem{-}t}\<%
\\
\>[21]\AgdaComment{{-}{-} gr applied to the membership element of the tail max}\<%
\\
\>[21]\AgdaSymbol{(}\AgdaFunction{subst}\AgdaSpace{}%
\AgdaSymbol{(λ}\AgdaSpace{}%
\AgdaBound{ll}\AgdaSpace{}%
\AgdaSymbol{→}\AgdaSpace{}%
\AgdaBound{wit{-}aux}\AgdaSpace{}%
\AgdaOperator{\AgdaDatatype{memberOf}}\AgdaSpace{}%
\AgdaBound{ll}\AgdaSymbol{)}\AgdaSpace{}%
\AgdaSymbol{(}\AgdaFunction{sym}\AgdaSpace{}%
\AgdaBound{eq}\AgdaSymbol{)}\AgdaSpace{}%
\AgdaSymbol{(}\AgdaFunction{∧{-}left}\AgdaSpace{}%
\AgdaBound{pr{-}aux}\AgdaSymbol{)))))}\<%
\\
\>[0]\<%
\\
\>[0]\AgdaFunction{consistent}\AgdaSpace{}%
\AgdaSymbol{\{}\AgdaBound{x}%
\>[949I]\AgdaOperator{\AgdaInductiveConstructor{∷}}\AgdaSpace{}%
\AgdaBound{xs}\AgdaSymbol{\}}\AgdaSpace{}%
\AgdaSymbol{\{}\AgdaBound{n}\AgdaSymbol{\}}\AgdaSpace{}%
\AgdaOperator{\AgdaInductiveConstructor{⟨}}\AgdaSpace{}%
\AgdaInductiveConstructor{mem{-}t}\AgdaSpace{}%
\AgdaBound{mem}\AgdaSpace{}%
\AgdaOperator{\AgdaInductiveConstructor{,}}\AgdaSpace{}%
\AgdaBound{gr}\AgdaSpace{}%
\AgdaOperator{\AgdaInductiveConstructor{⟩}}\AgdaSpace{}%
\AgdaSymbol{|}\AgdaSpace{}%
\AgdaSymbol{(}\AgdaBound{y}\AgdaSpace{}%
\AgdaOperator{\AgdaInductiveConstructor{∷}}\AgdaSpace{}%
\AgdaBound{ys}\AgdaSymbol{)}\AgdaSpace{}%
\AgdaOperator{\AgdaInductiveConstructor{with≡}}\AgdaSpace{}%
\AgdaBound{eq}\AgdaSpace{}%
\AgdaSymbol{=}\<%
\\
\>[.]\<[949I]%
\>[14]\AgdaInductiveConstructor{max{-}t}\AgdaSpace{}%
\AgdaSymbol{(}\AgdaOperator{\AgdaInductiveConstructor{⟨}}\AgdaSpace{}%
\AgdaBound{mem}\AgdaSpace{}%
\AgdaOperator{\AgdaInductiveConstructor{,}}\AgdaSpace{}%
\AgdaSymbol{(λ}\AgdaSpace{}%
\AgdaBound{m}\AgdaSpace{}%
\AgdaSymbol{→}\AgdaSpace{}%
\AgdaBound{gr}\AgdaSpace{}%
\AgdaSymbol{(}\AgdaInductiveConstructor{mem{-}t}\AgdaSpace{}%
\AgdaBound{m}\AgdaSymbol{))}\AgdaSpace{}%
\AgdaOperator{\AgdaInductiveConstructor{⟩}}\AgdaSymbol{)}\AgdaSpace{}%
\AgdaSymbol{(}\AgdaBound{gr}\AgdaSpace{}%
\AgdaInductiveConstructor{mem{-}h}\AgdaSymbol{)}\<%
\end{code}
\end{center}

The proof is on how the tail is observed and then on the position of the maximum again in terms of \textit{memberOf}. The code is organized according to the different cases and follows the same path of the theoretical proof. \agda{inspect} is the function of the module \textit{Singleton}.
\begin{itemize}
\item If the list is empty, then no constructors of \textit{memberOf} can be applied. This is an absurd case and it is automatically managed by Agda when trying to pattern match on the element of type \textit{memberOf}.
\item If the tail is empty, that is \agda{Thunk.force xs = []}:
\begin{itemize}
\item If the maximum is the head, then the axiom is applied.
\item If the maximum is not the head, then we are again in an absurd case that this time has to be managed through \agda{abs}.
\end{itemize}
\item If the tail is not empty such that it is built using \agda{y :: ys}:
\begin{itemize}
\item If the maximum is the head it means that there exists a number that is the maximum of the tail but it is lower that $n$. So the rule is applied with this two proofs as input parameters. This elements are obtained through \agda{cons-aux}. In a more detailed way, the proof that $n$ is the maximum between itself and the previous maximum is the result of \agda{gr} applied to the membership proof of this last maximum. Clearly the constructor \agda{mem-t} has to be used since the position of the maximum of the list is referred to the tail and not to the whole list.
\item If the maximum is not the head, it means that it is also the maximum of the tail $xs$. Thus the rule is applied. The proof that $n$ is the maximum between the maximum of the tail and $x$ is obtained by applying \agda{gr} to the constructor of \textit{memberOf} that identifies the head. On the other hand, we can give to the constructor \agda{meof-t} the couple whose first element basically says that $n$ is inside the tail while the second one is a restriction of \agda{gr} to the elements of $xs$.
\end{itemize}
\end{itemize}

\LC{Notice that the statement \agda{cons-aux} that tells that there exists a maximum in the tail is postulated. This is because it cannot be proved since proofs by absurd cannot be made in Agda. In fact, in intuitionistic logic the existence of a value is proved by providing that value; in this case this cannot be achieved since to get the maximum of an infinite list we should check each element.}
Finally we can prove the completeness of the predicate through the bounded coinduction principle that we already proved.

\begin{center}
\begin{code}
\>[0]\AgdaFunction{max{-}complete}\AgdaSpace{}%
\AgdaSymbol{:}\AgdaSpace{}%
\AgdaSymbol{∀}\AgdaSpace{}%
\AgdaSymbol{\{}\AgdaBound{i}\AgdaSymbol{\}\{}\AgdaBound{n}\AgdaSymbol{\}\{}\AgdaBound{xs}\AgdaSymbol{\}}\AgdaSpace{}%
\AgdaSymbol{→}\AgdaSpace{}%
\AgdaFunction{S}\AgdaSpace{}%
\AgdaBound{n}\AgdaSpace{}%
\AgdaBound{xs}\AgdaSpace{}%
\AgdaSymbol{→}\AgdaSpace{}%
\AgdaSymbol{(}\AgdaBound{n}\AgdaSpace{}%
\AgdaOperator{\AgdaDatatype{maxElem}}\AgdaSpace{}%
\AgdaBound{xs}\AgdaSymbol{)}\AgdaSpace{}%
\AgdaBound{i}\<%
\\
\>[0]\AgdaFunction{max{-}complete}\AgdaSpace{}%
\AgdaSymbol{\{}\AgdaBound{i}\AgdaSymbol{\}}\AgdaSpace{}%
\AgdaSymbol{\{}\AgdaBound{n}\AgdaSymbol{\}}\AgdaSpace{}%
\AgdaSymbol{\{}\AgdaBound{xs}\AgdaSymbol{\}}\AgdaSpace{}%
\AgdaBound{Sxs}\AgdaSpace{}%
\AgdaSymbol{=}\AgdaSpace{}%
\AgdaFunction{bd{-}c}\AgdaSpace{}%
\AgdaFunction{S}\AgdaSpace{}%
\AgdaFunction{bound}\AgdaSpace{}%
\AgdaFunction{consistent}\AgdaSpace{}%
\AgdaBound{Sxs}\<%
\end{code}
\end{center}

Where $S$ is the specification given before.\\
To sum up, we presented a guideline to deal with generalized inference systems in Agda. For the codes we referred only to the standard library implementation of colists in order to avoid the issues that we met in the previous chapters. Nevertheless it can be noticed that sometimes we faced some difficulties. As an example the constructor of the axiom of our types take in input the proof that the tail of the list is empty. If we had directly written the empty list as right argument to \agda{::} we would have met errors in the proofs stating that two empty lists that are built using \agda{[]} are not the same. The input proof on how the tail is built allows us to refer to all the lists that can be observed as \agda{[]}.
Another problem was the usage of the \textit{max} function instead of a specific data type. We had to implement \agda{max-refl} in order to keep the connections between the inputs and the output of \textit{max}. These relations are considered in terms of equality proofs between variables and adopted in order to make proofs by substitution.

\chapter{Temporal operators}
\label{chapter:temp-op}
In this chapter, we consider an interesting case study, that is, three parametric predicates on (possibly infinite) lists which correspond to three operators of linear temporal logic:
\emph{Eventually} (also called \textit{F}), \emph{Always} (also called \textit{G}),  and \emph{Infinitely Often}. The first two predicates can be seen as a generalization of $\memberOf{\x}{\_}$ and \textit{allPos}, respectively, and indeed their correctness proofs are very similar to those presented in \refToChapter{inductive} and \refToChapter{coinductive}. We propose their implementation both as library colists and as coinductive records. The predicate \emph{Infinitely Often} could be defined by stratification of the inference systems for the two basic operators. We show an alternative approach where \textit{Infinitely Often} is directly defined by an inference system with corules, needed since this predicate is neither inductive, nor purely coinductive, and we apply the techniques developed in \refToChapter{flex} to prove its correctness.
In this case we provide the code only for library colists.

\section{Eventually}\label{sect:eventually}

{Given an arbitrary predicate $\Pred$ on list elements, $\Eventually{\Pred} {\xs}$ holds if there is at least one element $\x$ in $\xs$ such that $\Pred(\x)$ holds.} 
This parametric predicate can be defined by the following inference system interpreted inductively.

\begin{small}
\begin{quote}
$\NamedRule{ev-h}{}{\Eventually {\Pred} {\x{:}\xs}}\Pred(\x)$\Space
$\NamedRule{ev-t}{\Eventually{\Pred}{\xs}}{\Eventually {\Pred} {\x{:}\xs}}$
\end{quote}
\end{small}

{Note that the previously considered predicate $\memberOf{\y}{\_}$ can be obtained as the instance of \textit{Eventually} on $\Pred(\x)=\y$.}

{The (parametric) specification of the predicate is $\Spec_\Pred = \{ \xs \mid \exists \x. \memberOf {\x} {\xs} \land \Pred(\x) \}$.}

\begin{statement}
The inductive definition of \textit{Eventually} is sound with respect to its specification.
\end{statement}
\begin{proof}
We have to prove that, if $\Eventually{\Pred}{\xs}\in\Ind{\is}$, then there exists $\x$ such that $\memberOf{\x}{\xs}$ and $\Pred(\x)$ holds. The proof is by induction on the definition of $\Eventually{\Pred}{\xs}$. 
\begin{description}
\item[\refToRule{ev-h}] We have $\Eventually {\Pred} {\x{:}\xs}$, and $\Pred(\x)$. Moreover, $\memberOf{\x}{\xs}$ holds by \refToRule{mem-h}, hence the thesis holds. 
\item[\refToRule{ev-t}] We have $\Eventually {\Pred} {\x{:}\xs}$, and $\Eventually {\Pred} {\xs}$. Hence, by inductive hypothesis, there exists $\y$ such that $\memberOf{\y}{\xs}$ and $\Pred(\y)$ holds. Then, we can apply rule \refToRule{mem-t} and get the thesis.
\end{description}
\end{proof}

\begin{statement}
The inductive definition of \textit{Eventually} is complete with respect to its specification.
\end{statement}
\begin{proof}
We have to prove that, if there exists $\x$ such that $\memberOf{\x}{\xs}$ and $\Pred(\x)$ hold, then $\Eventually {\Pred} {\xs}$ can be derived.
The proof is by induction on the definition of $\memberOf{\x}{\xs}$.
\begin{description}
\item[\refToRule{mem-h}] We have $\memberOf{\x}{\x{:}\xs}$, and $\Pred(\x)$ holds. Then, $\Eventually {\Pred} {\x{:}\xs}$ can be derived by rule \refToRule{ev-h}.
\item[\refToRule{mem-t}] We have $\memberOf{\x}{\y{:}\xs}$, $\memberOf{\x}{\xs}$, and $\Pred(\x)$ holds. By inductive hypothesis we get that $\Eventually {\Pred} {\xs}$ can be derived. 
Hence, $\Eventually {\Pred} {\y{:}\xs}$ can be derived by \refToRule{ev-t}.
\end{description}
\end{proof}

Now we describe the Agda implementation.
This first operator tells that the involved predicate must hold at least once thus, the implementations follow an inductive scheme. We begin by showing the codes for standard library colists. These are the specification and the type representing the operator.

\begin{center}
\begin{code}%
\>[0]\AgdaFunction{S}\AgdaSpace{}%
\AgdaSymbol{:}\AgdaSpace{}%
\AgdaSymbol{\{}\AgdaBound{A}\AgdaSpace{}%
\AgdaSymbol{:}\AgdaSpace{}%
\AgdaPrimitiveType{Set}\AgdaSymbol{\}}\AgdaSpace{}%
\AgdaSymbol{→}\AgdaSpace{}%
\AgdaSymbol{(}\AgdaBound{A}\AgdaSpace{}%
\AgdaSymbol{→}\AgdaSpace{}%
\AgdaPrimitiveType{Set}\AgdaSymbol{)}\AgdaSpace{}%
\AgdaSymbol{→}\AgdaSpace{}%
\AgdaDatatype{Colist}\AgdaSpace{}%
\AgdaBound{A}\AgdaSpace{}%
\AgdaPostulate{∞}\AgdaSpace{}%
\AgdaSymbol{→}\AgdaSpace{}%
\AgdaPrimitiveType{Set}\<%
\\
\>[0]\AgdaFunction{S}\AgdaSpace{}%
\AgdaSymbol{\{}\AgdaBound{A}\AgdaSymbol{\}}\AgdaSpace{}%
\AgdaBound{P}\AgdaSpace{}%
\AgdaBound{xs}\AgdaSpace{}%
\AgdaSymbol{=}\AgdaSpace{}%
\AgdaDatatype{∃}\AgdaSpace{}%
\AgdaBound{A}\AgdaSpace{}%
\AgdaSymbol{(λ}\AgdaSpace{}%
\AgdaBound{x}\AgdaSpace{}%
\AgdaSymbol{→}\AgdaSpace{}%
\AgdaSymbol{(}\AgdaBound{x}\AgdaSpace{}%
\AgdaOperator{\AgdaDatatype{memberOf}}\AgdaSpace{}%
\AgdaBound{xs}\AgdaSymbol{)}\AgdaSpace{}%
\AgdaOperator{\AgdaDatatype{∧}}\AgdaSpace{}%
\AgdaSymbol{(}\AgdaBound{P}\AgdaSpace{}%
\AgdaBound{x}\AgdaSymbol{))}\<%
\\
\\[\AgdaEmptyExtraSkip]%
\>[0]\AgdaKeyword{data}\AgdaSpace{}%
\AgdaOperator{\AgdaDatatype{\AgdaUnderscore{}Eventually\AgdaUnderscore{}}}\AgdaSpace{}%
\AgdaSymbol{\{}\AgdaBound{A}\AgdaSpace{}%
\AgdaSymbol{:}\AgdaSpace{}%
\AgdaPrimitiveType{Set}\AgdaSymbol{\}}\AgdaSpace{}%
\AgdaSymbol{:}\AgdaSpace{}%
\AgdaSymbol{(}\AgdaBound{A}\AgdaSpace{}%
\AgdaSymbol{→}\AgdaSpace{}%
\AgdaPrimitiveType{Set}\AgdaSymbol{)}\AgdaSpace{}%
\AgdaSymbol{→}\AgdaSpace{}%
\AgdaDatatype{Colist}\AgdaSpace{}%
\AgdaBound{A}\AgdaSpace{}%
\AgdaPostulate{∞}\AgdaSpace{}%
\AgdaSymbol{→}\AgdaSpace{}%
\AgdaPrimitiveType{Set}\AgdaSpace{}%
\AgdaKeyword{where}\<%
\\
\>[0][@{}l@{\AgdaIndent{0}}]%
\>[2]\AgdaInductiveConstructor{ev{-}h}\AgdaSpace{}%
\AgdaSymbol{:}\AgdaSpace{}%
\AgdaSymbol{∀}\AgdaSpace{}%
\AgdaSymbol{\{}\AgdaBound{x}\AgdaSpace{}%
\AgdaBound{xs}\AgdaSpace{}%
\AgdaBound{P}\AgdaSymbol{\}}\AgdaSpace{}%
\AgdaSymbol{→}\AgdaSpace{}%
\AgdaBound{P}\AgdaSpace{}%
\AgdaBound{x}\AgdaSpace{}%
\AgdaSymbol{→}\AgdaSpace{}%
\AgdaBound{P}\AgdaSpace{}%
\AgdaOperator{\AgdaDatatype{Eventually}}\AgdaSpace{}%
\AgdaSymbol{(}\AgdaBound{x}\AgdaSpace{}%
\AgdaOperator{\AgdaInductiveConstructor{∷}}\AgdaSpace{}%
\AgdaBound{xs}\AgdaSymbol{)}\<%
\\
\>[2]\AgdaInductiveConstructor{ev{-}t}\AgdaSpace{}%
\AgdaSymbol{:}\AgdaSpace{}%
\AgdaSymbol{∀}\AgdaSpace{}%
\AgdaSymbol{\{}\AgdaBound{x}\AgdaSpace{}%
\AgdaBound{xs}\AgdaSpace{}%
\AgdaBound{P}\AgdaSymbol{\}}\AgdaSpace{}%
\AgdaSymbol{→}\AgdaSpace{}%
\AgdaBound{P}\AgdaSpace{}%
\AgdaOperator{\AgdaDatatype{Eventually}}\AgdaSpace{}%
\AgdaSymbol{(}\AgdaField{Thunk.force}\AgdaSpace{}%
\AgdaBound{xs}\AgdaSymbol{)}\AgdaSpace{}%
\AgdaSymbol{→}\AgdaSpace{}%
\AgdaBound{P}\AgdaSpace{}%
\AgdaOperator{\AgdaDatatype{Eventually}}\AgdaSpace{}%
\AgdaSymbol{(}\AgdaBound{x}\AgdaSpace{}%
\AgdaOperator{\AgdaInductiveConstructor{∷}}\AgdaSpace{}%
\AgdaBound{xs}\AgdaSymbol{)}\<%
\end{code}

\end{center}

Notice that the resulting type does not depend on \agda{Size} and \agda{Thunk} because it is inductive and that it shows some similarities with \textit{memberOf}. In fact, if an element is inside a list this means that it can be found at least once.

\begin{center}
\begin{code}
\>[0]\AgdaFunction{ev{-}sound}%
\>[88I]\AgdaSymbol{:}\AgdaSpace{}%
\AgdaSymbol{\{}\AgdaBound{A}\AgdaSpace{}%
\AgdaSymbol{:}\AgdaSpace{}%
\AgdaPrimitiveType{Set}\AgdaSymbol{\}\{}\AgdaBound{P}\AgdaSpace{}%
\AgdaSymbol{:}\AgdaSpace{}%
\AgdaBound{A}\AgdaSpace{}%
\AgdaSymbol{→}\AgdaSpace{}%
\AgdaPrimitiveType{Set}\AgdaSymbol{\}\{}\AgdaBound{ys}\AgdaSpace{}%
\AgdaSymbol{:}\AgdaSpace{}%
\AgdaDatatype{Colist}\AgdaSpace{}%
\AgdaBound{A}\AgdaSpace{}%
\AgdaPostulate{∞}\AgdaSymbol{\}}\AgdaSpace{}%
\AgdaSymbol{→}\AgdaSpace{}%
\AgdaBound{P}\AgdaSpace{}%
\AgdaOperator{\AgdaDatatype{Eventually}}\AgdaSpace{}%
\AgdaBound{ys}\AgdaSpace{}%
\AgdaSymbol{→}\<%
\\
\>[.]\<[88I]%
\>[9]\AgdaDatatype{∃}\AgdaSpace{}%
\AgdaBound{A}\AgdaSpace{}%
\AgdaSymbol{(λ}\AgdaSpace{}%
\AgdaBound{x}\AgdaSpace{}%
\AgdaSymbol{→}\AgdaSpace{}%
\AgdaSymbol{(}\AgdaBound{x}\AgdaSpace{}%
\AgdaOperator{\AgdaDatatype{memberOf}}\AgdaSpace{}%
\AgdaBound{ys}\AgdaSymbol{)}\AgdaSpace{}%
\AgdaOperator{\AgdaDatatype{∧}}\AgdaSpace{}%
\AgdaSymbol{(}\AgdaBound{P}\AgdaSpace{}%
\AgdaBound{x}\AgdaSymbol{))}\<%
\\
\>[0]\<%
\\
\>[0]\AgdaFunction{ev{-}sound}\AgdaSpace{}%
\AgdaSymbol{\{}\AgdaBound{A}\AgdaSymbol{\}}\AgdaSpace{}%
\AgdaSymbol{\{}\AgdaBound{P}\AgdaSymbol{\}}\AgdaSpace{}%
\AgdaSymbol{\{}\AgdaBound{x}\AgdaSpace{}%
\AgdaOperator{\AgdaInductiveConstructor{∷}}\AgdaSpace{}%
\AgdaBound{xs}\AgdaSymbol{\}}\AgdaSpace{}%
\AgdaSymbol{(}\AgdaInductiveConstructor{ev{-}h}\AgdaSpace{}%
\AgdaBound{Px}\AgdaSymbol{)}\AgdaSpace{}%
\AgdaSymbol{=}\AgdaSpace{}%
\AgdaOperator{\AgdaInductiveConstructor{<}}\AgdaSpace{}%
\AgdaBound{x}\AgdaSpace{}%
\AgdaOperator{\AgdaInductiveConstructor{,}}\AgdaSpace{}%
\AgdaOperator{\AgdaInductiveConstructor{⟨}}\AgdaSpace{}%
\AgdaInductiveConstructor{mem{-}h}\AgdaSpace{}%
\AgdaOperator{\AgdaInductiveConstructor{,}}\AgdaSpace{}%
\AgdaBound{Px}\AgdaSpace{}%
\AgdaOperator{\AgdaInductiveConstructor{⟩}}\AgdaSpace{}%
\AgdaOperator{\AgdaInductiveConstructor{>}}\<%
\\
\>[0]\AgdaFunction{ev{-}sound}%
\>[132I]\AgdaSymbol{\{}\AgdaBound{A}\AgdaSymbol{\}}\AgdaSpace{}%
\AgdaSymbol{\{}\AgdaBound{P}\AgdaSymbol{\}}\AgdaSpace{}%
\AgdaSymbol{\{}\AgdaBound{x}\AgdaSpace{}%
\AgdaOperator{\AgdaInductiveConstructor{∷}}\AgdaSpace{}%
\AgdaBound{xs}\AgdaSymbol{\}}\AgdaSpace{}%
\AgdaSymbol{(}\AgdaInductiveConstructor{ev{-}t}\AgdaSpace{}%
\AgdaBound{ev{-}xs}\AgdaSymbol{)}\AgdaSpace{}%
\AgdaSymbol{=}\AgdaSpace{}%
\AgdaOperator{\AgdaInductiveConstructor{<}}\<%
\\
\>[.]\<[132I]%
\>[9]\AgdaSymbol{(}\AgdaFunction{witness}\AgdaSpace{}%
\AgdaSymbol{(}\AgdaFunction{ev{-}sound}\AgdaSpace{}%
\AgdaSymbol{\{}\AgdaArgument{ys}\AgdaSpace{}%
\AgdaSymbol{=}\AgdaSpace{}%
\AgdaField{Thunk.force}\AgdaSpace{}%
\AgdaBound{xs}\AgdaSymbol{\}}\AgdaSpace{}%
\AgdaBound{ev{-}xs}\AgdaSymbol{))}\AgdaSpace{}%
\AgdaOperator{\AgdaInductiveConstructor{,}}\<%
\\
\>[9]\AgdaOperator{\AgdaInductiveConstructor{⟨}}\<%
\\
\>[9][@{}l@{\AgdaIndent{0}}]%
\>[11]\AgdaInductiveConstructor{mem{-}t}\AgdaSpace{}%
\AgdaSymbol{(}\AgdaFunction{∧{-}left}\AgdaSpace{}%
\AgdaSymbol{(}\AgdaFunction{proof}\AgdaSpace{}%
\AgdaSymbol{(}\AgdaFunction{ev{-}sound}\AgdaSpace{}%
\AgdaBound{ev{-}xs}\AgdaSymbol{)))}\AgdaSpace{}%
\AgdaOperator{\AgdaInductiveConstructor{,}}\<%
\\
\>[11]\AgdaFunction{∧{-}right}\AgdaSpace{}%
\AgdaSymbol{(}\AgdaFunction{proof}\AgdaSpace{}%
\AgdaSymbol{(}\AgdaFunction{ev{-}sound}\AgdaSpace{}%
\AgdaBound{ev{-}xs}\AgdaSymbol{))}\<%
\\
\>[9]\AgdaOperator{\AgdaInductiveConstructor{⟩}}\AgdaSpace{}%
\AgdaOperator{\AgdaInductiveConstructor{>}}\<%
\end{code}
\end{center}

The soundness proof is not as simple as the \textit{memberOf} one because if we use the \textit{lookup} function we have to manage the \agda{Maybe}. We decided not to use the function above but the \textit{memberOf} predicate that is correct since we provided the proofs. There can beother different strategies like lifting $P$ or giving a definition of \textit{memberOf} that involves also the indexes.
The proof is by induction over the rules of \textit{Eventually} and it can be noticed again in the first case that \agda{mem-h} can be used and Agda knows that we are referring to $x$.

\begin{center}
\begin{code}
\>[0]\AgdaFunction{ev{-}complete}%
\>[157I]\AgdaSymbol{:}\AgdaSpace{}%
\AgdaSymbol{\{}\AgdaBound{A}\AgdaSpace{}%
\AgdaSymbol{:}\AgdaSpace{}%
\AgdaPrimitiveType{Set}\AgdaSymbol{\}\{}\AgdaBound{P}\AgdaSpace{}%
\AgdaSymbol{:}\AgdaSpace{}%
\AgdaBound{A}\AgdaSpace{}%
\AgdaSymbol{→}\AgdaSpace{}%
\AgdaPrimitiveType{Set}\AgdaSymbol{\}\{}\AgdaBound{xs}\AgdaSpace{}%
\AgdaSymbol{:}\AgdaSpace{}%
\AgdaDatatype{Colist}\AgdaSpace{}%
\AgdaBound{A}\AgdaSpace{}%
\AgdaPostulate{∞}\AgdaSymbol{\}}\AgdaSpace{}%
\AgdaSymbol{→}\<%
\\
\>[.]\<[157I]%
\>[12]\AgdaDatatype{∃}\AgdaSpace{}%
\AgdaBound{A}\AgdaSpace{}%
\AgdaSymbol{(λ}\AgdaSpace{}%
\AgdaBound{x}\AgdaSpace{}%
\AgdaSymbol{→}\AgdaSpace{}%
\AgdaSymbol{(}\AgdaBound{x}\AgdaSpace{}%
\AgdaOperator{\AgdaDatatype{memberOf}}\AgdaSpace{}%
\AgdaBound{xs}\AgdaSymbol{)}\AgdaSpace{}%
\AgdaOperator{\AgdaDatatype{∧}}\AgdaSpace{}%
\AgdaSymbol{(}\AgdaBound{P}\AgdaSpace{}%
\AgdaBound{x}\AgdaSymbol{))}\AgdaSpace{}%
\AgdaSymbol{→}\AgdaSpace{}%
\AgdaBound{P}\AgdaSpace{}%
\AgdaOperator{\AgdaDatatype{Eventually}}\AgdaSpace{}%
\AgdaBound{xs}\<%
\\
\>[0]\<%
\\
\>[0]\AgdaFunction{ev{-}complete}\AgdaSpace{}%
\AgdaOperator{\AgdaInductiveConstructor{<}}\AgdaSpace{}%
\AgdaBound{x}\AgdaSpace{}%
\AgdaOperator{\AgdaInductiveConstructor{,}}\AgdaSpace{}%
\AgdaOperator{\AgdaInductiveConstructor{⟨}}\AgdaSpace{}%
\AgdaInductiveConstructor{mem{-}h}\AgdaSpace{}%
\AgdaOperator{\AgdaInductiveConstructor{,}}\AgdaSpace{}%
\AgdaBound{Px}\AgdaSpace{}%
\AgdaOperator{\AgdaInductiveConstructor{⟩}}\AgdaSpace{}%
\AgdaOperator{\AgdaInductiveConstructor{>}}\AgdaSpace{}%
\AgdaSymbol{=}\AgdaSpace{}%
\AgdaInductiveConstructor{ev{-}h}\AgdaSpace{}%
\AgdaBound{Px}\<%
\\
\>[0]\AgdaFunction{ev{-}complete}%
\>[196I]\AgdaOperator{\AgdaInductiveConstructor{<}}\AgdaSpace{}%
\AgdaBound{x}\AgdaSpace{}%
\AgdaOperator{\AgdaInductiveConstructor{,}}\AgdaSpace{}%
\AgdaOperator{\AgdaInductiveConstructor{⟨}}\AgdaSpace{}%
\AgdaInductiveConstructor{mem{-}t}\AgdaSpace{}%
\AgdaBound{mem}\AgdaSpace{}%
\AgdaOperator{\AgdaInductiveConstructor{,}}\AgdaSpace{}%
\AgdaBound{Px}\AgdaSpace{}%
\AgdaOperator{\AgdaInductiveConstructor{⟩}}\AgdaSpace{}%
\AgdaOperator{\AgdaInductiveConstructor{>}}\AgdaSpace{}%
\AgdaSymbol{=}\<%
\\
\>[.]\<[196I]%
\>[12]\AgdaInductiveConstructor{ev{-}t}\AgdaSpace{}%
\AgdaSymbol{(}\AgdaFunction{ev{-}complete}\AgdaSpace{}%
\AgdaOperator{\AgdaInductiveConstructor{<}}\AgdaSpace{}%
\AgdaBound{x}\AgdaSpace{}%
\AgdaOperator{\AgdaInductiveConstructor{,}}\AgdaSpace{}%
\AgdaOperator{\AgdaInductiveConstructor{⟨}}\AgdaSpace{}%
\AgdaBound{mem}\AgdaSpace{}%
\AgdaOperator{\AgdaInductiveConstructor{,}}\AgdaSpace{}%
\AgdaBound{Px}\AgdaSpace{}%
\AgdaOperator{\AgdaInductiveConstructor{⟩}}\AgdaSpace{}%
\AgdaOperator{\AgdaInductiveConstructor{>}}\AgdaSymbol{)}\<%
\end{code}
\end{center}

The proof is by induction on the rules of \textit{memberOf} as we stated in the theoretical discussion. In the first case $Px$ is the proof that the predicate holds for the head of the list. Now we show the implementations using our approach to colists.

\begin{center}
\begin{code}%
\>[0]\AgdaFunction{S}\AgdaSpace{}%
\AgdaSymbol{:}\AgdaSpace{}%
\AgdaSymbol{\{}\AgdaBound{A}\AgdaSpace{}%
\AgdaSymbol{:}\AgdaSpace{}%
\AgdaPrimitiveType{Set}\AgdaSymbol{\}}\AgdaSpace{}%
\AgdaSymbol{→}\AgdaSpace{}%
\AgdaSymbol{(}\AgdaBound{A}\AgdaSpace{}%
\AgdaSymbol{→}\AgdaSpace{}%
\AgdaPrimitiveType{Set}\AgdaSymbol{)}\AgdaSpace{}%
\AgdaSymbol{→}\AgdaSpace{}%
\AgdaRecord{MyColist}\AgdaSpace{}%
\AgdaBound{A}\AgdaSpace{}%
\AgdaSymbol{→}\AgdaSpace{}%
\AgdaPrimitiveType{Set}\<%
\\
\>[0]\AgdaFunction{S}\AgdaSpace{}%
\AgdaSymbol{\{}\AgdaBound{A}\AgdaSymbol{\}}\AgdaSpace{}%
\AgdaBound{P}\AgdaSpace{}%
\AgdaBound{xs}\AgdaSpace{}%
\AgdaSymbol{=}\AgdaSpace{}%
\AgdaDatatype{∃}\AgdaSpace{}%
\AgdaBound{A}\AgdaSpace{}%
\AgdaSymbol{(λ}\AgdaSpace{}%
\AgdaBound{x}\AgdaSpace{}%
\AgdaSymbol{→}\AgdaSpace{}%
\AgdaSymbol{(}\AgdaBound{x}\AgdaSpace{}%
\AgdaOperator{\AgdaDatatype{memberOf}}\AgdaSpace{}%
\AgdaBound{xs}\AgdaSymbol{)}\AgdaSpace{}%
\AgdaOperator{\AgdaDatatype{∧}}\AgdaSpace{}%
\AgdaSymbol{(}\AgdaBound{P}\AgdaSpace{}%
\AgdaBound{x}\AgdaSymbol{))}\<%
\\
\\[\AgdaEmptyExtraSkip]%
\>[0]\AgdaKeyword{data}\AgdaSpace{}%
\AgdaOperator{\AgdaDatatype{\AgdaUnderscore{}Eventually\AgdaUnderscore{}}}\AgdaSpace{}%
\AgdaSymbol{\{}\AgdaBound{A}\AgdaSpace{}%
\AgdaSymbol{:}\AgdaSpace{}%
\AgdaPrimitiveType{Set}\AgdaSymbol{\}(}\AgdaBound{P}\AgdaSpace{}%
\AgdaSymbol{:}\AgdaSpace{}%
\AgdaBound{A}\AgdaSpace{}%
\AgdaSymbol{→}\AgdaSpace{}%
\AgdaPrimitiveType{Set}\AgdaSymbol{)(}\AgdaBound{ys}\AgdaSpace{}%
\AgdaSymbol{:}\AgdaSpace{}%
\AgdaRecord{MyColist}\AgdaSpace{}%
\AgdaBound{A}\AgdaSymbol{)}\AgdaSpace{}%
\AgdaSymbol{:}\AgdaSpace{}%
\AgdaPrimitiveType{Set}\AgdaSpace{}%
\AgdaKeyword{where}\<%
\\
\>[0][@{}l@{\AgdaIndent{0}}]%
\>[2]\AgdaInductiveConstructor{ev{-}h}\AgdaSpace{}%
\AgdaSymbol{:}%
\>[67I]\AgdaSymbol{\{}\AgdaBound{x}\AgdaSpace{}%
\AgdaSymbol{:}\AgdaSpace{}%
\AgdaBound{A}\AgdaSymbol{\}\{}\AgdaBound{xs}\AgdaSpace{}%
\AgdaSymbol{:}\AgdaSpace{}%
\AgdaRecord{MyColist}\AgdaSpace{}%
\AgdaBound{A}\AgdaSymbol{\}}\AgdaSpace{}%
\AgdaSymbol{→}\AgdaSpace{}%
\AgdaSymbol{(}\AgdaField{MyColist.list}\AgdaSpace{}%
\AgdaBound{ys}\AgdaSymbol{)}\AgdaSpace{}%
\AgdaOperator{\AgdaDatatype{≡}}\AgdaSpace{}%
\AgdaSymbol{(}\AgdaInductiveConstructor{just}\AgdaSpace{}%
\AgdaOperator{\AgdaInductiveConstructor{⟨}}\AgdaSpace{}%
\AgdaBound{x}\AgdaSpace{}%
\AgdaOperator{\AgdaInductiveConstructor{,}}\AgdaSpace{}%
\AgdaBound{xs}\AgdaSpace{}%
\AgdaOperator{\AgdaInductiveConstructor{⟩}}\AgdaSymbol{)}\AgdaSpace{}%
\AgdaSymbol{→}\<%
\\
\>[67I][@{}l@{\AgdaIndent{0}}]%
\>[10]\AgdaBound{P}\AgdaSpace{}%
\AgdaBound{x}\AgdaSpace{}%
\AgdaSymbol{→}\AgdaSpace{}%
\AgdaBound{P}\AgdaSpace{}%
\AgdaOperator{\AgdaDatatype{Eventually}}\AgdaSpace{}%
\AgdaBound{ys}\<%
\\
\>[2]\AgdaInductiveConstructor{ev{-}t}\AgdaSpace{}%
\AgdaSymbol{:}%
\>[90I]\AgdaSymbol{\{}\AgdaBound{x}\AgdaSpace{}%
\AgdaSymbol{:}\AgdaSpace{}%
\AgdaBound{A}\AgdaSymbol{\}\{}\AgdaBound{xs}\AgdaSpace{}%
\AgdaSymbol{:}\AgdaSpace{}%
\AgdaRecord{MyColist}\AgdaSpace{}%
\AgdaBound{A}\AgdaSymbol{\}}\AgdaSpace{}%
\AgdaSymbol{→}\AgdaSpace{}%
\AgdaSymbol{(}\AgdaField{MyColist.list}\AgdaSpace{}%
\AgdaBound{ys}\AgdaSymbol{)}\AgdaSpace{}%
\AgdaOperator{\AgdaDatatype{≡}}\AgdaSpace{}%
\AgdaSymbol{(}\AgdaInductiveConstructor{just}\AgdaSpace{}%
\AgdaOperator{\AgdaInductiveConstructor{⟨}}\AgdaSpace{}%
\AgdaBound{x}\AgdaSpace{}%
\AgdaOperator{\AgdaInductiveConstructor{,}}\AgdaSpace{}%
\AgdaBound{xs}\AgdaSpace{}%
\AgdaOperator{\AgdaInductiveConstructor{⟩}}\AgdaSymbol{)}\AgdaSpace{}%
\AgdaSymbol{→}\<%
\\
\>[90I][@{}l@{\AgdaIndent{0}}]%
\>[10]\AgdaBound{P}\AgdaSpace{}%
\AgdaOperator{\AgdaDatatype{Eventually}}\AgdaSpace{}%
\AgdaBound{xs}\AgdaSpace{}%
\AgdaSymbol{→}\AgdaSpace{}%
\AgdaBound{P}\AgdaSpace{}%
\AgdaOperator{\AgdaDatatype{Eventually}}\AgdaSpace{}%
\AgdaBound{ys}\<%
\end{code}

\end{center}

The proofs on how the reference list is built must be added. Also in this case the code is similar to the implementation of \textit{memberOf} for our colists.

\begin{center}
\begin{code}
\>[0]\AgdaFunction{ev{-}sound}\AgdaSpace{}%
\AgdaSymbol{:}%
\>[114I]\AgdaSymbol{\{}\AgdaBound{A}\AgdaSpace{}%
\AgdaSymbol{:}\AgdaSpace{}%
\AgdaPrimitiveType{Set}\AgdaSymbol{\}\{}\AgdaBound{xs}\AgdaSpace{}%
\AgdaSymbol{:}\AgdaSpace{}%
\AgdaRecord{MyColist}\AgdaSpace{}%
\AgdaBound{A}\AgdaSymbol{\}\{}\AgdaBound{P}\AgdaSpace{}%
\AgdaSymbol{:}\AgdaSpace{}%
\AgdaBound{A}\AgdaSpace{}%
\AgdaSymbol{→}\AgdaSpace{}%
\AgdaPrimitiveType{Set}\AgdaSymbol{\}}\AgdaSpace{}%
\AgdaSymbol{→}\AgdaSpace{}%
\AgdaBound{P}\AgdaSpace{}%
\AgdaOperator{\AgdaDatatype{Eventually}}\AgdaSpace{}%
\AgdaBound{xs}\AgdaSpace{}%
\AgdaSymbol{→}\<%
\\
\>[.]\<[114I]%
\>[11]\AgdaDatatype{∃}\AgdaSpace{}%
\AgdaBound{A}\AgdaSpace{}%
\AgdaSymbol{(λ}\AgdaSpace{}%
\AgdaBound{x}\AgdaSpace{}%
\AgdaSymbol{→}\AgdaSpace{}%
\AgdaSymbol{(}\AgdaBound{x}\AgdaSpace{}%
\AgdaOperator{\AgdaDatatype{memberOf}}\AgdaSpace{}%
\AgdaBound{xs}\AgdaSymbol{)}\AgdaSpace{}%
\AgdaOperator{\AgdaDatatype{∧}}\AgdaSpace{}%
\AgdaSymbol{(}\AgdaBound{P}\AgdaSpace{}%
\AgdaBound{x}\AgdaSymbol{))}\<%
\\
\>[0]\<%
\\
\>[0]\AgdaFunction{ev{-}sound}\AgdaSpace{}%
\AgdaSymbol{(}\AgdaInductiveConstructor{ev{-}h}\AgdaSpace{}%
\AgdaSymbol{\{}\AgdaBound{x}\AgdaSymbol{\}\{}\AgdaBound{xs}\AgdaSymbol{\}}\AgdaSpace{}%
\AgdaBound{eq}\AgdaSpace{}%
\AgdaBound{Px}\AgdaSymbol{)}\AgdaSpace{}%
\AgdaSymbol{=}\AgdaSpace{}%
\AgdaOperator{\AgdaInductiveConstructor{<}}\AgdaSpace{}%
\AgdaBound{x}\AgdaSpace{}%
\AgdaOperator{\AgdaInductiveConstructor{,}}\AgdaSpace{}%
\AgdaOperator{\AgdaInductiveConstructor{⟨}}\AgdaSpace{}%
\AgdaInductiveConstructor{mem{-}h}\AgdaSpace{}%
\AgdaBound{eq}\AgdaSpace{}%
\AgdaOperator{\AgdaInductiveConstructor{,}}\AgdaSpace{}%
\AgdaBound{Px}\AgdaSpace{}%
\AgdaOperator{\AgdaInductiveConstructor{⟩}}\AgdaSpace{}%
\AgdaOperator{\AgdaInductiveConstructor{>}}\<%
\\
\>[0]\AgdaFunction{ev{-}sound}%
\>[154I]\AgdaSymbol{(}\AgdaInductiveConstructor{ev{-}t}\AgdaSpace{}%
\AgdaBound{eq}\AgdaSpace{}%
\AgdaBound{ev{-}xs}\AgdaSymbol{)}\AgdaSpace{}%
\AgdaSymbol{=}\<%
\\
\>[.]\<[154I]%
\>[9]\AgdaOperator{\AgdaInductiveConstructor{<}}\<%
\\
\>[9]\AgdaFunction{witness}\AgdaSpace{}%
\AgdaSymbol{(}\AgdaFunction{ev{-}sound}\AgdaSpace{}%
\AgdaBound{ev{-}xs}\AgdaSymbol{)}\AgdaSpace{}%
\AgdaOperator{\AgdaInductiveConstructor{,}}\<%
\\
\>[9][@{}l@{\AgdaIndent{0}}]%
\>[11]\AgdaOperator{\AgdaInductiveConstructor{⟨}}\<%
\\
\>[11][@{}l@{\AgdaIndent{0}}]%
\>[13]\AgdaInductiveConstructor{mem{-}t}\AgdaSpace{}%
\AgdaBound{eq}\AgdaSpace{}%
\AgdaSymbol{(}\AgdaFunction{∧{-}left}\AgdaSpace{}%
\AgdaSymbol{(}\AgdaFunction{proof}\AgdaSpace{}%
\AgdaSymbol{(}\AgdaFunction{ev{-}sound}\AgdaSpace{}%
\AgdaBound{ev{-}xs}\AgdaSymbol{)))}\AgdaSpace{}%
\AgdaOperator{\AgdaInductiveConstructor{,}}\<%
\\
\>[13]\AgdaFunction{∧{-}right}\AgdaSpace{}%
\AgdaSymbol{(}\AgdaFunction{proof}\AgdaSpace{}%
\AgdaSymbol{(}\AgdaFunction{ev{-}sound}\AgdaSpace{}%
\AgdaBound{ev{-}xs}\AgdaSymbol{))}\<%
\\
\>[11]\AgdaOperator{\AgdaInductiveConstructor{⟩}}\<%
\\
\>[9]\AgdaOperator{\AgdaInductiveConstructor{>}}\<%
\end{code}
\end{center}

The proof is by induction over the rules of \textit{Eventually}. Differently from the code about standard library colists, the proof on how the list is built is crucial. In the first case Agda knows that $x$ is the head thanks to the equality proof \LC{$eq$. For what concerns the second case, we need to use the constructor \agda{mem-t} since the output of the recursive call is referred to the tail of the list}.

\begin{center}
\begin{code}
\>[0]\AgdaFunction{ev{-}complete}%
\>[170I]\AgdaSymbol{:}\AgdaSpace{}%
\AgdaSymbol{\{}\AgdaBound{A}\AgdaSpace{}%
\AgdaSymbol{:}\AgdaSpace{}%
\AgdaPrimitiveType{Set}\AgdaSymbol{\}\{}\AgdaBound{xs}\AgdaSpace{}%
\AgdaSymbol{:}\AgdaSpace{}%
\AgdaRecord{MyColist}\AgdaSpace{}%
\AgdaBound{A}\AgdaSymbol{\}\{}\AgdaBound{P}\AgdaSpace{}%
\AgdaSymbol{:}\AgdaSpace{}%
\AgdaBound{A}\AgdaSpace{}%
\AgdaSymbol{→}\AgdaSpace{}%
\AgdaPrimitiveType{Set}\AgdaSymbol{\}}\AgdaSpace{}%
\AgdaSymbol{→}\<%
\\
\>[.]\<[170I]%
\>[12]\AgdaDatatype{∃}\AgdaSpace{}%
\AgdaBound{A}\AgdaSpace{}%
\AgdaSymbol{(λ}\AgdaSpace{}%
\AgdaBound{x}\AgdaSpace{}%
\AgdaSymbol{→}\AgdaSpace{}%
\AgdaSymbol{(}\AgdaBound{x}\AgdaSpace{}%
\AgdaOperator{\AgdaDatatype{memberOf}}\AgdaSpace{}%
\AgdaBound{xs}\AgdaSymbol{)}\AgdaSpace{}%
\AgdaOperator{\AgdaDatatype{∧}}\AgdaSpace{}%
\AgdaSymbol{(}\AgdaBound{P}\AgdaSpace{}%
\AgdaBound{x}\AgdaSymbol{))}\AgdaSpace{}%
\AgdaSymbol{→}\AgdaSpace{}%
\AgdaBound{P}\AgdaSpace{}%
\AgdaOperator{\AgdaDatatype{Eventually}}\AgdaSpace{}%
\AgdaBound{xs}\<%
\\
\\
\>[0]\AgdaFunction{ev{-}complete}\AgdaSpace{}%
\AgdaOperator{\AgdaInductiveConstructor{<}}\AgdaSpace{}%
\AgdaBound{x}\AgdaSpace{}%
\AgdaOperator{\AgdaInductiveConstructor{,}}\AgdaSpace{}%
\AgdaOperator{\AgdaInductiveConstructor{⟨}}\AgdaSpace{}%
\AgdaInductiveConstructor{mem{-}h}\AgdaSpace{}%
\AgdaBound{eq}\AgdaSpace{}%
\AgdaOperator{\AgdaInductiveConstructor{,}}\AgdaSpace{}%
\AgdaBound{Px}\AgdaSpace{}%
\AgdaOperator{\AgdaInductiveConstructor{⟩}}\AgdaSpace{}%
\AgdaOperator{\AgdaInductiveConstructor{>}}\AgdaSpace{}%
\AgdaSymbol{=}\AgdaSpace{}%
\AgdaInductiveConstructor{ev{-}h}\AgdaSpace{}%
\AgdaBound{eq}\AgdaSpace{}%
\AgdaBound{Px}\<%
\\
\>[0]\AgdaFunction{ev{-}complete}%
\>[210I]\AgdaOperator{\AgdaInductiveConstructor{<}}\AgdaSpace{}%
\AgdaBound{x}\AgdaSpace{}%
\AgdaOperator{\AgdaInductiveConstructor{,}}\AgdaSpace{}%
\AgdaOperator{\AgdaInductiveConstructor{⟨}}\AgdaSpace{}%
\AgdaInductiveConstructor{mem{-}t}\AgdaSpace{}%
\AgdaBound{eq}\AgdaSpace{}%
\AgdaBound{mem}\AgdaSpace{}%
\AgdaOperator{\AgdaInductiveConstructor{,}}\AgdaSpace{}%
\AgdaBound{Px}\AgdaSpace{}%
\AgdaOperator{\AgdaInductiveConstructor{⟩}}\AgdaSpace{}%
\AgdaOperator{\AgdaInductiveConstructor{>}}\AgdaSpace{}%
\AgdaSymbol{=}\AgdaSpace{}%
\AgdaInductiveConstructor{ev{-}t}\AgdaSpace{}%
\AgdaBound{eq}\<%
\\
\>[.]\<[210I]%
\>[12]\AgdaSymbol{(}\AgdaFunction{ev{-}complete}\AgdaSpace{}%
\AgdaOperator{\AgdaInductiveConstructor{<}}\AgdaSpace{}%
\AgdaBound{x}\AgdaSpace{}%
\AgdaOperator{\AgdaInductiveConstructor{,}}\AgdaSpace{}%
\AgdaOperator{\AgdaInductiveConstructor{⟨}}\AgdaSpace{}%
\AgdaBound{mem}\AgdaSpace{}%
\AgdaOperator{\AgdaInductiveConstructor{,}}\AgdaSpace{}%
\AgdaBound{Px}\AgdaSpace{}%
\AgdaOperator{\AgdaInductiveConstructor{⟩}}\AgdaSpace{}%
\AgdaOperator{\AgdaInductiveConstructor{>}}\AgdaSymbol{)}\<%
\end{code}
\end{center}

The proof is by induction over the rules of \textit{memberOf} and the same considerations on $eq$ hold. There are no additional equalities because they are hidden inside \textit{memberOf}. It would have been different if we had decided to use the \textit{get} function together with a lift function on $P$. Probably we would have met those issues that we had to solve in \refToChapter{inductive}. 

\section{Always}\label{sect:always}

{Given an arbitrary predicate $\Pred$ on list elements, $\Always{\Pred} {\xs}$ holds if $\Pred(\x)$ holds for each $\x$ element of $\xs$.} 
This parametric predicate can be defined by the following inference system interpreted coinductively.

\begin{small}
\begin{quote}
$\NamedRule{alw-$\Lambda$} {}{\Always{\Pred} {\Lambda}}\Space
\NamedRule{alw-t} {\Always{\Pred}{\xs}}{\Always{\Pred} {\x{:}\xs}}\Pred(\x)$
\end{quote}
\end{small}

Note that the previously considered predicate \textit{allPos} can be obtained as the instance of \textit{Always} on $\Pred(\x)=\x>0$.

The (parametric) specification of the predicate is {$\Spec_\Pred = \{ \xs \mid \forall \x. \memberOf {\x} {\xs}\ \mbox{implies}\ \Pred(\x) \}$.}

\begin{statement}
The coinductive definition of \textit{Always} is sound with respect to its specification.
\end{statement}
\begin{proof}
We have to prove that, if $\Always{\Pred}{\xs}\in\CoInd{\is}$, then $\memberOf{\x}{\xs}$ implies that $\Pred(\x)$ holds. If $\memberOf {\x} {\xs}$ holds, then $\xs$ cannot be empty, hence $\xs=\y{:}\ys$. Hence, to derive that $\xs\in\CoInd{\is}$, we have used rule \refToRule{alw-t}, thus $\y>0$ and $\ys\in\CoInd{\is}$.
The proof is by induction on the definition of  $\memberOf{\x}{\xs}$.
Clearly, if the list is empty we have that $\Lambda \in S$.
\begin{description}
\item[\refToRule{mem-h}] We have $\memberOf{\x}{\x{:}\xs}$, hence the thesis holds by the side condition of \refToRule{alw-t}.
\item[\refToRule{mem-t}] If we look for an element in $x{:}xs$ which is not the head, we know that it surely lies inside $xs$. By inductive hypothesis we have that $\Pred$ holds on all the elements in $xs$, thus also on the one we are looking for.
\end{description}
\end{proof}

\begin{statement}
The coinductive definition of \textit{Always} is complete with respect to its specification.
\end{statement}
\begin{proof}
We have to prove that, if $\xs\in\Spec$, that is, $\Pred$ holds on all the elements of $\xs$, then $\Always{\Pred}{\xs}$ can be derived. 
The proof is by coinduction on the definition of $\Always{\Pred}{\xs}$. That is, we have to show that $\Spec$ is consistent: if $\xs$ is in $\Spec$, then it is the consequence of a rule with premises which are in $\Spec$ as well. We consider two cases.
\begin{itemize}
\item If $\xs \in \Spec$ and $\xs$ is empty, then it is the consequence of \refToRule{alw-$\Lambda$}.
\item If $\ys\in\Spec$ and $\ys = \x{:}\xs$, then it is the consequence of \refToRule{alw-t} with premise $\xs$, and we know that $\xs\in\Spec$. 
\end{itemize}
\end{proof}

If a predicate always holds, this means that it must hold for each element of a potentially infinite structure, so its implementation follows a coinductive schema. We begin showing the codes involving standard library colists as we did in \refToSection{eventually}.
These are the specification and the implementation of the type.

\begin{center}
\begin{code}%
\>[0]\AgdaFunction{S}\AgdaSpace{}%
\AgdaSymbol{:}\AgdaSpace{}%
\AgdaSymbol{\{}\AgdaBound{A}\AgdaSpace{}%
\AgdaSymbol{:}\AgdaSpace{}%
\AgdaPrimitiveType{Set}\AgdaSymbol{\}}\AgdaSpace{}%
\AgdaSymbol{→}\AgdaSpace{}%
\AgdaSymbol{(}\AgdaBound{A}\AgdaSpace{}%
\AgdaSymbol{→}\AgdaSpace{}%
\AgdaPrimitiveType{Set}\AgdaSymbol{)}\AgdaSpace{}%
\AgdaSymbol{→}\AgdaSpace{}%
\AgdaDatatype{Colist}\AgdaSpace{}%
\AgdaBound{A}\AgdaSpace{}%
\AgdaPostulate{∞}\AgdaSpace{}%
\AgdaSymbol{→}\AgdaSpace{}%
\AgdaPrimitiveType{Set}\<%
\\
\>[0]\AgdaFunction{S}\AgdaSpace{}%
\AgdaSymbol{\{}\AgdaBound{A}\AgdaSymbol{\}}\AgdaSpace{}%
\AgdaBound{P}\AgdaSpace{}%
\AgdaBound{xs}\AgdaSpace{}%
\AgdaSymbol{=}\AgdaSpace{}%
\AgdaSymbol{\{}\AgdaBound{x}\AgdaSpace{}%
\AgdaSymbol{:}\AgdaSpace{}%
\AgdaBound{A}\AgdaSymbol{\}}\AgdaSpace{}%
\AgdaSymbol{→}\AgdaSpace{}%
\AgdaBound{x}\AgdaSpace{}%
\AgdaOperator{\AgdaDatatype{memberOf}}\AgdaSpace{}%
\AgdaBound{xs}\AgdaSpace{}%
\AgdaSymbol{→}\AgdaSpace{}%
\AgdaBound{P}\AgdaSpace{}%
\AgdaBound{x}\<%
\\
\\[\AgdaEmptyExtraSkip]%
\>[0]\AgdaKeyword{data}\AgdaSpace{}%
\AgdaDatatype{Always}\AgdaSpace{}%
\AgdaSymbol{\{}\AgdaBound{A}\AgdaSpace{}%
\AgdaSymbol{:}\AgdaSpace{}%
\AgdaPrimitiveType{Set}\AgdaSymbol{\}}\AgdaSpace{}%
\AgdaSymbol{:}\AgdaSpace{}%
\AgdaSymbol{(}\AgdaBound{A}\AgdaSpace{}%
\AgdaSymbol{→}\AgdaSpace{}%
\AgdaPrimitiveType{Set}\AgdaSymbol{)}\AgdaSpace{}%
\AgdaSymbol{→}\AgdaSpace{}%
\AgdaDatatype{Colist}\AgdaSpace{}%
\AgdaBound{A}\AgdaSpace{}%
\AgdaPostulate{∞}\AgdaSpace{}%
\AgdaSymbol{→}\AgdaSpace{}%
\AgdaPostulate{Size}\AgdaSpace{}%
\AgdaSymbol{→}\AgdaSpace{}%
\AgdaPrimitiveType{Set}\AgdaSpace{}%
\AgdaKeyword{where}\<%
\\
\>[0][@{}l@{\AgdaIndent{0}}]%
\>[2]\AgdaInductiveConstructor{alw{-}Λ}\AgdaSpace{}%
\AgdaSymbol{:}\AgdaSpace{}%
\AgdaSymbol{∀}\AgdaSpace{}%
\AgdaSymbol{\{}\AgdaBound{P}\AgdaSpace{}%
\AgdaSymbol{:}\AgdaSpace{}%
\AgdaBound{A}\AgdaSpace{}%
\AgdaSymbol{→}\AgdaSpace{}%
\AgdaPrimitiveType{Set}\AgdaSymbol{\}\{}\AgdaBound{i}\AgdaSymbol{\}}\AgdaSpace{}%
\AgdaSymbol{→}\AgdaSpace{}%
\AgdaDatatype{Always}\AgdaSpace{}%
\AgdaBound{P}\AgdaSpace{}%
\AgdaInductiveConstructor{[]}\AgdaSpace{}%
\AgdaBound{i}\<%
\\
\>[2]\AgdaInductiveConstructor{alw{-}t}\AgdaSpace{}%
\AgdaSymbol{:}\AgdaSpace{}%
\AgdaSymbol{∀}\AgdaSpace{}%
\AgdaSymbol{\{}\AgdaBound{x}\AgdaSpace{}%
\AgdaBound{i}\AgdaSpace{}%
\AgdaBound{xs}\AgdaSpace{}%
\AgdaBound{P}\AgdaSymbol{\}}\AgdaSpace{}%
\AgdaSymbol{→}\AgdaSpace{}%
\AgdaBound{P}\AgdaSpace{}%
\AgdaBound{x}\AgdaSpace{}%
\AgdaSymbol{→}\<%
\\
\>[2][@{}l@{\AgdaIndent{0}}]%
\>[6]\AgdaRecord{Thunk}\AgdaSpace{}%
\AgdaSymbol{(}\AgdaDatatype{Always}\AgdaSpace{}%
\AgdaBound{P}\AgdaSpace{}%
\AgdaSymbol{(}\AgdaField{Thunk.force}\AgdaSpace{}%
\AgdaBound{xs}\AgdaSymbol{))}\AgdaSpace{}%
\AgdaBound{i}\AgdaSpace{}%
\AgdaSymbol{→}\<%
\\
\>[6]\AgdaDatatype{Always}\AgdaSpace{}%
\AgdaBound{P}\AgdaSpace{}%
\AgdaSymbol{(}\AgdaBound{x}\AgdaSpace{}%
\AgdaOperator{\AgdaInductiveConstructor{∷}}\AgdaSpace{}%
\AgdaBound{xs}\AgdaSymbol{)}\AgdaSpace{}%
\AgdaBound{i}\<%
\end{code}

\end{center}

It is clear that the type depends on \agda{Size} and \agda{Thunk} and it reminds the definition of \textit{allPos}. In this case the side condition that asks that the head is positive is replaced by a generic predicate $\Pred$.

\begin{center}
\begin{code}
\>[0]\AgdaFunction{alw{-}sound}%
\>[88I]\AgdaSymbol{:}\AgdaSpace{}%
\AgdaSymbol{\{}\AgdaBound{A}\AgdaSpace{}%
\AgdaSymbol{:}\AgdaSpace{}%
\AgdaPrimitiveType{Set}\AgdaSymbol{\}\{}\AgdaBound{P}\AgdaSpace{}%
\AgdaSymbol{:}\AgdaSpace{}%
\AgdaBound{A}\AgdaSpace{}%
\AgdaSymbol{→}\AgdaSpace{}%
\AgdaPrimitiveType{Set}\AgdaSymbol{\}\{}\AgdaBound{xs}\AgdaSpace{}%
\AgdaSymbol{:}\AgdaSpace{}%
\AgdaDatatype{Colist}\AgdaSpace{}%
\AgdaBound{A}\AgdaSpace{}%
\AgdaPostulate{∞}\AgdaSymbol{\}}\AgdaSpace{}%
\AgdaSymbol{→}\<%
\\
\>[.]\<[88I]%
\>[10]\AgdaSymbol{(∀}\AgdaSpace{}%
\AgdaSymbol{\{}\AgdaBound{i}\AgdaSymbol{\}}\AgdaSpace{}%
\AgdaSymbol{→}\AgdaSpace{}%
\AgdaDatatype{Always}\AgdaSpace{}%
\AgdaBound{P}\AgdaSpace{}%
\AgdaBound{xs}\AgdaSpace{}%
\AgdaBound{i}\AgdaSymbol{)}\AgdaSpace{}%
\AgdaSymbol{→}\<%
\\
\>[10]\AgdaSymbol{(∀}\AgdaSpace{}%
\AgdaSymbol{\{}\AgdaBound{x}\AgdaSymbol{\}}\AgdaSpace{}%
\AgdaSymbol{→}\AgdaSpace{}%
\AgdaBound{x}\AgdaSpace{}%
\AgdaOperator{\AgdaDatatype{memberOf}}\AgdaSpace{}%
\AgdaBound{xs}\AgdaSpace{}%
\AgdaSymbol{→}\AgdaSpace{}%
\AgdaBound{P}\AgdaSpace{}%
\AgdaBound{x}\AgdaSymbol{)}\<%
\\
\>[0]\<%
\\
\>[0]\AgdaFunction{alw{-}sound}\AgdaSpace{}%
\AgdaBound{alw}\AgdaSpace{}%
\AgdaInductiveConstructor{mem{-}h}\AgdaSpace{}%
\AgdaKeyword{with}\AgdaSpace{}%
\AgdaBound{alw}\<%
\\
\>[0]\AgdaSymbol{...}\AgdaSpace{}%
\AgdaSymbol{|}\AgdaSpace{}%
\AgdaInductiveConstructor{alw{-}t}\AgdaSpace{}%
\AgdaBound{Px}\AgdaSpace{}%
\AgdaBound{alw{-}xs}\AgdaSpace{}%
\AgdaSymbol{=}\AgdaSpace{}%
\AgdaBound{Px}\<%
\\
\>[0]\AgdaFunction{alw{-}sound}\AgdaSpace{}%
\AgdaBound{alw}\AgdaSpace{}%
\AgdaSymbol{(}\AgdaInductiveConstructor{mem{-}t}\AgdaSpace{}%
\AgdaBound{mem}\AgdaSymbol{)}\AgdaSpace{}%
\AgdaKeyword{with}\AgdaSpace{}%
\AgdaBound{alw}\<%
\\
\>[0]\AgdaSymbol{...}\AgdaSpace{}%
\AgdaSymbol{|}\AgdaSpace{}%
\AgdaInductiveConstructor{alw{-}t}\AgdaSpace{}%
\AgdaBound{Px}\AgdaSpace{}%
\AgdaBound{alw{-}xs}\AgdaSpace{}%
\AgdaSymbol{=}\AgdaSpace{}%
\AgdaFunction{alw{-}sound}\AgdaSpace{}%
\AgdaSymbol{(}\AgdaField{Thunk.force}\AgdaSpace{}%
\AgdaBound{alw{-}xs}\AgdaSymbol{)}\AgdaSpace{}%
\AgdaBound{mem}\<%
\end{code}
\end{center}

The proof is the same as the \textit{allPos} and also in this case its type requires that the predicate holds for all approximations $i$. Agda knows that the list cannot be empty when pattern matching on \textit{memberOf} so splitting on $alw$ produces only one case.

\begin{center}
\begin{code}
\>[0]\AgdaFunction{alw{-}complete}%
\>[140I]\AgdaSymbol{:}\AgdaSpace{}%
\AgdaSymbol{\{}\AgdaBound{A}\AgdaSpace{}%
\AgdaSymbol{:}\AgdaSpace{}%
\AgdaPrimitiveType{Set}\AgdaSymbol{\}\{}\AgdaBound{P}\AgdaSpace{}%
\AgdaSymbol{:}\AgdaSpace{}%
\AgdaBound{A}\AgdaSpace{}%
\AgdaSymbol{→}\AgdaSpace{}%
\AgdaPrimitiveType{Set}\AgdaSymbol{\}\{}\AgdaBound{ys}\AgdaSpace{}%
\AgdaSymbol{:}\AgdaSpace{}%
\AgdaDatatype{Colist}\AgdaSpace{}%
\AgdaBound{A}\AgdaSpace{}%
\AgdaPostulate{∞}\AgdaSymbol{\}}\AgdaSpace{}%
\AgdaSymbol{→}\<%
\\
\>[.]\<[140I]%
\>[13]\AgdaSymbol{(∀}\AgdaSpace{}%
\AgdaSymbol{\{}\AgdaBound{x}\AgdaSymbol{\}}\AgdaSpace{}%
\AgdaSymbol{→}\AgdaSpace{}%
\AgdaBound{x}\AgdaSpace{}%
\AgdaOperator{\AgdaDatatype{memberOf}}\AgdaSpace{}%
\AgdaBound{ys}\AgdaSpace{}%
\AgdaSymbol{→}\AgdaSpace{}%
\AgdaBound{P}\AgdaSpace{}%
\AgdaBound{x}\AgdaSymbol{)}\AgdaSpace{}%
\AgdaSymbol{→}\<%
\\
\>[13]\AgdaSymbol{(∀}\AgdaSpace{}%
\AgdaSymbol{\{}\AgdaBound{i}\AgdaSymbol{\}}\AgdaSpace{}%
\AgdaSymbol{→}\AgdaSpace{}%
\AgdaDatatype{Always}\AgdaSpace{}%
\AgdaBound{P}\AgdaSpace{}%
\AgdaBound{ys}\AgdaSpace{}%
\AgdaBound{i}\AgdaSymbol{)}\<%
\\
\>[0]\<%
\\
\>[0]\AgdaFunction{alw{-}complete}\AgdaSpace{}%
\AgdaSymbol{\{}\AgdaArgument{ys}\AgdaSpace{}%
\AgdaSymbol{=}\AgdaSpace{}%
\AgdaInductiveConstructor{[]}\AgdaSymbol{\}}\AgdaSpace{}%
\AgdaBound{f}\AgdaSpace{}%
\AgdaSymbol{=}\AgdaSpace{}%
\AgdaInductiveConstructor{alw{-}Λ}\<%
\\
\>[0]\AgdaFunction{alw{-}complete}%
\>[174I]\AgdaSymbol{\{}\AgdaArgument{ys}\AgdaSpace{}%
\AgdaSymbol{=}\AgdaSpace{}%
\AgdaBound{x}\AgdaSpace{}%
\AgdaOperator{\AgdaInductiveConstructor{∷}}\AgdaSpace{}%
\AgdaBound{xs}\AgdaSymbol{\}}\AgdaSpace{}%
\AgdaBound{f}\AgdaSpace{}%
\AgdaSymbol{=}\AgdaSpace{}%
\AgdaInductiveConstructor{alw{-}t}\AgdaSpace{}%
\AgdaSymbol{(}\AgdaBound{f}\AgdaSpace{}%
\AgdaInductiveConstructor{mem{-}h}\AgdaSymbol{)}\<%
\\
\>[.]\<[174I]%
\>[13]\AgdaSymbol{(λ}\AgdaSpace{}%
\AgdaKeyword{where}\AgdaSpace{}%
\AgdaSymbol{.}\AgdaField{force}\AgdaSpace{}%
\AgdaSymbol{→}\AgdaSpace{}%
\AgdaFunction{alw{-}complete}\AgdaSpace{}%
\AgdaSymbol{(λ}\AgdaSpace{}%
\AgdaBound{m}\AgdaSpace{}%
\AgdaSymbol{→}\AgdaSpace{}%
\AgdaBound{f}\AgdaSpace{}%
\AgdaSymbol{(}\AgdaInductiveConstructor{mem{-}t}\AgdaSpace{}%
\AgdaBound{m}\AgdaSymbol{)))}\<%
\end{code}
\end{center}

Where the last parameter in the recursive call is the restriction of $f$ to the tail. Notice that the proofs are very similar to those we discussed in the theoretical part. No we can move to our approach to colists. 

\begin{center}
\begin{code}%
\>[0]\AgdaFunction{S}\AgdaSpace{}%
\AgdaSymbol{:}\AgdaSpace{}%
\AgdaSymbol{\{}\AgdaBound{A}\AgdaSpace{}%
\AgdaSymbol{:}\AgdaSpace{}%
\AgdaPrimitiveType{Set}\AgdaSymbol{\}}\AgdaSpace{}%
\AgdaSymbol{→}\AgdaSpace{}%
\AgdaSymbol{(}\AgdaBound{A}\AgdaSpace{}%
\AgdaSymbol{→}\AgdaSpace{}%
\AgdaPrimitiveType{Set}\AgdaSymbol{)}\AgdaSpace{}%
\AgdaSymbol{→}\AgdaSpace{}%
\AgdaRecord{MyColist}\AgdaSpace{}%
\AgdaBound{A}\AgdaSpace{}%
\AgdaSymbol{→}\AgdaSpace{}%
\AgdaPrimitiveType{Set}\<%
\\
\>[0]\AgdaFunction{S}\AgdaSpace{}%
\AgdaSymbol{\{}\AgdaBound{A}\AgdaSymbol{\}}\AgdaSpace{}%
\AgdaBound{P}\AgdaSpace{}%
\AgdaBound{xs}\AgdaSpace{}%
\AgdaSymbol{=}\AgdaSpace{}%
\AgdaSymbol{\{}\AgdaBound{x}\AgdaSpace{}%
\AgdaSymbol{:}\AgdaSpace{}%
\AgdaBound{A}\AgdaSymbol{\}}\AgdaSpace{}%
\AgdaSymbol{→}\AgdaSpace{}%
\AgdaBound{x}\AgdaSpace{}%
\AgdaOperator{\AgdaDatatype{memberOf}}\AgdaSpace{}%
\AgdaBound{xs}\AgdaSpace{}%
\AgdaSymbol{→}\AgdaSpace{}%
\AgdaBound{P}\AgdaSpace{}%
\AgdaBound{x}\<%
\\
\\[\AgdaEmptyExtraSkip]%
\>[0]\AgdaKeyword{record}\AgdaSpace{}%
\AgdaRecord{Always}\AgdaSpace{}%
\AgdaSymbol{\{}\AgdaBound{A}\AgdaSpace{}%
\AgdaSymbol{:}\AgdaSpace{}%
\AgdaPrimitiveType{Set}\AgdaSymbol{\}(}\AgdaBound{P}\AgdaSpace{}%
\AgdaSymbol{:}\AgdaSpace{}%
\AgdaBound{A}\AgdaSpace{}%
\AgdaSymbol{→}\AgdaSpace{}%
\AgdaPrimitiveType{Set}\AgdaSymbol{)(}\AgdaBound{xs}\AgdaSpace{}%
\AgdaSymbol{:}\AgdaSpace{}%
\AgdaRecord{MyColist}\AgdaSpace{}%
\AgdaBound{A}\AgdaSymbol{)}\AgdaSpace{}%
\AgdaSymbol{:}\AgdaSpace{}%
\AgdaPrimitiveType{Set}\AgdaSpace{}%
\AgdaKeyword{where}\<%
\\
\>[0][@{}l@{\AgdaIndent{0}}]%
\>[2]\AgdaKeyword{coinductive}\<%
\\
\>[2]\AgdaKeyword{field}\<%
\\
\>[2][@{}l@{\AgdaIndent{0}}]%
\>[4]\AgdaField{list}\AgdaSpace{}%
\AgdaSymbol{:}\AgdaSpace{}%
\AgdaSymbol{(}\AgdaField{MyColist.list}\AgdaSpace{}%
\AgdaBound{xs}\AgdaSpace{}%
\AgdaOperator{\AgdaDatatype{≡}}\AgdaSpace{}%
\AgdaInductiveConstructor{nothing}\AgdaSymbol{)}\AgdaSpace{}%
\AgdaOperator{\AgdaDatatype{∨}}\<%
\\
\>[4][@{}l@{\AgdaIndent{0}}]%
\>[6]\AgdaSymbol{(}\AgdaDatatype{∃}\AgdaSpace{}%
\AgdaSymbol{(}\AgdaBound{A}\AgdaSpace{}%
\AgdaOperator{\AgdaDatatype{×}}\AgdaSpace{}%
\AgdaRecord{MyColist}\AgdaSpace{}%
\AgdaBound{A}\AgdaSymbol{)}\<%
\\
\>[6]\AgdaSymbol{(λ}\AgdaSpace{}%
\AgdaBound{c}\AgdaSpace{}%
\AgdaSymbol{→}\<%
\\
\>[6][@{}l@{\AgdaIndent{0}}]%
\>[8]\AgdaSymbol{((}\AgdaField{MyColist.list}\AgdaSpace{}%
\AgdaBound{xs}\AgdaSymbol{)}\AgdaSpace{}%
\AgdaOperator{\AgdaDatatype{≡}}\AgdaSpace{}%
\AgdaSymbol{(}\AgdaInductiveConstructor{just}\AgdaSpace{}%
\AgdaBound{c}\AgdaSymbol{))}\AgdaSpace{}%
\AgdaOperator{\AgdaDatatype{∧}}\<%
\\
\>[8]\AgdaSymbol{((}\AgdaBound{P}\AgdaSpace{}%
\AgdaSymbol{(}\AgdaFunction{∧{-}left}\AgdaSpace{}%
\AgdaBound{c}\AgdaSymbol{))}\AgdaSpace{}%
\AgdaOperator{\AgdaDatatype{∧}}\AgdaSpace{}%
\AgdaSymbol{(}\AgdaRecord{Always}\AgdaSpace{}%
\AgdaBound{P}\AgdaSpace{}%
\AgdaSymbol{(}\AgdaFunction{∧{-}right}\AgdaSpace{}%
\AgdaBound{c}\AgdaSymbol{)))}\<%
\\
\>[6]\AgdaSymbol{))}\<%
\end{code}

\end{center}

As we presented for \textit{allPos} on our implementation of colists, we face again the same difficulties. The field brings the proof on how the colist is made. The variable \agda{c} inside the lambda function represents a couple such that the first element is the head of the list and the second one the tail; in fact its type is \agda{A $\times$ MyColist A}.

\begin{center}
\begin{code}
\>[0]\AgdaFunction{alw{-}sound}\AgdaSpace{}%
\AgdaSymbol{:}%
\>[92I]\AgdaSymbol{\{}\AgdaBound{A}\AgdaSpace{}%
\AgdaSymbol{:}\AgdaSpace{}%
\AgdaPrimitiveType{Set}\AgdaSymbol{\}\{}\AgdaBound{ys}\AgdaSpace{}%
\AgdaSymbol{:}\AgdaSpace{}%
\AgdaRecord{MyColist}\AgdaSpace{}%
\AgdaBound{A}\AgdaSymbol{\}\{}\AgdaBound{P}\AgdaSpace{}%
\AgdaSymbol{:}\AgdaSpace{}%
\AgdaBound{A}\AgdaSpace{}%
\AgdaSymbol{→}\AgdaSpace{}%
\AgdaPrimitiveType{Set}\AgdaSymbol{\}}\AgdaSpace{}%
\AgdaSymbol{→}\<%
\\
\>[92I][@{}l@{\AgdaIndent{0}}]%
\>[13]\AgdaRecord{Always}\AgdaSpace{}%
\AgdaBound{P}\AgdaSpace{}%
\AgdaBound{ys}\AgdaSpace{}%
\AgdaSymbol{→}\<%
\\
\>[13]\AgdaSymbol{(\{}\AgdaBound{n}\AgdaSpace{}%
\AgdaSymbol{:}\AgdaSpace{}%
\AgdaBound{A}\AgdaSymbol{\}}\AgdaSpace{}%
\AgdaSymbol{→}\AgdaSpace{}%
\AgdaBound{n}\AgdaSpace{}%
\AgdaOperator{\AgdaDatatype{memberOf}}\AgdaSpace{}%
\AgdaBound{ys}\AgdaSpace{}%
\AgdaSymbol{→}\AgdaSpace{}%
\AgdaBound{P}\AgdaSpace{}%
\AgdaBound{n}\AgdaSymbol{)}\<%
\\
\>[0]\<%
\\
\>[0]\AgdaFunction{alw{-}sound}\AgdaSpace{}%
\AgdaSymbol{\{}\AgdaArgument{ys}\AgdaSpace{}%
\AgdaSymbol{=}\AgdaSpace{}%
\AgdaBound{ys}\AgdaSymbol{\}\{}\AgdaArgument{P}\AgdaSpace{}%
\AgdaSymbol{=}\AgdaSpace{}%
\AgdaBound{P}\AgdaSymbol{\}}\AgdaSpace{}%
\AgdaBound{alw}\AgdaSpace{}%
\AgdaSymbol{(}\AgdaInductiveConstructor{mem{-}h}\AgdaSpace{}%
\AgdaBound{eq}\AgdaSymbol{)}\AgdaSpace{}%
\AgdaKeyword{with}\AgdaSpace{}%
\AgdaField{Always.list}\AgdaSpace{}%
\AgdaBound{alw}\<%
\\
\>[0]\AgdaSymbol{...}\AgdaSpace{}%
\AgdaSymbol{|}\AgdaSpace{}%
\AgdaInductiveConstructor{inl}\AgdaSpace{}%
\AgdaBound{p}\AgdaSpace{}%
\AgdaSymbol{=}\AgdaSpace{}%
\AgdaFunction{⊥{-}elim}\AgdaSpace{}%
\AgdaSymbol{(}\AgdaFunction{mycolist{-}abs}\AgdaSpace{}%
\AgdaSymbol{\{}\AgdaArgument{l}\AgdaSpace{}%
\AgdaSymbol{=}\AgdaSpace{}%
\AgdaBound{ys}\AgdaSymbol{\}}\AgdaSpace{}%
\AgdaBound{p}\AgdaSpace{}%
\AgdaBound{eq}\AgdaSymbol{)}\<%
\\
\>[0]\AgdaSymbol{...}\AgdaSpace{}%
\AgdaSymbol{|}\AgdaSpace{}%
\AgdaInductiveConstructor{inr}%
\>[139I]\AgdaOperator{\AgdaInductiveConstructor{<}}\AgdaSpace{}%
\AgdaOperator{\AgdaInductiveConstructor{⟨}}\AgdaSpace{}%
\AgdaBound{x}\AgdaSpace{}%
\AgdaOperator{\AgdaInductiveConstructor{,}}\AgdaSpace{}%
\AgdaSymbol{\AgdaUnderscore{}}\AgdaSpace{}%
\AgdaOperator{\AgdaInductiveConstructor{⟩}}\AgdaSpace{}%
\AgdaOperator{\AgdaInductiveConstructor{,}}\AgdaSpace{}%
\AgdaOperator{\AgdaInductiveConstructor{⟨}}\AgdaSpace{}%
\AgdaBound{eq₁}\AgdaSpace{}%
\AgdaOperator{\AgdaInductiveConstructor{,}}\AgdaSpace{}%
\AgdaOperator{\AgdaInductiveConstructor{⟨}}\AgdaSpace{}%
\AgdaBound{Px}\AgdaSpace{}%
\AgdaOperator{\AgdaInductiveConstructor{,}}\AgdaSpace{}%
\AgdaSymbol{\AgdaUnderscore{}}\AgdaSpace{}%
\AgdaOperator{\AgdaInductiveConstructor{⟩}}\AgdaSpace{}%
\AgdaOperator{\AgdaInductiveConstructor{⟩}}\AgdaSpace{}%
\AgdaOperator{\AgdaInductiveConstructor{>}}\AgdaSpace{}%
\AgdaSymbol{=}\<%
\\
\>[.]\<[139I]%
\>[10]\AgdaFunction{subst}\AgdaSpace{}%
\AgdaSymbol{(λ}\AgdaSpace{}%
\AgdaBound{n}\AgdaSpace{}%
\AgdaSymbol{→}\AgdaSpace{}%
\AgdaBound{P}\AgdaSpace{}%
\AgdaBound{n}\AgdaSymbol{)}\AgdaSpace{}%
\AgdaSymbol{(}\AgdaFunction{eq2sx}\AgdaSpace{}%
\AgdaSymbol{(}\AgdaFunction{just{-}elim}\AgdaSpace{}%
\AgdaSymbol{(}\AgdaFunction{trans}\AgdaSpace{}%
\AgdaSymbol{(}\AgdaFunction{sym}\AgdaSpace{}%
\AgdaBound{eq₁}\AgdaSymbol{)}\AgdaSpace{}%
\AgdaBound{eq}\AgdaSymbol{))}\AgdaSpace{}%
\AgdaSymbol{)}\AgdaSpace{}%
\AgdaBound{Px}\<%
\\
\\[\AgdaEmptyExtraSkip]%
\>[0]\AgdaFunction{alw{-}sound}\AgdaSpace{}%
\AgdaSymbol{\{}\AgdaArgument{ys}\AgdaSpace{}%
\AgdaSymbol{=}\AgdaSpace{}%
\AgdaBound{ys}\AgdaSymbol{\}\{}\AgdaArgument{P}\AgdaSpace{}%
\AgdaSymbol{=}\AgdaSpace{}%
\AgdaBound{P}\AgdaSymbol{\}}\AgdaSpace{}%
\AgdaBound{alw}\AgdaSpace{}%
\AgdaSymbol{(}\AgdaInductiveConstructor{mem{-}t}\AgdaSpace{}%
\AgdaBound{eq}\AgdaSpace{}%
\AgdaBound{mem}\AgdaSymbol{)}\AgdaSpace{}%
\AgdaKeyword{with}\AgdaSpace{}%
\AgdaField{Always.list}\AgdaSpace{}%
\AgdaBound{alw}\<%
\\
\>[0]\AgdaSymbol{...}\AgdaSpace{}%
\AgdaSymbol{|}\AgdaSpace{}%
\AgdaInductiveConstructor{inl}\AgdaSpace{}%
\AgdaBound{p}\AgdaSpace{}%
\AgdaSymbol{=}\AgdaSpace{}%
\AgdaFunction{⊥{-}elim}\AgdaSpace{}%
\AgdaSymbol{(}\AgdaFunction{mycolist{-}abs}\AgdaSpace{}%
\AgdaSymbol{\{}\AgdaArgument{l}\AgdaSpace{}%
\AgdaSymbol{=}\AgdaSpace{}%
\AgdaBound{ys}\AgdaSymbol{\}}\AgdaSpace{}%
\AgdaBound{p}\AgdaSpace{}%
\AgdaBound{eq}\AgdaSymbol{)}\<%
\\
\>[0]\AgdaSymbol{...}\AgdaSpace{}%
\AgdaSymbol{|}%
\>[194I]\AgdaInductiveConstructor{inr}\AgdaSpace{}%
\AgdaOperator{\AgdaInductiveConstructor{<}}\AgdaSpace{}%
\AgdaOperator{\AgdaInductiveConstructor{⟨}}\AgdaSpace{}%
\AgdaSymbol{\AgdaUnderscore{}}\AgdaSpace{}%
\AgdaOperator{\AgdaInductiveConstructor{,}}\AgdaSpace{}%
\AgdaBound{xs}\AgdaSpace{}%
\AgdaOperator{\AgdaInductiveConstructor{⟩}}\AgdaSpace{}%
\AgdaOperator{\AgdaInductiveConstructor{,}}\AgdaSpace{}%
\AgdaOperator{\AgdaInductiveConstructor{⟨}}\AgdaSpace{}%
\AgdaBound{eq₁}\AgdaSpace{}%
\AgdaOperator{\AgdaInductiveConstructor{,}}\AgdaSpace{}%
\AgdaOperator{\AgdaInductiveConstructor{⟨}}\AgdaSpace{}%
\AgdaSymbol{\AgdaUnderscore{}}\AgdaSpace{}%
\AgdaOperator{\AgdaInductiveConstructor{,}}\AgdaSpace{}%
\AgdaBound{alw{-}xs}\AgdaSpace{}%
\AgdaOperator{\AgdaInductiveConstructor{⟩}}\AgdaSpace{}%
\AgdaOperator{\AgdaInductiveConstructor{⟩}}\AgdaSpace{}%
\AgdaOperator{\AgdaInductiveConstructor{>}}\AgdaSpace{}%
\AgdaSymbol{=}\<%
\\
\>[.]\<[194I]%
\>[6]\AgdaFunction{alw{-}sound}\<%
\\
\>[6]\AgdaSymbol{(}\<%
\\
\>[6][@{}l@{\AgdaIndent{0}}]%
\>[8]\AgdaFunction{subst}\<%
\\
\>[8]\AgdaSymbol{(λ}\AgdaSpace{}%
\AgdaBound{v}\AgdaSpace{}%
\AgdaSymbol{→}\AgdaSpace{}%
\AgdaRecord{Always}\AgdaSpace{}%
\AgdaBound{P}\AgdaSpace{}%
\AgdaBound{v}\AgdaSymbol{)}\<%
\\
\>[8]\AgdaSymbol{(}\AgdaFunction{eq2dx}\AgdaSpace{}%
\AgdaSymbol{(}\AgdaFunction{just{-}elim}\AgdaSpace{}%
\AgdaSymbol{(}\AgdaFunction{trans}\AgdaSpace{}%
\AgdaSymbol{(}\AgdaFunction{sym}\AgdaSpace{}%
\AgdaBound{eq₁}\AgdaSymbol{)}\AgdaSpace{}%
\AgdaBound{eq}\AgdaSymbol{)))}\<%
\\
\>[8]\AgdaBound{alw{-}xs}\<%
\\
\>[6]\AgdaSymbol{)}\AgdaSpace{}%
\AgdaBound{mem}\<%
\end{code}
\end{center}

Absurd case must be managed again. Then the proof is by induction over \textit{memberOf}. In the first case \agda{$\Pred$x} is the proof that $\Pred$ holds for \agda{x} but the proof that it holds for the head specified in \agda{eq} is required: this is proved by substitution. The second case is similar but the problem is shifted to \agda{xs} where \agda{alw-xs} is the proof that \textit{Always} holds on the tail. The solutions are provided through the tools presented in the previous chapter.

\begin{center}
\begin{code}
\>[0]\AgdaFunction{alw{-}complete}\AgdaSpace{}%
\AgdaSymbol{:}%
\>[16]\AgdaSymbol{\{}\AgdaBound{A}\AgdaSpace{}%
\AgdaSymbol{:}\AgdaSpace{}%
\AgdaPrimitiveType{Set}\AgdaSymbol{\}(}\AgdaBound{P}\AgdaSpace{}%
\AgdaSymbol{:}\AgdaSpace{}%
\AgdaBound{A}\AgdaSpace{}%
\AgdaSymbol{→}\AgdaSpace{}%
\AgdaPrimitiveType{Set}\AgdaSymbol{)(}\AgdaBound{ys}\AgdaSpace{}%
\AgdaSymbol{:}\AgdaSpace{}%
\AgdaRecord{MyColist}\AgdaSpace{}%
\AgdaBound{A}\AgdaSymbol{)}\AgdaSpace{}%
\AgdaSymbol{→}\<%
\\
\>[16]\AgdaSymbol{(\{}\AgdaBound{n}\AgdaSpace{}%
\AgdaSymbol{:}\AgdaSpace{}%
\AgdaBound{A}\AgdaSymbol{\}}\AgdaSpace{}%
\AgdaSymbol{→}\AgdaSpace{}%
\AgdaBound{n}\AgdaSpace{}%
\AgdaOperator{\AgdaDatatype{memberOf}}\AgdaSpace{}%
\AgdaBound{ys}\AgdaSpace{}%
\AgdaSymbol{→}\AgdaSpace{}%
\AgdaBound{P}\AgdaSpace{}%
\AgdaBound{n}\AgdaSymbol{)}\AgdaSpace{}%
\AgdaSymbol{→}\<%
\\
\>[16]\AgdaRecord{Always}\AgdaSpace{}%
\AgdaBound{P}\AgdaSpace{}%
\AgdaBound{ys}\<%
\\
\\[\AgdaEmptyExtraSkip]%
\>[0]\AgdaField{Always.list}\AgdaSpace{}%
\AgdaSymbol{(}\AgdaFunction{alw{-}complete}\AgdaSpace{}%
\AgdaBound{P}\AgdaSpace{}%
\AgdaBound{ys}\AgdaSpace{}%
\AgdaBound{f}\AgdaSymbol{)}\AgdaSpace{}%
\AgdaKeyword{with}\AgdaSpace{}%
\AgdaFunction{inspect}\AgdaSpace{}%
\AgdaSymbol{(}\AgdaField{MyColist.list}\AgdaSpace{}%
\AgdaBound{ys}\AgdaSymbol{)}\<%
\\
\>[0]\AgdaSymbol{...}\AgdaSpace{}%
\AgdaSymbol{|}\AgdaSpace{}%
\AgdaInductiveConstructor{nothing}\AgdaSpace{}%
\AgdaOperator{\AgdaInductiveConstructor{with≡}}\AgdaSpace{}%
\AgdaBound{eq}\AgdaSpace{}%
\AgdaSymbol{=}\AgdaSpace{}%
\AgdaInductiveConstructor{inl}\AgdaSpace{}%
\AgdaBound{eq}\<%
\\
\>[0]\AgdaSymbol{...}%
\>[262I]\AgdaSymbol{|}\AgdaSpace{}%
\AgdaInductiveConstructor{just}\AgdaSpace{}%
\AgdaOperator{\AgdaInductiveConstructor{⟨}}\AgdaSpace{}%
\AgdaBound{x}\AgdaSpace{}%
\AgdaOperator{\AgdaInductiveConstructor{,}}\AgdaSpace{}%
\AgdaBound{xs}\AgdaSpace{}%
\AgdaOperator{\AgdaInductiveConstructor{⟩}}\AgdaSpace{}%
\AgdaOperator{\AgdaInductiveConstructor{with≡}}\AgdaSpace{}%
\AgdaBound{eq}\AgdaSpace{}%
\AgdaSymbol{=}\<%
\\
\>[.]\<[262I]%
\>[4]\AgdaInductiveConstructor{inr}\AgdaSpace{}%
\AgdaOperator{\AgdaInductiveConstructor{<}}\<%
\\
\>[4][@{}l@{\AgdaIndent{0}}]%
\>[6]\AgdaOperator{\AgdaInductiveConstructor{⟨}}\AgdaSpace{}%
\AgdaBound{x}\AgdaSpace{}%
\AgdaOperator{\AgdaInductiveConstructor{,}}\AgdaSpace{}%
\AgdaBound{xs}\AgdaSpace{}%
\AgdaOperator{\AgdaInductiveConstructor{⟩}}\AgdaSpace{}%
\AgdaOperator{\AgdaInductiveConstructor{,}}\<%
\\
\>[6]\AgdaOperator{\AgdaInductiveConstructor{⟨}}\AgdaSpace{}%
\AgdaBound{eq}\AgdaSpace{}%
\AgdaOperator{\AgdaInductiveConstructor{,}}\AgdaSpace{}%
\AgdaOperator{\AgdaInductiveConstructor{⟨}}\AgdaSpace{}%
\AgdaBound{f}\AgdaSpace{}%
\AgdaSymbol{(}\AgdaInductiveConstructor{mem{-}h}\AgdaSpace{}%
\AgdaBound{eq}\AgdaSymbol{)}\AgdaSpace{}%
\AgdaOperator{\AgdaInductiveConstructor{,}}\AgdaSpace{}%
\AgdaFunction{alw{-}complete}\AgdaSpace{}%
\AgdaBound{P}\AgdaSpace{}%
\AgdaBound{xs}\AgdaSpace{}%
\AgdaSymbol{(λ}\AgdaSpace{}%
\AgdaBound{m}\AgdaSpace{}%
\AgdaSymbol{→}\AgdaSpace{}%
\AgdaBound{f}\AgdaSpace{}%
\AgdaSymbol{(}\AgdaInductiveConstructor{mem{-}t}\AgdaSpace{}%
\AgdaBound{eq}\AgdaSpace{}%
\AgdaBound{m}\AgdaSymbol{))}\AgdaSpace{}%
\AgdaOperator{\AgdaInductiveConstructor{⟩}}\AgdaSpace{}%
\AgdaOperator{\AgdaInductiveConstructor{⟩}}\<%
\\
\>[4]\AgdaOperator{\AgdaInductiveConstructor{>}}\<%
\end{code}
\end{center}

The proof is by copattern matching on the result of \agda{alw-complete} and inside by induction on how the list is observed. In this case the code has no challenges to face due to the fact that \agda{inspect} (from the module \textit{Singleton}, see \refToSection{agda-modules}) provides the required proof \agda{eq} to feed the constructors of \textit{Always}.

\section{Infinitely Often}
\label{sect:infinitely-often}
Given a predicate $P$, $\InfinitelyOften {P} {\xs}$ holds if there exist infinite elements of  $\xs$ such that $P$ holds on them. 
This implies that the {list} must be infinite. 
The inference system is similar to that of \textit{Eventually}, but in this case the axiom is replaced by a coaxiom. \LC{Notice that \textit{InfinitelyOften} must be defined through corules, otherwise we would accept judgments in which the predicate $P$ does not hold on any of the elements of the list.}

\EZComm{cambiati nomi regole, unformare in Agda}

\begin{small}
\begin{quote}
$\NamedCoRule{co-io-h} {}{\InfinitelyOften {\Pred} {\x{:}\xs}} \Pred(\x)\Space
\NamedRule{io-t} {\InfinitelyOften{\Pred}{\xs}}{\InfinitelyOften {\Pred} {\x{:}\xs}}$
\end{quote}
\end{small}

In this section we will follow the same steps of \refToChapter{flex}. First of all, we show the specification for this predicate:

\begin{center}
$\Spec_\Pred = \{ \xs \mid \forall i\geq 0. \exists n.n > i \land \Pred (\get {\xs} {n})$
\end{center}

Notice that the specification needs to be expressed in terms of the index of the element in the list, in order to use the $>$ relation. Indeed, the judgments $\memberOf {\x} {\xs}$ and $\memberOf {\y} {\xs}$ provide no information on whether $\x$ comes before or after $\y$. Alternatively we could have defined a variant of \textit{memberOf} that takes into account also the index.
\EZComm{non c'entra qui: Recalling what we said in \refToChapter{inductive}, \textit{get} for colists is not guaranteed to return a value (more details in \refToSection{get}). In this case this problem is avoided since the predicate implies that the list is only infinite.}

We begin by proving the soundness of the predicate.

\begin{statement}
The definition of \textit{InfinitelyOften} by the inference system with corules is sound with respect to its specification.
\end{statement}
\begin{proof}
We have to prove that $\Generated{\is}{\cois}\subseteq\Spec$. That is, assuming that $\InfinitelyOften{\Pred} {\xs}\in\Generated{\is}{\cois}$, we have to prove that, for all $i\geq 0$, there exists $n>i$ such that $\Pred (\get {\xs} {n})$ holds. The proof is by arithmetic induction on $i$.\EZComm{ricontrollare!}
First of all, from $\InfinitelyOften{\Pred} {\xs}\in\Generated{\is}{\cois}$, we get $\xs=\x{:}\ys$.
\begin{description}
\item[i = 0] From $\InfinitelyOften{\Pred} {\xs}\in\Generated{\is}{\cois}$, we get $\InfinitelyOften{\Pred} {\ys}\in\Generated{\is}{\cois}$.
We prove that there exists $n \geq 0$ such that\footnote{Note that we have to reason on the tail because we are using $>$ rather than $\geq$. Otherwise the proof would not work in the \refToRule{co-io-h} case, since it is not true that $0 > 0$.} $\Pred(\get {\ys} {n})$, which implies $\Pred(\get {\xs} {n+1})$. Since $\Generated{\is}{\cois}\subseteq\Ind{\is\cup\cois}$ holds by definition, we can reason by induction on the definition of $\InfinitelyOften{\Pred}{\ys}$ in $\Ind{\is\cup\cois}$.
\begin{description}
\item[\refToRule{co-io-h}] We have $\ys=\y{:}\zs$, and $\Pred(\y)$. Hence, the thesis holds for $n=0$.
\item[\refToRule{io-t}] We have $\ys=\y{:}\zs$, and $\Eventually {\Pred} {\zs}$. Hence, by inductive hypothesis, there exists $n\geq 0$ such that $\Pred(\get {\ys} {n})$, which implies $\Pred(\get {\xs} {n+1})$.
\end{description}
\item[i + 1] We have to prove that there exists $n > i + 1$ such that $\Pred(\get {\xs} {n})$. By inductive hypothesis we know that there exists $m > i$ such that $\Pred(\get {\ys} {m})$ holds, hence  $\Pred(\get {\xs} {m+1})$ holds and $m+1>i+1$.
\end{description}
\end{proof}

\begin{statement}
The definition of \textit{InfinitelyOften} by the inference system with corules is complete with respect to its specification.
\end{statement}

To prove the completeness statement above, we separately prove \textit{boundedness} and \textit{consistency} of $\Spec$.

\begin{statement}
\label{stmt:max-bound}
$\Spec$ is bounded with respect to the inference system with corules of \textit{InfinitelyOften}.
\end{statement}
\begin{proof}
We have to prove that $\Spec \subseteq \Ind{\is\cup\cois}$, that is, for each $\xs\in\Spec$, the judgment $\InfinitelyOften{\Pred}{\xs}$ can be derived in $\is\cup\cois$. 
Since $\xs\in\Spec$ implies $\Spec'=\{\exists n \geq 0 \mid \Pred(\get{\xs}{n})\}$,  we prove that, for each $\xs\in\Spec'$, the judgment $\InfinitelyOften{\Pred}{\xs}$ can be derived.
 The proof is by arithmetic induction on $n$.
\begin{description}
\item[n = 0] We have $\xs=\x{:}\ys$, and the judgment can be derived by rule \refToRule{co-io-h}.
\item[n + 1] We have $\Pred(\get {\xs} {n+1})$, hence $\xs=\x{:}\ys$, and $\Pred(\get {\xs} {n})$ holds.  By inductive hypothesis we can derive $\InfinitelyOften{\Pred}{\ys}$, and we get the thesis by applying rule \refToRule{io-t}.
\end{description}
\end{proof}

\begin{statement}
$\Spec$ is consistent with respect to $\is$.
\end{statement}
\begin{proof}
We have to prove that, for each $\xs\in\Spec$, $\InfinitelyOften{\Pred}{\xs}$ is the consequence of a rule with premises which are in $\Spec$ as well.
Since $\xs\in\Spec$, $\xs$ cannot be empty, hence it is of shape $\x{:}\ys$. Hence, $\InfinitelyOften{\Pred}{\xs}$ is the consequence of rule \refToRule{io-t} with premise $\InfinitelyOften{\Pred}{\xs}$.
Hence, we have to prove (the rather obvious fact) that $\x{:}\ys\in\Spec$ implies $\ys \in \Spec$. To be formal: we know that, for all  $i \geq 0$, there exists $n>i$ such that $\Pred(\get {\x{:}\ys} {n}) $ holds, 
 and we want to prove that, for all  $j\geq 0$, there exists $m>j$ such that $\Pred(\get {\ys} {m} $ holds. From the hypothesis we know that there exists $n>j+1$ such that $\Pred(\get {\x{:}\ys} {n}) $ holds, which is equivalent to $\Pred(\get {\ys} {n-1})$. Since $n>j+1>0$, we have that $n-1>j\geq 0$.
\end{proof}

Now that we proved the completeness of our predicate we can move to 
Agda. First we show the codes of the types and the auxiliary functions that are the basic ingredients of the correctness proofs.

\begin{center}
\begin{code}
\>[0]\AgdaKeyword{data}\AgdaSpace{}%
\AgdaOperator{\AgdaDatatype{\AgdaUnderscore{}InfinitelyOften{-}ind\AgdaUnderscore{}}}\AgdaSpace{}%
\AgdaSymbol{\{}\AgdaBound{A}\AgdaSpace{}%
\AgdaSymbol{:}\AgdaSpace{}%
\AgdaPrimitiveType{Set}\AgdaSymbol{\}}\AgdaSpace{}%
\AgdaSymbol{:}\AgdaSpace{}%
\AgdaSymbol{(}\AgdaBound{A}\AgdaSpace{}%
\AgdaSymbol{→}\AgdaSpace{}%
\AgdaPrimitiveType{Set}\AgdaSymbol{)}\AgdaSpace{}%
\AgdaSymbol{→}\AgdaSpace{}%
\AgdaDatatype{Colist}\AgdaSpace{}%
\AgdaBound{A}\AgdaSpace{}%
\AgdaPostulate{∞}\AgdaSpace{}%
\AgdaSymbol{→}\AgdaSpace{}%
\AgdaPrimitiveType{Set}\AgdaSpace{}%
\AgdaKeyword{where}\<%
\\
\>[0][@{}l@{\AgdaIndent{0}}]%
\>[2]\AgdaInductiveConstructor{co{-}io{-}h}\AgdaSpace{}%
\AgdaSymbol{:}\AgdaSpace{}%
\AgdaSymbol{∀}\AgdaSpace{}%
\AgdaSymbol{\{}\AgdaBound{x}\AgdaSpace{}%
\AgdaBound{xs}\AgdaSpace{}%
\AgdaBound{P}\AgdaSymbol{\}}\AgdaSpace{}%
\AgdaSymbol{→}\AgdaSpace{}%
\AgdaBound{P}\AgdaSpace{}%
\AgdaBound{x}\AgdaSpace{}%
\AgdaSymbol{→}\AgdaSpace{}%
\AgdaBound{P}\AgdaSpace{}%
\AgdaOperator{\AgdaDatatype{InfinitelyOften{-}ind}}\AgdaSpace{}%
\AgdaSymbol{(}\AgdaBound{x}\AgdaSpace{}%
\AgdaOperator{\AgdaInductiveConstructor{∷}}\AgdaSpace{}%
\AgdaBound{xs}\AgdaSymbol{)}\<%
\\
\>[2]\AgdaInductiveConstructor{io{-}t{-}ind}\AgdaSpace{}%
\AgdaSymbol{:}\AgdaSpace{}%
\AgdaSymbol{∀}\AgdaSpace{}%
\AgdaSymbol{\{}\AgdaBound{x}%
\>[56I]\AgdaBound{xs}\AgdaSpace{}%
\AgdaBound{P}\AgdaSymbol{\}}\AgdaSpace{}%
\AgdaSymbol{→}\AgdaSpace{}%
\AgdaBound{P}\AgdaSpace{}%
\AgdaOperator{\AgdaDatatype{InfinitelyOften{-}ind}}\AgdaSpace{}%
\AgdaSymbol{(}\AgdaField{Thunk.force}\AgdaSpace{}%
\AgdaBound{xs}\AgdaSymbol{)}\AgdaSpace{}%
\AgdaSymbol{→}\<%
\\
\>[.]\<[56I]%
\>[18]\AgdaBound{P}\AgdaSpace{}%
\AgdaOperator{\AgdaDatatype{InfinitelyOften{-}ind}}\AgdaSpace{}%
\AgdaSymbol{(}\AgdaBound{x}\AgdaSpace{}%
\AgdaOperator{\AgdaInductiveConstructor{∷}}\AgdaSpace{}%
\AgdaBound{xs}\AgdaSymbol{)}\<%
\\
\\[\AgdaEmptyExtraSkip]%
\>[0]\AgdaKeyword{data}\AgdaSpace{}%
\AgdaDatatype{InfinitelyOften}\AgdaSpace{}%
\AgdaSymbol{\{}\AgdaBound{A}\AgdaSpace{}%
\AgdaSymbol{:}\AgdaSpace{}%
\AgdaPrimitiveType{Set}\AgdaSymbol{\}}\AgdaSpace{}%
\AgdaSymbol{:}\AgdaSpace{}%
\AgdaSymbol{(}\AgdaBound{A}\AgdaSpace{}%
\AgdaSymbol{→}\AgdaSpace{}%
\AgdaPrimitiveType{Set}\AgdaSymbol{)}\AgdaSpace{}%
\AgdaSymbol{→}\AgdaSpace{}%
\AgdaDatatype{Colist}\AgdaSpace{}%
\AgdaBound{A}\AgdaSpace{}%
\AgdaPostulate{∞}\AgdaSpace{}%
\AgdaSymbol{→}\AgdaSpace{}%
\AgdaPostulate{Size}\AgdaSpace{}%
\AgdaSymbol{→}\AgdaSpace{}%
\AgdaPrimitiveType{Set}\AgdaSpace{}%
\AgdaKeyword{where}\<%
\\
\>[0][@{}l@{\AgdaIndent{0}}]%
\>[2]\AgdaInductiveConstructor{io{-}t}%
\>[85I]\AgdaSymbol{:}\AgdaSpace{}%
\AgdaSymbol{∀}\AgdaSpace{}%
\AgdaSymbol{\{}\AgdaBound{x}\AgdaSpace{}%
\AgdaBound{xs}\AgdaSpace{}%
\AgdaBound{P}\AgdaSpace{}%
\AgdaBound{i}\AgdaSymbol{\}}\AgdaSpace{}%
\AgdaSymbol{→}\AgdaSpace{}%
\AgdaRecord{Thunk}\AgdaSpace{}%
\AgdaSymbol{(}\AgdaDatatype{InfinitelyOften}\AgdaSpace{}%
\AgdaBound{P}\AgdaSpace{}%
\AgdaSymbol{(}\AgdaField{Thunk.force}\AgdaSpace{}%
\AgdaBound{xs}\AgdaSymbol{))}\AgdaSpace{}%
\AgdaBound{i}\AgdaSpace{}%
\AgdaSymbol{→}\<%
\\
\>[.]\<[85I]%
\>[7]\AgdaBound{P}\AgdaSpace{}%
\AgdaOperator{\AgdaDatatype{InfinitelyOften{-}ind}}\AgdaSpace{}%
\AgdaSymbol{(}\AgdaBound{x}\AgdaSpace{}%
\AgdaOperator{\AgdaInductiveConstructor{∷}}\AgdaSpace{}%
\AgdaBound{xs}\AgdaSymbol{)}\AgdaSpace{}%
\AgdaSymbol{→}\AgdaSpace{}%
\AgdaDatatype{InfinitelyOften}\AgdaSpace{}%
\AgdaBound{P}\AgdaSpace{}%
\AgdaSymbol{(}\AgdaBound{x}\AgdaSpace{}%
\AgdaOperator{\AgdaInductiveConstructor{∷}}\AgdaSpace{}%
\AgdaBound{xs}\AgdaSymbol{)}\AgdaSpace{}%
\AgdaBound{i}\<%
\\
\\[\AgdaEmptyExtraSkip]%
\>[0]\AgdaKeyword{data}\AgdaSpace{}%
\AgdaDatatype{Step}\AgdaSpace{}%
\AgdaSymbol{\{}\AgdaBound{A}\AgdaSpace{}%
\AgdaSymbol{:}\AgdaSpace{}%
\AgdaPrimitiveType{Set}\AgdaSymbol{\}(}\AgdaBound{S}\AgdaSpace{}%
\AgdaSymbol{:}\AgdaSpace{}%
\AgdaDatatype{Colist}\AgdaSpace{}%
\AgdaBound{A}\AgdaSpace{}%
\AgdaPostulate{∞}\AgdaSpace{}%
\AgdaSymbol{→}\AgdaSpace{}%
\AgdaPrimitiveType{Set}\AgdaSymbol{)}\AgdaSpace{}%
\AgdaSymbol{:}\AgdaSpace{}%
\AgdaDatatype{Colist}\AgdaSpace{}%
\AgdaBound{A}\AgdaSpace{}%
\AgdaPostulate{∞}\AgdaSpace{}%
\AgdaSymbol{→}\AgdaSpace{}%
\AgdaPrimitiveType{Set}\AgdaSpace{}%
\AgdaKeyword{where}\<%
\\
\>[0][@{}l@{\AgdaIndent{0}}]%
\>[2]\AgdaInductiveConstructor{io{-}t}\AgdaSpace{}%
\AgdaSymbol{:}\AgdaSpace{}%
\AgdaSymbol{∀}\AgdaSpace{}%
\AgdaSymbol{\{}\AgdaBound{x}\AgdaSpace{}%
\AgdaBound{xs}\AgdaSymbol{\}}\AgdaSpace{}%
\AgdaSymbol{→}\AgdaSpace{}%
\AgdaBound{S}\AgdaSpace{}%
\AgdaSymbol{(}\AgdaField{Thunk.force}\AgdaSpace{}%
\AgdaBound{xs}\AgdaSymbol{)}\AgdaSpace{}%
\AgdaSymbol{→}\AgdaSpace{}%
\AgdaDatatype{Step}\AgdaSpace{}%
\AgdaBound{S}\AgdaSpace{}%
\AgdaSymbol{(}\AgdaBound{x}\AgdaSpace{}%
\AgdaOperator{\AgdaInductiveConstructor{∷}}\AgdaSpace{}%
\AgdaBound{xs}\AgdaSymbol{)}\<%
\\
\\[\AgdaEmptyExtraSkip]%
\>[0]\AgdaFunction{lift}\AgdaSpace{}%
\AgdaSymbol{:}\AgdaSpace{}%
\AgdaSymbol{\{}\AgdaBound{A}\AgdaSpace{}%
\AgdaSymbol{:}\AgdaSpace{}%
\AgdaPrimitiveType{Set}\AgdaSymbol{\}}\AgdaSpace{}%
\AgdaSymbol{→}\AgdaSpace{}%
\AgdaSymbol{(}\AgdaBound{A}\AgdaSpace{}%
\AgdaSymbol{→}\AgdaSpace{}%
\AgdaPrimitiveType{Set}\AgdaSymbol{)}\AgdaSpace{}%
\AgdaSymbol{→}\AgdaSpace{}%
\AgdaSymbol{(}\AgdaDatatype{Maybe}\AgdaSpace{}%
\AgdaBound{A}\AgdaSpace{}%
\AgdaSymbol{→}\AgdaSpace{}%
\AgdaPrimitiveType{Set}\AgdaSymbol{)}\<%
\\
\>[0]\AgdaFunction{lift}\AgdaSpace{}%
\AgdaBound{P}\AgdaSpace{}%
\AgdaSymbol{(}\AgdaInductiveConstructor{just}\AgdaSpace{}%
\AgdaBound{x}\AgdaSymbol{)}\AgdaSpace{}%
\AgdaSymbol{=}\AgdaSpace{}%
\AgdaBound{P}\AgdaSpace{}%
\AgdaBound{x}\<%
\\
\>[0]\AgdaFunction{lift}\AgdaSpace{}%
\AgdaBound{P}\AgdaSpace{}%
\AgdaInductiveConstructor{nothing}\AgdaSpace{}%
\AgdaSymbol{=}\AgdaSpace{}%
\AgdaDatatype{⊥}\<%
\\
\\[\AgdaEmptyExtraSkip]%
\>[0]\AgdaFunction{S}\AgdaSpace{}%
\AgdaSymbol{:}\AgdaSpace{}%
\AgdaSymbol{\{}\AgdaBound{A}\AgdaSpace{}%
\AgdaSymbol{:}\AgdaSpace{}%
\AgdaPrimitiveType{Set}\AgdaSymbol{\}}\AgdaSpace{}%
\AgdaSymbol{→}\AgdaSpace{}%
\AgdaSymbol{(}\AgdaBound{A}\AgdaSpace{}%
\AgdaSymbol{→}\AgdaSpace{}%
\AgdaPrimitiveType{Set}\AgdaSymbol{)}\AgdaSpace{}%
\AgdaSymbol{→}\AgdaSpace{}%
\AgdaDatatype{Colist}\AgdaSpace{}%
\AgdaBound{A}\AgdaSpace{}%
\AgdaPostulate{∞}\AgdaSpace{}%
\AgdaSymbol{→}\AgdaSpace{}%
\AgdaPrimitiveType{Set}\<%
\\
\>[0]\AgdaFunction{S}\AgdaSpace{}%
\AgdaBound{P}\AgdaSpace{}%
\AgdaBound{xs}\AgdaSpace{}%
\AgdaSymbol{=}\AgdaSpace{}%
\AgdaSymbol{(∀}\AgdaSpace{}%
\AgdaSymbol{(}\AgdaBound{i}\AgdaSpace{}%
\AgdaSymbol{:}\AgdaSpace{}%
\AgdaDatatype{Nat}\AgdaSymbol{)}\AgdaSpace{}%
\AgdaSymbol{→}\AgdaSpace{}%
\AgdaDatatype{∃}\AgdaSpace{}%
\AgdaDatatype{Nat}\AgdaSpace{}%
\AgdaSymbol{(λ}\AgdaSpace{}%
\AgdaBound{n}\AgdaSpace{}%
\AgdaSymbol{→}\AgdaSpace{}%
\AgdaSymbol{(}\AgdaBound{n}\AgdaSpace{}%
\AgdaOperator{\AgdaDatatype{>}}\AgdaSpace{}%
\AgdaBound{i}\AgdaSymbol{)}\AgdaSpace{}%
\AgdaOperator{\AgdaDatatype{∧}}\AgdaSpace{}%
\AgdaSymbol{((}\AgdaFunction{lift}\AgdaSpace{}%
\AgdaBound{P}\AgdaSymbol{)}\AgdaSpace{}%
\AgdaSymbol{(}\AgdaFunction{lookup}\AgdaSpace{}%
\AgdaBound{n}\AgdaSpace{}%
\AgdaBound{xs}\AgdaSymbol{))))}\<%
\end{code}
\end{center}

Some comments follow.
\begin{itemize}
\item \agda{InfinitelyOften-ind} represents the inference system with corules inductively interpreted. In fact the \agda{data} constructor is used without the need of \textit{thunks} and \textit{size}.
\item \agda{InfinitelyOften} represents $\Generated {\is} {\cois}$.
\item \agda{Step} is the type used to prove \textit{consistency} of the specification. As we already said in \refToChapter{flex}, it means that there exists a rule such that the premises are in the specification. There is only a constructor since there is only a rule in $\is$.
\item In the theoretical part we discussed about the fact that the involved predicate works only on infinite lists. In Agda we are giving an implementation of such predicate on colists, that are possibly infinite  data structures. Thus, the \agda{lift} function allows us to apply the predicate $P$ directly on the result of the \agda{lookup} function which is not guaranteed to always return a value.
\item \agda{S} is the specification. You can see that the usage of the \textit{lift} function leads to a compact code. Otherwise we should have specified that there exists an element at position $n$ and that $P$ holds on it.
\end{itemize}

Notice that we meet again some notions that we have already introduced in \refToChapter{flex}. Now we can show the proof of \textit{soundness}.

\begin{center}
\begin{code}
\>[0]\AgdaFunction{sound{-}aux}%
\>[200I]\AgdaSymbol{:}\AgdaSpace{}%
\AgdaSymbol{\{}\AgdaBound{A}\AgdaSpace{}%
\AgdaSymbol{:}\AgdaSpace{}%
\AgdaPrimitiveType{Set}\AgdaSymbol{\}\{}\AgdaBound{ys}\AgdaSpace{}%
\AgdaSymbol{:}\AgdaSpace{}%
\AgdaDatatype{Colist}\AgdaSpace{}%
\AgdaBound{A}\AgdaSpace{}%
\AgdaPostulate{∞}\AgdaSymbol{\}\{}\AgdaBound{P}\AgdaSpace{}%
\AgdaSymbol{:}\AgdaSpace{}%
\AgdaBound{A}\AgdaSpace{}%
\AgdaSymbol{→}\AgdaSpace{}%
\AgdaPrimitiveType{Set}\AgdaSymbol{\}}\AgdaSpace{}%
\AgdaSymbol{→}\<%
\\
\>[.]\<[200I]%
\>[10]\AgdaBound{P}\AgdaSpace{}%
\AgdaOperator{\AgdaDatatype{InfinitelyOften{-}ind}}\AgdaSpace{}%
\AgdaBound{ys}\AgdaSpace{}%
\AgdaSymbol{→}\AgdaSpace{}%
\AgdaDatatype{∃}\AgdaSpace{}%
\AgdaDatatype{Nat}\AgdaSpace{}%
\AgdaSymbol{(λ}\AgdaSpace{}%
\AgdaBound{n}\AgdaSpace{}%
\AgdaSymbol{→}\AgdaSpace{}%
\AgdaSymbol{(}\AgdaFunction{lift}\AgdaSpace{}%
\AgdaBound{P}\AgdaSymbol{)}\AgdaSpace{}%
\AgdaSymbol{(}\AgdaFunction{lookup}\AgdaSpace{}%
\AgdaBound{n}\AgdaSpace{}%
\AgdaBound{ys}\AgdaSymbol{))}\<%
\\
\>[0]\AgdaFunction{sound{-}aux}\AgdaSpace{}%
\AgdaSymbol{\{}\AgdaBound{A}\AgdaSymbol{\}}\AgdaSpace{}%
\AgdaSymbol{\{\AgdaUnderscore{}}\AgdaSpace{}%
\AgdaOperator{\AgdaInductiveConstructor{∷}}\AgdaSpace{}%
\AgdaBound{xs}\AgdaSymbol{\}}\AgdaSpace{}%
\AgdaSymbol{\{}\AgdaBound{P}\AgdaSymbol{\}}\AgdaSpace{}%
\AgdaSymbol{(}\AgdaInductiveConstructor{co{-}io{-}h}\AgdaSpace{}%
\AgdaBound{Px}\AgdaSymbol{)}\AgdaSpace{}%
\AgdaSymbol{=}\AgdaSpace{}%
\AgdaOperator{\AgdaInductiveConstructor{<}}\AgdaSpace{}%
\AgdaInductiveConstructor{zero}\AgdaSpace{}%
\AgdaOperator{\AgdaInductiveConstructor{,}}\AgdaSpace{}%
\AgdaBound{Px}\AgdaSpace{}%
\AgdaOperator{\AgdaInductiveConstructor{>}}\<%
\\
\>[0]\AgdaFunction{sound{-}aux}\AgdaSpace{}%
\AgdaSymbol{\{}\AgdaBound{A}\AgdaSymbol{\}}%
\>[240I]\AgdaSymbol{\{\AgdaUnderscore{}}\AgdaSpace{}%
\AgdaOperator{\AgdaInductiveConstructor{∷}}\AgdaSpace{}%
\AgdaBound{xs}\AgdaSymbol{\}}\AgdaSpace{}%
\AgdaSymbol{\{}\AgdaBound{P}\AgdaSymbol{\}}\AgdaSpace{}%
\AgdaSymbol{(}\AgdaInductiveConstructor{io{-}t{-}ind}\AgdaSpace{}%
\AgdaBound{io{-}xs}\AgdaSymbol{)}\AgdaSpace{}%
\AgdaSymbol{=}\<%
\\
\>[.]\<[240I]%
\>[14]\AgdaOperator{\AgdaInductiveConstructor{<}}\AgdaSpace{}%
\AgdaSymbol{(}\AgdaInductiveConstructor{suc}\AgdaSpace{}%
\AgdaSymbol{(}\AgdaFunction{witness}\AgdaSpace{}%
\AgdaSymbol{(}\AgdaFunction{sound{-}aux}\AgdaSpace{}%
\AgdaBound{io{-}xs}\AgdaSymbol{)))}\AgdaSpace{}%
\AgdaOperator{\AgdaInductiveConstructor{,}}\AgdaSpace{}%
\AgdaFunction{proof}\AgdaSpace{}%
\AgdaSymbol{(}\AgdaFunction{sound{-}aux}\AgdaSpace{}%
\AgdaBound{io{-}xs}\AgdaSymbol{)}\AgdaSpace{}%
\AgdaOperator{\AgdaInductiveConstructor{>}}\<%
\\
\\[\AgdaEmptyExtraSkip]%
\\[\AgdaEmptyExtraSkip]%
\>[0]\AgdaFunction{io2ioind}\AgdaSpace{}%
\AgdaSymbol{:}\AgdaSpace{}%
\AgdaSymbol{∀\{}\AgdaBound{A}\AgdaSpace{}%
\AgdaBound{ys}\AgdaSymbol{\}\{}\AgdaBound{P}%
\>[259I]\AgdaSymbol{:}\AgdaSpace{}%
\AgdaBound{A}\AgdaSpace{}%
\AgdaSymbol{→}\AgdaSpace{}%
\AgdaPrimitiveType{Set}\AgdaSymbol{\}}\AgdaSpace{}%
\AgdaSymbol{→}\AgdaSpace{}%
\AgdaSymbol{(∀}\AgdaSpace{}%
\AgdaSymbol{\{}\AgdaBound{i}\AgdaSymbol{\}}\AgdaSpace{}%
\AgdaSymbol{→}\AgdaSpace{}%
\AgdaDatatype{InfinitelyOften}\AgdaSpace{}%
\AgdaBound{P}\AgdaSpace{}%
\AgdaBound{ys}\AgdaSpace{}%
\AgdaBound{i}\AgdaSymbol{)}\AgdaSpace{}%
\AgdaSymbol{→}\<%
\\
\>[.]\<[259I]%
\>[21]\AgdaBound{P}\AgdaSpace{}%
\AgdaOperator{\AgdaDatatype{InfinitelyOften{-}ind}}\AgdaSpace{}%
\AgdaBound{ys}\<%
\\
\>[0]\AgdaFunction{io2ioind}\AgdaSpace{}%
\AgdaBound{inf{-}o}\AgdaSpace{}%
\AgdaKeyword{with}\AgdaSpace{}%
\AgdaBound{inf{-}o}\<%
\\
\>[0]\AgdaFunction{io2ioind}\AgdaSpace{}%
\AgdaBound{inf{-}o}\AgdaSpace{}%
\AgdaSymbol{|}\AgdaSpace{}%
\AgdaInductiveConstructor{io{-}t}\AgdaSpace{}%
\AgdaBound{inf{-}o{-}xs}\AgdaSpace{}%
\AgdaBound{io{-}ys{-}ind}\AgdaSpace{}%
\AgdaSymbol{=}\AgdaSpace{}%
\AgdaBound{io{-}ys{-}ind}\<%
\\
\\[\AgdaEmptyExtraSkip]%
\\[\AgdaEmptyExtraSkip]%
\>[0]\AgdaFunction{io{-}sound}\AgdaSpace{}%
\AgdaSymbol{:}%
\>[285I]\AgdaSymbol{\{}\AgdaBound{A}\AgdaSpace{}%
\AgdaSymbol{:}\AgdaSpace{}%
\AgdaPrimitiveType{Set}\AgdaSymbol{\}\{}\AgdaBound{P}\AgdaSpace{}%
\AgdaSymbol{:}\AgdaSpace{}%
\AgdaBound{A}\AgdaSpace{}%
\AgdaSymbol{→}\AgdaSpace{}%
\AgdaPrimitiveType{Set}\AgdaSymbol{\}\{}\AgdaBound{ys}\AgdaSpace{}%
\AgdaSymbol{:}\AgdaSpace{}%
\AgdaDatatype{Colist}\AgdaSpace{}%
\AgdaBound{A}\AgdaSpace{}%
\AgdaPostulate{∞}\AgdaSymbol{\}}\AgdaSpace{}%
\AgdaSymbol{→}\<%
\\
\>[.]\<[285I]%
\>[11]\AgdaSymbol{(∀}\AgdaSpace{}%
\AgdaSymbol{\{}\AgdaBound{i}\AgdaSymbol{\}}\AgdaSpace{}%
\AgdaSymbol{→}\AgdaSpace{}%
\AgdaDatatype{InfinitelyOften}\AgdaSpace{}%
\AgdaBound{P}\AgdaSpace{}%
\AgdaBound{ys}\AgdaSpace{}%
\AgdaBound{i}\AgdaSymbol{)}\AgdaSpace{}%
\AgdaSymbol{→}\AgdaSpace{}%
\AgdaSymbol{((}\AgdaBound{i}\AgdaSpace{}%
\AgdaSymbol{:}\AgdaSpace{}%
\AgdaDatatype{Nat}\AgdaSymbol{)}\AgdaSpace{}%
\AgdaSymbol{→}\<%
\\
\>[11]\AgdaDatatype{∃}\AgdaSpace{}%
\AgdaDatatype{Nat}\AgdaSpace{}%
\AgdaSymbol{(λ}\AgdaSpace{}%
\AgdaBound{n}\AgdaSpace{}%
\AgdaSymbol{→}\AgdaSpace{}%
\AgdaSymbol{(}\AgdaBound{n}\AgdaSpace{}%
\AgdaOperator{\AgdaDatatype{>}}\AgdaSpace{}%
\AgdaBound{i}\AgdaSymbol{)}\AgdaSpace{}%
\AgdaOperator{\AgdaDatatype{∧}}\AgdaSpace{}%
\AgdaSymbol{((}\AgdaFunction{lift}\AgdaSpace{}%
\AgdaBound{P}\AgdaSymbol{)}\AgdaSpace{}%
\AgdaSymbol{(}\AgdaFunction{lookup}\AgdaSpace{}%
\AgdaBound{n}\AgdaSpace{}%
\AgdaBound{ys}\AgdaSymbol{))))}\<%
\\
\\[\AgdaEmptyExtraSkip]%
\>[0]\AgdaFunction{io{-}sound}\AgdaSpace{}%
\AgdaSymbol{\{}\AgdaBound{A}\AgdaSymbol{\}}\AgdaSpace{}%
\AgdaSymbol{\{}\AgdaBound{P}\AgdaSymbol{\}}\AgdaSpace{}%
\AgdaSymbol{\{}\AgdaBound{ys}\AgdaSymbol{\}}\AgdaSpace{}%
\AgdaBound{inf{-}o}\AgdaSpace{}%
\AgdaInductiveConstructor{zero}\AgdaSpace{}%
\AgdaKeyword{with}\AgdaSpace{}%
\AgdaBound{inf{-}o}\<%
\\
\>[0]\AgdaFunction{io{-}sound}%
\>[328I]\AgdaSymbol{\{}\AgdaBound{A}\AgdaSymbol{\}}\AgdaSpace{}%
\AgdaSymbol{\{}\AgdaBound{P}\AgdaSymbol{\}}\AgdaSpace{}%
\AgdaSymbol{\{\AgdaUnderscore{}}\AgdaSpace{}%
\AgdaOperator{\AgdaInductiveConstructor{∷}}\AgdaSpace{}%
\AgdaBound{xs}\AgdaSymbol{\}}\AgdaSpace{}%
\AgdaBound{inf{-}o}\AgdaSpace{}%
\AgdaInductiveConstructor{zero}\AgdaSpace{}%
\AgdaSymbol{|}\AgdaSpace{}%
\AgdaInductiveConstructor{io{-}t}\AgdaSpace{}%
\AgdaBound{inf{-}o{-}xs}\AgdaSpace{}%
\AgdaSymbol{\AgdaUnderscore{}}\AgdaSpace{}%
\AgdaSymbol{=}\<%
\\
\>[328I][@{}l@{\AgdaIndent{0}}]%
\>[10]\AgdaKeyword{let}\AgdaSpace{}%
\AgdaBound{ex}\AgdaSpace{}%
\AgdaSymbol{=}\AgdaSpace{}%
\AgdaSymbol{(}\AgdaFunction{sound{-}aux}\AgdaSpace{}%
\AgdaSymbol{(}\AgdaFunction{io2ioind}\AgdaSpace{}%
\AgdaSymbol{(}\AgdaBound{inf{-}o{-}xs}\AgdaSpace{}%
\AgdaSymbol{.}\AgdaField{force}\AgdaSymbol{)))}\AgdaSpace{}%
\AgdaKeyword{in}\<%
\\
\>[.]\<[328I]%
\>[9]\AgdaOperator{\AgdaInductiveConstructor{<}}\AgdaSpace{}%
\AgdaInductiveConstructor{suc}\AgdaSpace{}%
\AgdaSymbol{(}\AgdaFunction{witness}\AgdaSpace{}%
\AgdaBound{ex}\AgdaSymbol{)}\AgdaSpace{}%
\AgdaOperator{\AgdaInductiveConstructor{,}}\AgdaSpace{}%
\AgdaOperator{\AgdaInductiveConstructor{⟨}}\AgdaSpace{}%
\AgdaInductiveConstructor{g{-}zero}\AgdaSpace{}%
\AgdaOperator{\AgdaInductiveConstructor{,}}%
\>[42]\AgdaSymbol{(}\AgdaFunction{proof}\AgdaSpace{}%
\AgdaBound{ex}\AgdaSymbol{)}\AgdaSpace{}%
\AgdaOperator{\AgdaInductiveConstructor{⟩}}\AgdaSpace{}%
\AgdaOperator{\AgdaInductiveConstructor{>}}\<%
\\
\\[\AgdaEmptyExtraSkip]%
\>[0]\AgdaFunction{io{-}sound}\AgdaSpace{}%
\AgdaSymbol{\{}\AgdaBound{A}\AgdaSymbol{\}}\AgdaSpace{}%
\AgdaSymbol{\{}\AgdaBound{P}\AgdaSymbol{\}}\AgdaSpace{}%
\AgdaSymbol{\{}\AgdaBound{ys}\AgdaSymbol{\}}\AgdaSpace{}%
\AgdaBound{inf{-}o}\AgdaSpace{}%
\AgdaSymbol{(}\AgdaInductiveConstructor{suc}\AgdaSpace{}%
\AgdaBound{i}\AgdaSymbol{)}\AgdaSpace{}%
\AgdaKeyword{with}\AgdaSpace{}%
\AgdaBound{inf{-}o}\<%
\\
\>[0]\AgdaFunction{io{-}sound}\AgdaSpace{}%
\AgdaSymbol{\{}\AgdaBound{A}\AgdaSymbol{\}}%
\>[366I]\AgdaSymbol{\{}\AgdaBound{P}\AgdaSymbol{\}}\AgdaSpace{}%
\AgdaSymbol{\{\AgdaUnderscore{}}\AgdaSpace{}%
\AgdaOperator{\AgdaInductiveConstructor{∷}}\AgdaSpace{}%
\AgdaBound{xs}\AgdaSymbol{\}}\AgdaSpace{}%
\AgdaBound{inf{-}o}\AgdaSpace{}%
\AgdaSymbol{(}\AgdaInductiveConstructor{suc}\AgdaSpace{}%
\AgdaBound{i}\AgdaSymbol{)}\AgdaSpace{}%
\AgdaSymbol{|}\AgdaSpace{}%
\AgdaInductiveConstructor{io{-}t}\AgdaSpace{}%
\AgdaBound{info{-}xs}\AgdaSpace{}%
\AgdaSymbol{\AgdaUnderscore{}}\AgdaSpace{}%
\AgdaSymbol{=}\<%
\\
\>[.]\<[366I]%
\>[13]\AgdaKeyword{let}\AgdaSpace{}%
\AgdaBound{rec}\AgdaSpace{}%
\AgdaSymbol{=}\AgdaSpace{}%
\AgdaFunction{io{-}sound}\AgdaSpace{}%
\AgdaSymbol{(}\AgdaBound{info{-}xs}\AgdaSpace{}%
\AgdaSymbol{.}\AgdaField{force}\AgdaSymbol{)}\AgdaSpace{}%
\AgdaBound{i}\AgdaSpace{}%
\AgdaKeyword{in}\<%
\\
\>[13]\AgdaOperator{\AgdaInductiveConstructor{<}}%
\>[385I]\AgdaSymbol{(}\AgdaInductiveConstructor{suc}\AgdaSpace{}%
\AgdaSymbol{(}\AgdaFunction{witness}\AgdaSpace{}%
\AgdaBound{rec}\AgdaSymbol{))}\AgdaSpace{}%
\AgdaOperator{\AgdaInductiveConstructor{,}}\<%
\\
\>[385I][@{}l@{\AgdaIndent{0}}]%
\>[17]\AgdaOperator{\AgdaInductiveConstructor{⟨}}\AgdaSpace{}%
\AgdaInductiveConstructor{g{-}suc}\AgdaSpace{}%
\AgdaSymbol{(}\AgdaFunction{∧{-}left}\AgdaSpace{}%
\AgdaSymbol{(}\AgdaFunction{proof}\AgdaSpace{}%
\AgdaBound{rec}\AgdaSymbol{))}\AgdaSpace{}%
\AgdaOperator{\AgdaInductiveConstructor{,}}\AgdaSpace{}%
\AgdaFunction{∧{-}right}\AgdaSpace{}%
\AgdaSymbol{(}\AgdaFunction{proof}\AgdaSpace{}%
\AgdaBound{rec}\AgdaSymbol{)}\AgdaOperator{\AgdaInductiveConstructor{⟩}}\AgdaSpace{}%
\AgdaOperator{\AgdaInductiveConstructor{>}}\<%
\end{code}
\end{center}

The proof is by induction over the index. Notice that there are two auxiliary functions that allow to:
\begin{itemize}
\item Get the finite proof tree, that is the proof in \agda{InfinitelyOften-ind}.
\item Get the index of the element on which the predicate holds given the proof in\\ \agda{InfinitelyOften-ind}. This is done by induction over the rules of the inference system with corules.
\end{itemize}
When the index $i$ is \agda{zero} we call this two function keeping in mind that we have to find an index greater than zero. In fact, we call them with the proof that \textit{InfinitelyOften} holds on the tail as input and then we take the successor of the result in order to refer the output to the whole list. \agda{g-zero} can be used because Agda recognizes that we are looking for an $n$ such that \agda{n > zero} and so also \agda{suc n > zero}.
When \agda{suc i} is considered we call the proof recursively. We need to find $n > i + 1$ and the call tells that there exists $j > i$ inside the tail. But the element at position $j$ in the tail is at position $j + 1$ in the whole list, thus $j + 1 > i + 1$. In Agda this means that we need to take the successor of the witness of the recursive call and that we need to  apply the constructor \agda{g-suc}.
In order to prove the completeness of \textit{InfinitelyOften} we report the code of the \textbf{bounded coinduction principle} as we already did in \refToChapter{flex}.

\begin{center}
\begin{code}%
\>[0]\AgdaFunction{bd{-}c}\AgdaSpace{}%
\AgdaSymbol{:}%
\>[399I]\AgdaSymbol{\{}\AgdaBound{A}\AgdaSpace{}%
\AgdaSymbol{:}\AgdaSpace{}%
\AgdaPrimitiveType{Set}\AgdaSymbol{\}\{}\AgdaBound{P}\AgdaSpace{}%
\AgdaSymbol{:}\AgdaSpace{}%
\AgdaBound{A}\AgdaSpace{}%
\AgdaSymbol{→}\AgdaSpace{}%
\AgdaPrimitiveType{Set}\AgdaSymbol{\}}\AgdaSpace{}%
\AgdaSymbol{→}\<%
\\
\>[.]\<[399I]%
\>[7]\AgdaSymbol{(}\AgdaBound{S}\AgdaSpace{}%
\AgdaSymbol{:}\AgdaSpace{}%
\AgdaSymbol{(}\AgdaBound{A}\AgdaSpace{}%
\AgdaSymbol{→}\AgdaSpace{}%
\AgdaPrimitiveType{Set}\AgdaSymbol{)}\AgdaSpace{}%
\AgdaSymbol{→}\AgdaSpace{}%
\AgdaDatatype{Colist}\AgdaSpace{}%
\AgdaBound{A}\AgdaSpace{}%
\AgdaPostulate{∞}\AgdaSpace{}%
\AgdaSymbol{→}\AgdaSpace{}%
\AgdaPrimitiveType{Set}\AgdaSymbol{)}\AgdaSpace{}%
\AgdaSymbol{→}\<%
\\
\>[7]\AgdaSymbol{(\{}\AgdaBound{ys}\AgdaSpace{}%
\AgdaSymbol{:}\AgdaSpace{}%
\AgdaDatatype{Colist}\AgdaSpace{}%
\AgdaBound{A}\AgdaSpace{}%
\AgdaPostulate{∞}\AgdaSymbol{\}}\AgdaSpace{}%
\AgdaSymbol{→}\AgdaSpace{}%
\AgdaBound{S}\AgdaSpace{}%
\AgdaBound{P}\AgdaSpace{}%
\AgdaBound{ys}\AgdaSpace{}%
\AgdaSymbol{→}\AgdaSpace{}%
\AgdaBound{P}\AgdaSpace{}%
\AgdaOperator{\AgdaDatatype{InfinitelyOften{-}ind}}\AgdaSpace{}%
\AgdaBound{ys}\AgdaSymbol{)}\AgdaSpace{}%
\AgdaSymbol{→}\<%
\\
\>[7]\AgdaSymbol{(\{}\AgdaBound{ys}\AgdaSpace{}%
\AgdaSymbol{:}\AgdaSpace{}%
\AgdaDatatype{Colist}\AgdaSpace{}%
\AgdaBound{A}\AgdaSpace{}%
\AgdaPostulate{∞}\AgdaSymbol{\}}\AgdaSpace{}%
\AgdaSymbol{→}\AgdaSpace{}%
\AgdaBound{S}\AgdaSpace{}%
\AgdaBound{P}\AgdaSpace{}%
\AgdaBound{ys}\AgdaSpace{}%
\AgdaSymbol{→}\AgdaSpace{}%
\AgdaDatatype{Step}\AgdaSpace{}%
\AgdaSymbol{(}\AgdaBound{S}\AgdaSpace{}%
\AgdaBound{P}\AgdaSymbol{)}\AgdaSpace{}%
\AgdaBound{ys}\AgdaSymbol{)}\AgdaSpace{}%
\AgdaSymbol{→}\<%
\\
\>[7]\AgdaComment{{-}{-}{-}{-}{-}{-}{-}{-}{-}{-}{-}{-}{-}{-}{-}{-}{-}{-}{-}{-}{-}{-}{-}{-}{-}{-}{-}{-}{-}{-}{-}{-}{-}{-}{-}{-}{-}{-}{-}{-}{-}{-}{-}{-}{-}{-}{-}{-}{-}{-}{-}{-}{-}{-}{-}{-}{-}{-}{-}{-}{-}{-}}\<%
\\
\>[7]\AgdaSymbol{(∀}\AgdaSpace{}%
\AgdaSymbol{\{}\AgdaBound{i}\AgdaSpace{}%
\AgdaBound{ys}\AgdaSymbol{\}}\AgdaSpace{}%
\AgdaSymbol{→}\AgdaSpace{}%
\AgdaBound{S}\AgdaSpace{}%
\AgdaBound{P}\AgdaSpace{}%
\AgdaBound{ys}\AgdaSpace{}%
\AgdaSymbol{→}\AgdaSpace{}%
\AgdaDatatype{InfinitelyOften}\AgdaSpace{}%
\AgdaBound{P}\AgdaSpace{}%
\AgdaBound{ys}\AgdaSpace{}%
\AgdaBound{i}\AgdaSymbol{)}\<%
\\
\\[\AgdaEmptyExtraSkip]%
\>[0]\AgdaFunction{bd{-}c}\AgdaSpace{}%
\AgdaSymbol{\{}\AgdaBound{A}\AgdaSymbol{\}}\AgdaSpace{}%
\AgdaSymbol{\{}\AgdaBound{P}\AgdaSymbol{\}}\AgdaSpace{}%
\AgdaBound{S}\AgdaSpace{}%
\AgdaBound{bd}\AgdaSpace{}%
\AgdaBound{cons}\AgdaSpace{}%
\AgdaBound{Sys}\AgdaSpace{}%
\AgdaKeyword{with}\AgdaSpace{}%
\AgdaBound{cons}\AgdaSpace{}%
\AgdaBound{Sys}\<%
\\
\>[0]\AgdaSymbol{...}\AgdaSpace{}%
\AgdaSymbol{|}\AgdaSpace{}%
\AgdaInductiveConstructor{io{-}t}\AgdaSpace{}%
\AgdaBound{Sxs}\AgdaSpace{}%
\AgdaSymbol{=}\AgdaSpace{}%
\AgdaInductiveConstructor{io{-}t}\AgdaSpace{}%
\AgdaSymbol{(λ}\AgdaSpace{}%
\AgdaKeyword{where}\AgdaSpace{}%
\AgdaSymbol{.}\AgdaField{force}\AgdaSpace{}%
\AgdaSymbol{→}\AgdaSpace{}%
\AgdaFunction{bd{-}c}\AgdaSpace{}%
\AgdaBound{S}\AgdaSpace{}%
\AgdaBound{bd}\AgdaSpace{}%
\AgdaBound{cons}\AgdaSpace{}%
\AgdaBound{Sxs}\AgdaSymbol{)}\AgdaSpace{}%
\AgdaSymbol{(}\AgdaBound{bd}\AgdaSpace{}%
\AgdaBound{Sys}\AgdaSymbol{)}\<%
\end{code}
\end{center}

Now we can move to the proofs of the required properties.

\begin{center}
\begin{code}
\>[0]\AgdaFunction{bound{-}aux}\AgdaSpace{}%
\AgdaSymbol{:}%
\>[484I]\AgdaSymbol{\{}\AgdaBound{A}\AgdaSpace{}%
\AgdaSymbol{:}\AgdaSpace{}%
\AgdaPrimitiveType{Set}\AgdaSymbol{\}\{}\AgdaBound{ys}\AgdaSpace{}%
\AgdaSymbol{:}\AgdaSpace{}%
\AgdaDatatype{Colist}\AgdaSpace{}%
\AgdaBound{A}\AgdaSpace{}%
\AgdaPostulate{∞}\AgdaSymbol{\}\{}\AgdaBound{P}\AgdaSpace{}%
\AgdaSymbol{:}\AgdaSpace{}%
\AgdaBound{A}\AgdaSpace{}%
\AgdaSymbol{→}\AgdaSpace{}%
\AgdaPrimitiveType{Set}\AgdaSymbol{\}}\AgdaSpace{}%
\AgdaSymbol{→}\<%
\\
\>[.]\<[484I]%
\>[12]\AgdaDatatype{∃}\AgdaSpace{}%
\AgdaDatatype{Nat}\AgdaSpace{}%
\AgdaSymbol{(λ}\AgdaSpace{}%
\AgdaBound{n}\AgdaSpace{}%
\AgdaSymbol{→}\AgdaSpace{}%
\AgdaSymbol{(}\AgdaFunction{lift}\AgdaSpace{}%
\AgdaBound{P}\AgdaSymbol{)}\AgdaSpace{}%
\AgdaSymbol{(}\AgdaFunction{lookup}\AgdaSpace{}%
\AgdaBound{n}\AgdaSpace{}%
\AgdaBound{ys}\AgdaSymbol{))}\AgdaSpace{}%
\AgdaSymbol{→}\AgdaSpace{}%
\AgdaBound{P}\AgdaSpace{}%
\AgdaOperator{\AgdaDatatype{InfinitelyOften{-}ind}}\AgdaSpace{}%
\AgdaBound{ys}\<%
\\
\>[0]\AgdaFunction{bound{-}aux}\AgdaSpace{}%
\AgdaSymbol{\{}\AgdaArgument{ys}\AgdaSpace{}%
\AgdaSymbol{=}\AgdaSpace{}%
\AgdaInductiveConstructor{[]}\AgdaSymbol{\}}\AgdaSpace{}%
\AgdaOperator{\AgdaInductiveConstructor{<}}\AgdaSpace{}%
\AgdaBound{n}\AgdaSpace{}%
\AgdaOperator{\AgdaInductiveConstructor{,}}\AgdaSpace{}%
\AgdaSymbol{()}\AgdaSpace{}%
\AgdaOperator{\AgdaInductiveConstructor{>}}\<%
\\
\>[0]\AgdaFunction{bound{-}aux}\AgdaSpace{}%
\AgdaSymbol{\{}\AgdaArgument{ys}\AgdaSpace{}%
\AgdaSymbol{=}\AgdaSpace{}%
\AgdaBound{x}\AgdaSpace{}%
\AgdaOperator{\AgdaInductiveConstructor{∷}}\AgdaSpace{}%
\AgdaBound{xs}\AgdaSymbol{\}}\AgdaSpace{}%
\AgdaOperator{\AgdaInductiveConstructor{<}}\AgdaSpace{}%
\AgdaInductiveConstructor{zero}\AgdaSpace{}%
\AgdaOperator{\AgdaInductiveConstructor{,}}\AgdaSpace{}%
\AgdaBound{Px}\AgdaSpace{}%
\AgdaOperator{\AgdaInductiveConstructor{>}}\AgdaSpace{}%
\AgdaSymbol{=}\AgdaSpace{}%
\AgdaInductiveConstructor{co{-}io{-}h}\AgdaSpace{}%
\AgdaBound{Px}\<%
\\
\>[0]\AgdaFunction{bound{-}aux}\AgdaSpace{}%
\AgdaSymbol{\{}\AgdaArgument{ys}\AgdaSpace{}%
\AgdaSymbol{=}\AgdaSpace{}%
\AgdaBound{x}\AgdaSpace{}%
\AgdaOperator{\AgdaInductiveConstructor{∷}}\AgdaSpace{}%
\AgdaBound{xs}\AgdaSymbol{\}}\AgdaSpace{}%
\AgdaOperator{\AgdaInductiveConstructor{<}}\AgdaSpace{}%
\AgdaInductiveConstructor{suc}\AgdaSpace{}%
\AgdaBound{n}\AgdaSpace{}%
\AgdaOperator{\AgdaInductiveConstructor{,}}\AgdaSpace{}%
\AgdaBound{Px}\AgdaSpace{}%
\AgdaOperator{\AgdaInductiveConstructor{>}}\AgdaSpace{}%
\AgdaSymbol{=}\AgdaSpace{}%
\AgdaInductiveConstructor{io{-}t{-}ind}\AgdaSpace{}%
\AgdaSymbol{(}\AgdaFunction{bound{-}aux}\AgdaSpace{}%
\AgdaOperator{\AgdaInductiveConstructor{<}}\AgdaSpace{}%
\AgdaBound{n}\AgdaSpace{}%
\AgdaOperator{\AgdaInductiveConstructor{,}}\AgdaSpace{}%
\AgdaBound{Px}\AgdaSpace{}%
\AgdaOperator{\AgdaInductiveConstructor{>}}\AgdaSymbol{)}\<%
\\
\\[\AgdaEmptyExtraSkip]%
\\[\AgdaEmptyExtraSkip]%
\>[0]\AgdaFunction{bound{-}aux2}\AgdaSpace{}%
\AgdaSymbol{:}%
\>[550I]\AgdaSymbol{\{}\AgdaBound{A}\AgdaSpace{}%
\AgdaSymbol{:}\AgdaSpace{}%
\AgdaPrimitiveType{Set}\AgdaSymbol{\}\{}\AgdaBound{ys}\AgdaSpace{}%
\AgdaSymbol{:}\AgdaSpace{}%
\AgdaDatatype{Colist}\AgdaSpace{}%
\AgdaBound{A}\AgdaSpace{}%
\AgdaPostulate{∞}\AgdaSymbol{\}\{}\AgdaBound{P}\AgdaSpace{}%
\AgdaSymbol{:}\AgdaSpace{}%
\AgdaBound{A}\AgdaSpace{}%
\AgdaSymbol{→}\AgdaSpace{}%
\AgdaPrimitiveType{Set}\AgdaSymbol{\}}\AgdaSpace{}%
\AgdaSymbol{→}\<%
\\
\>[.]\<[550I]%
\>[13]\AgdaFunction{S}\AgdaSpace{}%
\AgdaBound{P}\AgdaSpace{}%
\AgdaBound{ys}\AgdaSpace{}%
\AgdaSymbol{→}\AgdaSpace{}%
\AgdaDatatype{∃}\AgdaSpace{}%
\AgdaDatatype{Nat}\AgdaSpace{}%
\AgdaSymbol{(λ}\AgdaSpace{}%
\AgdaBound{n}\AgdaSpace{}%
\AgdaSymbol{→}\AgdaSpace{}%
\AgdaSymbol{(}\AgdaFunction{lift}\AgdaSpace{}%
\AgdaBound{P}\AgdaSymbol{)}\AgdaSpace{}%
\AgdaSymbol{(}\AgdaFunction{lookup}\AgdaSpace{}%
\AgdaBound{n}\AgdaSpace{}%
\AgdaBound{ys}\AgdaSymbol{))}\<%
\\
\>[0]\AgdaFunction{bound{-}aux2}\AgdaSpace{}%
\AgdaBound{Sys}\AgdaSpace{}%
\AgdaKeyword{with}\AgdaSpace{}%
\AgdaBound{Sys}\AgdaSpace{}%
\AgdaInductiveConstructor{zero}\<%
\\
\>[0]\AgdaFunction{bound{-}aux2}\AgdaSpace{}%
\AgdaBound{Sys}\AgdaSpace{}%
\AgdaSymbol{|}\AgdaSpace{}%
\AgdaOperator{\AgdaInductiveConstructor{<}}\AgdaSpace{}%
\AgdaBound{n}\AgdaSpace{}%
\AgdaOperator{\AgdaInductiveConstructor{,}}\AgdaSpace{}%
\AgdaOperator{\AgdaInductiveConstructor{⟨}}\AgdaSpace{}%
\AgdaInductiveConstructor{g{-}zero}\AgdaSpace{}%
\AgdaOperator{\AgdaInductiveConstructor{,}}\AgdaSpace{}%
\AgdaBound{Px}\AgdaSpace{}%
\AgdaOperator{\AgdaInductiveConstructor{⟩}}\AgdaSpace{}%
\AgdaOperator{\AgdaInductiveConstructor{>}}\AgdaSpace{}%
\AgdaSymbol{=}\AgdaSpace{}%
\AgdaOperator{\AgdaInductiveConstructor{<}}\AgdaSpace{}%
\AgdaBound{n}\AgdaSpace{}%
\AgdaOperator{\AgdaInductiveConstructor{,}}\AgdaSpace{}%
\AgdaBound{Px}\AgdaSpace{}%
\AgdaOperator{\AgdaInductiveConstructor{>}}\<%
\\
\\[\AgdaEmptyExtraSkip]%
\\[\AgdaEmptyExtraSkip]%
\>[0]\AgdaFunction{bound}%
\>[596I]\AgdaSymbol{:\{}\AgdaBound{A}\AgdaSpace{}%
\AgdaSymbol{:}\AgdaSpace{}%
\AgdaPrimitiveType{Set}\AgdaSymbol{\}\{}\AgdaBound{P}\AgdaSpace{}%
\AgdaSymbol{:}\AgdaSpace{}%
\AgdaBound{A}\AgdaSpace{}%
\AgdaSymbol{→}\AgdaSpace{}%
\AgdaPrimitiveType{Set}\AgdaSymbol{\}\{}\AgdaBound{ys}\AgdaSpace{}%
\AgdaSymbol{:}\AgdaSpace{}%
\AgdaDatatype{Colist}\AgdaSpace{}%
\AgdaBound{A}\AgdaSpace{}%
\AgdaPostulate{∞}\AgdaSymbol{\}}\AgdaSpace{}%
\AgdaSymbol{→}\<%
\\
\>[596I][@{}l@{\AgdaIndent{0}}]%
\>[8]\AgdaFunction{S}\AgdaSpace{}%
\AgdaBound{P}\AgdaSpace{}%
\AgdaBound{ys}\AgdaSpace{}%
\AgdaSymbol{→}\AgdaSpace{}%
\AgdaBound{P}\AgdaSpace{}%
\AgdaOperator{\AgdaDatatype{InfinitelyOften{-}ind}}\AgdaSpace{}%
\AgdaBound{ys}\<%
\\
\>[0]\AgdaFunction{bound}\AgdaSpace{}%
\AgdaSymbol{\{}\AgdaBound{A}\AgdaSymbol{\}}\AgdaSpace{}%
\AgdaSymbol{\{}\AgdaBound{P}\AgdaSymbol{\}}\AgdaSpace{}%
\AgdaSymbol{\{}\AgdaBound{ys}\AgdaSymbol{\}}\AgdaSpace{}%
\AgdaBound{fun}\AgdaSpace{}%
\AgdaSymbol{=}\AgdaSpace{}%
\AgdaFunction{bound{-}aux}\AgdaSpace{}%
\AgdaSymbol{(}\AgdaFunction{bound{-}aux2}\AgdaSpace{}%
\AgdaSymbol{\{}\AgdaBound{A}\AgdaSymbol{\}}\AgdaSpace{}%
\AgdaSymbol{\{}\AgdaBound{ys}\AgdaSymbol{\}}\AgdaSpace{}%
\AgdaSymbol{\{}\AgdaBound{P}\AgdaSymbol{\}}\AgdaSpace{}%
\AgdaBound{fun}\AgdaSymbol{)}\<%
\end{code}
\end{center}

The proof is very similar to the one we provided in theoretical part. We pass through an auxiliary specification that basically says that there exists at least one index such that $P$ holds on the element at that position.

\begin{center}
$\Spec_{1} = \{xs \mid \exists n. \Pred (\get {xs} {n})\}$
\end{center}  
An auxiliary function helps us to obtain this proof given that our specification holds as input parameter.
Then we proceed by induction over the index and this is done by another auxiliary function. If the index $n$ is zero then we can apply the coaxiom \agda{co-io-h} constructor inside \agda{InfinitelyOften-ind}.  If the index is not zero, that is \agda{suc n}, we know that we can derive \agda{P IO xs} by recursively calling the function and thus we can apply the constructor \agda{io-t-ind}. The proof of boundedness is obtained by combining the two functions above.
Now we show the code of the consistency proof.

\begin{center}
\begin{code}
\>[0]\AgdaFunction{pred}\AgdaSpace{}%
\AgdaSymbol{:}\AgdaSpace{}%
\AgdaSymbol{∀}\AgdaSpace{}%
\AgdaBound{n}\AgdaSpace{}%
\AgdaSymbol{→}\AgdaSpace{}%
\AgdaBound{n}\AgdaSpace{}%
\AgdaOperator{\AgdaDatatype{>}}\AgdaSpace{}%
\AgdaNumber{0}\AgdaSpace{}%
\AgdaSymbol{→}\AgdaSpace{}%
\AgdaDatatype{Nat}\<%
\\
\>[0]\AgdaFunction{pred}\AgdaSpace{}%
\AgdaSymbol{(}\AgdaInductiveConstructor{suc}\AgdaSpace{}%
\AgdaBound{n}\AgdaSymbol{)}\AgdaSpace{}%
\AgdaInductiveConstructor{g{-}zero}\AgdaSpace{}%
\AgdaSymbol{=}\AgdaSpace{}%
\AgdaBound{n}\<%
\\
\\[\AgdaEmptyExtraSkip]%
\>[0]\AgdaFunction{lookup{-}tail}%
\>[639I]\AgdaSymbol{:}\AgdaSpace{}%
\AgdaSymbol{\{}\AgdaBound{A}\AgdaSpace{}%
\AgdaSymbol{:}\AgdaSpace{}%
\AgdaPrimitiveType{Set}\AgdaSymbol{\}}\AgdaSpace{}%
\AgdaSymbol{\{}\AgdaBound{x}\AgdaSpace{}%
\AgdaSymbol{:}\AgdaSpace{}%
\AgdaBound{A}\AgdaSymbol{\}}\AgdaSpace{}%
\AgdaSymbol{\{}\AgdaBound{xs}\AgdaSpace{}%
\AgdaSymbol{:}\AgdaSpace{}%
\AgdaRecord{Thunk}\AgdaSpace{}%
\AgdaSymbol{(}\AgdaDatatype{Colist}\AgdaSpace{}%
\AgdaBound{A}\AgdaSymbol{)}\AgdaSpace{}%
\AgdaPostulate{∞}\AgdaSymbol{\}}\AgdaSpace{}%
\AgdaSymbol{→}\<%
\\
\>[.]\<[639I]%
\>[12]\AgdaSymbol{∀}\AgdaSpace{}%
\AgdaSymbol{\{}\AgdaBound{i}\AgdaSymbol{\}}\AgdaSpace{}%
\AgdaSymbol{→}\AgdaSpace{}%
\AgdaSymbol{(}\AgdaBound{gt}\AgdaSpace{}%
\AgdaSymbol{:}\AgdaSpace{}%
\AgdaBound{i}\AgdaSpace{}%
\AgdaOperator{\AgdaDatatype{>}}\AgdaSpace{}%
\AgdaNumber{0}\AgdaSymbol{)}\AgdaSpace{}%
\AgdaSymbol{→}\AgdaSpace{}%
\AgdaFunction{lookup}\AgdaSpace{}%
\AgdaBound{i}\AgdaSpace{}%
\AgdaSymbol{(}\AgdaBound{x}\AgdaSpace{}%
\AgdaOperator{\AgdaInductiveConstructor{∷}}\AgdaSpace{}%
\AgdaBound{xs}\AgdaSymbol{)}\AgdaSpace{}%
\AgdaOperator{\AgdaDatatype{≡}}\AgdaSpace{}%
\AgdaFunction{lookup}\AgdaSpace{}%
\AgdaSymbol{(}\AgdaFunction{pred}\AgdaSpace{}%
\AgdaBound{i}\AgdaSpace{}%
\AgdaBound{gt}\AgdaSymbol{)}\AgdaSpace{}%
\AgdaSymbol{(}\AgdaField{force}\AgdaSpace{}%
\AgdaBound{xs}\AgdaSymbol{)}\<%
\\
\>[0]\AgdaFunction{lookup{-}tail}\AgdaSpace{}%
\AgdaInductiveConstructor{g{-}zero}\AgdaSpace{}%
\AgdaSymbol{=}\AgdaSpace{}%
\AgdaInductiveConstructor{refl}\<%
\\
\\[\AgdaEmptyExtraSkip]%
\>[0]\AgdaFunction{pred{-}eq}\AgdaSpace{}%
\AgdaSymbol{:}\AgdaSpace{}%
\AgdaSymbol{∀}\AgdaSpace{}%
\AgdaSymbol{\{}\AgdaBound{n}\AgdaSpace{}%
\AgdaBound{i}\AgdaSymbol{\}}\AgdaSpace{}%
\AgdaSymbol{→}\AgdaSpace{}%
\AgdaSymbol{(}\AgdaBound{g}\AgdaSpace{}%
\AgdaSymbol{:}\AgdaSpace{}%
\AgdaBound{n}\AgdaSpace{}%
\AgdaOperator{\AgdaDatatype{>}}\AgdaSpace{}%
\AgdaInductiveConstructor{zero}\AgdaSymbol{)}\AgdaSpace{}%
\AgdaSymbol{→}\AgdaSpace{}%
\AgdaSymbol{(}\AgdaBound{g₁}\AgdaSpace{}%
\AgdaSymbol{:}\AgdaSpace{}%
\AgdaBound{n}\AgdaSpace{}%
\AgdaOperator{\AgdaDatatype{>}}\AgdaSpace{}%
\AgdaInductiveConstructor{suc}\AgdaSpace{}%
\AgdaBound{i}\AgdaSymbol{)}\AgdaSpace{}%
\AgdaSymbol{→}\AgdaSpace{}%
\AgdaSymbol{((}\AgdaFunction{pred}\AgdaSpace{}%
\AgdaBound{n}\AgdaSymbol{)}\AgdaSpace{}%
\AgdaBound{g}\AgdaSymbol{)}\AgdaSpace{}%
\AgdaOperator{\AgdaDatatype{>}}\AgdaSpace{}%
\AgdaBound{i}\<%
\\
\>[0]\AgdaFunction{pred{-}eq}\AgdaSpace{}%
\AgdaInductiveConstructor{g{-}zero}\AgdaSpace{}%
\AgdaSymbol{(}\AgdaInductiveConstructor{g{-}suc}\AgdaSpace{}%
\AgdaBound{g₁}\AgdaSymbol{)}\AgdaSpace{}%
\AgdaSymbol{=}\AgdaSpace{}%
\AgdaBound{g₁}\<%
\\
\\[\AgdaEmptyExtraSkip]%
\\[\AgdaEmptyExtraSkip]%
\>[0]\AgdaFunction{consistent}%
\>[704I]\AgdaSymbol{:\{}\AgdaBound{A}\AgdaSpace{}%
\AgdaSymbol{:}\AgdaSpace{}%
\AgdaPrimitiveType{Set}\AgdaSymbol{\}\{}\AgdaBound{P}\AgdaSpace{}%
\AgdaSymbol{:}\AgdaSpace{}%
\AgdaBound{A}\AgdaSpace{}%
\AgdaSymbol{→}\AgdaSpace{}%
\AgdaPrimitiveType{Set}\AgdaSymbol{\}\{}\AgdaBound{ys}\AgdaSpace{}%
\AgdaSymbol{:}\AgdaSpace{}%
\AgdaDatatype{Colist}\AgdaSpace{}%
\AgdaBound{A}\AgdaSpace{}%
\AgdaPostulate{∞}\AgdaSymbol{\}}\AgdaSpace{}%
\AgdaSymbol{→}\<%
\\
\>[.]\<[704I]%
\>[11]\AgdaFunction{S}\AgdaSpace{}%
\AgdaBound{P}\AgdaSpace{}%
\AgdaBound{ys}\AgdaSpace{}%
\AgdaSymbol{→}\AgdaSpace{}%
\AgdaDatatype{Step}\AgdaSpace{}%
\AgdaSymbol{(}\AgdaFunction{S}\AgdaSpace{}%
\AgdaBound{P}\AgdaSymbol{)}\AgdaSpace{}%
\AgdaBound{ys}\<%
\\
\>[0]\<%
\\
\>[0]\AgdaFunction{consistent}\AgdaSpace{}%
\AgdaSymbol{\{}\AgdaBound{A}\AgdaSymbol{\}}\AgdaSpace{}%
\AgdaSymbol{\{}\AgdaBound{P}\AgdaSymbol{\}}\AgdaSpace{}%
\AgdaSymbol{\{}\AgdaInductiveConstructor{[]}\AgdaSymbol{\}}\AgdaSpace{}%
\AgdaBound{Sys}\AgdaSpace{}%
\AgdaSymbol{=}\AgdaSpace{}%
\AgdaFunction{⊥{-}elim}\AgdaSpace{}%
\AgdaSymbol{(}\AgdaFunction{∧{-}right}\AgdaSpace{}%
\AgdaSymbol{(}\AgdaFunction{proof}\AgdaSpace{}%
\AgdaSymbol{(}\AgdaBound{Sys}\AgdaSpace{}%
\AgdaInductiveConstructor{zero}\AgdaSymbol{)))}\<%
\\
\>[0]\AgdaFunction{consistent}\AgdaSpace{}%
\AgdaSymbol{\{}\AgdaBound{A}\AgdaSymbol{\}}%
\>[734I]\AgdaSymbol{\{}\AgdaBound{P}\AgdaSymbol{\}}\AgdaSpace{}%
\AgdaSymbol{\{}\AgdaBound{x}\AgdaSpace{}%
\AgdaOperator{\AgdaInductiveConstructor{∷}}\AgdaSpace{}%
\AgdaBound{xs}\AgdaSymbol{\}}\AgdaSpace{}%
\AgdaBound{Sys}\AgdaSpace{}%
\AgdaSymbol{=}\<%
\\
\>[.]\<[734I]%
\>[15]\AgdaInductiveConstructor{io{-}t}\AgdaSpace{}%
\AgdaSymbol{(λ}%
\>[741I]\AgdaBound{i}\AgdaSpace{}%
\AgdaSymbol{→}\AgdaSpace{}%
\AgdaKeyword{let}\AgdaSpace{}%
\AgdaBound{ex}\AgdaSpace{}%
\AgdaSymbol{=}\AgdaSpace{}%
\AgdaBound{Sys}\AgdaSpace{}%
\AgdaSymbol{(}\AgdaInductiveConstructor{suc}\AgdaSpace{}%
\AgdaBound{i}\AgdaSymbol{)}\AgdaSpace{}%
\AgdaKeyword{in}\<%
\\
\>[.]\<[741I]%
\>[23]\AgdaOperator{\AgdaInductiveConstructor{<}}%
\>[750I]\AgdaFunction{pred}\AgdaSpace{}%
\AgdaSymbol{(}\AgdaFunction{witness}\AgdaSpace{}%
\AgdaBound{ex}\AgdaSymbol{)}\AgdaSpace{}%
\AgdaSymbol{(}\AgdaFunction{gt{-}tr}\AgdaSpace{}%
\AgdaSymbol{(}\AgdaFunction{∧{-}left}\AgdaSpace{}%
\AgdaSymbol{(}\AgdaFunction{proof}\AgdaSpace{}%
\AgdaBound{ex}\AgdaSymbol{))}%
\>[71]\AgdaInductiveConstructor{g{-}zero}\AgdaSymbol{)}\AgdaSpace{}%
\AgdaOperator{\AgdaInductiveConstructor{,}}\<%
\\
\>[.]\<[750I]%
\>[25]\AgdaOperator{\AgdaInductiveConstructor{⟨}}%
\>[758I]\AgdaFunction{pred{-}eq}\AgdaSpace{}%
\AgdaSymbol{(}\AgdaFunction{gt{-}tr}\AgdaSpace{}%
\AgdaSymbol{(}\AgdaFunction{∧{-}left}\AgdaSpace{}%
\AgdaSymbol{(}\AgdaFunction{proof}\AgdaSpace{}%
\AgdaBound{ex}\AgdaSymbol{))}\AgdaSpace{}%
\AgdaInductiveConstructor{g{-}zero}\AgdaSymbol{)}\AgdaSpace{}%
\AgdaSymbol{(}\AgdaFunction{∧{-}left}\AgdaSpace{}%
\AgdaSymbol{(}\AgdaFunction{proof}\AgdaSpace{}%
\AgdaBound{ex}\AgdaSymbol{))}\AgdaSpace{}%
\AgdaOperator{\AgdaInductiveConstructor{,}}\<%
\\
\>[.]\<[758I]%
\>[27]\AgdaFunction{subst}\AgdaSpace{}%
\AgdaSymbol{(}\AgdaFunction{lift}\AgdaSpace{}%
\AgdaBound{P}\AgdaSymbol{)}\<%
\\
\>[27][@{}l@{\AgdaIndent{0}}]%
\>[29]\AgdaSymbol{(}\AgdaFunction{lookup{-}tail}\AgdaSpace{}%
\AgdaSymbol{\{}\AgdaBound{A}\AgdaSymbol{\}}\AgdaSpace{}%
\AgdaSymbol{\{}\AgdaBound{x}\AgdaSymbol{\}}\AgdaSpace{}%
\AgdaSymbol{\{}\AgdaBound{xs}\AgdaSymbol{\}}\AgdaSpace{}%
\AgdaSymbol{\{}\AgdaFunction{witness}\AgdaSpace{}%
\AgdaBound{ex}\AgdaSymbol{\}}\<%
\\
\>[29]\AgdaSymbol{(}\AgdaFunction{gt{-}tr}\AgdaSpace{}%
\AgdaSymbol{(}\AgdaFunction{∧{-}left}\AgdaSpace{}%
\AgdaSymbol{(}\AgdaFunction{proof}\AgdaSpace{}%
\AgdaBound{ex}\AgdaSymbol{))}%
\>[57]\AgdaInductiveConstructor{g{-}zero}\AgdaSymbol{))}\<%
\\
\>[29]\AgdaSymbol{(}\AgdaFunction{∧{-}right}\AgdaSpace{}%
\AgdaSymbol{(}\AgdaFunction{proof}\AgdaSpace{}%
\AgdaBound{ex}\AgdaSymbol{))}\<%
\\
\>[25][@{}l@{\AgdaIndent{0}}]%
\>[26]\AgdaOperator{\AgdaInductiveConstructor{⟩}}\AgdaSpace{}%
\AgdaOperator{\AgdaInductiveConstructor{>}}\AgdaSymbol{)}\<%
\end{code}
\end{center}

In this case the auxiliary functions are equality properties but the first one which simply returns the predecessor of a number that must be greater than zero. There is only one rule in \agda{Step} which is \agda{InfinitelyOften-ind} and in order to apply it we need the proof that the specification holds on \agda{xs} which is the tail. All the remaining code gets this proof from \agda{Sys} (which is the proof that the specification holds on the whole list) by restricting it to the tail. The predecessor function \agda{pred} is needed since the output of \agda{Sys} refers to the whole list while we need it to refer to the tail \agda{xs}. Notice that in the call \agda{Sys (suc i)} the successor is needed to avoid considering the head of \agda{xs}. \agda{gt-tr} is the transitive property of $>$, see \refToSection{agda-modules} for more details.
There is an absurd case which can automatically managed by Agda but we decided to simplify the code by directly calling the elimination rule of $\bot$ on the absurd term. In fact this function is called on the proof the $\Pred$ holds on the output of \agda{lookup} but this leads to an absurd since the list in empty.
Finally we can apply the bounded coinduction principle in order to prove the completeness of \agda{InfinitelyOften}.

\begin{center}
\begin{code}
\>[0]\AgdaFunction{io{-}complete}\AgdaSpace{}%
\AgdaSymbol{:}\AgdaSpace{}%
\AgdaSymbol{\{}\AgdaBound{A}\AgdaSpace{}%
\AgdaSymbol{:}\AgdaSpace{}%
\AgdaPrimitiveType{Set}\AgdaSymbol{\}\{}\AgdaBound{P}\AgdaSpace{}%
\AgdaSymbol{:}\AgdaSpace{}%
\AgdaBound{A}\AgdaSpace{}%
\AgdaSymbol{→}\AgdaSpace{}%
\AgdaPrimitiveType{Set}\AgdaSymbol{\}\{}\AgdaBound{ys}\AgdaSpace{}%
\AgdaSymbol{:}\AgdaSpace{}%
\AgdaDatatype{Colist}\AgdaSpace{}%
\AgdaBound{A}\AgdaSpace{}%
\AgdaPostulate{∞}\AgdaSymbol{\}\{}\AgdaBound{i}\AgdaSpace{}%
\AgdaSymbol{:}\AgdaSpace{}%
\AgdaPostulate{Size}\AgdaSymbol{\}}\AgdaSpace{}%
\AgdaSymbol{→}\<%
\\
\>[0][@{}l@{\AgdaIndent{0}}]%
\>[3]\AgdaFunction{S}\AgdaSpace{}%
\AgdaBound{P}\AgdaSpace{}%
\AgdaBound{ys}\AgdaSpace{}%
\AgdaSymbol{→}\AgdaSpace{}%
\AgdaDatatype{InfinitelyOften}\AgdaSpace{}%
\AgdaBound{P}\AgdaSpace{}%
\AgdaBound{ys}\AgdaSpace{}%
\AgdaBound{i}\<%
\\
\>[0]\AgdaFunction{io{-}complete}\AgdaSpace{}%
\AgdaSymbol{\{}\AgdaBound{A}\AgdaSymbol{\}}\AgdaSpace{}%
\AgdaSymbol{\{}\AgdaBound{P}\AgdaSymbol{\}}\AgdaSpace{}%
\AgdaSymbol{\{}\AgdaBound{ys}\AgdaSymbol{\}}\AgdaSpace{}%
\AgdaBound{Sys}\AgdaSpace{}%
\AgdaSymbol{=}\AgdaSpace{}%
\AgdaFunction{bd{-}c}\AgdaSpace{}%
\AgdaFunction{S}\AgdaSpace{}%
\AgdaFunction{bound}\AgdaSpace{}%
\AgdaFunction{consistent}\AgdaSpace{}%
\AgdaBound{Sys}\<%
\end{code}
\end{center}

Summing up, we presented another predicate that cannot be interpreted inductively or coinductively. From the Agda point of view, we followed again the theoretical notions and the scheme to prove the completeness as we did in \refToChapter{flex}. Differently from \textit{maxElem}, in this case we had an advantage since the inference system of \textit{InfinitelyOften} has a single rule. Despite this, we didn't face many issues in general. In \textit{maxElem} there were problems related to the \textit{max} function and to pattern matching over the tail of a colist which is a thunk. Clearly all the codes in this chapter could have been simplified by considering only streams but the comparison would not have been possible.

\newpage

\chapter{Conclusion}\label{chapter:conclu}
The aim of this work was to provide guidelines to express in Agda inference systems, starting from standard inference systems interpreted either inductively of coinductively, to focus then on \emph{generalized inference systems} which are a recent formalism to define predicates which are neither inductive, nor purely coinductive.

To prove \emph{soundness} and \emph{completeness} of predicates defined by (generalized) inference systems with respect to a given specification, the following proof principles are available.
\begin{itemize}
\item \textit{Induction principle} to prove the soundness of inductive predicates.
\item \textit{Coinduction principle} to prove the completeness of coinductive predicates.
\item \textit{Bounded coinduction principle} to prove the completeness of predicates defined through generalized inference systems.
\end{itemize}

We described how inductive and coinductive data structures can be implemented by two Agda constructs: \emph{data types} and \emph{coinductive records}. 
Generally, a function cannot be defined in the same way on data types and coinductive records. Indeed, in the first case the definition is typically by \emph{pattern matching}, whereas, on infinite structures, it is not possible to inspect the whole structure (e.g., a stream), and \emph{copattern matching} must be used. This second technique allows one to reason on how the result will be observed.

For what concerns data structures which can be either finite or infinite, we considered the paradigmatic example of \emph{colists} (possibly infinite lists).
The standard library offers an implementation of colists relying on the auxiliary types \agda{Size} and \agda{Thunk}. The first one models the approximation level (hence it is $\infty$ for lists which need to be approximated at any level) while the second one is the type of \emph{suspended computations}. The resulting code has pros and cons. On the one hand, the syntax is the same as for finite lists and pattern matching is allowed.  On the other hand, the usage of the auxiliary types is not always clear, especially the boxing and unboxing of values inside thunks.
We proposed a different solution based on coinductive records with a single field, that does not require to understand additional mechanisms but can become complicated to handle. 

Thanks to the key feature of Agda that predicates are types, our investigation on the different ways to implement data structures smoothly extends to implement also predicates on them.
We considered \textit{memberOf} and \textit{allPos} as paradigmatic examples of predicates that must be interpreted inductively and coinductively, respectively.
Then, we generalized them by considering \textit{Eventually} and \textit{Always} which are linear temporal logic operators.

For such predicates, we developed the Agda code using both the implementations discussed above, with the aim of comparing them. While the mentioned advantages and disadvantages still apply to predicates, we had to face additional challenges in the approach of coinductive records. In fact predicates implemented in this way involve variables that are existentially quantified, since we reason on how the data structure can be observed. This requires some machinery with equality properties to keep the relation between these variables and those obtained inside proofs, for example using \agda{with}.
We also had to manage absurd cases that were avoided using the library colists.
Finally, we investigated how the two approaches scale on predicates with more than one argument. In this case, the approach as coinductive records becomes problematic, since the existentially quantified variables must be multiplied by the number of involved colists. 

Coming then to main focus of this work, we considered \textit{maxElem} and \textit{InfinitelyOften} as examples of predicates on which neither the inductive nor the coinductive interpretation provides the intended meaning, whereas this can be obtained by an inference system with corules. In this case, to express the predicate and to prove its correctness in Agda is \EZ{more challenging}. Indeed:
\begin{itemize}
\item There is no built-in type corresponding to the predicate.
\item The bounded induction principle needs to be expressed and proved.
\end{itemize}

\EZ{Our investigation leads to the following guidelines to handle these issues. Consider a predicate defined by an inference system with corules $\Pair{\is}{\cois}$.}

\EZ{To implement the predicate, two Agda types can be used:
\begin{itemize}
\item A coinductive type that represents the interpretation of the inference system with corules, that is, $\Generated{\is}{\cois}$. Accordingly with the definition, this type internally uses the following other type:
\item An inductive type that represents the inductive interpretation of the inference system extended with corules, that is, $\Ind{\is\cup\cois}$.
\end{itemize}}

\EZ{Moreover, an auxiliary type  (called \agda{Step} in \refToChapter{flex} and \refToSection{infinitely-often}) can be defined, expressing the parametric predicate that, for a given specification $\Spec$, a judgment is the consequence of a rule whose premises are in the specification. 
This type is not recursive and uses the same meta-rules of the coinductive one.}

\EZ{Finally, the bounded coinduction principle corresponding to the predicate must be proved, using the type \agda{Step} defined before to express  consistency. This provides a canonical technique to prove completeness.}

In this case, we only developed the code for library colists, which we expected to be more suitable after the previous investigation. However, we had to solve some non trivial problems. For example, pattern matching on the tail of a colist is not allowed since it is boxed inside a thunk, making the usage of \agda{with} necessary,  \EZ{and} again we had to manage absurd cases and auxiliary equalities.

Summing up, we provided guidelines driven by examples for writing in Agda (generalized) inference systems and to check their correctness. 
A nice outcome of the thesis is that it shows that it is possible to write Agda definitions and proofs which exactly follow the hand-written schemes. However, as should be expected, for an automatic theorem prover nothing is obvious.
Hence, besides the high-level proof, users have to implement on their own all the needed auxiliary properties such as transitivity of a relation, in terms of functions that given proofs return proofs. This additional aspect can become very important in evaluating advantages and disadvantages of different approaches. 

\EZ{A natural and important further development of the work done in this thesis is to transform the methodology in an automatic translation. That is, a user should be allowed to write an inference system (with corules) in a natural syntax, and the corresponding Agda types should be generated automatically. This could be achieved by an external tool, that is, user definitions could be given to a parser producing the Agda code. An alternative, more interesting and challenging, solution, is to write the inference system as an Agda type, and then use \emph{reflection}, which was recently added in Agda. In this case, 
the Agda type should express the syntax of the inference system. Notably, we should use a data type with constructors for rules and corules.}

\EZ{Another interesting direction for further work is the specification and verification of concurrent systems, extending what we did in \refToChapter{temp-op}. In fact, often properties of such systems can be classified as \textit{safety} or \textit{liveness} ones. Roughly speaking, a safety property states that something bad will never happen, while a liveness one states that something good will eventually happen. 
Generally, we have to reason inductively to prove a liveness property, and coinductively to prove a safety property, as shown by the examples of temporal operators we studied in \refToSection{eventually} and \refToSection{always}. However, there are more complex properties obtained by mixing safety and liveness, as shown by the example in \refToSection{infinitely-often}. In these cases, which require
a combination of inductive and coinductive techniques, inference systems with corules, and the experience we gained in their Agda implementation, look a promising approach.}

\bibliographystyle{alphaurl}
\bibliography{bib}

\appendix

\chapter{}

\section{Agda modules}\label{sect:agda-modules}

\EZ{Here we report all the auxiliary modules used in the previous chapters, and not imported from a  built-in library.}

\subsection{Getting the $i$-th element of a list}\label{sect:get}

\EZ{This function returns the element of a list at given index.}
Here we report its implementations for streams and colists as coinductive records (\agda{MyColist}). The implementation for library colists is provided in the \agda{Colist} module and is called \agda{lookup}.

\begin{center}
\begin{code}
\>[0]\AgdaFunction{get}\AgdaSpace{}%
\AgdaSymbol{:}\AgdaSpace{}%
\AgdaSymbol{\{}\AgdaBound{A}\AgdaSpace{}%
\AgdaSymbol{:}\AgdaSpace{}%
\AgdaPrimitiveType{Set}\AgdaSymbol{\}}\AgdaSpace{}%
\AgdaSymbol{→}\AgdaSpace{}%
\AgdaRecord{MyStream}\AgdaSpace{}%
\AgdaBound{A}\AgdaSpace{}%
\AgdaSymbol{→}\AgdaSpace{}%
\AgdaDatatype{Nat}\AgdaSpace{}%
\AgdaSymbol{→}\AgdaSpace{}%
\AgdaBound{A}\<%
\\
\>[0]\AgdaFunction{get}\AgdaSpace{}%
\AgdaBound{xs}\AgdaSpace{}%
\AgdaInductiveConstructor{zero}\AgdaSpace{}%
\AgdaSymbol{=}\AgdaSpace{}%
\AgdaField{MyStream.hd}\AgdaSpace{}%
\AgdaBound{xs}\<%
\\
\>[0]\AgdaFunction{get}\AgdaSpace{}%
\AgdaBound{xs}\AgdaSpace{}%
\AgdaSymbol{(}\AgdaInductiveConstructor{suc}\AgdaSpace{}%
\AgdaBound{i}\AgdaSymbol{)}\AgdaSpace{}%
\AgdaSymbol{=}\AgdaSpace{}%
\AgdaFunction{get}\AgdaSpace{}%
\AgdaSymbol{(}\AgdaField{MyStream.tl}\AgdaSpace{}%
\AgdaBound{xs}\AgdaSymbol{)}\AgdaSpace{}%
\AgdaBound{i}\<%
\\
\\
\>[0]\AgdaFunction{get}\AgdaSpace{}%
\AgdaSymbol{:}\AgdaSpace{}%
\AgdaSymbol{\{}\AgdaBound{A}\AgdaSpace{}%
\AgdaSymbol{:}\AgdaSpace{}%
\AgdaPrimitiveType{Set}\AgdaSymbol{\}}\AgdaSpace{}%
\AgdaSymbol{→}\AgdaSpace{}%
\AgdaRecord{MyColist}\AgdaSpace{}%
\AgdaBound{A}\AgdaSpace{}%
\AgdaSymbol{→}\AgdaSpace{}%
\AgdaDatatype{Nat}\AgdaSpace{}%
\AgdaSymbol{→}\AgdaSpace{}%
\AgdaDatatype{Maybe}\AgdaSpace{}%
\AgdaBound{A}\<%
\\
\>[0]\AgdaFunction{get}\AgdaSpace{}%
\AgdaBound{ys}\AgdaSpace{}%
\AgdaBound{i}\AgdaSpace{}%
\AgdaKeyword{with}\AgdaSpace{}%
\AgdaFunction{inspect}\AgdaSpace{}%
\AgdaSymbol{(}\AgdaField{MyColist.list}\AgdaSpace{}%
\AgdaBound{ys}\AgdaSymbol{)}\<%
\\
\>[0]\AgdaSymbol{...}\AgdaSpace{}%
\AgdaSymbol{|}\AgdaSpace{}%
\AgdaInductiveConstructor{nothing}\AgdaSpace{}%
\AgdaOperator{\AgdaInductiveConstructor{with≡}}\AgdaSpace{}%
\AgdaBound{eq}\AgdaSpace{}%
\AgdaSymbol{=}\AgdaSpace{}%
\AgdaInductiveConstructor{nothing}\<%
\\
\>[0]\AgdaFunction{get}\AgdaSpace{}%
\AgdaBound{l}\AgdaSpace{}%
\AgdaInductiveConstructor{zero}\AgdaSpace{}%
\AgdaSymbol{|}\AgdaSpace{}%
\AgdaInductiveConstructor{just}\AgdaSpace{}%
\AgdaOperator{\AgdaInductiveConstructor{⟨}}\AgdaSpace{}%
\AgdaBound{x}\AgdaSpace{}%
\AgdaOperator{\AgdaInductiveConstructor{,}}\AgdaSpace{}%
\AgdaBound{xs}\AgdaSpace{}%
\AgdaOperator{\AgdaInductiveConstructor{⟩}}\AgdaSpace{}%
\AgdaOperator{\AgdaInductiveConstructor{with≡}}\AgdaSpace{}%
\AgdaBound{eq}\AgdaSpace{}%
\AgdaSymbol{=}\AgdaSpace{}%
\AgdaInductiveConstructor{just}\AgdaSpace{}%
\AgdaBound{x}\<%
\\
\>[0]\AgdaFunction{get}\AgdaSpace{}%
\AgdaBound{l}\AgdaSpace{}%
\AgdaSymbol{(}\AgdaInductiveConstructor{suc}\AgdaSpace{}%
\AgdaBound{i}\AgdaSymbol{)}\AgdaSpace{}%
\AgdaSymbol{|}\AgdaSpace{}%
\AgdaInductiveConstructor{just}\AgdaSpace{}%
\AgdaOperator{\AgdaInductiveConstructor{⟨}}\AgdaSpace{}%
\AgdaBound{x}\AgdaSpace{}%
\AgdaOperator{\AgdaInductiveConstructor{,}}\AgdaSpace{}%
\AgdaBound{xs}\AgdaSpace{}%
\AgdaOperator{\AgdaInductiveConstructor{⟩}}\AgdaSpace{}%
\AgdaOperator{\AgdaInductiveConstructor{with≡}}\AgdaSpace{}%
\AgdaBound{eq}\AgdaSpace{}%
\AgdaSymbol{=}\AgdaSpace{}%
\AgdaFunction{get}\AgdaSpace{}%
\AgdaBound{xs}\AgdaSpace{}%
\AgdaBound{i}\<%
\end{code}
\end{center}

\subsection{Equality properties}

We show the implementation of the equality properties listed in \refToSection{equality}. Each property is a function that takes proofs as input and returns a new proof.

\begin{center}
\begin{code}%
\>[0]\AgdaFunction{sym}\AgdaSpace{}%
\AgdaSymbol{:}\AgdaSpace{}%
\AgdaSymbol{∀}\AgdaSpace{}%
\AgdaSymbol{\{}\AgdaBound{A}\AgdaSpace{}%
\AgdaSymbol{:}\AgdaSpace{}%
\AgdaPrimitiveType{Set}\AgdaSymbol{\}}\AgdaSpace{}%
\AgdaSymbol{\{}\AgdaBound{x}\AgdaSpace{}%
\AgdaBound{y}\AgdaSpace{}%
\AgdaSymbol{:}\AgdaSpace{}%
\AgdaBound{A}\AgdaSymbol{\}}\AgdaSpace{}%
\AgdaSymbol{→}\AgdaSpace{}%
\AgdaBound{x}\AgdaSpace{}%
\AgdaOperator{\AgdaDatatype{≡}}\AgdaSpace{}%
\AgdaBound{y}\AgdaSpace{}%
\AgdaSymbol{→}%
\>[39]\AgdaBound{y}\AgdaSpace{}%
\AgdaOperator{\AgdaDatatype{≡}}\AgdaSpace{}%
\AgdaBound{x}\<%
\\
\>[0]\AgdaFunction{sym}\AgdaSpace{}%
\AgdaInductiveConstructor{refl}\AgdaSpace{}%
\AgdaSymbol{=}\AgdaSpace{}%
\AgdaInductiveConstructor{refl}\<%
\\
\\[\AgdaEmptyExtraSkip]%
\>[0]\AgdaFunction{trans}\AgdaSpace{}%
\AgdaSymbol{:}\AgdaSpace{}%
\AgdaSymbol{∀}\AgdaSpace{}%
\AgdaSymbol{\{}\AgdaBound{A}\AgdaSpace{}%
\AgdaSymbol{:}\AgdaSpace{}%
\AgdaPrimitiveType{Set}\AgdaSymbol{\}}\AgdaSpace{}%
\AgdaSymbol{\{}\AgdaBound{x}\AgdaSpace{}%
\AgdaBound{y}\AgdaSpace{}%
\AgdaBound{z}\AgdaSpace{}%
\AgdaSymbol{:}\AgdaSpace{}%
\AgdaBound{A}\AgdaSymbol{\}}\AgdaSpace{}%
\AgdaSymbol{→}\AgdaSpace{}%
\AgdaBound{x}\AgdaSpace{}%
\AgdaOperator{\AgdaDatatype{≡}}\AgdaSpace{}%
\AgdaBound{y}\AgdaSpace{}%
\AgdaSymbol{→}\AgdaSpace{}%
\AgdaBound{y}\AgdaSpace{}%
\AgdaOperator{\AgdaDatatype{≡}}\AgdaSpace{}%
\AgdaBound{z}\AgdaSpace{}%
\AgdaSymbol{→}\AgdaSpace{}%
\AgdaBound{x}\AgdaSpace{}%
\AgdaOperator{\AgdaDatatype{≡}}\AgdaSpace{}%
\AgdaBound{z}\<%
\\
\>[0]\AgdaFunction{trans}\AgdaSpace{}%
\AgdaInductiveConstructor{refl}\AgdaSpace{}%
\AgdaInductiveConstructor{refl}%
\>[17]\AgdaSymbol{=}%
\>[20]\AgdaInductiveConstructor{refl}\<%
\\
\\[\AgdaEmptyExtraSkip]%
\>[0]\AgdaFunction{cong}\AgdaSpace{}%
\AgdaSymbol{:}\AgdaSpace{}%
\AgdaSymbol{∀}\AgdaSpace{}%
\AgdaSymbol{\{}\AgdaBound{A}\AgdaSpace{}%
\AgdaBound{B}\AgdaSpace{}%
\AgdaSymbol{:}\AgdaSpace{}%
\AgdaPrimitiveType{Set}\AgdaSymbol{\}}\AgdaSpace{}%
\AgdaSymbol{(}\AgdaBound{f}\AgdaSpace{}%
\AgdaSymbol{:}\AgdaSpace{}%
\AgdaBound{A}\AgdaSpace{}%
\AgdaSymbol{→}\AgdaSpace{}%
\AgdaBound{B}\AgdaSymbol{)}\AgdaSpace{}%
\AgdaSymbol{\{}\AgdaBound{x}\AgdaSpace{}%
\AgdaBound{y}\AgdaSpace{}%
\AgdaSymbol{:}\AgdaSpace{}%
\AgdaBound{A}\AgdaSymbol{\}}\AgdaSpace{}%
\AgdaSymbol{→}\AgdaSpace{}%
\AgdaBound{x}\AgdaSpace{}%
\AgdaOperator{\AgdaDatatype{≡}}\AgdaSpace{}%
\AgdaBound{y}\AgdaSpace{}%
\AgdaSymbol{→}\AgdaSpace{}%
\AgdaBound{f}\AgdaSpace{}%
\AgdaBound{x}\AgdaSpace{}%
\AgdaOperator{\AgdaDatatype{≡}}\AgdaSpace{}%
\AgdaBound{f}\AgdaSpace{}%
\AgdaBound{y}\<%
\\
\>[0]\AgdaFunction{cong}\AgdaSpace{}%
\AgdaBound{f}\AgdaSpace{}%
\AgdaInductiveConstructor{refl}%
\>[13]\AgdaSymbol{=}%
\>[16]\AgdaInductiveConstructor{refl}\<%
\\
\\[\AgdaEmptyExtraSkip]%
\>[0]\AgdaFunction{subst}\AgdaSpace{}%
\AgdaSymbol{:}\AgdaSpace{}%
\AgdaSymbol{∀}\AgdaSpace{}%
\AgdaSymbol{\{}\AgdaBound{A}\AgdaSpace{}%
\AgdaSymbol{:}\AgdaSpace{}%
\AgdaPrimitiveType{Set}\AgdaSymbol{\}}\AgdaSpace{}%
\AgdaSymbol{\{}\AgdaBound{x}\AgdaSpace{}%
\AgdaBound{y}\AgdaSpace{}%
\AgdaSymbol{:}\AgdaSpace{}%
\AgdaBound{A}\AgdaSymbol{\}}\AgdaSpace{}%
\AgdaSymbol{(}\AgdaBound{P}\AgdaSpace{}%
\AgdaSymbol{:}\AgdaSpace{}%
\AgdaBound{A}\AgdaSpace{}%
\AgdaSymbol{→}\AgdaSpace{}%
\AgdaPrimitiveType{Set}\AgdaSymbol{)}\AgdaSpace{}%
\AgdaSymbol{→}\AgdaSpace{}%
\AgdaBound{x}\AgdaSpace{}%
\AgdaOperator{\AgdaDatatype{≡}}\AgdaSpace{}%
\AgdaBound{y}\AgdaSpace{}%
\AgdaSymbol{→}\AgdaSpace{}%
\AgdaBound{P}\AgdaSpace{}%
\AgdaBound{x}\AgdaSpace{}%
\AgdaSymbol{→}\AgdaSpace{}%
\AgdaBound{P}\AgdaSpace{}%
\AgdaBound{y}\<%
\\
\>[0]\AgdaFunction{subst}\AgdaSpace{}%
\AgdaBound{P}\AgdaSpace{}%
\AgdaInductiveConstructor{refl}\AgdaSpace{}%
\AgdaBound{px}\AgdaSpace{}%
\AgdaSymbol{=}\AgdaSpace{}%
\AgdaBound{px}\<%
\end{code}

\end{center}

The proofs of the properties are very simple: they take in input equality proofs and there is only one constructor that can be used.

\subsection{Equality reasoning}

This module is taken from \cite{plfa2019}, and used in auxiliary equality proofs. The advantage of using these constructs is that we can chain different equalities without defining each step separately as a single property.
\EZComm{non capisco come risulta scritto il codice}
\begin{center}
\begin{code}
\>[0][@{}l@{\AgdaIndent{0}}]%
\>[2]\AgdaOperator{\AgdaFunction{begin\AgdaUnderscore{}}}\AgdaSpace{}%
\AgdaSymbol{:}\AgdaSpace{}%
\AgdaSymbol{∀}\AgdaSpace{}%
\AgdaSymbol{\{}\AgdaBound{x}\AgdaSpace{}%
\AgdaBound{y}\AgdaSpace{}%
\AgdaSymbol{:}\AgdaSpace{}%
\AgdaBound{A}\AgdaSymbol{\}}\AgdaSpace{}%
\AgdaSymbol{→}\AgdaSpace{}%
\AgdaBound{x}\AgdaSpace{}%
\AgdaOperator{\AgdaDatatype{≡}}\AgdaSpace{}%
\AgdaBound{y}\AgdaSpace{}%
\AgdaSymbol{→}\AgdaSpace{}%
\AgdaBound{x}\AgdaSpace{}%
\AgdaOperator{\AgdaDatatype{≡}}\AgdaSpace{}%
\AgdaBound{y}\<%
\\
\>[2]\AgdaOperator{\AgdaFunction{begin}}\AgdaSpace{}%
\AgdaBound{x≡y}%
\>[13]\AgdaSymbol{=}%
\>[16]\AgdaBound{x≡y}\<%
\\
\\[\AgdaEmptyExtraSkip]%
\>[2]\AgdaOperator{\AgdaFunction{\AgdaUnderscore{}≡⟨⟩\AgdaUnderscore{}}}\AgdaSpace{}%
\AgdaSymbol{:}\AgdaSpace{}%
\AgdaSymbol{∀}\AgdaSpace{}%
\AgdaSymbol{(}\AgdaBound{x}\AgdaSpace{}%
\AgdaSymbol{:}\AgdaSpace{}%
\AgdaBound{A}\AgdaSymbol{)}\AgdaSpace{}%
\AgdaSymbol{\{}\AgdaBound{y}\AgdaSpace{}%
\AgdaSymbol{:}\AgdaSpace{}%
\AgdaBound{A}\AgdaSymbol{\}}\AgdaSpace{}%
\AgdaSymbol{→}\AgdaSpace{}%
\AgdaBound{x}\AgdaSpace{}%
\AgdaOperator{\AgdaDatatype{≡}}\AgdaSpace{}%
\AgdaBound{y}\AgdaSpace{}%
\AgdaSymbol{→}\AgdaSpace{}%
\AgdaBound{x}\AgdaSpace{}%
\AgdaOperator{\AgdaDatatype{≡}}\AgdaSpace{}%
\AgdaBound{y}\<%
\\
\>[2]\AgdaBound{x}\AgdaSpace{}%
\AgdaOperator{\AgdaFunction{≡⟨⟩}}\AgdaSpace{}%
\AgdaBound{x≡y}%
\>[13]\AgdaSymbol{=}%
\>[16]\AgdaBound{x≡y}\<%
\\
\\[\AgdaEmptyExtraSkip]%
\>[2]\AgdaOperator{\AgdaFunction{\AgdaUnderscore{}≡⟨\AgdaUnderscore{}⟩\AgdaUnderscore{}}}\AgdaSpace{}%
\AgdaSymbol{:}\AgdaSpace{}%
\AgdaSymbol{∀}\AgdaSpace{}%
\AgdaSymbol{(}\AgdaBound{x}\AgdaSpace{}%
\AgdaSymbol{:}\AgdaSpace{}%
\AgdaBound{A}\AgdaSymbol{)}\AgdaSpace{}%
\AgdaSymbol{\{}\AgdaBound{y}\AgdaSpace{}%
\AgdaBound{z}\AgdaSpace{}%
\AgdaSymbol{:}\AgdaSpace{}%
\AgdaBound{A}\AgdaSymbol{\}}\AgdaSpace{}%
\AgdaSymbol{→}\AgdaSpace{}%
\AgdaBound{x}\AgdaSpace{}%
\AgdaOperator{\AgdaDatatype{≡}}\AgdaSpace{}%
\AgdaBound{y}\AgdaSpace{}%
\AgdaSymbol{→}\AgdaSpace{}%
\AgdaBound{y}\AgdaSpace{}%
\AgdaOperator{\AgdaDatatype{≡}}\AgdaSpace{}%
\AgdaBound{z}\AgdaSpace{}%
\AgdaSymbol{→}\AgdaSpace{}%
\AgdaBound{x}\AgdaSpace{}%
\AgdaOperator{\AgdaDatatype{≡}}\AgdaSpace{}%
\AgdaBound{z}\<%
\\
\>[2]\AgdaBound{x}\AgdaSpace{}%
\AgdaOperator{\AgdaFunction{≡⟨}}\AgdaSpace{}%
\AgdaBound{x≡y}\AgdaSpace{}%
\AgdaOperator{\AgdaFunction{⟩}}\AgdaSpace{}%
\AgdaBound{y≡z}%
\>[18]\AgdaSymbol{=}%
\>[21]\AgdaFunction{trans}\AgdaSpace{}%
\AgdaBound{x≡y}\AgdaSpace{}%
\AgdaBound{y≡z}\<%
\\
\\[\AgdaEmptyExtraSkip]%
\>[2]\AgdaOperator{\AgdaFunction{\AgdaUnderscore{}∎}}\AgdaSpace{}%
\AgdaSymbol{:}\AgdaSpace{}%
\AgdaSymbol{∀}\AgdaSpace{}%
\AgdaSymbol{(}\AgdaBound{x}\AgdaSpace{}%
\AgdaSymbol{:}\AgdaSpace{}%
\AgdaBound{A}\AgdaSymbol{)}%
\>[18]\AgdaSymbol{→}\AgdaSpace{}%
\AgdaBound{x}\AgdaSpace{}%
\AgdaOperator{\AgdaDatatype{≡}}\AgdaSpace{}%
\AgdaBound{x}\<%
\\
\>[2]\AgdaBound{x}\AgdaSpace{}%
\AgdaOperator{\AgdaFunction{∎}}%
\>[7]\AgdaSymbol{=}%
\>[10]\AgdaInductiveConstructor{refl}\<%
\end{code}
\end{center}

\subsection{Conjunction/Product type}
\EZ{The logical conjunction is implemented as a data type with a single constructor which returns a pair.}

\begin{center}
\begin{code}%
\>[0]\AgdaKeyword{data}\AgdaSpace{}%
\AgdaOperator{\AgdaDatatype{\AgdaUnderscore{}∧\AgdaUnderscore{}}}\AgdaSpace{}%
\AgdaSymbol{(}\AgdaBound{A}\AgdaSpace{}%
\AgdaBound{B}\AgdaSpace{}%
\AgdaSymbol{:}\AgdaSpace{}%
\AgdaPrimitiveType{Set}\AgdaSymbol{)}\AgdaSpace{}%
\AgdaSymbol{:}\AgdaSpace{}%
\AgdaPrimitiveType{Set}\AgdaSpace{}%
\AgdaKeyword{where}\<%
\\
\>[0][@{}l@{\AgdaIndent{0}}]%
\>[2]\AgdaOperator{\AgdaInductiveConstructor{⟨\AgdaUnderscore{},\AgdaUnderscore{}⟩}}\AgdaSpace{}%
\AgdaSymbol{:}%
\>[11]\AgdaBound{A}\AgdaSpace{}%
\AgdaSymbol{→}\AgdaSpace{}%
\AgdaBound{B}\AgdaSpace{}%
\AgdaSymbol{→}\AgdaSpace{}%
\AgdaBound{A}\AgdaSpace{}%
\AgdaOperator{\AgdaDatatype{∧}}\AgdaSpace{}%
\AgdaBound{B}\<%
\\
\\[\AgdaEmptyExtraSkip]%
\>[0]\AgdaOperator{\AgdaFunction{\AgdaUnderscore{}×\AgdaUnderscore{}}}\AgdaSpace{}%
\AgdaSymbol{:}\AgdaSpace{}%
\AgdaSymbol{(}\AgdaBound{A}\AgdaSpace{}%
\AgdaBound{B}\AgdaSpace{}%
\AgdaSymbol{:}\AgdaSpace{}%
\AgdaPrimitiveType{Set}\AgdaSymbol{)}\AgdaSpace{}%
\AgdaSymbol{→}\AgdaSpace{}%
\AgdaPrimitiveType{Set}\<%
\\
\>[0]\AgdaBound{A}\AgdaSpace{}%
\AgdaOperator{\AgdaFunction{×}}\AgdaSpace{}%
\AgdaBound{B}\AgdaSpace{}%
\AgdaSymbol{=}\AgdaSpace{}%
\AgdaBound{A}\AgdaSpace{}%
\AgdaOperator{\AgdaDatatype{∧}}\AgdaSpace{}%
\AgdaBound{B}\<%
\\
\\[\AgdaEmptyExtraSkip]%
\>[0]\AgdaComment{{-}{-}Projections}\<%
\\
\>[0]\AgdaFunction{∧{-}left}\AgdaSpace{}%
\AgdaSymbol{:}\AgdaSpace{}%
\AgdaSymbol{\{}\AgdaBound{A}\AgdaSpace{}%
\AgdaBound{B}\AgdaSpace{}%
\AgdaSymbol{:}\AgdaSpace{}%
\AgdaPrimitiveType{Set}\AgdaSymbol{\}}\AgdaSpace{}%
\AgdaSymbol{→}\AgdaSpace{}%
\AgdaSymbol{(}\AgdaBound{A}\AgdaSpace{}%
\AgdaOperator{\AgdaDatatype{∧}}\AgdaSpace{}%
\AgdaBound{B}\AgdaSymbol{)}\AgdaSpace{}%
\AgdaSymbol{→}\AgdaSpace{}%
\AgdaBound{A}\<%
\\
\>[0]\AgdaFunction{∧{-}left}\AgdaSpace{}%
\AgdaOperator{\AgdaInductiveConstructor{⟨}}\AgdaSpace{}%
\AgdaBound{l}\AgdaSpace{}%
\AgdaOperator{\AgdaInductiveConstructor{,}}\AgdaSpace{}%
\AgdaBound{r}\AgdaSpace{}%
\AgdaOperator{\AgdaInductiveConstructor{⟩}}\AgdaSpace{}%
\AgdaSymbol{=}\AgdaSpace{}%
\AgdaBound{l}\<%
\\
\\[\AgdaEmptyExtraSkip]%
\>[0]\AgdaFunction{∧{-}right}\AgdaSpace{}%
\AgdaSymbol{:}\AgdaSpace{}%
\AgdaSymbol{\{}\AgdaBound{A}\AgdaSpace{}%
\AgdaBound{B}\AgdaSpace{}%
\AgdaSymbol{:}\AgdaSpace{}%
\AgdaPrimitiveType{Set}\AgdaSymbol{\}}\AgdaSpace{}%
\AgdaSymbol{→}\AgdaSpace{}%
\AgdaSymbol{(}\AgdaBound{A}\AgdaSpace{}%
\AgdaOperator{\AgdaDatatype{∧}}\AgdaSpace{}%
\AgdaBound{B}\AgdaSymbol{)}\AgdaSpace{}%
\AgdaSymbol{→}\AgdaSpace{}%
\AgdaBound{B}\<%
\\
\>[0]\AgdaFunction{∧{-}right}\AgdaSpace{}%
\AgdaOperator{\AgdaInductiveConstructor{⟨}}\AgdaSpace{}%
\AgdaBound{l}\AgdaSpace{}%
\AgdaOperator{\AgdaInductiveConstructor{,}}\AgdaSpace{}%
\AgdaBound{r}\AgdaSpace{}%
\AgdaOperator{\AgdaInductiveConstructor{⟩}}\AgdaSpace{}%
\AgdaSymbol{=}\AgdaSpace{}%
\AgdaBound{r}\<%
\\
\\[\AgdaEmptyExtraSkip]%
\\[\AgdaEmptyExtraSkip]%
\>[0]\AgdaComment{{-}{-}Elimination}\<%
\\
\>[0]\AgdaFunction{eq2sx}\AgdaSpace{}%
\AgdaSymbol{:}\AgdaSpace{}%
\AgdaSymbol{\{}\AgdaBound{A}\AgdaSpace{}%
\AgdaBound{B}\AgdaSpace{}%
\AgdaSymbol{:}\AgdaSpace{}%
\AgdaPrimitiveType{Set}\AgdaSymbol{\}\{}\AgdaBound{h1}\AgdaSpace{}%
\AgdaBound{h2}\AgdaSpace{}%
\AgdaSymbol{:}\AgdaSpace{}%
\AgdaBound{A}\AgdaSymbol{\}\{}\AgdaBound{t1}\AgdaSpace{}%
\AgdaBound{t2}\AgdaSpace{}%
\AgdaSymbol{:}\AgdaSpace{}%
\AgdaBound{B}\AgdaSymbol{\}}\AgdaSpace{}%
\AgdaSymbol{→}\AgdaSpace{}%
\AgdaOperator{\AgdaInductiveConstructor{⟨}}\AgdaSpace{}%
\AgdaBound{h1}\AgdaSpace{}%
\AgdaOperator{\AgdaInductiveConstructor{,}}\AgdaSpace{}%
\AgdaBound{t1}\AgdaSpace{}%
\AgdaOperator{\AgdaInductiveConstructor{⟩}}\AgdaSpace{}%
\AgdaOperator{\AgdaDatatype{≡}}\AgdaSpace{}%
\AgdaOperator{\AgdaInductiveConstructor{⟨}}\AgdaSpace{}%
\AgdaBound{h2}\AgdaSpace{}%
\AgdaOperator{\AgdaInductiveConstructor{,}}\AgdaSpace{}%
\AgdaBound{t2}\AgdaSpace{}%
\AgdaOperator{\AgdaInductiveConstructor{⟩}}\AgdaSpace{}%
\AgdaSymbol{→}\AgdaSpace{}%
\AgdaBound{h1}\AgdaSpace{}%
\AgdaOperator{\AgdaDatatype{≡}}\AgdaSpace{}%
\AgdaBound{h2}\<%
\\
\>[0]\AgdaFunction{eq2sx}\AgdaSpace{}%
\AgdaInductiveConstructor{refl}\AgdaSpace{}%
\AgdaSymbol{=}\AgdaSpace{}%
\AgdaInductiveConstructor{refl}\<%
\\
\\[\AgdaEmptyExtraSkip]%
\>[0]\AgdaFunction{eq2dx}\AgdaSpace{}%
\AgdaSymbol{:}\AgdaSpace{}%
\AgdaSymbol{\{}\AgdaBound{A}\AgdaSpace{}%
\AgdaBound{B}\AgdaSpace{}%
\AgdaSymbol{:}\AgdaSpace{}%
\AgdaPrimitiveType{Set}\AgdaSymbol{\}\{}\AgdaBound{h1}\AgdaSpace{}%
\AgdaBound{h2}\AgdaSpace{}%
\AgdaSymbol{:}\AgdaSpace{}%
\AgdaBound{A}\AgdaSymbol{\}\{}\AgdaBound{t1}\AgdaSpace{}%
\AgdaBound{t2}\AgdaSpace{}%
\AgdaSymbol{:}\AgdaSpace{}%
\AgdaBound{B}\AgdaSymbol{\}}\AgdaSpace{}%
\AgdaSymbol{→}\AgdaSpace{}%
\AgdaOperator{\AgdaInductiveConstructor{⟨}}\AgdaSpace{}%
\AgdaBound{h1}\AgdaSpace{}%
\AgdaOperator{\AgdaInductiveConstructor{,}}\AgdaSpace{}%
\AgdaBound{t1}\AgdaSpace{}%
\AgdaOperator{\AgdaInductiveConstructor{⟩}}\AgdaSpace{}%
\AgdaOperator{\AgdaDatatype{≡}}\AgdaSpace{}%
\AgdaOperator{\AgdaInductiveConstructor{⟨}}\AgdaSpace{}%
\AgdaBound{h2}\AgdaSpace{}%
\AgdaOperator{\AgdaInductiveConstructor{,}}\AgdaSpace{}%
\AgdaBound{t2}\AgdaSpace{}%
\AgdaOperator{\AgdaInductiveConstructor{⟩}}\AgdaSpace{}%
\AgdaSymbol{→}\AgdaSpace{}%
\AgdaBound{t1}\AgdaSpace{}%
\AgdaOperator{\AgdaDatatype{≡}}\AgdaSpace{}%
\AgdaBound{t2}\<%
\\
\>[0]\AgdaFunction{eq2dx}\AgdaSpace{}%
\AgdaInductiveConstructor{refl}\AgdaSpace{}%
\AgdaSymbol{=}\AgdaSpace{}%
\AgdaInductiveConstructor{refl}\<%
\end{code}

\end{center}

\EZ{Depending on the context we may use the alias \agda{$\times$}. The projections extract the left and right element of the pair. The functions} \textit{eq2sx} and \textit{eq2dx} provide the equality proofs of the left and right elements, respectively, starting from the proof that two pairs are equal. 

\subsection{Disjunction type}
\EZ{The logical disjunction is implemented as a data type with two constructors that correspond to the injections.}

\begin{center}
\begin{code}%
\>[0]\AgdaKeyword{data}\AgdaSpace{}%
\AgdaOperator{\AgdaDatatype{\AgdaUnderscore{}∨\AgdaUnderscore{}}}%
\>[10]\AgdaSymbol{\{}\AgdaBound{ι}\AgdaSymbol{\}}\AgdaSpace{}%
\AgdaSymbol{(}\AgdaBound{A}\AgdaSpace{}%
\AgdaBound{B}\AgdaSpace{}%
\AgdaSymbol{:}\AgdaSpace{}%
\AgdaPrimitiveType{Set}\AgdaSpace{}%
\AgdaBound{ι}\AgdaSymbol{)}\AgdaSpace{}%
\AgdaSymbol{:}\AgdaSpace{}%
\AgdaPrimitiveType{Set}\AgdaSpace{}%
\AgdaBound{ι}\AgdaSpace{}%
\AgdaKeyword{where}\<%
\\
\>[0][@{}l@{\AgdaIndent{0}}]%
\>[2]\AgdaInductiveConstructor{inl}\AgdaSpace{}%
\AgdaSymbol{:}\AgdaSpace{}%
\AgdaBound{A}\AgdaSpace{}%
\AgdaSymbol{→}\AgdaSpace{}%
\AgdaBound{A}\AgdaSpace{}%
\AgdaOperator{\AgdaDatatype{∨}}\AgdaSpace{}%
\AgdaBound{B}\<%
\\
\>[2]\AgdaInductiveConstructor{inr}\AgdaSpace{}%
\AgdaSymbol{:}\AgdaSpace{}%
\AgdaBound{B}\AgdaSpace{}%
\AgdaSymbol{→}\AgdaSpace{}%
\AgdaBound{A}\AgdaSpace{}%
\AgdaOperator{\AgdaDatatype{∨}}\AgdaSpace{}%
\AgdaBound{B}\<%
\\
\\[\AgdaEmptyExtraSkip]%
\\[\AgdaEmptyExtraSkip]%
\>[0]\AgdaFunction{∨{-}elim}%
\>[22I]\AgdaSymbol{:}\AgdaSpace{}%
\AgdaSymbol{∀}\AgdaSpace{}%
\AgdaSymbol{\{}\AgdaBound{ι₁}\AgdaSpace{}%
\AgdaBound{ι₂}\AgdaSymbol{\}}\AgdaSpace{}%
\AgdaSymbol{\{}\AgdaBound{A}\AgdaSpace{}%
\AgdaBound{B}\AgdaSpace{}%
\AgdaSymbol{:}\AgdaSpace{}%
\AgdaPrimitiveType{Set}\AgdaSpace{}%
\AgdaBound{ι₁}\AgdaSymbol{\}}\AgdaSpace{}%
\AgdaSymbol{\{}\AgdaBound{Q}\AgdaSpace{}%
\AgdaSymbol{:}\AgdaSpace{}%
\AgdaPrimitiveType{Set}\AgdaSpace{}%
\AgdaBound{ι₂}\AgdaSymbol{\}}\AgdaSpace{}%
\AgdaSymbol{→}\<%
\\
\>[.]\<[22I]%
\>[7]\AgdaSymbol{(}\AgdaBound{A}\AgdaSpace{}%
\AgdaSymbol{→}\AgdaSpace{}%
\AgdaBound{Q}\AgdaSymbol{)}\AgdaSpace{}%
\AgdaSymbol{→}\<%
\\
\>[7]\AgdaSymbol{(}\AgdaBound{B}\AgdaSpace{}%
\AgdaSymbol{→}\AgdaSpace{}%
\AgdaBound{Q}\AgdaSymbol{)}\AgdaSpace{}%
\AgdaSymbol{→}\<%
\\
\>[7]\AgdaBound{A}\AgdaSpace{}%
\AgdaOperator{\AgdaDatatype{∨}}\AgdaSpace{}%
\AgdaBound{B}\AgdaSpace{}%
\AgdaSymbol{→}\AgdaSpace{}%
\AgdaBound{Q}\<%
\\
\>[0]\<%
\\
\>[0]\AgdaFunction{∨{-}elim}\AgdaSpace{}%
\AgdaBound{f}\AgdaSpace{}%
\AgdaBound{g}\AgdaSpace{}%
\AgdaSymbol{(}\AgdaInductiveConstructor{inl}\AgdaSpace{}%
\AgdaBound{x}\AgdaSymbol{)}\AgdaSpace{}%
\AgdaSymbol{=}\AgdaSpace{}%
\AgdaBound{f}\AgdaSpace{}%
\AgdaBound{x}\<%
\\
\>[0]\AgdaFunction{∨{-}elim}\AgdaSpace{}%
\AgdaBound{f}\AgdaSpace{}%
\AgdaBound{g}\AgdaSpace{}%
\AgdaSymbol{(}\AgdaInductiveConstructor{inr}\AgdaSpace{}%
\AgdaBound{x}\AgdaSymbol{)}\AgdaSpace{}%
\AgdaSymbol{=}\AgdaSpace{}%
\AgdaBound{g}\AgdaSpace{}%
\AgdaBound{x}\<%
\end{code}

\end{center}

The elimination rule allows a different behavior depending on how the term of type disjunction \EZ{is constructed}, by pattern matching on it. The function is written using also universe levels in order to make it usable when dealing with types. 

\subsection{Existential type}\label{sect:exists}
\EZ{The existential type is implemented as a data type with a single pair constructor, analogous to that of the conjiunction type}

\begin{center}
\begin{code}%
\>[0]\AgdaKeyword{data}\AgdaSpace{}%
\AgdaDatatype{∃}\AgdaSpace{}%
\AgdaSymbol{(}\AgdaBound{A}\AgdaSpace{}%
\AgdaSymbol{:}\AgdaSpace{}%
\AgdaPrimitiveType{Set}\AgdaSymbol{)}\AgdaSpace{}%
\AgdaSymbol{(}\AgdaBound{P}\AgdaSpace{}%
\AgdaSymbol{:}\AgdaSpace{}%
\AgdaBound{A}\AgdaSpace{}%
\AgdaSymbol{→}\AgdaSpace{}%
\AgdaPrimitiveType{Set}\AgdaSymbol{)}\AgdaSpace{}%
\AgdaSymbol{:}\AgdaSpace{}%
\AgdaPrimitiveType{Set}\AgdaSpace{}%
\AgdaKeyword{where}\<%
\\
\>[0][@{}l@{\AgdaIndent{0}}]%
\>[2]\AgdaOperator{\AgdaInductiveConstructor{<\AgdaUnderscore{},\AgdaUnderscore{}>}}\AgdaSpace{}%
\AgdaSymbol{:}\AgdaSpace{}%
\AgdaSymbol{(}\AgdaBound{a}\AgdaSpace{}%
\AgdaSymbol{:}\AgdaSpace{}%
\AgdaBound{A}\AgdaSymbol{)}\AgdaSpace{}%
\AgdaSymbol{→}\AgdaSpace{}%
\AgdaBound{P}\AgdaSpace{}%
\AgdaBound{a}\AgdaSpace{}%
\AgdaSymbol{→}\AgdaSpace{}%
\AgdaDatatype{∃}\AgdaSpace{}%
\AgdaBound{A}\AgdaSpace{}%
\AgdaBound{P}\<%
\\
\\[\AgdaEmptyExtraSkip]%
\\[\AgdaEmptyExtraSkip]%
\>[0]\AgdaFunction{witness}\AgdaSpace{}%
\AgdaSymbol{:}\AgdaSpace{}%
\AgdaSymbol{\{}\AgdaBound{A}\AgdaSpace{}%
\AgdaSymbol{:}\AgdaSpace{}%
\AgdaPrimitiveType{Set}\AgdaSymbol{\}\{}\AgdaBound{P}\AgdaSpace{}%
\AgdaSymbol{:}\AgdaSpace{}%
\AgdaBound{A}\AgdaSpace{}%
\AgdaSymbol{→}\AgdaSpace{}%
\AgdaPrimitiveType{Set}\AgdaSymbol{\}}\AgdaSpace{}%
\AgdaSymbol{→}\AgdaSpace{}%
\AgdaDatatype{∃}\AgdaSpace{}%
\AgdaBound{A}\AgdaSpace{}%
\AgdaBound{P}\AgdaSpace{}%
\AgdaSymbol{→}\AgdaSpace{}%
\AgdaBound{A}\<%
\\
\>[0]\AgdaFunction{witness}\AgdaSpace{}%
\AgdaOperator{\AgdaInductiveConstructor{<}}\AgdaSpace{}%
\AgdaBound{a}\AgdaSpace{}%
\AgdaOperator{\AgdaInductiveConstructor{,}}\AgdaSpace{}%
\AgdaBound{p}\AgdaSpace{}%
\AgdaOperator{\AgdaInductiveConstructor{>}}\AgdaSpace{}%
\AgdaSymbol{=}\AgdaSpace{}%
\AgdaBound{a}\<%
\\
\\[\AgdaEmptyExtraSkip]%
\>[0]\AgdaFunction{proof}\AgdaSpace{}%
\AgdaSymbol{:}\AgdaSpace{}%
\AgdaSymbol{\{}\AgdaBound{A}\AgdaSpace{}%
\AgdaSymbol{:}\AgdaSpace{}%
\AgdaPrimitiveType{Set}\AgdaSymbol{\}\{}\AgdaBound{P}\AgdaSpace{}%
\AgdaSymbol{:}\AgdaSpace{}%
\AgdaBound{A}\AgdaSpace{}%
\AgdaSymbol{→}\AgdaSpace{}%
\AgdaPrimitiveType{Set}\AgdaSymbol{\}}\AgdaSpace{}%
\AgdaSymbol{→}\AgdaSpace{}%
\AgdaSymbol{(}\AgdaBound{c}\AgdaSpace{}%
\AgdaSymbol{:}\AgdaSpace{}%
\AgdaDatatype{∃}\AgdaSpace{}%
\AgdaBound{A}\AgdaSpace{}%
\AgdaBound{P}\AgdaSymbol{)}\AgdaSpace{}%
\AgdaSymbol{→}\AgdaSpace{}%
\AgdaBound{P}\AgdaSpace{}%
\AgdaSymbol{(}\AgdaFunction{witness}\AgdaSpace{}%
\AgdaBound{c}\AgdaSymbol{)}\<%
\\
\>[0]\AgdaFunction{proof}\AgdaSpace{}%
\AgdaOperator{\AgdaInductiveConstructor{<}}\AgdaSpace{}%
\AgdaBound{a}\AgdaSpace{}%
\AgdaOperator{\AgdaInductiveConstructor{,}}\AgdaSpace{}%
\AgdaBound{p}\AgdaSpace{}%
\AgdaOperator{\AgdaInductiveConstructor{>}}\AgdaSpace{}%
\AgdaSymbol{=}\AgdaSpace{}%
\AgdaBound{p}\<%
\end{code}

\end{center}

Here the left element is the value on which the predicate holds, and the right one is the proof. Accordingly, the two projections are called \agda{witness} and \agda{proof}. 
When we use the existential, we write the predicate in the form of a lambda function.

\subsection{\agda{Max} function and properties}
This module has a huge importance in \refToChapter{flex} since we gave the definition of the predicate \textit{maxElem} using the binary function \textit{max} instead of a specific data type.
First we report the implementation of the function and then the properties that we used. For the sake of brevity we limit those properties to the types since their proofs are by pattern matching on each input element.

\begin{center}
\begin{code}%
\>[0]\AgdaFunction{max}\AgdaSpace{}%
\AgdaSymbol{:}\AgdaSpace{}%
\AgdaDatatype{Nat}\AgdaSpace{}%
\AgdaSymbol{→}\AgdaSpace{}%
\AgdaDatatype{Nat}\AgdaSpace{}%
\AgdaSymbol{→}\AgdaSpace{}%
\AgdaDatatype{Nat}\<%
\\
\>[0]\AgdaFunction{max}\AgdaSpace{}%
\AgdaInductiveConstructor{zero}\AgdaSpace{}%
\AgdaBound{b}\AgdaSpace{}%
\AgdaSymbol{=}\AgdaSpace{}%
\AgdaBound{b}\<%
\\
\>[0]\AgdaFunction{max}\AgdaSpace{}%
\AgdaSymbol{(}\AgdaInductiveConstructor{suc}\AgdaSpace{}%
\AgdaBound{a}\AgdaSymbol{)}\AgdaSpace{}%
\AgdaInductiveConstructor{zero}\AgdaSpace{}%
\AgdaSymbol{=}\AgdaSpace{}%
\AgdaInductiveConstructor{suc}\AgdaSpace{}%
\AgdaBound{a}\<%
\\
\>[0]\AgdaFunction{max}\AgdaSpace{}%
\AgdaSymbol{(}\AgdaInductiveConstructor{suc}\AgdaSpace{}%
\AgdaBound{a}\AgdaSymbol{)}\AgdaSpace{}%
\AgdaSymbol{(}\AgdaInductiveConstructor{suc}\AgdaSpace{}%
\AgdaBound{b}\AgdaSymbol{)}\AgdaSpace{}%
\AgdaSymbol{=}\AgdaSpace{}%
\AgdaInductiveConstructor{suc}\AgdaSpace{}%
\AgdaSymbol{(}\AgdaFunction{max}\AgdaSpace{}%
\AgdaBound{a}\AgdaSpace{}%
\AgdaBound{b}\AgdaSymbol{)}\<%
\\
\\[\AgdaEmptyExtraSkip]%
\>[0]\AgdaComment{{-}{-}{-}{-}{-}{-}{-}{-}{-}{-}{-}{-}{-}{-}{-}{-}{-}{-}{-}{-}{-}{-}{-}{-}PROPERTIES{-}{-}{-}{-}{-}{-}{-}{-}{-}{-}{-}{-}{-}{-}{-}{-}{-}{-}{-}{-}{-}{-}{-}{-}{-}{-}{-}{-}{-}{-}{-}{-}{-}}\<%
\\[\AgdaEmptyExtraSkip]%
\>[0]\AgdaFunction{max{-}comm}\AgdaSpace{}%
\AgdaSymbol{:}\AgdaSpace{}%
\AgdaSymbol{∀}\AgdaSpace{}%
\AgdaSymbol{\{}\AgdaBound{a}\AgdaSpace{}%
\AgdaBound{b}\AgdaSpace{}%
\AgdaBound{c}\AgdaSymbol{\}}\AgdaSpace{}%
\AgdaSymbol{→}\AgdaSpace{}%
\AgdaBound{c}\AgdaSpace{}%
\AgdaOperator{\AgdaDatatype{≡}}\AgdaSpace{}%
\AgdaFunction{max}\AgdaSpace{}%
\AgdaBound{a}\AgdaSpace{}%
\AgdaBound{b}\AgdaSpace{}%
\AgdaSymbol{→}\AgdaSpace{}%
\AgdaBound{c}\AgdaSpace{}%
\AgdaOperator{\AgdaDatatype{≡}}\AgdaSpace{}%
\AgdaFunction{max}\AgdaSpace{}%
\AgdaBound{b}\AgdaSpace{}%
\AgdaBound{a}\<%
\\
\\[\AgdaEmptyExtraSkip]%
\>[0]\AgdaFunction{max{-}suc}\AgdaSpace{}%
\AgdaSymbol{:}\AgdaSpace{}%
\AgdaSymbol{∀}\AgdaSpace{}%
\AgdaSymbol{\{}\AgdaBound{x}\AgdaSpace{}%
\AgdaBound{y}\AgdaSymbol{\}}\AgdaSpace{}%
\AgdaSymbol{→}\AgdaSpace{}%
\AgdaInductiveConstructor{suc}\AgdaSpace{}%
\AgdaSymbol{(}\AgdaFunction{max}\AgdaSpace{}%
\AgdaBound{x}\AgdaSpace{}%
\AgdaBound{y}\AgdaSymbol{)}\AgdaSpace{}%
\AgdaOperator{\AgdaDatatype{≡}}\AgdaSpace{}%
\AgdaFunction{max}\AgdaSpace{}%
\AgdaSymbol{(}\AgdaInductiveConstructor{suc}\AgdaSpace{}%
\AgdaBound{x}\AgdaSymbol{)}\AgdaSpace{}%
\AgdaSymbol{(}\AgdaInductiveConstructor{suc}\AgdaSpace{}%
\AgdaBound{y}\AgdaSymbol{)}\<%
\\
\\[\AgdaEmptyExtraSkip]%
\>[0]\AgdaFunction{pred{-}max}\AgdaSpace{}%
\AgdaSymbol{:}\AgdaSpace{}%
\AgdaSymbol{∀}\AgdaSpace{}%
\AgdaSymbol{\{}\AgdaBound{n}\AgdaSpace{}%
\AgdaBound{x}\AgdaSpace{}%
\AgdaBound{y}\AgdaSymbol{\}}\AgdaSpace{}%
\AgdaSymbol{→}\AgdaSpace{}%
\AgdaInductiveConstructor{suc}\AgdaSpace{}%
\AgdaBound{n}\AgdaSpace{}%
\AgdaOperator{\AgdaDatatype{≡}}\AgdaSpace{}%
\AgdaInductiveConstructor{suc}\AgdaSpace{}%
\AgdaSymbol{(}\AgdaFunction{max}\AgdaSpace{}%
\AgdaBound{x}\AgdaSpace{}%
\AgdaBound{y}\AgdaSymbol{)}\AgdaSpace{}%
\AgdaSymbol{→}\AgdaSpace{}%
\AgdaBound{n}\AgdaSpace{}%
\AgdaOperator{\AgdaDatatype{≡}}\AgdaSpace{}%
\AgdaFunction{max}\AgdaSpace{}%
\AgdaBound{x}\AgdaSpace{}%
\AgdaBound{y}\<%
\\
\\[\AgdaEmptyExtraSkip]%
\>[0]\AgdaFunction{max{-}refl}\AgdaSpace{}%
\AgdaSymbol{:}\AgdaSpace{}%
\AgdaSymbol{∀}\AgdaSpace{}%
\AgdaSymbol{\{}\AgdaBound{n}\AgdaSpace{}%
\AgdaBound{x}\AgdaSpace{}%
\AgdaBound{y}\AgdaSymbol{\}}\AgdaSpace{}%
\AgdaSymbol{→}\AgdaSpace{}%
\AgdaBound{n}\AgdaSpace{}%
\AgdaOperator{\AgdaDatatype{≡}}\AgdaSpace{}%
\AgdaFunction{max}\AgdaSpace{}%
\AgdaBound{x}\AgdaSpace{}%
\AgdaBound{y}\AgdaSpace{}%
\AgdaSymbol{→}\AgdaSpace{}%
\AgdaSymbol{(}\AgdaBound{n}\AgdaSpace{}%
\AgdaOperator{\AgdaDatatype{≡}}\AgdaSpace{}%
\AgdaBound{x}\AgdaSymbol{)}\AgdaSpace{}%
\AgdaOperator{\AgdaDatatype{∨}}\AgdaSpace{}%
\AgdaSymbol{(}\AgdaBound{n}\AgdaSpace{}%
\AgdaOperator{\AgdaDatatype{≡}}\AgdaSpace{}%
\AgdaBound{y}\AgdaSymbol{)}\<%
\\
\\[\AgdaEmptyExtraSkip]%
\>[0]\AgdaFunction{max{-}eq}\AgdaSpace{}%
\AgdaSymbol{:}\AgdaSpace{}%
\AgdaSymbol{∀}\AgdaSpace{}%
\AgdaSymbol{\{}\AgdaBound{n}\AgdaSpace{}%
\AgdaSymbol{:}\AgdaSpace{}%
\AgdaDatatype{Nat}\AgdaSymbol{\}}\AgdaSpace{}%
\AgdaSymbol{→}\AgdaSpace{}%
\AgdaBound{n}\AgdaSpace{}%
\AgdaOperator{\AgdaDatatype{≡}}\AgdaSpace{}%
\AgdaFunction{max}\AgdaSpace{}%
\AgdaBound{n}\AgdaSpace{}%
\AgdaBound{n}\<%
\\
\\[\AgdaEmptyExtraSkip]%
\>[0]\AgdaFunction{max{-}trans}\AgdaSpace{}%
\AgdaSymbol{:}\AgdaSpace{}%
\AgdaSymbol{∀}\AgdaSpace{}%
\AgdaSymbol{\{}\AgdaBound{z}\AgdaSpace{}%
\AgdaBound{n}\AgdaSpace{}%
\AgdaBound{i}\AgdaSymbol{\}}\AgdaSpace{}%
\AgdaSymbol{→}\AgdaSpace{}%
\AgdaBound{z}\AgdaSpace{}%
\AgdaOperator{\AgdaDatatype{≡}}\AgdaSpace{}%
\AgdaFunction{max}\AgdaSpace{}%
\AgdaBound{z}\AgdaSpace{}%
\AgdaBound{n}\AgdaSpace{}%
\AgdaSymbol{→}\AgdaSpace{}%
\AgdaBound{n}\AgdaSpace{}%
\AgdaOperator{\AgdaDatatype{≡}}\AgdaSpace{}%
\AgdaFunction{max}\AgdaSpace{}%
\AgdaBound{n}\AgdaSpace{}%
\AgdaBound{i}\AgdaSpace{}%
\AgdaSymbol{→}\AgdaSpace{}%
\AgdaBound{z}\AgdaSpace{}%
\AgdaOperator{\AgdaDatatype{≡}}\AgdaSpace{}%
\AgdaFunction{max}\AgdaSpace{}%
\AgdaBound{z}\AgdaSpace{}%
\AgdaBound{i}\<%
\end{code}

\end{center}

Where the most important function is \textit{max-refl}. If we consider \textit{maxElem} as example, when we say that $z = \Max {x} {y}$ we lose the connection between $z$ and $x, y$ even though it must be equal to $x$ or $y$. The function above allows us to pattern match over the result of it applied to the side condition of the rule of the predicate thanks also to the usage of the disjunction type.

\subsection{Singleton type}
While looking at Agda codes, you can see that \agda{inspect} is often called. This function comes from the \textit{Singleton} module which is one of the most important tools we needed.

\begin{center}
\begin{code}%
\>[0]\AgdaKeyword{data}\AgdaSpace{}%
\AgdaDatatype{Singleton}\AgdaSpace{}%
\AgdaSymbol{\{}\AgdaBound{a}\AgdaSymbol{\}}\AgdaSpace{}%
\AgdaSymbol{\{}\AgdaBound{A}\AgdaSpace{}%
\AgdaSymbol{:}\AgdaSpace{}%
\AgdaPrimitiveType{Set}\AgdaSpace{}%
\AgdaBound{a}\AgdaSymbol{\}}\AgdaSpace{}%
\AgdaSymbol{(}\AgdaBound{x}\AgdaSpace{}%
\AgdaSymbol{:}\AgdaSpace{}%
\AgdaBound{A}\AgdaSymbol{)}\AgdaSpace{}%
\AgdaSymbol{:}\AgdaSpace{}%
\AgdaPrimitiveType{Set}\AgdaSpace{}%
\AgdaBound{a}\AgdaSpace{}%
\AgdaKeyword{where}\<%
\\
\>[0][@{}l@{\AgdaIndent{0}}]%
\>[2]\AgdaOperator{\AgdaInductiveConstructor{\AgdaUnderscore{}with≡\AgdaUnderscore{}}}\AgdaSpace{}%
\AgdaSymbol{:}\AgdaSpace{}%
\AgdaSymbol{(}\AgdaBound{y}\AgdaSpace{}%
\AgdaSymbol{:}\AgdaSpace{}%
\AgdaBound{A}\AgdaSymbol{)}\AgdaSpace{}%
\AgdaSymbol{→}\AgdaSpace{}%
\AgdaBound{x}\AgdaSpace{}%
\AgdaOperator{\AgdaDatatype{≡}}\AgdaSpace{}%
\AgdaBound{y}\AgdaSpace{}%
\AgdaSymbol{→}\AgdaSpace{}%
\AgdaDatatype{Singleton}\AgdaSpace{}%
\AgdaBound{x}\<%
\\
\\[\AgdaEmptyExtraSkip]%
\>[0]\AgdaFunction{inspect}\AgdaSpace{}%
\AgdaSymbol{:}\AgdaSpace{}%
\AgdaSymbol{∀}\AgdaSpace{}%
\AgdaSymbol{\{}\AgdaBound{a}\AgdaSymbol{\}}\AgdaSpace{}%
\AgdaSymbol{\{}\AgdaBound{A}\AgdaSpace{}%
\AgdaSymbol{:}\AgdaSpace{}%
\AgdaPrimitiveType{Set}\AgdaSpace{}%
\AgdaBound{a}\AgdaSymbol{\}}\AgdaSpace{}%
\AgdaSymbol{(}\AgdaBound{x}\AgdaSpace{}%
\AgdaSymbol{:}\AgdaSpace{}%
\AgdaBound{A}\AgdaSymbol{)}\AgdaSpace{}%
\AgdaSymbol{→}\AgdaSpace{}%
\AgdaDatatype{Singleton}\AgdaSpace{}%
\AgdaBound{x}\<%
\\
\>[0]\AgdaFunction{inspect}\AgdaSpace{}%
\AgdaBound{x}\AgdaSpace{}%
\AgdaSymbol{=}\AgdaSpace{}%
\AgdaBound{x}\AgdaSpace{}%
\AgdaOperator{\AgdaInductiveConstructor{with≡}}\AgdaSpace{}%
\AgdaInductiveConstructor{refl}\<%
\end{code}

\end{center}

Given an element $x$, \textit{Singleton x} represents the set of all the elements that are equal to $x$.
It is used because the reference to the old variables are lost when pattern matching through \textit{with}. Pattern matching on the result of \agda{inspect} allows to bring the equality proofs inside each case.
We implemented \textit{Singleton} according to the directives in the documentation.

\subsection{Empty type}
Another very important module is \textit{Empty}. We used it many times in order to manage absurd cases. As it is reported in the codes, sometimes this cases are obvious and Agda automatically recognizes them. On the other hand, especially when we used colists implemented as coinductive records, we had to manually solve such situations. For example there were cases in which we had the proof of the membership of an element to a list and at the same time a proof that the list was empty. 

\begin{center}
\begin{code}%
\>[0]\AgdaKeyword{data}\AgdaSpace{}%
\AgdaDatatype{⊥}\AgdaSpace{}%
\AgdaSymbol{:}\AgdaSpace{}%
\AgdaPrimitiveType{Set}\AgdaSpace{}%
\AgdaKeyword{where}\<%
\\
\\[\AgdaEmptyExtraSkip]%
\>[0]\AgdaFunction{⊥{-}elim}\AgdaSpace{}%
\AgdaSymbol{:}\AgdaSpace{}%
\AgdaSymbol{∀}\AgdaSpace{}%
\AgdaSymbol{\{}\AgdaBound{w}\AgdaSymbol{\}}\AgdaSpace{}%
\AgdaSymbol{\{}\AgdaBound{Whatever}\AgdaSpace{}%
\AgdaSymbol{:}\AgdaSpace{}%
\AgdaPrimitiveType{Set}\AgdaSpace{}%
\AgdaBound{w}\AgdaSymbol{\}}\AgdaSpace{}%
\AgdaSymbol{→}\AgdaSpace{}%
\AgdaDatatype{⊥}\AgdaSpace{}%
\AgdaSymbol{→}\AgdaSpace{}%
\AgdaBound{Whatever}\<%
\\
\>[0]\AgdaFunction{⊥{-}elim}\AgdaSpace{}%
\AgdaSymbol{()}\<%
\end{code}

\end{center}

$\bot$ is a data type without constructors and the elimination rule allows to manage the mentioned absurd cases by returning whatever type is needed.
In order to say to Agda that there is an absurd, a function that takes some proofs in input and returns $\bot$ must be implemented. Then it can be given as a parameter to \agda{$\bot$-elim}.
We did not implement this module since it is already contained in the standard library.

\section{\agda{MyColist} equalities and absurd cases}
\label{sect:mycolisteq}
Here we show some technicalities about the main equality proofs that are used when dealing with properties of colists implemented as coinductive records. These proofs are needed because we reason by observations in such codes. For example, existentially quantified variables must be related between themselves in some way according to the context.

\begin{center}
\begin{code}%
\>[0]\AgdaFunction{just{-}elim}%
\>[18I]\AgdaSymbol{:}\AgdaSpace{}%
\AgdaSymbol{\{}\AgdaBound{A}\AgdaSpace{}%
\AgdaSymbol{:}\AgdaSpace{}%
\AgdaPrimitiveType{Set}\AgdaSymbol{\}\{}\AgdaBound{c1}\AgdaSpace{}%
\AgdaBound{c2}\AgdaSpace{}%
\AgdaSymbol{:}\AgdaSpace{}%
\AgdaBound{A}\AgdaSymbol{\}}\AgdaSpace{}%
\AgdaSymbol{→}\<%
\\
\>[.]\<[18I]%
\>[10]\AgdaOperator{\AgdaDatatype{\AgdaUnderscore{}≡\AgdaUnderscore{}}}\AgdaSpace{}%
\AgdaSymbol{\{}\AgdaArgument{A}\AgdaSpace{}%
\AgdaSymbol{=}\AgdaSpace{}%
\AgdaDatatype{Maybe}\AgdaSpace{}%
\AgdaBound{A}\AgdaSymbol{\}}\AgdaSpace{}%
\AgdaSymbol{(}\AgdaInductiveConstructor{just}\AgdaSpace{}%
\AgdaBound{c1}\AgdaSymbol{)}\AgdaSpace{}%
\AgdaSymbol{(}\AgdaInductiveConstructor{just}\AgdaSpace{}%
\AgdaBound{c2}\AgdaSymbol{)}\AgdaSpace{}%
\AgdaSymbol{→}\<%
\\
\>[10]\AgdaBound{c1}\AgdaSpace{}%
\AgdaOperator{\AgdaDatatype{≡}}\AgdaSpace{}%
\AgdaBound{c2}\<%
\\
\>[0]\AgdaFunction{just{-}elim}\AgdaSpace{}%
\AgdaInductiveConstructor{refl}\AgdaSpace{}%
\AgdaSymbol{=}\AgdaSpace{}%
\AgdaInductiveConstructor{refl}\<%
\\
\\[\AgdaEmptyExtraSkip]%
\>[0]\AgdaFunction{eq{-}hd}%
\>[40I]\AgdaSymbol{:}\AgdaSpace{}%
\AgdaSymbol{\{}\AgdaBound{A}\AgdaSpace{}%
\AgdaSymbol{:}\AgdaSpace{}%
\AgdaPrimitiveType{Set}\AgdaSymbol{\}}\AgdaSpace{}%
\AgdaSymbol{\{}\AgdaBound{h1}\AgdaSpace{}%
\AgdaBound{h2}\AgdaSpace{}%
\AgdaSymbol{:}\AgdaSpace{}%
\AgdaBound{A}\AgdaSymbol{\}}\AgdaSpace{}%
\AgdaSymbol{\{}\AgdaBound{l}\AgdaSpace{}%
\AgdaBound{t1}\AgdaSpace{}%
\AgdaBound{t2}\AgdaSpace{}%
\AgdaSymbol{:}\AgdaSpace{}%
\AgdaRecord{MyColist}\AgdaSpace{}%
\AgdaBound{A}\AgdaSymbol{\}}\AgdaSpace{}%
\AgdaSymbol{→}\<%
\\
\>[.]\<[40I]%
\>[6]\AgdaField{MyColist.list}\AgdaSpace{}%
\AgdaBound{l}\AgdaSpace{}%
\AgdaOperator{\AgdaDatatype{≡}}\AgdaSpace{}%
\AgdaInductiveConstructor{just}\AgdaSpace{}%
\AgdaOperator{\AgdaInductiveConstructor{⟨}}\AgdaSpace{}%
\AgdaBound{h1}\AgdaSpace{}%
\AgdaOperator{\AgdaInductiveConstructor{,}}\AgdaSpace{}%
\AgdaBound{t1}\AgdaSpace{}%
\AgdaOperator{\AgdaInductiveConstructor{⟩}}\AgdaSpace{}%
\AgdaSymbol{→}\<%
\\
\>[6]\AgdaField{MyColist.list}\AgdaSpace{}%
\AgdaBound{l}\AgdaSpace{}%
\AgdaOperator{\AgdaDatatype{≡}}\AgdaSpace{}%
\AgdaInductiveConstructor{just}\AgdaSpace{}%
\AgdaOperator{\AgdaInductiveConstructor{⟨}}\AgdaSpace{}%
\AgdaBound{h2}\AgdaSpace{}%
\AgdaOperator{\AgdaInductiveConstructor{,}}\AgdaSpace{}%
\AgdaBound{t2}\AgdaSpace{}%
\AgdaOperator{\AgdaInductiveConstructor{⟩}}\AgdaSpace{}%
\AgdaSymbol{→}\<%
\\
\>[6]\AgdaBound{h1}\AgdaSpace{}%
\AgdaOperator{\AgdaDatatype{≡}}\AgdaSpace{}%
\AgdaBound{h2}\<%
\\
\>[0]\AgdaFunction{eq{-}hd}\AgdaSpace{}%
\AgdaBound{eq1}\AgdaSpace{}%
\AgdaBound{eq2}\AgdaSpace{}%
\AgdaSymbol{=}%
\>[17]\AgdaFunction{eq2sx}\AgdaSpace{}%
\AgdaSymbol{((}\AgdaFunction{just{-}elim}\AgdaSpace{}%
\AgdaSymbol{(}\AgdaFunction{trans}\AgdaSpace{}%
\AgdaSymbol{(}\AgdaFunction{sym}\AgdaSpace{}%
\AgdaBound{eq1}\AgdaSymbol{)}\AgdaSpace{}%
\AgdaBound{eq2}\AgdaSymbol{)))}\<%
\\
\\[\AgdaEmptyExtraSkip]%
\>[0]\AgdaFunction{eq{-}tl}%
\>[83I]\AgdaSymbol{:}\AgdaSpace{}%
\AgdaSymbol{\{}\AgdaBound{A}\AgdaSpace{}%
\AgdaSymbol{:}\AgdaSpace{}%
\AgdaPrimitiveType{Set}\AgdaSymbol{\}}\AgdaSpace{}%
\AgdaSymbol{\{}\AgdaBound{h1}\AgdaSpace{}%
\AgdaBound{h2}\AgdaSpace{}%
\AgdaSymbol{:}\AgdaSpace{}%
\AgdaBound{A}\AgdaSymbol{\}}\AgdaSpace{}%
\AgdaSymbol{\{}\AgdaBound{l}\AgdaSpace{}%
\AgdaBound{t1}\AgdaSpace{}%
\AgdaBound{t2}\AgdaSpace{}%
\AgdaSymbol{:}\AgdaSpace{}%
\AgdaRecord{MyColist}\AgdaSpace{}%
\AgdaBound{A}\AgdaSymbol{\}}\AgdaSpace{}%
\AgdaSymbol{→}\<%
\\
\>[.]\<[83I]%
\>[6]\AgdaField{MyColist.list}\AgdaSpace{}%
\AgdaBound{l}\AgdaSpace{}%
\AgdaOperator{\AgdaDatatype{≡}}\AgdaSpace{}%
\AgdaInductiveConstructor{just}\AgdaSpace{}%
\AgdaOperator{\AgdaInductiveConstructor{⟨}}\AgdaSpace{}%
\AgdaBound{h1}\AgdaSpace{}%
\AgdaOperator{\AgdaInductiveConstructor{,}}\AgdaSpace{}%
\AgdaBound{t1}\AgdaSpace{}%
\AgdaOperator{\AgdaInductiveConstructor{⟩}}\AgdaSpace{}%
\AgdaSymbol{→}\<%
\\
\>[6]\AgdaField{MyColist.list}\AgdaSpace{}%
\AgdaBound{l}\AgdaSpace{}%
\AgdaOperator{\AgdaDatatype{≡}}\AgdaSpace{}%
\AgdaInductiveConstructor{just}\AgdaSpace{}%
\AgdaOperator{\AgdaInductiveConstructor{⟨}}\AgdaSpace{}%
\AgdaBound{h2}\AgdaSpace{}%
\AgdaOperator{\AgdaInductiveConstructor{,}}\AgdaSpace{}%
\AgdaBound{t2}\AgdaSpace{}%
\AgdaOperator{\AgdaInductiveConstructor{⟩}}\AgdaSpace{}%
\AgdaSymbol{→}\<%
\\
\>[6]\AgdaBound{t1}\AgdaSpace{}%
\AgdaOperator{\AgdaDatatype{≡}}\AgdaSpace{}%
\AgdaBound{t2}\<%
\\
\>[0]\AgdaFunction{eq{-}tl}\AgdaSpace{}%
\AgdaBound{eq1}\AgdaSpace{}%
\AgdaBound{eq2}\AgdaSpace{}%
\AgdaSymbol{=}%
\>[17]\AgdaFunction{eq2dx}\AgdaSpace{}%
\AgdaSymbol{((}\AgdaFunction{just{-}elim}\AgdaSpace{}%
\AgdaSymbol{(}\AgdaFunction{trans}\AgdaSpace{}%
\AgdaSymbol{(}\AgdaFunction{sym}\AgdaSpace{}%
\AgdaBound{eq1}\AgdaSymbol{)}\AgdaSpace{}%
\AgdaBound{eq2}\AgdaSymbol{)))}\<%
\\
\\[\AgdaEmptyExtraSkip]%
\>[0]\AgdaFunction{maybe{-}abs}%
\>[126I]\AgdaSymbol{:}\AgdaSpace{}%
\AgdaSymbol{\{}\AgdaBound{A}\AgdaSpace{}%
\AgdaSymbol{:}\AgdaSpace{}%
\AgdaPrimitiveType{Set}\AgdaSymbol{\}\{}\AgdaBound{m}\AgdaSpace{}%
\AgdaSymbol{:}\AgdaSpace{}%
\AgdaDatatype{Maybe}\AgdaSpace{}%
\AgdaBound{A}\AgdaSymbol{\}\{}\AgdaBound{something}\AgdaSpace{}%
\AgdaSymbol{:}\AgdaSpace{}%
\AgdaBound{A}\AgdaSymbol{\}}\AgdaSpace{}%
\AgdaSymbol{→}\<%
\\
\>[.]\<[126I]%
\>[10]\AgdaBound{m}\AgdaSpace{}%
\AgdaOperator{\AgdaDatatype{≡}}\AgdaSpace{}%
\AgdaInductiveConstructor{nothing}\AgdaSpace{}%
\AgdaSymbol{→}\AgdaSpace{}%
\AgdaBound{m}\AgdaSpace{}%
\AgdaOperator{\AgdaDatatype{≡}}\AgdaSpace{}%
\AgdaSymbol{(}\AgdaInductiveConstructor{just}\AgdaSpace{}%
\AgdaBound{something}\AgdaSymbol{)}\AgdaSpace{}%
\AgdaSymbol{→}\AgdaSpace{}%
\AgdaDatatype{⊥}\<%
\\
\>[0]\AgdaFunction{maybe{-}abs}\AgdaSpace{}%
\AgdaInductiveConstructor{refl}\AgdaSpace{}%
\AgdaSymbol{()}\<%
\\
\\[\AgdaEmptyExtraSkip]%
\>[0]\AgdaFunction{maybe{-}abs{-}2}\AgdaSpace{}%
\AgdaSymbol{:}\AgdaSpace{}%
\AgdaSymbol{\{}\AgdaBound{A}\AgdaSpace{}%
\AgdaSymbol{:}\AgdaSpace{}%
\AgdaPrimitiveType{Set}\AgdaSymbol{\}}\AgdaSpace{}%
\AgdaSymbol{\{}\AgdaBound{x}\AgdaSpace{}%
\AgdaSymbol{:}\AgdaSpace{}%
\AgdaBound{A}\AgdaSymbol{\}}\AgdaSpace{}%
\AgdaSymbol{→}\AgdaSpace{}%
\AgdaOperator{\AgdaDatatype{\AgdaUnderscore{}≡\AgdaUnderscore{}}}\AgdaSpace{}%
\AgdaSymbol{\{}\AgdaArgument{A}\AgdaSpace{}%
\AgdaSymbol{=}\AgdaSpace{}%
\AgdaDatatype{Maybe}\AgdaSpace{}%
\AgdaBound{A}\AgdaSymbol{\}}\AgdaSpace{}%
\AgdaInductiveConstructor{nothing}\AgdaSpace{}%
\AgdaSymbol{(}\AgdaInductiveConstructor{just}\AgdaSpace{}%
\AgdaBound{x}\AgdaSymbol{)}\AgdaSpace{}%
\AgdaSymbol{→}\AgdaSpace{}%
\AgdaDatatype{⊥}\<%
\\
\>[0]\AgdaFunction{maybe{-}abs{-}2}\AgdaSpace{}%
\AgdaSymbol{\{}\AgdaBound{x}\AgdaSymbol{\}}\AgdaSpace{}%
\AgdaSymbol{()}\<%
\\
\\[\AgdaEmptyExtraSkip]%
\>[0]\AgdaFunction{mycolist{-}abs}%
\>[167I]\AgdaSymbol{:}\AgdaSpace{}%
\AgdaSymbol{\{}\AgdaBound{A}\AgdaSpace{}%
\AgdaSymbol{:}\AgdaSpace{}%
\AgdaPrimitiveType{Set}\AgdaSymbol{\}\{}\AgdaBound{l}\AgdaSpace{}%
\AgdaSymbol{:}\AgdaSpace{}%
\AgdaRecord{MyColist}\AgdaSpace{}%
\AgdaBound{A}\AgdaSymbol{\}\{}\AgdaBound{something}\AgdaSpace{}%
\AgdaSymbol{:}\AgdaSpace{}%
\AgdaBound{A}\AgdaSpace{}%
\AgdaOperator{\AgdaDatatype{∧}}\AgdaSpace{}%
\AgdaRecord{MyColist}\AgdaSpace{}%
\AgdaBound{A}\AgdaSymbol{\}}\AgdaSpace{}%
\AgdaSymbol{→}\<%
\\
\>[.]\<[167I]%
\>[13]\AgdaSymbol{(}\AgdaField{MyColist.list}\AgdaSpace{}%
\AgdaBound{l}\AgdaSymbol{)}\AgdaSpace{}%
\AgdaOperator{\AgdaDatatype{≡}}\AgdaSpace{}%
\AgdaInductiveConstructor{nothing}\AgdaSpace{}%
\AgdaSymbol{→}\<%
\\
\>[13]\AgdaSymbol{(}\AgdaField{MyColist.list}\AgdaSpace{}%
\AgdaBound{l}\AgdaSymbol{)}\AgdaSpace{}%
\AgdaOperator{\AgdaDatatype{≡}}\AgdaSpace{}%
\AgdaInductiveConstructor{just}\AgdaSpace{}%
\AgdaBound{something}\AgdaSpace{}%
\AgdaSymbol{→}\AgdaSpace{}%
\AgdaDatatype{⊥}\<%
\\
\>[0]\AgdaFunction{mycolist{-}abs}\AgdaSpace{}%
\AgdaBound{p}\AgdaSpace{}%
\AgdaBound{q}\AgdaSpace{}%
\AgdaSymbol{=}\AgdaSpace{}%
\AgdaFunction{maybe{-}abs}\AgdaSpace{}%
\AgdaBound{p}\AgdaSpace{}%
\AgdaBound{q}\<%
\\
\\[\AgdaEmptyExtraSkip]%
\>[0]\AgdaFunction{get{-}eq{-}0}%
\>[196I]\AgdaSymbol{:}\AgdaSpace{}%
\AgdaSymbol{\{}\AgdaBound{A}\AgdaSpace{}%
\AgdaSymbol{:}\AgdaSpace{}%
\AgdaPrimitiveType{Set}\AgdaSymbol{\}}\AgdaSpace{}%
\AgdaSymbol{\{}\AgdaBound{h}\AgdaSpace{}%
\AgdaSymbol{:}\AgdaSpace{}%
\AgdaBound{A}\AgdaSymbol{\}}\AgdaSpace{}%
\AgdaSymbol{\{}\AgdaBound{l}\AgdaSpace{}%
\AgdaBound{t}\AgdaSpace{}%
\AgdaSymbol{:}\AgdaSpace{}%
\AgdaRecord{MyColist}\AgdaSpace{}%
\AgdaBound{A}\AgdaSymbol{\}}\AgdaSpace{}%
\AgdaSymbol{→}\<%
\\
\>[.]\<[196I]%
\>[9]\AgdaField{MyColist.list}\AgdaSpace{}%
\AgdaBound{l}\AgdaSpace{}%
\AgdaOperator{\AgdaDatatype{≡}}\AgdaSpace{}%
\AgdaInductiveConstructor{just}\AgdaSpace{}%
\AgdaOperator{\AgdaInductiveConstructor{⟨}}\AgdaSpace{}%
\AgdaBound{h}\AgdaSpace{}%
\AgdaOperator{\AgdaInductiveConstructor{,}}\AgdaSpace{}%
\AgdaBound{t}\AgdaSpace{}%
\AgdaOperator{\AgdaInductiveConstructor{⟩}}\AgdaSpace{}%
\AgdaSymbol{→}\<%
\\
\>[9]\AgdaFunction{get}\AgdaSpace{}%
\AgdaBound{l}\AgdaSpace{}%
\AgdaInductiveConstructor{zero}\AgdaSpace{}%
\AgdaOperator{\AgdaDatatype{≡}}\AgdaSpace{}%
\AgdaInductiveConstructor{just}\AgdaSpace{}%
\AgdaBound{h}\<%
\\
\>[0]\<%
\\
\>[0]\AgdaFunction{get{-}eq{-}0}\AgdaSpace{}%
\AgdaSymbol{\{}\AgdaArgument{h}\AgdaSpace{}%
\AgdaSymbol{=}\AgdaSpace{}%
\AgdaBound{h}\AgdaSymbol{\}}\AgdaSpace{}%
\AgdaSymbol{\{}\AgdaBound{l}\AgdaSymbol{\}}\AgdaSpace{}%
\AgdaSymbol{\{}\AgdaBound{t}\AgdaSymbol{\}}\AgdaSpace{}%
\AgdaBound{eq}\AgdaSpace{}%
\AgdaKeyword{with}\AgdaSpace{}%
\AgdaFunction{inspect}\AgdaSpace{}%
\AgdaSymbol{(}\AgdaField{MyColist.list}\AgdaSpace{}%
\AgdaBound{l}\AgdaSymbol{)}\<%
\\
\>[0]\AgdaSymbol{...}\AgdaSpace{}%
\AgdaSymbol{|}\AgdaSpace{}%
\AgdaInductiveConstructor{nothing}\AgdaSpace{}%
\AgdaOperator{\AgdaInductiveConstructor{with≡}}\AgdaSpace{}%
\AgdaBound{eq1}\AgdaSpace{}%
\AgdaSymbol{=}\AgdaSpace{}%
\AgdaFunction{⊥{-}elim}\AgdaSpace{}%
\AgdaSymbol{(}\AgdaFunction{mycolist{-}abs}\AgdaSpace{}%
\AgdaSymbol{\{}\AgdaArgument{l}\AgdaSpace{}%
\AgdaSymbol{=}\AgdaSpace{}%
\AgdaBound{l}\AgdaSymbol{\}}\AgdaSpace{}%
\AgdaBound{eq1}\AgdaSpace{}%
\AgdaBound{eq}\AgdaSymbol{)}\<%
\\
\>[0]\AgdaSymbol{...}\AgdaSpace{}%
\AgdaSymbol{|}\AgdaSpace{}%
\AgdaInductiveConstructor{just}\AgdaSpace{}%
\AgdaOperator{\AgdaInductiveConstructor{⟨}}\AgdaSpace{}%
\AgdaBound{h1}\AgdaSpace{}%
\AgdaOperator{\AgdaInductiveConstructor{,}}\AgdaSpace{}%
\AgdaBound{t1}\AgdaSpace{}%
\AgdaOperator{\AgdaInductiveConstructor{⟩}}\AgdaSpace{}%
\AgdaOperator{\AgdaInductiveConstructor{with≡}}\AgdaSpace{}%
\AgdaBound{eq1}\AgdaSpace{}%
\AgdaSymbol{=}\AgdaSpace{}%
\AgdaFunction{cong}\AgdaSpace{}%
\AgdaSymbol{(λ}\AgdaSpace{}%
\AgdaBound{k}\AgdaSpace{}%
\AgdaSymbol{→}\AgdaSpace{}%
\AgdaInductiveConstructor{just}\AgdaSpace{}%
\AgdaBound{k}\AgdaSymbol{)}\AgdaSpace{}%
\AgdaSymbol{(}\AgdaFunction{eq{-}hd}\AgdaSpace{}%
\AgdaSymbol{\{}\AgdaArgument{l}\AgdaSpace{}%
\AgdaSymbol{=}\AgdaSpace{}%
\AgdaBound{l}\AgdaSymbol{\}}\AgdaSpace{}%
\AgdaBound{eq1}\AgdaSpace{}%
\AgdaBound{eq}\AgdaSymbol{)}\<%
\\
\\[\AgdaEmptyExtraSkip]%
\>[0]\AgdaFunction{get{-}eq{-}tl}%
\>[267I]\AgdaSymbol{:}\AgdaSpace{}%
\AgdaSymbol{\{}\AgdaBound{A}\AgdaSpace{}%
\AgdaSymbol{:}\AgdaSpace{}%
\AgdaPrimitiveType{Set}\AgdaSymbol{\}}\AgdaSpace{}%
\AgdaSymbol{\{}\AgdaBound{h}\AgdaSpace{}%
\AgdaBound{x}\AgdaSpace{}%
\AgdaSymbol{:}\AgdaSpace{}%
\AgdaBound{A}\AgdaSymbol{\}}\AgdaSpace{}%
\AgdaSymbol{\{}\AgdaBound{l}\AgdaSpace{}%
\AgdaBound{t}\AgdaSpace{}%
\AgdaSymbol{:}\AgdaSpace{}%
\AgdaRecord{MyColist}\AgdaSpace{}%
\AgdaBound{A}\AgdaSymbol{\}}\AgdaSpace{}%
\AgdaSymbol{\{}\AgdaBound{i}\AgdaSpace{}%
\AgdaSymbol{:}\AgdaSpace{}%
\AgdaDatatype{Nat}\AgdaSymbol{\}}\AgdaSpace{}%
\AgdaSymbol{→}\<%
\\
\>[.]\<[267I]%
\>[10]\AgdaField{MyColist.list}\AgdaSpace{}%
\AgdaBound{l}\AgdaSpace{}%
\AgdaOperator{\AgdaDatatype{≡}}\AgdaSpace{}%
\AgdaInductiveConstructor{just}\AgdaSpace{}%
\AgdaOperator{\AgdaInductiveConstructor{⟨}}\AgdaSpace{}%
\AgdaBound{h}\AgdaSpace{}%
\AgdaOperator{\AgdaInductiveConstructor{,}}\AgdaSpace{}%
\AgdaBound{t}\AgdaSpace{}%
\AgdaOperator{\AgdaInductiveConstructor{⟩}}\AgdaSpace{}%
\AgdaSymbol{→}\<%
\\
\>[10]\AgdaFunction{get}\AgdaSpace{}%
\AgdaBound{t}\AgdaSpace{}%
\AgdaBound{i}\AgdaSpace{}%
\AgdaOperator{\AgdaDatatype{≡}}\AgdaSpace{}%
\AgdaInductiveConstructor{just}\AgdaSpace{}%
\AgdaBound{x}\AgdaSpace{}%
\AgdaSymbol{→}\<%
\\
\>[10]\AgdaFunction{get}\AgdaSpace{}%
\AgdaBound{l}\AgdaSpace{}%
\AgdaSymbol{(}\AgdaInductiveConstructor{suc}\AgdaSpace{}%
\AgdaBound{i}\AgdaSymbol{)}\AgdaSpace{}%
\AgdaOperator{\AgdaDatatype{≡}}\AgdaSpace{}%
\AgdaInductiveConstructor{just}\AgdaSpace{}%
\AgdaBound{x}\<%
\\
\>[0]\<%
\\
\>[0]\AgdaFunction{get{-}eq{-}tl}\AgdaSpace{}%
\AgdaSymbol{\{}\AgdaArgument{x}\AgdaSpace{}%
\AgdaSymbol{=}\AgdaSpace{}%
\AgdaBound{x}\AgdaSymbol{\}}\AgdaSpace{}%
\AgdaSymbol{\{}\AgdaBound{l}\AgdaSymbol{\}}\AgdaSpace{}%
\AgdaSymbol{\{}\AgdaBound{t}\AgdaSymbol{\}}\AgdaSpace{}%
\AgdaSymbol{\{}\AgdaBound{i}\AgdaSymbol{\}}\AgdaSpace{}%
\AgdaBound{eq{-}l}\AgdaSpace{}%
\AgdaBound{eq{-}get}\AgdaSpace{}%
\AgdaKeyword{with}\AgdaSpace{}%
\AgdaFunction{inspect}\AgdaSpace{}%
\AgdaSymbol{(}\AgdaField{MyColist.list}\AgdaSpace{}%
\AgdaBound{l}\AgdaSymbol{)}\<%
\\
\>[0]\AgdaSymbol{...}\AgdaSpace{}%
\AgdaSymbol{|}\AgdaSpace{}%
\AgdaInductiveConstructor{nothing}\AgdaSpace{}%
\AgdaOperator{\AgdaInductiveConstructor{with≡}}\AgdaSpace{}%
\AgdaBound{eq}\AgdaSpace{}%
\AgdaSymbol{=}\AgdaSpace{}%
\AgdaFunction{⊥{-}elim}\AgdaSpace{}%
\AgdaSymbol{(}\AgdaFunction{mycolist{-}abs}\AgdaSpace{}%
\AgdaSymbol{\{}\AgdaArgument{l}\AgdaSpace{}%
\AgdaSymbol{=}\AgdaSpace{}%
\AgdaBound{l}\AgdaSymbol{\}}\AgdaSpace{}%
\AgdaBound{eq}\AgdaSpace{}%
\AgdaBound{eq{-}l}\AgdaSymbol{)}\<%
\\
\>[0]\AgdaSymbol{...}\AgdaSpace{}%
\AgdaSymbol{|}\AgdaSpace{}%
\AgdaInductiveConstructor{just}\AgdaSpace{}%
\AgdaOperator{\AgdaInductiveConstructor{⟨}}\AgdaSpace{}%
\AgdaBound{h1}\AgdaSpace{}%
\AgdaOperator{\AgdaInductiveConstructor{,}}\AgdaSpace{}%
\AgdaBound{t1}\AgdaSpace{}%
\AgdaOperator{\AgdaInductiveConstructor{⟩}}\AgdaSpace{}%
\AgdaOperator{\AgdaInductiveConstructor{with≡}}\AgdaSpace{}%
\AgdaBound{eq}\AgdaSpace{}%
\AgdaSymbol{=}%
\>[339I]\AgdaOperator{\AgdaFunction{begin}}\<%
\\
\>[339I][@{}l@{\AgdaIndent{0}}]%
\>[36]\AgdaFunction{get}\AgdaSpace{}%
\AgdaBound{t1}\AgdaSpace{}%
\AgdaBound{i}\<%
\\
\>[.]\<[339I]%
\>[34]\AgdaOperator{\AgdaFunction{≡⟨}}\AgdaSpace{}%
\AgdaFunction{cong}\AgdaSpace{}%
\AgdaSymbol{(λ}\AgdaSpace{}%
\AgdaBound{k}\AgdaSpace{}%
\AgdaSymbol{→}\AgdaSpace{}%
\AgdaFunction{get}\AgdaSpace{}%
\AgdaBound{k}\AgdaSpace{}%
\AgdaBound{i}\AgdaSymbol{)}\AgdaSpace{}%
\AgdaSymbol{(}\AgdaFunction{eq{-}tl}\AgdaSpace{}%
\AgdaSymbol{\{}\AgdaArgument{l}\AgdaSpace{}%
\AgdaSymbol{=}\AgdaSpace{}%
\AgdaBound{l}\AgdaSymbol{\}}\AgdaSpace{}%
\AgdaBound{eq}\AgdaSpace{}%
\AgdaBound{eq{-}l}\AgdaSymbol{)}\AgdaSpace{}%
\AgdaOperator{\AgdaFunction{⟩}}\<%
\\
\>[34][@{}l@{\AgdaIndent{0}}]%
\>[36]\AgdaFunction{get}\AgdaSpace{}%
\AgdaBound{t}\AgdaSpace{}%
\AgdaBound{i}\<%
\\
\>[34]\AgdaOperator{\AgdaFunction{≡⟨}}\AgdaSpace{}%
\AgdaBound{eq{-}get}\AgdaSpace{}%
\AgdaOperator{\AgdaFunction{⟩}}\<%
\\
\>[34][@{}l@{\AgdaIndent{0}}]%
\>[36]\AgdaInductiveConstructor{just}\AgdaSpace{}%
\AgdaBound{x}\AgdaSpace{}%
\AgdaOperator{\AgdaFunction{∎}}\<%
\end{code}

\end{center}

\section{Binary infinite trees}
\label{sect:trees}
As an additional example of coinductive records usage we show infinite-only binary trees.

\begin{center}
\begin{code}%
\>[0]\AgdaKeyword{record}\AgdaSpace{}%
\AgdaRecord{infTree}\AgdaSpace{}%
\AgdaSymbol{(}\AgdaBound{A}\AgdaSpace{}%
\AgdaSymbol{:}\AgdaSpace{}%
\AgdaPrimitiveType{Set}\AgdaSymbol{)}\AgdaSpace{}%
\AgdaSymbol{:}\AgdaSpace{}%
\AgdaPrimitiveType{Set}\AgdaSpace{}%
\AgdaKeyword{where}\<%
\\
\>[0][@{}l@{\AgdaIndent{0}}]%
\>[2]\AgdaKeyword{coinductive}\<%
\\
\>[2]\AgdaKeyword{field}\<%
\\
\>[2][@{}l@{\AgdaIndent{0}}]%
\>[4]\AgdaField{root}\AgdaSpace{}%
\AgdaSymbol{:}\AgdaSpace{}%
\AgdaBound{A}\<%
\\
\>[4]\AgdaField{left}\AgdaSpace{}%
\AgdaSymbol{:}\AgdaSpace{}%
\AgdaRecord{infTree}\AgdaSpace{}%
\AgdaBound{A}\<%
\\
\>[4]\AgdaField{right}\AgdaSpace{}%
\AgdaSymbol{:}\AgdaSpace{}%
\AgdaRecord{infTree}\AgdaSpace{}%
\AgdaBound{A}\<%
\end{code}

\end{center}

The type is parametrized over $A$ leading to a flexible data structure as we did also for infinite lists. \LC{\agda{infTree} can be implemented also using thunks as for the library colists}. In that case the two subtrees would be boxed inside suspended computations. Now we show the implementations of \textit{allPos}. Again we use \textit{memberOf} as auxiliary predicate but we do not make comments on it since the proof scheme is the same of infinite lists.

\begin{center}
\begin{code}%
\>[0]\AgdaKeyword{data}\AgdaSpace{}%
\AgdaOperator{\AgdaDatatype{\AgdaUnderscore{}memberOf\AgdaUnderscore{}}}\AgdaSpace{}%
\AgdaSymbol{\{}\AgdaBound{A}\AgdaSpace{}%
\AgdaSymbol{:}\AgdaSpace{}%
\AgdaPrimitiveType{Set}\AgdaSymbol{\}(}\AgdaBound{x}\AgdaSpace{}%
\AgdaSymbol{:}\AgdaSpace{}%
\AgdaBound{A}\AgdaSymbol{)(}\AgdaBound{t}\AgdaSpace{}%
\AgdaSymbol{:}\AgdaSpace{}%
\AgdaRecord{infTree}\AgdaSpace{}%
\AgdaBound{A}\AgdaSymbol{)}\AgdaSpace{}%
\AgdaSymbol{:}\AgdaSpace{}%
\AgdaPrimitiveType{Set}\AgdaSpace{}%
\AgdaKeyword{where}\<%
\\
\>[0][@{}l@{\AgdaIndent{0}}]%
\>[2]\AgdaInductiveConstructor{mem{-}h}\AgdaSpace{}%
\AgdaSymbol{:}\AgdaSpace{}%
\AgdaBound{x}\AgdaSpace{}%
\AgdaOperator{\AgdaDatatype{≡}}\AgdaSpace{}%
\AgdaSymbol{(}\AgdaField{infTree.root}\AgdaSpace{}%
\AgdaBound{t}\AgdaSymbol{)}\AgdaSpace{}%
\AgdaSymbol{→}\AgdaSpace{}%
\AgdaBound{x}\AgdaSpace{}%
\AgdaOperator{\AgdaDatatype{memberOf}}\AgdaSpace{}%
\AgdaBound{t}\<%
\\
\>[2]\AgdaInductiveConstructor{mem{-}l}\AgdaSpace{}%
\AgdaSymbol{:}\AgdaSpace{}%
\AgdaBound{x}\AgdaSpace{}%
\AgdaOperator{\AgdaDatatype{memberOf}}\AgdaSpace{}%
\AgdaSymbol{(}\AgdaField{infTree.left}\AgdaSpace{}%
\AgdaBound{t}\AgdaSymbol{)}\AgdaSpace{}%
\AgdaSymbol{→}%
\>[41]\AgdaBound{x}\AgdaSpace{}%
\AgdaOperator{\AgdaDatatype{memberOf}}\AgdaSpace{}%
\AgdaBound{t}\<%
\\
\>[2]\AgdaInductiveConstructor{mem{-}r}\AgdaSpace{}%
\AgdaSymbol{:}\AgdaSpace{}%
\AgdaBound{x}\AgdaSpace{}%
\AgdaOperator{\AgdaDatatype{memberOf}}\AgdaSpace{}%
\AgdaSymbol{(}\AgdaField{infTree.right}\AgdaSpace{}%
\AgdaBound{t}\AgdaSymbol{)}\AgdaSpace{}%
\AgdaSymbol{→}%
\>[42]\AgdaBound{x}\AgdaSpace{}%
\AgdaOperator{\AgdaDatatype{memberOf}}\AgdaSpace{}%
\AgdaBound{t}\<%
\end{code}

\end{center}

The code about \textit{allPos} follows.

\begin{center}
\begin{code}%
\>[0]\AgdaKeyword{record}\AgdaSpace{}%
\AgdaRecord{allPos}\AgdaSpace{}%
\AgdaSymbol{(}\AgdaBound{t}\AgdaSpace{}%
\AgdaSymbol{:}\AgdaSpace{}%
\AgdaRecord{infTree}\AgdaSpace{}%
\AgdaDatatype{Nat}\AgdaSymbol{)}\AgdaSpace{}%
\AgdaSymbol{:}\AgdaSpace{}%
\AgdaPrimitiveType{Set}\AgdaSpace{}%
\AgdaKeyword{where}\<%
\\
\>[0][@{}l@{\AgdaIndent{0}}]%
\>[2]\AgdaKeyword{coinductive}\<%
\\
\>[2]\AgdaKeyword{field}\<%
\\
\>[2][@{}l@{\AgdaIndent{0}}]%
\>[4]\AgdaField{root}\AgdaSpace{}%
\AgdaSymbol{:}\AgdaSpace{}%
\AgdaSymbol{(}\AgdaField{infTree.root}\AgdaSpace{}%
\AgdaBound{t}\AgdaSymbol{)}\AgdaSpace{}%
\AgdaOperator{\AgdaDatatype{>}}\AgdaSpace{}%
\AgdaInductiveConstructor{zero}\<%
\\
\>[4]\AgdaField{left}\AgdaSpace{}%
\AgdaSymbol{:}\AgdaSpace{}%
\AgdaRecord{allPos}\AgdaSpace{}%
\AgdaSymbol{(}\AgdaField{infTree.left}\AgdaSpace{}%
\AgdaBound{t}\AgdaSymbol{)}\<%
\\
\>[4]\AgdaField{right}\AgdaSpace{}%
\AgdaSymbol{:}\AgdaSpace{}%
\AgdaRecord{allPos}\AgdaSpace{}%
\AgdaSymbol{(}\AgdaField{infTree.right}\AgdaSpace{}%
\AgdaBound{t}\AgdaSymbol{)}\<%
\\
\\[\AgdaEmptyExtraSkip]%
\>[0]\AgdaFunction{sound{-}allpos}\AgdaSpace{}%
\AgdaSymbol{:}%
\>[34I]\AgdaSymbol{\{}\AgdaBound{t}\AgdaSpace{}%
\AgdaSymbol{:}\AgdaSpace{}%
\AgdaRecord{infTree}\AgdaSpace{}%
\AgdaDatatype{Nat}\AgdaSymbol{\}}\AgdaSpace{}%
\AgdaSymbol{→}\AgdaSpace{}%
\AgdaRecord{allPos}\AgdaSpace{}%
\AgdaBound{t}\AgdaSpace{}%
\AgdaSymbol{→}\<%
\\
\>[.]\<[34I]%
\>[15]\AgdaSymbol{(\{}\AgdaBound{n}\AgdaSpace{}%
\AgdaSymbol{:}\AgdaSpace{}%
\AgdaDatatype{Nat}\AgdaSymbol{\}}\AgdaSpace{}%
\AgdaSymbol{→}\AgdaSpace{}%
\AgdaBound{n}\AgdaSpace{}%
\AgdaOperator{\AgdaDatatype{memberOf}}\AgdaSpace{}%
\AgdaBound{t}\AgdaSpace{}%
\AgdaSymbol{→}\AgdaSpace{}%
\AgdaBound{n}\AgdaSpace{}%
\AgdaOperator{\AgdaDatatype{>}}\AgdaSpace{}%
\AgdaInductiveConstructor{zero}\AgdaSymbol{)}\<%
\\
\\[\AgdaEmptyExtraSkip]%
\>[0]\AgdaFunction{sound{-}allpos}\AgdaSpace{}%
\AgdaBound{ap}\AgdaSpace{}%
\AgdaSymbol{(}\AgdaInductiveConstructor{mem{-}h}\AgdaSpace{}%
\AgdaInductiveConstructor{refl}\AgdaSymbol{)}\AgdaSpace{}%
\AgdaSymbol{=}\AgdaSpace{}%
\AgdaField{allPos.root}\AgdaSpace{}%
\AgdaBound{ap}\<%
\\
\>[0]\AgdaFunction{sound{-}allpos}\AgdaSpace{}%
\AgdaBound{ap}\AgdaSpace{}%
\AgdaSymbol{(}\AgdaInductiveConstructor{mem{-}l}\AgdaSpace{}%
\AgdaBound{mem}\AgdaSymbol{)}\AgdaSpace{}%
\AgdaSymbol{=}\AgdaSpace{}%
\AgdaFunction{sound{-}allpos}\AgdaSpace{}%
\AgdaSymbol{(}\AgdaField{allPos.left}\AgdaSpace{}%
\AgdaBound{ap}\AgdaSymbol{)}\AgdaSpace{}%
\AgdaBound{mem}\<%
\\
\>[0]\AgdaFunction{sound{-}allpos}\AgdaSpace{}%
\AgdaBound{ap}\AgdaSpace{}%
\AgdaSymbol{(}\AgdaInductiveConstructor{mem{-}r}\AgdaSpace{}%
\AgdaBound{mem}\AgdaSymbol{)}\AgdaSpace{}%
\AgdaSymbol{=}\AgdaSpace{}%
\AgdaFunction{sound{-}allpos}\AgdaSpace{}%
\AgdaSymbol{(}\AgdaField{allPos.right}\AgdaSpace{}%
\AgdaBound{ap}\AgdaSymbol{)}\AgdaSpace{}%
\AgdaBound{mem}\<%
\\
\\[\AgdaEmptyExtraSkip]%
\\[\AgdaEmptyExtraSkip]%
\>[0]\AgdaFunction{complete{-}allpos}%
\>[74I]\AgdaSymbol{:}%
\>[19]\AgdaSymbol{(}\AgdaBound{t}\AgdaSpace{}%
\AgdaSymbol{:}\AgdaSpace{}%
\AgdaRecord{infTree}\AgdaSpace{}%
\AgdaDatatype{Nat}\AgdaSymbol{)}\AgdaSpace{}%
\AgdaSymbol{→}\<%
\\
\>[.]\<[74I]%
\>[16]\AgdaSymbol{(\{}\AgdaBound{n}\AgdaSpace{}%
\AgdaSymbol{:}\AgdaSpace{}%
\AgdaDatatype{Nat}\AgdaSymbol{\}}\AgdaSpace{}%
\AgdaSymbol{→}\AgdaSpace{}%
\AgdaBound{n}\AgdaSpace{}%
\AgdaOperator{\AgdaDatatype{memberOf}}\AgdaSpace{}%
\AgdaBound{t}\AgdaSpace{}%
\AgdaSymbol{→}\AgdaSpace{}%
\AgdaBound{n}\AgdaSpace{}%
\AgdaOperator{\AgdaDatatype{>}}\AgdaSpace{}%
\AgdaNumber{0}\AgdaSymbol{)}\AgdaSpace{}%
\AgdaSymbol{→}\AgdaSpace{}%
\AgdaRecord{allPos}\AgdaSpace{}%
\AgdaBound{t}\<%
\\
\>[0]\<%
\\
\>[0]\AgdaField{allPos.root}\AgdaSpace{}%
\AgdaSymbol{(}\AgdaFunction{complete{-}allpos}\AgdaSpace{}%
\AgdaBound{t}\AgdaSpace{}%
\AgdaBound{p}\AgdaSymbol{)}\AgdaSpace{}%
\AgdaSymbol{=}\AgdaSpace{}%
\AgdaBound{p}\AgdaSpace{}%
\AgdaSymbol{(}\AgdaInductiveConstructor{mem{-}h}\AgdaSpace{}%
\AgdaInductiveConstructor{refl}\AgdaSymbol{)}\<%
\\
\>[0]\AgdaField{allPos.left}%
\>[99I]\AgdaSymbol{(}\AgdaFunction{complete{-}allpos}\AgdaSpace{}%
\AgdaBound{t}\AgdaSpace{}%
\AgdaBound{p}\AgdaSymbol{)}\AgdaSpace{}%
\AgdaSymbol{=}\<%
\\
\>[.]\<[99I]%
\>[12]\AgdaFunction{complete{-}allpos}\AgdaSpace{}%
\AgdaSymbol{(}\AgdaField{infTree.left}\AgdaSpace{}%
\AgdaBound{t}\AgdaSymbol{)}\AgdaSpace{}%
\AgdaSymbol{(λ}\AgdaSpace{}%
\AgdaBound{m}\AgdaSpace{}%
\AgdaSymbol{→}\AgdaSpace{}%
\AgdaBound{p}\AgdaSpace{}%
\AgdaSymbol{(}\AgdaInductiveConstructor{mem{-}l}\AgdaSpace{}%
\AgdaBound{m}\AgdaSymbol{))}\<%
\\
\>[0]\AgdaField{allPos.right}\AgdaSpace{}%
\AgdaSymbol{(}\AgdaFunction{complete{-}allpos}\AgdaSpace{}%
\AgdaBound{t}\AgdaSpace{}%
\AgdaBound{p}\AgdaSymbol{)}\AgdaSpace{}%
\AgdaSymbol{=}\<%
\\
\>[0][@{}l@{\AgdaIndent{0}}]%
\>[12]\AgdaFunction{complete{-}allpos}\AgdaSpace{}%
\AgdaSymbol{(}\AgdaField{infTree.right}\AgdaSpace{}%
\AgdaBound{t}\AgdaSymbol{)}\AgdaSpace{}%
\AgdaSymbol{(λ}\AgdaSpace{}%
\AgdaBound{m}\AgdaSpace{}%
\AgdaSymbol{→}\AgdaSpace{}%
\AgdaBound{p}\AgdaSpace{}%
\AgdaSymbol{(}\AgdaInductiveConstructor{mem{-}r}\AgdaSpace{}%
\AgdaBound{m}\AgdaSymbol{))}\<%
\end{code}

\end{center}

Notice that \textit{memberOf} must have three constructors. A more interest predicate over an infinite tree $t$ of natural numbers is \textit{t contains a path whose elements are all positive}. 

\begin{center}
\begin{code}%
\>[0]\AgdaKeyword{record}\AgdaSpace{}%
\AgdaRecord{pathPos}\AgdaSpace{}%
\AgdaSymbol{(}\AgdaBound{t}\AgdaSpace{}%
\AgdaSymbol{:}\AgdaSpace{}%
\AgdaRecord{infTree}\AgdaSpace{}%
\AgdaDatatype{Nat}\AgdaSymbol{)}\AgdaSpace{}%
\AgdaSymbol{:}\AgdaSpace{}%
\AgdaPrimitiveType{Set}\AgdaSpace{}%
\AgdaKeyword{where}\<%
\\
\>[0][@{}l@{\AgdaIndent{0}}]%
\>[2]\AgdaKeyword{coinductive}\<%
\\
\>[2]\AgdaKeyword{field}\<%
\\
\>[2][@{}l@{\AgdaIndent{0}}]%
\>[4]\AgdaField{hd}\AgdaSpace{}%
\AgdaSymbol{:}\AgdaSpace{}%
\AgdaSymbol{(}\AgdaField{infTree.root}\AgdaSpace{}%
\AgdaBound{t}\AgdaSymbol{)}\AgdaSpace{}%
\AgdaOperator{\AgdaDatatype{>}}\AgdaSpace{}%
\AgdaInductiveConstructor{zero}\<%
\\
\>[4]\AgdaField{tl}%
\>[29I]\AgdaSymbol{:}\AgdaSpace{}%
\AgdaDatatype{∃}\AgdaSpace{}%
\AgdaSymbol{(}\AgdaRecord{infTree}\AgdaSpace{}%
\AgdaDatatype{Nat}\AgdaSymbol{)}\<%
\\
\>[.]\<[29I]%
\>[7]\AgdaSymbol{(λ}\AgdaSpace{}%
\AgdaBound{s}\AgdaSpace{}%
\AgdaSymbol{→}\AgdaSpace{}%
\AgdaSymbol{((}\AgdaBound{s}\AgdaSpace{}%
\AgdaOperator{\AgdaDatatype{≡}}\AgdaSpace{}%
\AgdaSymbol{(}\AgdaField{infTree.left}\AgdaSpace{}%
\AgdaBound{t}\AgdaSymbol{))}\AgdaSpace{}%
\AgdaOperator{\AgdaDatatype{∧}}\AgdaSpace{}%
\AgdaRecord{pathPos}\AgdaSpace{}%
\AgdaBound{s}\AgdaSymbol{)}\AgdaSpace{}%
\AgdaOperator{\AgdaDatatype{∨}}\<%
\\
\>[7]\AgdaSymbol{((}\AgdaBound{s}\AgdaSpace{}%
\AgdaOperator{\AgdaDatatype{≡}}\AgdaSpace{}%
\AgdaSymbol{(}\AgdaField{infTree.right}\AgdaSpace{}%
\AgdaBound{t}\AgdaSymbol{))}\AgdaSpace{}%
\AgdaOperator{\AgdaDatatype{∧}}\AgdaSpace{}%
\AgdaRecord{pathPos}\AgdaSpace{}%
\AgdaBound{s}\AgdaSymbol{))}\<%
\end{code}

\end{center}

The predicate inside the $\exists$ type asks that \textit{pathPos} holds for the left or for the right subtree thus requiring also the proof that the interested subtree is inside $t$.

\end{document}